\documentclass[12pt]{report}

\usepackage[english]{babel}
\usepackage{mathbbold}
\usepackage{float}
\usepackage{amssymb}
\usepackage{amsmath}
\usepackage{babel}
\usepackage{a4}
\usepackage{headings}
\usepackage{latexsym}
\usepackage{graphics}
\usepackage{graphicx}
\usepackage{epsfig}
\usepackage{times}
\usepackage{amsfonts}
\usepackage[dvips]{color}
\usepackage{setspace}
\def\0{|0\rangle}
\def\Ep{E_{P}}

\def\O{\Omega}
\def\Lp{L_{P}}
\def\ee{Einstein's equation }
\def\pd{\partial}
\def\abs#1{\left|#1\right|}
\def\cd{\nabla}
\def\phit{\Phi}
\def\pl{\widetilde{p}}
\def\vl{\widetilde{v}}
\def\dv{\delta v}
\def\ex{\exp{\left(-iMc^2 t/\hbar\right)}}
\def\m#1{\left\langle#1\right\rangle}
\def\cI{C_1}
\def\cII{C_2}
\newcommand{\xv}{\mathbf{x}}
\newcommand{\kv}{\mathbf{k}}
\def\ms#1{\langle#1\rangle}
\def\lamm{L\"{a}mmerzahl }
\def\schr{Schr\"{o}dinger }
\def\Ho{\hat{H}}
\def\Uo{\hat{U}}
\def\I{\hat{\mathrm{I}}}
\def\Ko{\hat{U}}
\def\ket#1{|#1\rangle}
\def\kett#1{#1\rangle}
\def\bra#1{\langle#1|}
\def\brk#1{\langle#1\rangle}
\def\Pio{\hat{\Pi}}
\def\d{d}
\def\v#1{\mathbf{#1}}
\def\Tp{T_{\mathrm{P}}}
\newcommand{\kvs}{\hat{\mathbf{k}}}
\newcommand{\yv}{\mathbf{y}}
\newcommand{\Kvs}{\hat{\mathbf{K}}}
\def\Mp{M_{\mathrm{P}}}
\newcommand{\lvs}{\hat{\mathbf{z}}}
\newcommand{\ivs}{\hat{\mathbf{x}}}
\newcommand{\jvs}{\hat{\mathbf{y}}}
\def\lr{L_{\mathrm{\scriptsize{R}}}}
\def\llr{\lambda_{\mathrm{\scriptsize{R}}}}
\def\kcut{k_\lambda}
\def\Pr#1{\ket{#1}\bra{#1}}
\def\tex{T_{\mathrm{\scriptsize{ex}}}}
\def\mc70{M_{\text{C}_{70}}}

\def\gmb{\gamma^{\text{\scriptsize{B}}}}
\def\cfm{conformal metric }
\def\cfmp{conformal metric}
\def\l{\ell}
\def\lab{L_{\text{\scriptsize{lab}}}}

\def\Lobs{\Lambda^{\text{\scriptsize{obs}}}}
\def\T{T_{ab}}
\def\rhom{\rho_{\psi}^*}

\def\LM{\Lambda_M}
\def\LP{\Lambda_P}

\def\rV{\rho^{\text{\scriptsize{QFT}}}_{\text{\scriptsize{vacuum}}}}
\def\LRG{\Lambda_{\text{\scriptsize{RG}}}}
\def\rRG{\rho^{\text{\scriptsize{RG}}}_{\text{\scriptsize{vacuum}}}}

\def\Csb{{\Gamma^{\text{\scriptsize{B}}}}}
\def\cdb{\nabla^{\text{\scriptsize{B}}}}

\def\hgw{h^{\text{\scriptsize{GW}}}}
\def\Teff{T^{\text{\scriptsize{eff}}}}
\def\aa{a}
\def\bb{b}
\def\cc{c}
\def\dd{d}
\def\eee{e}
\def\ar{\tilde{A}}

\def\phbd{{\phi}}
\def\phbdo{{\phi_{0}}}
\def\L{\mathcal{L}}
\def\S{\mathcal{S}}
\def\Sb{\S_{\mf}}
\def\lb{\lambda}
\def\gc{\bar{g}}
\def\rg{\sqrt{-g}\,}
\def\rgc{\sqrt{-\bar{g}}\,}
\def\phic{\bar{\phi}}
\def\Rc{\bar{R}}
\def\Vc{\bar{V}}
\def\cdc{\bar{\cd}}
\def\ph{\varphi}
\def\B{\square}
\def\mf{{\mathfrak{M}}}
\def\Sjf{\S_{\text{\scriptsize{tot}}}^{\text{\scriptsize{JF}}}}
\def\Sef{\S_{\text{\scriptsize{tot}}}^{\text{\scriptsize{EF}}}}

\def\ggg{g^{c}}
\def\ki{k_1}
\def\kii{k_2}
\def\kiii{k_3}
\def\kiiii{k_4}
\def\cbd{C_{\text{\scriptsize{BD}}}}

\def\vg#1{\boldsymbol{#1}}
\def\a{\alpha}
\def\b{\beta}

\def\TKG{T_{ab}^{\mathrm{\scriptsize{KG}}}}
\def\Sphi{S_\phi}

\def\eep{Einstein's equation}
\def\gb{g^{\text{\scriptsize{B}}}}
\def\gbu{{g^{\text{\scriptsize{B}}}}}
\def\Tgw{T^{\text{\scriptsize{GW}}}}
\def\Tgwgo{T^{\text{\scriptsize{GW-GO}}}}
\def\Lav{L_{\m{\,}}}
\def\bx{\pd^c\pd_c}

\def\TKG{T_{ab}^{\text{\scriptsize{KG}}}}
\def\Sphi{S_\phi}
\def\P{\mathcal{P}}

\def\Ao{\hat{A}}
\def\xo{\hat{x}}

\def\Uo{\hat{U}}
\def\tr{\text{Tr\,}}
\def\H{\mathcal{H}}

\begin{document}

\newpage\thispagestyle{empty}
\begin{center}
\large \mbox\par\vspace{1cm}\par UNIVERSITY OF ABERDEEN

\vspace{0.2cm} Department of Physics
\\ \vspace{4cm} {\LARGE{\textbf{SPACETIME CONFORMAL FLUCTUATIONS AND QUANTUM
DEPHASING}}}\\
\vspace{1cm} A thesis presented for the degree of Doctor of
Philosophy\\ at the University of Aberdeen
\\ \vspace{3cm}
Paolo BONIFACIO\\
Master Degree Universit\`{a} degli studi di Milano, ITALY
\\ \vspace{1.5cm}
2008/09
\\ \vspace{2cm}
\end{center}


\newpage\thispagestyle{empty}
\mbox{ }

\begin{center}
\textbf{Abstract}
\end{center}

Any quantum system interacting with a complex environment undergoes
decoherence and loses the `ability' to show genuine quantum effects.
Empty space is filled with vacuum energy due to matter fields in
their ground state and represents an underlying environment that any
quantum particle has to cope with. In particular quantum gravity
vacuum fluctuations should represent a universal source of
decoherence. It is important to assess which parameters control such
an effect, also in relation to the issue of gaining experimental
access to Planck scale physics.

To this end we employ a stochastic approach that models spacetime
fluctuations close to the Planck scale by means of a classical,
randomly fluctuating metric (random gravity framework). We enrich
the classical scheme for metric perturbations over a curved
background by also including matter fields and metric conformal
fluctuations. We show in general that a conformally modulated metric
induces dephasing as a result of an effective nonlinear newtonian
potential obtained in the appropriate non-relativistic limit of a
minimally coupled Klein-Gordon field. The special case of vacuum
fluctuations is considered and a quantitative estimate of the
expected effect deduced.

Secondly we address the question of how conformal fluctuations could
physically arise. By applying the random gravity framework we first
show that standard GR seems to forbid spontaneous conformal metric
modulations. Finally we argue that a different result follows within
scalar-tensor theories of gravity such as e.g. Brans-Dicke theory.
In this case a conformal modulation of the metric arises naturally
as a result of the fluctuations in the Brans-Dicke field and quantum
dephasing of a test particle is expected to occur. For large
negative values of the coupling parameter the conformal fluctuations
may also contribute to alleviate the well known problem of the large
zero point energy due to quantum matter fields.

\newpage
\thispagestyle{empty}


\newpage\thispagestyle{empty}
\mbox{ }

\begin{center}
\textbf{Declaration}
\end{center}

This thesis has been composed by the candidate. This thesis has not
been accepted in any previous application for a degree. The work
described in this thesis has been done by the candidate. All
quotations have been distinguished by quotation marks and the
sources of information are
specifically acknowledged.\\

Chapter \ref{ch1} is revised and extended from: Paolo M. Bonifacio,
Charles H.-T. Wang, J. Tito Mendonca, Robert Bingham 2009 Dephasing
of a non-relativistic quantum particle due to a conformally
fluctuating spacetime, \emph{Class. Quantum Grav.} \textbf{26}
145013

\vspace{3.0cm}

Signed:

\vspace{0.5cm}

June 2009

\newpage
\thispagestyle{empty}


\newpage\thispagestyle{empty}
\mbox{ }

\begin{center}
\textbf{Acknowledgements}
\end{center}

First of all I want to thank my supervisor, Dr Charles Wang, for
giving me the chance to do a PhD and for all his support and most
valuable scientific advice. I also thank Dr Norval Strachan and the
department of Physics for giving me the opportunity to work as a
lecturer during the final phase of my PhD research. It has been
challenging but highly rewarding. Special thanks also goes to STFC
Centre for Fundamental Physics and to Prof Bingham. Various people
contributed to make this work possible through their knowledge and
stimulating discussions. Without any preferred order I wish to
thank: Tito Mendon\c{c}a, Luigi Galgani, Andrea Carati, Victor
Varela, Antonaldo Diaferio, Daniele Bertacca, Rui Reis.

I am deeply grateful to my parents for their support and
\emph{amore}, as well as to all the beautiful friends I met during
these years. For their love, company, help and understanding during
the good and the tough times. Thanks Dana, Christian, Costantino,
Valerio, Cesar \& Andrea (or the G5), Mauro, Margaux, Luca, Marko,
Jeremy, Fabien, Alice, Ida, Cristian, Louren\c{c}o, Annalise,
Lavanya, Kenny, Gabriel, Tiziano, Alessio. At last, a sweetly
refined thought goes to \emph{HannaH}, this palindromic special
creature that keeps on crossing my path. Thanks to the music,
eternal invisible friend and companion.

\newpage
\thispagestyle{empty}


\newpage\thispagestyle{empty}
\mbox{ } \singlespacing
\begin{quote}
\emph{Sbattuti e stanchi\\
di guerreggiar tant'anni, e risospinti\\
ancor da' fati, i greci condottieri\\
a l'insidie si di\^{e}ro; e da Minerva\\
divinamente instrutti, un gran cavallo\\
di ben contesti e ben confitti abeti\\
in sembianza d'un monte edificaro.\\
Poscia, finto che ci\`{o} fosse per v\'{o}to\\
del lor ritorno, di tornar sembiante\\
fecero tal, che se ne sparse il grido.\\
Dentro al suo cieco ventre e ne le grotte,\\
che molte erano e grandi, in s\'{i} gran mole,\\
rinchiuser di nascosto arme e guerrieri\\
a ci\`{o} per sorte e per valore eletti.\\}
\\
Virgil, Aeneid Book II, 29-19 BC
\end{quote}
\vspace{3.5cm}
\begin{flushright}
    \emph{To my parents}
\end{flushright}
\vspace{3.5cm}
\begin{quote}
\emph{E C. il ricercatore chiese a F. il musicista:\\
``Perche' ti interessi di Fisica e Filosofia della Natura?''\\
``Ma perche' e' fondamentale!'', fu la risposta.}
\end{quote}

\singlespacing

\newpage
\thispagestyle{empty}


\pagenumbering{roman} \setcounter{page}{1}
\tableofcontents
\newpage
\pagenumbering{arabic}

\chapter*{Notation and conventions}

\markboth{Notation and conventions}{Notation and conventions}

\addcontentsline{toc}{chapter}{Notation and conventions}

We use metric signature $(-,+,+,+)$ and, unless specified,
geometrized unites with $G = c = 1$. The coordinates in a given
system are denoted as $x = (x^0,x^1,x^2,x^3) \equiv (x^0, \xv),$ in
such a way that the $0$-th coordinate represents time.

Latin indices from the beginning of the alphabet such as $a,b,c,d
\ldots=0,1,2,3.$ are used for spacetime tensors. Coordinates
equations holding for spacial components will be written using latin
indices from the middle of the alphabet $i,j,k,l\ldots =1,2,3.$

The physical metric tensor is denoted as $g_{ab}$ and its inverse is
$g^{ab}$. When we need to highlight that a tensor, say $T_{ab}$,
depends functionally upon other tensors ${A_{a\ldots}}^{b\ldots},
{B_{a\ldots}}^{b\ldots}, \ldots$, we will write
$T_{ab}[A,B,\ldots]$. The notation for a totally symmetric and
totally antisymmetric tensor is
$$ T_{(ab)}:=\frac{1}{2}(T_{ab} + T_{ba}), $$
$$ T_{[ba]}:=\frac{1}{2}(T_{ab} - T_{ba}). $$

The Christoffel symbol is defined as
\begin{equation*}
{\Gamma^c}_{ab} = \frac{1}{2}g^{cd}\left\{ \pd_a g_{bd} + \pd_b
g_{ad} - \pd_d g_{ab} \right\}.
\end{equation*}
The sign definition for the Riemann tensor is the same as in Wald
\cite{wald84}
\begin{equation*}
\cd_a\cd_b \,\omega_c - \cd_b\cd_a\,\omega_c = {R_{abc}}^d \omega_d,
\end{equation*}
where $\omega_d$ is a covariant vector field. This implies the
following expression for the Ricci tensor:
\begin{equation*}
R_{ab}:={R_{acb}}^c = \pd_c{\Gamma^c}_{ab} - \pd_a{\Gamma^c}_{cb} +
{\Gamma^c}_{ab}{\Gamma^d}_{cd} - {\Gamma^d}_{cb}{\Gamma^c}_{da}.
\end{equation*}

The notation $h^{(n)}_{ab\ldots}$ indicates a $n$-th order quantity.
Accordingly, when expansions are made, e.g. of the Ricci tensor, the
notation $R^{(2)}_{ab}[h^{(1)}]$ indicates the part of the Ricci
tensor which is second order in the linear perturbation
$h^{(1)}_{ab}$.

\newpage
\thispagestyle{empty}


\chapter*{Introduction and motivations}

\markboth{Introduction and motivations}{Introduction and
motivations}

\addcontentsline{toc}{chapter}{Introduction and motivations}

\section*{Probing Quantum Gravity?}

\addcontentsline{toc}{section}{Probing Quantum Gravity?}

General relativity (GR) \cite{wald84} and quantum field theory (QFT)
\cite{ryder85} are both very successful theories of nature. Beside
providing a deeper understanding of how gravity works by relating it
to the curvature of spacetime, GR can be applied to a wide variety
of physical domains, ranging from the solar system to the whole
universe. So far it has `passed' various crucial experimental tests
\cite{clifford1981}, including anomalous perihelion shift of
Mercury, light deflection, as well as energy loss in pulsar binary
systems \cite{taylor1979} which provided indirect evidence for the
existence of gravitational waves (GWs). The large scale application
of GR culminated in the standard cosmological model. Through a
balanced interplay of theoretical inputs and observations, this is
converging in providing a credible description of the visible
universe and its evolution. Future direct detection of gravitational
radiation, e.g. from projects such as LIGO \cite{abbott08} or LISA
\cite{araujo07}, could provide further spectacular evidence in favor
of GR.

The developments of quantum theory into a relativistic theory of
quantum fields has converged into the standard model of particle
physics \cite{weinberg05}. This shed light onto the nature of
elementary particles and, though still plagued by rather technical
subtleties, provides quite a coherent framework which has been so
far spectacularly verified experimentally in basically all respects.
The only key element that still fails to be detected to date is the
Higgs boson, related to the mechanism at the origin of mass. The
fact that the newly built LHC \cite{han03} could provide future
direct evidence of this elusive particle is an exciting possibility
which would put the standard model on an even more solid basis.

Both GR and QFT are based on important symmetry principles. The
general relativistic framework extends Lorentz invariance between
inertial frames to a deeper invariance under arbitrary coordinates
transformation (covariance). The standard model is based, beyond the
principle of Lorentz invariance, upon a certain number of gauge
symmetries, i.e. it enjoys invariance under some special kind of
internal (non-spacetime) transformations. An example is the Dirac
field for the electron, whose local invariance under arbitrary phase
transformations necessarily implies the existence of an extra gauge
field with which it interacts. This is clearly the electromagnetic
field. The resulting QED theory describes electrons interacting with
the quantized radiation field and was the first example of
successful gauge theory based upon a $U(1)$ type symmetry. Other
examples are the Weinberg-Salam electroweak theory, where a $SU(2)$
type symmetry is used to unify the weak and electromagnetic
interactions, and the QCD describing the strong interaction between
quarks and gluons, based upon the $SU(3)$ symmetry.

Though successful in their various domains of application, GR and
QFT are deeply different at the conceptual level. GR is a classical
theory where the evolution of the spacetime metric $g_{ab}$ stems
deterministically from classical field equations. Even though
extremely complicated to do in any realistic situation, one can in
principle solve \ee once all the other \emph{classical} matter
fields and their coupling to gravity are specified. The evolution is
determined precisely from the initial data and there are no
intrinsic limits to one's ability to know precisely the matter
fields or gravity configuration at a given spacetime point. An
essential feature of the theory is that of being background
independent: the spacetime metric is not fixed a priori but it is
itself a physical actor, both affected by and affecting the matter
fields distribution. In any practically simple situation in which an
exact solution of \ee can be found analytically, together with the
included matter fields, then the past, present and future of the
physical system are completely determined and mathematically
described by a well defined manifold. These features imply that GR
is successful when applied to large scale systems, where the
underlying microphysics can be bypassed and described in classical
macroscopic terms.

The situation is radically different for QFT and the standard model.
These deal with physical entities that are believed to rank among
the most elementary constituent of the universe. Even though
technically complex, the basic principles at the heart of QFT are
those of quantum mechanics. This makes QFT a fundamentally
statistical theory, in which possible predictions are probabilities
for certain events (such as scattering, particles production from
the vacuum, field strengths, number of fields quanta) to happen. The
state of a given system is described by a state vector $\psi$ in a
Hilbert space satisfying a \schr-like equation with a suitable
Hamiltonian operator. Even though the evolution of the state vector
describing a closed system is deterministic, it is impossible, even
in principle, to foresee the outcome of a particular measurement.
The indeterminism inherent to any quantum theory occurs in a rather
subtle way involving the interaction of the physical system under
exam and some `measuring' apparatus. It appears that, whenever an
interaction and subsequent measurement take place, the state vector
collapses, \emph{unpredictably}, to a new state. The related
probability is the only information that the quantum theory
apparatus enables one to know in advance. Another important feature
of QFT that contrasts with GR is that the former is formulated
mathematically using a fixed non-dynamical flat spacetime geometry.
A quite successful formulation of QFT on a given \emph{fixed} curved
background geometry also exists. This allowed to predict important
effects such as e.g. particle creation from the vacuum, effects of
vacuum fluctuations and polarizations in the early universe and
Hawking radiation from black holes. However the passage from a flat
Minkowski to a general background involves lots of non trivial
issues \cite{wald94}. Moreover QFT on a curved background neglects
the backreaction effect of quantum matter onto the spacetime
geometry, which is treated as non-dynamical.

It is widely believed that the ultimate theory of nature should
unify GR and QFT into a more comprehensive scheme that could then be
applied both to the macroscopic classical world and to the
microscopic unpredictable world. The dominant belief is that, within
such a unified framework, gravity should be quantized as well. The
eventuality that gravity is a classical field coupled to, say, the
expectation value of the quantum matter fields leads to theoretical
inconsistences \cite{kiefer04,kiefer05} also supported by
experiments \cite{page81}. The resulting Quantum Gravity (QG) theory
\cite{giulini03} is currently the `holy grail' of physics. Beyond
esthetic and philosophical reasons, according to which a unified
description of nature is very desirable, a valid QG theory will
surely prove to be the most valuable tool to unveil new physics.
This will probably shed new light upon concepts such as
singularities, black holes, the big bang and the origin of the
universe, as well as the deeper nature of space and time.

Since the pioneering times of Einstein, Heisenberg and Dirac among
the others, people have been trying to build a coherent theory, so
far with no success. Various approaches have been attempted, from
the early covariant perturbation method (plagued by the issue of
non-renormalizability \cite{rosenfeld30,goroff85}) to effective
field theories \cite{donoghue94}, path integral method, Euclidean
path integral approach \cite{kiefer05}, Penrose twistor approach,
supergravity theories, lattice theories, dynamical triangulation
\cite{giulini03}, canonical geometrodynamics \cite{DeWitt1967}. The
two approaches that are being most widely pursued and are believed
by many to turn eventually into the right direction are String
Theory \cite{polchinski98} and Loop Quantum Gravity (LQG)
\cite{rovelli04}. String Theory is closer in spirit to a QFT (though
of a rather particular type since the physical particles arise here
as excitation states of the string fields). It must recur to
perturbative methods over a background in order to deduce the
particles spectrum of which the spin 2 massless graviton would be a
particular member. On the other hand LQG (based upon a development
of the canonical quantization method) sticks closer to the original
spirit of GR. It is a less ambitious program in that it does not
attempt to unify physics as a whole. However the quantization
program is standard in the sense that it is applied to GR once this
is recast into Hamiltonian form by suitable canonical
transformations. Once the so called Ashtekar variables
\cite{ashtekar86} are introduced to replace the metric by the affine
connection and the correct dynamical variables identified, the
quantization is relatively straightforward. Within LQG, lengths,
areas and volumes become themselves operator whose spectrum has been
calculated. This yields a microscopic vision of the spacetime
structure in terms of elementary `chunks' of geometry, i.e. areas
and volumes which can been described using the spin network theory.

Quite independently of the specific form of the correct QG theory,
it is generally believed that spacetime should have non-trivial
microscopic properties. These could include complex topologies or an
intrinsic discreteness (as emerging in the LQG approach) and,
definitely, a non classical behavior \cite{kimberly05}. By becoming
an operator, the spacetime metric is exptected to have unpredictable
configurations at the small scale; moreover any viable theory should
be able to specify how an appropriate classical limit yielding a
smooth metric obeying \ee can arise. On simple dimensional grounds
QG is expected to alter significantly the structure of spacetime and
the related physics at the Planck scale $\Lp \approx 10^{-35}$ m.
The corresponding Planck energy $\Ep \approx 10^{19}$ GeV is far
beyond direct reach of even the most powerful particle accelerators
such as LHC (even though it is still hoped that this could provide
hints for supersymmetry or hidden dimensions). Nonetheless finding
experimental access to QG phenomena is a matter of the greatest
importance. Various possibilities are being considered, e.g.
observing primordial black holes, QG induced anisotropy on the CMB,
varying coupling constants, violation of equivalence principle
and/or of Lorentz invariance, modified dispersion relations for
distant EM radiation due to the discrete microstructure of spacetime
\cite{giulini03,kiefer05}. A possibility that also receives
attention is that according to which vacuum fluctuations of the
spacetime geometry could affect the dynamics of a test quantum
particle coupled to gravity: it is expected that microscopic Planck
scale physics may be `amplified' at a macroscopic observable scale
as a \emph{decoherence} `signal'
\cite{kok03,powerpercival00,percival94,percival95}.

Decoherence is a well known and studied phenomena which occurs
whenever a quantum particle is coupled to an environment with many
degrees of freedom \cite{kiefer98,zeh06}. Extensive studies have
begun in the 70s. Today it is widely believed that decoherence can
play an important role in explaining the quantum to classical
transition, localization, the measurement problem and addressing the
issue of the wave function collapse \cite{giulini96}. It has also
been proposed that decoherence could play a role in the emergence of
classical spacetime from the full QG regime
\cite{kiefer98,whelan98}. In a nutshell, what happens is that the
closed system given by the particle plus the environment evolves in
the usual unitary way typical of quantum mechanics; this introduces
non trivial correlations between the particle and the environment
degrees of freedom; if the latter are ignored, technically by
performing a trace that effectively averages over the environment,
and the attention is focused on the open system constituted by the
particle alone, it is found in general that this loses coherence and
the `ability' to show genuine quantum effects. For example pure
superposition states decay to mixed states, basically equivalent to
simple statistical mixtures and the possibility of interference is
suppressed. Decoherence theory recently found brilliant confirmation
in interferometry experiments with mesoscopic particles such as
fullerene \cite{hackermuller03}. In these \emph{matter waves
interferometry experiments} heavy molecules are propagated through a
Talbot-Lau interferometry apparatus. The environment is mainly
represented by the air molecules. Unless the pressure is reduced
drastically and all other external influences appropriately
controlled, the molecular beam interference pattern's visibility is
reduced or canceled. The experimental results are in very good
agreement with decoherence theory calculations
\cite{hornberger03,hornberger04}.

Coming back to GQ and the decoherence induced by spacetime vacuum
fluctuations, the idea is that these would represent some sort of
inevitable environment that any quantum particle has to cope with.
It has first been suggested by Percival and collaborators that this
could be measured through matter waves interferometry experiments
\cite{percival94,percival95,percival97}. These ideas have been first
explored from the point of view of the Primary State Diffusion
theory \cite{percival94,percival98} and further investigated by
Power \cite{power99} and Power-Percival \cite{powerpercival00} for
the case of wave packets within a standard Quantum Mechanics
approach. When trying to study decoherence due to QG effects we face
a serious problem: a definite theory describing the quantum behavior
of spacetime is still lacking. Usually this is circumvented in the
literature by adopting some sort of stochastic treatment. Since we
do observe a classical spacetime down to at least the subnuclear
scale, it is clear that the true GQ theory must have a classical
limit out of which the smooth spacetime manifold emerges. It is
usually believed that there should be an intermediate regime,
between the Planck scale and the subnuclear scale, where the
residual signs of the spacetime quantum nature come in the form of
an effective stochastic geometry \cite{powerpercival00,wang2006}. In
other words the spacetime metric is expected to fluctuate randomly.
Given this assumption one can replace the `hard' fully quantum
decoherence problem with the problem of the interaction of a quantum
particle with a classical stochastic environment: it is then more
appropriate to speak of quantum \emph{dephasing} rather than
decoherence since the environment is not treated at the quantum
level. The problem of the interaction of a quantum particle with a
stochastic gravitational environment emerges even in other domains
involving fluctuations of a more general nature than due to vacuum
effects. For example Reynaud and collaborators have extensively
studied the dephasing induced by stochastic GWs of cosmic origin
\cite{reynaud04} and argued that these would have a negligible
effect on HYPER-like atomic interferometers. More recently the
interaction of a quantum particle with a stochastic metric and the
induced dephasing have been considered by Breuer \emph{et al}.
\cite{breuer08}: in the case of Planck scale fluctuations, this
study predicts far a too small effect to be detected with `ordinary'
quantum particles. However the authors do not rule out the
possibility of detecting dephasing experimentally by using large
quantum composite systems.

\section*{Brief description and salient points of this work}

\addcontentsline{toc}{section}{Brief description and salient points
of this work}

The investigations described in this thesis start out of the
previous considerations. The original question that we wished to
answer is whether decoherence suffered by a quantum particle due to
spacetime fluctuations can be detected at present, e.g. by means of
matter waves interferometry experiments like those described above
for fullerene. In our study we focus on a special class of spacetime
fluctuations, namely \emph{conformal fluctuations}. These represent
perturbations that change locally the scale of the spacetime
geometry. Conformal transformations of the metric play an important
role in GR and alternative gravity theories such as scalar-tensor
\cite{wagoner70}. The main reason why we want to study conformal
fluctuations is that these have been advocated in a recent series of
papers to be effective in providing a mechanism for quantum
dephasing \cite{powerpercival00,bingham05,wang2006}. Power and
Percival considered a simple and idealized toy model with conformal
fluctuations propagating in 1D over a flat spacetime geometry
\cite{powerpercival00}. Inspired by this work and by a related
proposal by Bingham \cite{bingham05}, Wang and collaborators further
investigated the issue trying to improve over Percival work by
incorporating GWs into the analysis \cite{wang2006}. The two main
conclusions of this work were: (1) conformal fluctuations can induce
dephasing (2) their `amount' is statistically set by a balancing
mechanism according to which for every graviton quantum there
corresponds a `conformal quantum' with a negative compensating
energy. Beyond being limited to conformal fluctuations traveling in
1D, the main limit of these analysis stands in the fact that the
presence of conformal fluctuations within the framework of GR is
somehow `postulated' but not shown to be consistent in a rigorous
way.

Our goal here is trying to clarify upon these and some related
issues by attempting to answer the following questions:
\begin{enumerate}
  \item if we model the conformal fluctuations more realistically
  in 3D, can they still induce dephasing of a quantum particle?
  \item if yes, how would this dephasing manifest itself
  qualitatively?
  \item how can we treat consistently the interesting case of vacuum
  conformal fluctuations?
  \item can we make a quantitative theoretical prediction for the amount of
  expected dephasing on a given quantum particle?
  \item are conformal fluctuations physical? In other words, can we
  identify a theory of gravity which is compatible with the presence of conformal
  fluctuations?
  \item can this theory be GR or shall we recur to alternative
  theories such as scalar-tensor?
  \item can conformal fluctuations really provide a mechanism to
  balance part or all of the vacuum energy due to GWs and/or matter
  fields?
\end{enumerate}
Questions 1. and 2. 3. and 4. are addressed in Chapter \ref{ch1},
which is based on the paper \cite{bonifacio2009I}. Here we study the
dynamics of a quantum particle in the metric $g_{ab} =
(1+A)^2\eta_{ab}$, where the \emph{conformal field} $A$ is a
stochastic field propagating in 3D space and satisfying the wave
equation. We consider this particular metric as given from the
outset without asking how it could practically arise: we limit
ourselves to studying the quantum mechanics problem of the
statistical evolution of the density matrix describing a non
relativistic quantum particle coupled to gravity. We derive a
dephasing formula generalizing to three space dimensions Percival
and Wang's results. The main result will be that, for an arbitrary
power spectrum $S(\omega)$ describing the statistical features of
the conformal field, dephasing does indeed occur and it is expected
to be roughly two orders of magnitude larger than in the 1D case.
Another important feature, in agreement with what found in previous
studies, is that dephasing occurs only as a nonlinear effect
resulting from a nonlinear effective newtonian potential $\propto
A^2$ to which the particle couples. A linear potential $\propto A$
\emph{cannot} induce dephasing.

In order to address question 4. and the problem of vacuum
fluctuations we find inspiration from the work of Boyer in the 70s
on random electrodynamics \cite{boyer75}. We simply assume that the
classical fluctuations in $A$ \emph{mimic} the zero point quantum
fluctuations of the hypothetical quantum field in its vacuum state.
This is implemented in practise by decomposing $A$ into normal modes
and by assigning an energy $\hbar\omega/2$ to each mode. This will
correspond to choosing $S(\omega) \propto 1/\omega$. In this case
the general results can be made explicit and the expected amount of
dephasing suffered by a particle of mass $M$ numerically estimated.
The conclusion is that the present day technology will not be able
to detect any effect, unless very heavy quantum particles are
employed, possibly in space based experiments, where the influence
of other external environmental factors is minimized.

The second part of the thesis addresses questions 5. 6. and 7. The
main goal is to identify a theory of gravity compatible with the
presence of conformal fluctuations induced by a conformal field $A$
satisfying the wave equation. In Chapter \ref{ch2} we establish the
main framework by developing a generalization of Isaacson's theory
for metric perturbations on a curved background
\cite{isaacson1968I,isaacson1968II} which can also incorporate
matter fields and conformal fluctuations. The treatment is
perturbative and can describe GWs at first order. Since we wish to
study phenomena related to vacuum we are ideally seeking for a
formalism in which any large scale curvature of the geometry of
empty spacetime is induced by the energy content of matter fields,
the backreaction due to GWs \emph{and}, possibly, to the conformal
fluctuations. In order to account for this it is necessary to push
the analysis to second order. Because of these facts the formalism
introduced in Chapter \ref{ch2} is referred to as the
\emph{nonlinear random gravity framework}. Like in Boyer random
electrodynamics the idea is that some of the vacuum properties,
namely those related to zero point energy, are modelled by using
classical stochastic fields with an appropriate spectrum as
described above. This formally allows to implement the passage from
the microscopic scale, where fields fluctuate, to the classical
scale, where geometry appears smooth, by means of a spacetime
averaging procedure. Treating vacuum in this way will raise the
issues of whether vacuum energy gravitates or not and how the
resulting vacuum energy due to matter fields, GWs and, possibly,
conformal fluctuations can be made compatible with the basically
vanishing observed value. Thus we will be led to discuss the well
known cosmological constant problem \cite{Carroll2001} and some
issues related to the still hypothetical mechanisms that can
regularize vacuum energy.

The nonlinear random gravity framework is applied to standard GR in
Chapter \ref{ch3}, where we try to find a solution of \ee in the
form $g_{ab} = \left[\Omega(A)\right]^2 \gamma_{ab}$ that may encode
GWs \emph{and} conformal fluctuations. Contrary to what claimed
previously in the literature we find that spontaneous conformal
fluctuations seem to be incompatible with the structure of GR, at
least in the case of the problem considered here dealing with empty
spacetime `filled' with fields in their vacuum state. We will show
that imposing a metric like $g_{ab} = \left[\Omega(A)\right]^2
\gamma_{ab}$ to \ee leads to constraint equations for the conformal
field $A$ that are unphysical and cannot be satisfied by a short
wavelength fluctuating field. Moreover and in relation to question
7. we find that conformal fluctuations within GR \emph{cannot} offer
a mean to provide vacuum energy regularization. This conclusion in
particular comes in correction of a technical imprecision in
\cite{wango8rg} where it had been wrongly found that conformal
fluctuations could offer a way to balance the large amount of vacuum
energy due to massless fields with a traceless stress energy tensor.

Chapter \ref{ch4} provides one possible answer to question 6. in
relation to alternative theories of gravity. First we correctly
re-interpret conformal transformations as a local change in the
physical units used in physics. Along these lines we illustrate
Bekenstein's theory of conformal invariant gravity
\cite{bekenstein1980}, where conformal fluctuations can be
introduced in a reasonable way through the fluctuations of
Bekenstein's mass gauge field. This theory is still GR and, though
enriched of a new symmetry, conformal fluctuations still cannot
induce any measurable effect. Finally we consider the more general
class of theories known as scalar-tensor \cite{wagoner70}, where
gravity is described by a rank 2 symmetric tensor and a scalar field
which sets the local value of the gravitational constant. We focus
in particular on the Brans-Dicke class of theories \cite{dicke62}
and apply the nonlinear random gravity framework. The main result is
that a vacuum solution in which the spacetime metric also presents a
conformal perturbation is indeed possible. The conformal field $A$
in this case represents a first order fluctuation of the Brans-Dicke
field and the wave equation is a simple consequence of the general
formalism. The main conclusion is that, \emph{independently} of the
value of the Brans-Dicke coupling parameter $\omega$, quantum
dephasing due to conformal fluctuations is predicted to occur within
the Brans-Dicke framework. A secondary conclusion regards the
coupling parameter $\omega$: at second order the structure of
spacetime is determined by matter fields, GWs \emph{and} $A$; this
is found to contribute through an usual Klein-Gordon stress energy
tensor for a massless field; for a certain range of values for
$\omega$, for which the Brans-Dicke field would be a equivalent to a
ghost, the conformal fluctuations could indeed provide a vacuum
energy balance mechanism. This would serve to balance the traceless
part of the matter fields stress energy tensor only. The
cosmological constant problem still cannot be addressed.
\\
\\

To conclude we just spend a few words on how the work is organized.
We discussed and included all our original results and contributions
in the first four chapters, representing the main body of this
thesis. We tried, when possible, not to overload the treatment with
long derivations or background knowledge. We thus included in the
appendices technical derivations and other related material that we
deemed would have interrupted the flow of the treatment. In
particular, Appendix \ref{ap1} introduces the general material for
the treatment of stochastic fields and is to be read in conjunction
with Chapter \ref{ch1}. There we derive an important theorem which
links the power spectral density of a spacetime stochastic process
to its autocorrelation function. All technical and lengthy
derivations related to Chapter \ref{ch1} are included in Appendix
\ref{ap2}, while Appendix \ref{ap3} shows how the spectral density
$S \propto 1/\omega$ in relation to vacuum can be deduced.

All the background material needed for this work is also shortly
discussed in the appendices. In Appendix \ref{ap4} we review in
great detail Isaacson's theory of metric perturbations over a curved
background. This is the starting point for the material presented in
Chapter \ref{ch2}. Finally, the necessary quantum mechanics
material, especially in relation to the density matric formalism and
decoherence is briefly illustrated in Appendix \ref{ap5}.

\newpage
\thispagestyle{empty}


\singlespacing
\chapter{Dephasing of a non-relativistic quantum
particle due to a conformally fluctuating spacetime}\label{ch1}

\begin{quote}
\begin{small}
In this chapter, based on \cite{bonifacio2009I}, we investigate the
dephasing suffered by a nonrelativistic quantum particle within a
conformally fluctuating spacetime geometry. Starting from a
minimally coupled massive Klein-Gordon field, we derive an effective
\schr equation in the non-relativistic limit. The wave function
couples to gravity through an effective nonlinear potential induced
by the conformal fluctuations. The quantum evolution is studied
through a Dyson expansion scheme up to second order. We show that
only the nonlinear part of the potential can induce dephasing. This
happens through an exponential decay of the off diagonal terms of
the particle density matrix. The bath of conformal radiation is
modeled in three-dimensions and its statistical properties are
described in terms of a general power spectral density. Vacuum
fluctuations at a low energy domain are investigated by introducing
an appropriate power spectral density and a general formula
describing the loss of coherence is derived. This depends
quadratically on the particle mass and on the inverse cube of a
particle dependent typical cutoff scale. Finally, the possibilities
for experimental verification are discussed. It is shown that
current interferometry experiments cannot detect such an effect.
However this conclusion may improve by using high mass entangled
quantum states.
\end{small}
\end{quote}

\singlespacing
\section{Introduction}

It is generally agreed that the underlying quantum nature of gravity
implies that the spacetime structure close to the Planck scale
departs from that predicted by GR. Unfortunately the QG domain is
still beyond modern particle accelerators such as LHC. Nonetheless,
finding experimental ways to test the quantum properties of
spacetime would be highly beneficial to the theoretical developments
of our fundamental theories of nature. In this respect it is has
been suggested that QG could induce decoherence on a quantum
particle through its underlying Planck scale spacetime fluctuations
\cite{powerpercival00,wang2006,breuer08,bingham05,camelia05,anastopoulos07}.

As the sensitivity and performance of matter wave interferometers is
increasing \cite{hackermuller03,hornberger03,hornberger04,
hackermuller04,stibor05}, it is important to assess the theoretical
possibility of a future experimental detection of intrinsic,
spacetime induced decoherence. The closely related dephasing effect
due to a random bath of classical GWs (e.g. of astrophysical origin)
has been extensively studied e.g. in \cite{reynaud04}. The problem
of the decoherence induced by spacetime fluctuations is difficult to
study: notwithstanding promising progress, mainly in loop quantum
gravity and superstring theory \cite{kiefer05}, a coherent and
established QG theoretical framework is still missing. Thus
theoretical attempts for a prediction of the decoherence induced by
spacetime fluctuations usually exploit some semiclassical framework.
Such approaches typically represent the spacetime metric close to
the Planck scale by means of fluctuating functions. These are
usually supposed to mimic the vacuum quantum properties of spacetime
down to some cutoff scale $\ell = \lambda \Lp,$ where $\Lp$ is the
Planck scale. The dimensionless parameter $\lambda$ marks the
benchmark between the fully quantum regime and the scale where the
classical properties of spacetime start to emerge
\cite{powerpercival00,wang2006}. The fact that classical fluctuating
fields can be used to reproduce various genuine quantum effects is
well known, e.g. from the work of Boyer \cite{boyer75,boyer1969} in
the case of the EM field. Frederick also applied the same technique
to model spacetime fluctuations \cite{frederick76}. A classical but
fluctuating metric is often exploited in the literature in relation
to problems related to the microscopic behavior of the spacetime;
e.g. a stochastic metric was employed in \cite{moffat97,miller99} to
study the problem of gravitational collapse and big bang
singularities, while in \cite{goklu08} spacetime metric fluctuations
were introduced and their ability to induce a weak equivalence
principle violation studied.

A pioneering analysis of the problem of spacetime induced
decoherence has been proposed by Power \& Percival (PP in the
following) \cite{powerpercival00} in the case of a conformally
modulated Minkowski spacetime with conformal fluctuations traveling
along one space dimension. This was improved by Wang \emph{et al}.
\cite{wang2006}, who extended upon PP work attempting to include the
effect of GWs. Conformal fluctuations are interesting as they are
relatively easy to treat and offer a convenient way to build `toy'
models to assess some of the problem's features. They have an
important role in theoretical physics \cite{kastrup08} and are
sometimes invoked in the literature also in relation to universal
scalar fields \cite{nelson85,miller99} that also arise naturally in
some modified theories of gravity such as scalar-tensor
\cite{brans61,dicke62,wagoner70}.

Within a semiclassical approach that `replaces' the true quantum
environment by classical fluctuating fields we should properly speak
of \emph{dephasing} of the quantum particle rather than decoherence.
In a remarkable paper about quantum interference in the presence of
an environment \cite{stern90}, Stern and co-authors showed that the
fully quantum approach where decoherence is studied by `tracing
away' the environment degrees of freedom and that in which the
dephasing of the quantum particle is due to a stochastic background
field give equivalent results.

Here we consider a conformally modulated four-dimensional spacetime
metric of the form $g_{ab} = (1 + A)^2 \eta_{ab}$. Such a metric has
been considered by PP \cite{powerpercival00}, where the dephasing
problem was studied in the simple idealized case of a particle
propagating in one-dimension. By imposing \ee on the metric
$g_{ab}$, PP deduced a wave equation for $A$. Their procedure to
derive an effective newtonian potential interacting with the quantum
probe started from the geodesic equation for a test particle. Even
though this did not take properly into account the nonlinearity in
the conformal factor $(1 + A)^2$, they found correctly that the
change in the density matrix is given by $\delta \rho \propto M^2 T
A_0^4 \tau_*$, where $M$ is the probing particle mass, $T$ the
flight time, $A_0$ the amplitude of the conformal fluctuations and
$\tau_*$ their correlation time. This formula was used to set limits
upon $\lambda$. However in doing so they did not treat the
statistical properties of the fluctuations properly and this
resulted in the wrong estimate $\lambda \propto (M^2 T / \delta
\rho)^{1/7}$, as already noted by Wang \emph{et al} in
\cite{wang2006}.

In their work, Wang and co-authors considered a metric of the kind
$g_{ab} = (1 + A)^2 \gamma_{ab}$, where the conformal metric
$\gamma_{ab}$ was supposed to encode GWs. This was done by
exploiting the results in \cite{wang05a,wang05b} where a canonical
geometrodynamics approach employing a conformal spacial 3-metric has
been studied. By exploiting an energy density balancing mechanism
between the conformal and GWs parts of the total gravitational
Hamiltonian, the statistical properties of the conformal
fluctuations where fixed. This corresponded to assume that each
`quantum' of the conformal field possessed a zero point energy
$-\hbar \omega$. Though an improvement over PP work, this approach
is still one-dimensional and too crude to make predictions.
Moreover, as it shall be discussed extensively in chapters \ref{ch3}
and \ref{ch4}, the issue of energy balance between conformal
fluctuations and GWs is a delicate one and, in fact, it seems to be
ruled out within the standard GR framework.

In the following sections we provide a coherent three-dimensional
treatment of the problem of a slow massive test particle coupled to
a conformally fluctuating spacetime. The conformal field $A$ is
assumed to satisfy a simple wave equation. This will allow for a
direct comparison with PP result. We also notice that such a
framework is expected to arise naturally within a scalar-tensor
theory of gravity, as it will be shown in chapter \ref{ch4}.

The material is organized as follows: in \emph{Section
\ref{P1_lvle}} the non-relativistic limit of a minimally coupled
Klein-Gordon field is deduced and an effective newtonian potential
depending nonlinearly on $A$ is identified in the resulting
effective \schr equation. In \emph{Section \ref{P1_aqe}} we set the
general formalism to study the average quantum evolution of the
particle density matrix $\rho$ through a Dyson expansion scheme. In
\emph{Section \ref{P1_cfcp}}, general results derived in Appendix
\ref{ap1} are used to model the statistical and correlation
properties of the fluctuations through a general, unspecified, power
spectral density. In \emph{Section \ref{P1_sec1} and \ref{P1_lemma}}
we compute the average quantum evolution and derive a general
expression for the evolved density matrix. We show in general that
only a nonlinear potential can induce dephasing. The resulting
dephasing formula implies an exponential decay of the density matrix
off-diagonal elements and is shown to hold in general and
independently of the specific spectral properties of the
fluctuations.  All we assume is that these obey a simple wave
equation and that they are a zero mean stochastic process. The
overall dephasing predicted within the present three-dimensional
model -equation \eqref{P1_P2_3d}- is seen to be about two orders of
magnitudes larger than in the one-dimensional case as derived by PP.
This result improves over both Percival's and Wang's work in that
its key ingredients are general enough to be potentially suited for
a variety of physical situations. Next we consider in \emph{Section
\ref{P1_pcf}} the problem of vacuum fluctuations. To this end a
power spectrum $S(\omega) \propto 1 / \omega$ is introduced and we
derive an explicit formula for the rate of change of the density
matrix. Finally the discussion in \emph{Section \ref{P1_disc}}
addresses the question of whether the dephasing due to conformal
vacuum spacetime fluctuations could be detected. A possibility would
be through matter wave interferometry employing heavy molecules. We
consider this issue in the final part of this chapter by estimating
the probing particle resolution scale, setting its ability to be
affected by the fluctuations. The resulting formula for the
dephasing indicates that the level of the effect is still likely to
be beyond experimental reach, even for heavy molecules such as
fullerene \cite{hackermuller04}. A measurable effect could possibly
result for larger masses, e.g. if entangles quantum states were
employed \cite{everitt08}.

\section{Low velocity limit and effective \schr
equation}\label{P1_lvle}

The problem we wish to solve is clearly defined: we consider a
scalar field $A$ inducing conformal fluctuations on an otherwise
flat spacetime geometry according to
\begin{equation}\label{P1_e1}
g_{ab} = \O^2\eta_{ab} = (1 + A)^2 \eta_{ab},
\end{equation}
where $\eta_{ab} = \mathrm{diag}(-1,1,1,1,)$ is the Minkowski
tensor. We will refer to $A$ as to the \emph{conformal field} and
this will be assumed to satisfy the wave equation $\pd^c\pd_c A =
0$. Solving this equation with random boundary conditions results in
a randomly fluctuating field propagating in three-dimensional space.
We assume this to be a small first order quantity, i.e. $\abs{A} =
O(\varepsilon \ll 1)$. Equation \eqref{P1_e1} expresses the
spacetime metric in the laboratory frame. We also suppose that the
typical wavelengths of $A$ are effectively cut off at a scale set by
$\ell = \lambda \Lp,$ where $\Lp = (\hbar G / c^3)^{1/2} \approx
10^{-35}$ m is the Planck length. The dimensionless parameter
$\lambda$ represents a structural property of spacetime: below
$\ell$ a full quantum treatment of gravity would be needed so that,
by definition, $\ell$ represents the scale at which a semiclassical
approach that treats quantum effects by means of classical randomly
fluctuating fields is supposed to be a valid approximation. The
value of $\lambda$ is model dependent but it is generally agreed
that $\lambda \gtrsim 10^2$ \cite{wang2006}, so that $\ell$ is
expected to be extremely small from a macroscopic point of view.
This motivates the assumption that classical macroscopic bodies,
including the objects making up the laboratory frame and also the
observers, are unaffected by the fluctuations in $A$. This
corresponds to the idea that a physical object is characterized by
some typical resolution scale $\lr$ setting its ability to `feel'
the fluctuations: if $\lr \gg \ell$ these average out and do not
affect the body, which then simply follows the geodesic of the flat
background metric. On the other hand a microscopic particle can
represent a successful probe of the conformal fluctuations if its
resolution scale is small enough.

We are interested in the phase change induced on the wave function
of a quantum particle by the fluctuating gravitational field.
Various approaches to the problem of how spacetime curvature affects
the propagation of a quantum wave exist in the literature; e.g. for
a stationary, weak field and a non relativistic particle a
\schr-like equation can be recovered \cite{dewitt66}. The more
interesting case of time varying gravitational fields can be treated
e.g. by eikonal methods that are usually restricted to weak fields
with $g_{ab} = \eta_{ab} + h_{ab}$ and $\abs{h_{ab}} \ll 1$
\cite{linet76,cai89}. Other approaches, e.g. in
\cite{goklu08,breuer08}, are based on the scheme developed by Kiefer
\cite{kiefer91} for the nonrelativistic reduction of a Klein-Gordon
field which is minimally coupled to a linearly perturbed metric.

The approach of PP in \cite{powerpercival00} and of Wang at el. in
\cite{wang2006} was to derive the geodesic equation in the weak
field limit. Their treatments were however employing, incorrectly,
the usual newtonian limit scheme which is valid \emph{only} for
weak, linear and static perturbations \cite{wald84}. This cannot be
done in the present case as $\O^2 = (1+A)^2$ induces a fast varying,
nonlinear perturbation. In alternative to the Newtonian limit
approach one could compute the geodesics of a conformally modulated
Minkowski metric \emph{exactly} and without making assumptions on
the conformal factor. However this is ideally suitable for a
\emph{zero size} test particle or, more precisely, for a particle
whose typical size is much less than the typical scale over which
the spacetime geometry varies. This is not suitable for the
situation we wish to study, where the spacetime fluctuations are
assumed to vary on the very short scale $\ell = \lambda \Lp$.

A wave approach that starts from a relativistic KG field does not
suffer from the limitations of geodesic approach: this could be
suitable for localized particles, while the KG approach does not
assume any wave profile. For these reasons we expect the two methods
to be inequivalent. Moreover the wave approach is conceptually
clearer also because the coupling between gravity and a scalar field
is well understood and in the appropriate non-relativistic weak
field limit an effective \schr equation emerges naturally. This will
be our approach below.

We describe the quantum particle of mass $M$ by means of a minimally
coupled Klein-Gordon (KG) field $\phi$:
\begin{equation*}
g^{ab}\cd_a\cd_b \phi = \frac{M^2 c^2}{\hbar^2} \phi,
\end{equation*}
where $\cd_a$ is the covariant derivative of the physical metric
$g_{ab}$. Using $g_{ab} = \O^2 \eta_{ab}$ this equation can easily
be made explicit \cite{wald84} and reads:
\begin{equation}\label{P1_KG1}
\left( -\frac{1}{c^2}\frac{\pd^2}{\pd t^2} + \nabla^2 \right) \phi =
\frac{\Omega^2 M^2 c^2}{\hbar^2} \phi - 2 \pd_a (\ln \Omega) \pd^a
\phi,
\end{equation}
i.e. the wave equation for a massive scalar field plus a
perturbation due to $A$ describing the coupling to the conformally
fluctuating spacetime. We remark that, had we considered the
alternative meaningful scenario of a conformally coupled scalar
field, then the equation $g^{ab}\cd_a\cd_b \phi - R\phi/6 - M^2 c^2
\phi / \hbar^2 = 0$ would read explicitly
\begin{equation*}
\left( -\frac{1}{c^2}\frac{\pd^2}{\pd t^2} + \nabla^2 \right) \phi =
\frac{\Omega^2 M^2 c^2}{\hbar^2} \phi - 2 \pd_a (\ln \Omega) \pd^a
\phi - \phi \O^{-1}\pd^c \pd_c \O.
\end{equation*}
The extra curvature term reads $\O^{-1} \pd^c \pd_c \O = (1 - A)
\pd^c \pd_c A + O(\varepsilon^3)$. We see that \emph{if $A$ is
assumed to satisfy the wave equation then it has no effect}: in this
case the minimally and conformally coupled KG equations are
equivalent up to second order in $A$. We also note that, by
introducing the \emph{auxiliary field} $\phit := \O \phi$, equation
\eqref{P1_KG1} turns out to be equivalent to:
\begin{equation*}
\left( -\frac{1}{c^2}\frac{\pd^2}{\pd t^2} + \nabla^2 \right) \phit
= \frac{\Omega^2 M^2 c^2}{\hbar^2} \phit.
\end{equation*}
In principle, if a solution for $\phit$ were known, then \emph{the
physical scalar field representing the particle} would follow
formally as $\phi = \O^{-1}\phit = (1 - A + A^2) \phit,$ up to
second order in $A$. However in studying the dephasing problem we
will \emph{only} find an averaged solution for the average density
matrix representing the quantum particle. Therefore, even if a
solution in this sense is know in relation to $\phit$, it would not
be obvious how to obtain the corresponding averaged density matrix
related to $\phi$, which is what we are interested in.

In view of the above considerations we work directly with equation
\eqref{P1_KG1} and now proceed in deriving its suitable
non-relativistic limit. We will make two assumptions:
\begin{enumerate}
  \item the particle is slow i.e., if $\pl = M \vl$ is its momentum in the
  laboratory, we have:
  \begin{equation*}
  \frac{\vl}{c} \ll 1;
  \end{equation*}
  \item the effect of the conformal fluctuations is small, i.e. the
  induced change in momentum $\delta p = M \dv$ is small compared to
  $M \vl$:
  \begin{equation*}
  \frac{\dv}{\vl} \ll 1.
  \end{equation*}
\end{enumerate}
In view of these assumptions we can write:
\begin{equation*}
\phi = \psi \ex,
\end{equation*}
where the field $\psi$ is close to be a plane wave of momentum
$\pl$. As a consequence we have:
\begin{equation*}
\frac{\pd^2 \psi}{\pd t^2} \approx
-\frac{1}{\hbar^2}\left(\frac{\pl^2}{2M}\right)^2\psi.
\end{equation*}
Using this and multiplying by $\hbar^2 / 2M$, equation
\eqref{P1_KG1} yields:
\begin{eqnarray}\label{P1_eqeqeq}
\left[ \underbrace{i\hbar\frac{\pd}{\pd t} +
\frac{\hbar^2}{2M}\nabla^2}_{T_1}
-\underbrace{\frac{Mc^2}{8}\left(\frac{\vl}{c}\right)^4}_{T_2}
\right] \psi = \nonumber\\
\hspace{-1cm}=\underbrace{\left(A +
\frac{A^2}{2}\right)Mc^2}_{T_3}\psi -
\underbrace{\frac{\hbar^2}{M}\pd^a \phi\pd_a
\ln(1+A)\times\exp{\left(iMc^2 t/\hbar\right)}}_{T_4}.
\end{eqnarray}

Leaving the term $T_4$ aside for the moment, the orders of the three
underlined terms must be carefully assessed. We have:
\begin{equation*}
T_1 \sim M\vl^2,\quad T_2 \sim Mc^2\left(\frac{\vl}{c}\right)^4\quad
\m{T_3} \sim \varepsilon^2 M c^2,
\end{equation*}
where an average has been inserted since $T_3$ is fluctuating. It
follows that
\begin{equation*}
\frac{T_2}{T_1} \sim \left(\frac{\vl}{c}\right)^2 \ll 1.
\end{equation*}
Thus, in the non-relativistic limit, $T_2$ is negligible in
comparison to $T_1$. This is the case in a typical interferometry
experiment where it can be $\vl \approx 10^2$ m s$^{-1}$
\cite{stibor05}, so that $(\vl / c)^2 \sim 10^{-12}$. Next we have:
\begin{equation*}
\frac{\m{T_3}}{T_1} \sim \left(\frac{\varepsilon c}{\vl}\right)^2.
\end{equation*}
The request that the conformal fluctuations have a small effect thus
gives the condition
\begin{equation}\label{P1_C2}
\frac{\m{T_3}}{T_1} \ll 1 \quad \Leftrightarrow \quad \varepsilon^2
\sim \m{A^2} \ll \left(\frac{\vl}{c}\right)^2.
\end{equation}
That this condition is effectively satisfied can be checked a
posteriori after the model is complete. It depends on the
statistical properties of the conformal field and the particle
ability to probe them. This will be related to a particle
\emph{resolution scale}. At the end of the discussion in Section
\ref{P1_sneo} we will show that \eqref{P1_C2} is satisfied if, e.g.,
the particle resolution scale is given by its Compton length.

Under these conditions the non-relativistic limit of equation
\eqref{P1_eqeqeq} yields:
\begin{equation}\label{P1_eqeq}
-\frac{\hbar^2}{2M}\nabla^2\psi + \left(A + \frac{A^2}{2}\right)Mc^2
\psi + T_4 = i\hbar\frac{\pd \psi}{\pd t},
\end{equation}
where
\begin{equation*}
T_4 := -\frac{\hbar^2}{M}\pd^a \phi\pd_a \ln(1+A) \times
\exp{\left(iMc^2 t/\hbar\right)}.
\end{equation*}
In order to assess the correction due to this term we split it into
two contributions by writing separately the time and space
derivatives. Using the fact that
\begin{equation*}
\frac{\pd \phi}{\pd t} \approx -\frac{i M c^2}{\hbar}\psi\ex
\end{equation*}
it is easy to see that:
\begin{equation}\label{P1_t4}
T_4 = -i\hbar\left( \dot{A} - A\dot{A} \right)\psi - i\hbar \vl
\left( A_{,x} - A A_{,x} \right)\psi,
\end{equation}
where $\dot{A} := \pd A / \pd t$, $A_{,x} := \pd A / \pd x$ and
where we assumed that the particle velocity in the laboratory is
along the $x$ axis. In Appendix \ref{ap2XXX} we show that \emph{if
$A$ is (i) a stochastic isotropic perturbation and (ii) effectively
fast varying over a typical length $ \lambda_A = \kappa h / (Mc)$
related to the particle resolution scale}, then $T_4$ reduces to:
\begin{equation}\label{P1_t4bis}
T_4 = \left( - A + A^2 \right) \frac{Mc^2}{\kappa}\psi.
\end{equation}
Here $\kappa \sim 1$ is dimensionless and its precise value is
unimportant. The important point is that $T_4$ yields a
\emph{positive} extra nonlinear term in $A$ that adds up to what we
already have in \eqref{P1_eqeq}. Finally we get the \emph{effective
\schr equation}
\begin{equation*}
-\frac{\hbar^2}{2M}\nabla^2\psi + V \psi = i\hbar\frac{\pd \psi}{\pd
t},
\end{equation*}
where the \emph{nonlinear fluctuating potential} $V$ is defined by
\begin{equation}\label{P1_pot}
V := \left(\cI A + \cII A^2\right)Mc^2.
\end{equation}
The values of the constants $\cI$ and $\cII$ depend on $\kappa$. For
$\kappa = 1$ it would be $\cI = 0$ and $\cII = 3/2.$ For generality
we will leave them unspecified and consider $\kappa$ as a constant
of order one.

\section{Average quantum evolution}\label{P1_aqe}

\subsection{Dyson expansion for short evolution
time}\label{P1_P2_ipde}

We now have a rather well defined problem: that of the dynamics of a
non relativistic quantum particle under the influence of the
nonlinear stochastic potential \eqref{P1_pot}. The \schr equation
describing the dynamics of a free particle is suitable to describe
the interference patterns that could result e.g. in an
interferometry experiment employing cold molecular beams. When the
particle in the beam propagates through an environment, we are
dealing with an open quantum system. This in general suffers
decoherence, resulting in a loss of visibility in the fringes
pattern \cite{hornberger04,stibor05}. This is a well defined
macroscopic quantity. In the present semiclassical treatment the
environment due to spacetime fluctuations is represented, down to
the semiclassical scale $\ell$, by a sea of random radiation encoded
in $A$ and resulting in the fluctuating potential $V$. An estimate
of the overall dephasing can be obtained by considering the
statistical averaged dynamics of a single quantum particle
interacting with $V$. In practise we will need (i) to solve for the
dynamics of a single particle of mass $M$ and (ii) calculate the
averaged wavefunction by averaging over the fluctuations. The
outcome of (i) would be some sort of `fluctuating' wavefunction
carrying, beyond the information related to the innate quantum
behavior of the system, that related to the fluctuations in the
potential. The outcome of (ii) is to yield a general statistical
result describing what would be obtained in an experiment where many
identical particles propagate through the same fluctuating
potential.

We thus consider the Hamiltonian operator $\Ho(t)=\Ho^0+\Ho^1(t),$
where $\Ho^0$ is the kinetic part while
\begin{equation*}
\Ho^1(t)=\int \d^3 x V(\xv,t) \ket{\xv}\bra{\xv},
\end{equation*}
is the perturbation due to the fluctuating potential energy. Here
$\ket{\xv}\bra{\xv}$ is the projection operator on the space spanned
by the position operator eigenstate $\ket{\xv}$. Indicating the
state vector at time $t$ with $\psi_t$, the related \schr equation
reads
\begin{equation*}
\Ho(t)\psi_t = i\hbar\frac{\pd\psi_t}{\pd t}.
\end{equation*}
Using the density matrix formalism and as shown in Appendix
\ref{ap5}, the general solution can be expressed through a Dyson
series as \cite{roman65}
\begin{equation}\label{dy}
\rho_T=\rho_0 + \Ko_1(T)\rho_0+\rho_0
\Ko_1^{\dag}(T)+\Ko_2(T)\rho_0+\Ko_1(T)\rho_0 \Ko_1^{\dag}(T)+\rho_0
\Ko_2^{\dag}(T)+\ldots,
\end{equation}
where $\rho_0$ is the initial density matrix and the propagators
$\Ko_1(T)$ and $\Ko_2(T)$ are given by
\begin{equation*}
\Ko_1(T):=-\frac{i}{\hbar}\int_{0}^{T}{\Ho}(t'){\d}t',
\end{equation*}
\begin{equation*}
\Ko_2(T):=-\frac{1}{\hbar^2}
\int_{0}^{T}{\d}t'\int_{0}^{t'}{\d}t''{\Ho}(t'){\Ho}(t'').
\end{equation*}
In truncating the series to second order we assume that the system
evolves for a time $T$ such that $ T\ll T^*$, where $T^*$ is defined
as the typical time scale required to have a significant change in
the density matrix $\rho$.

The effect of the environment upon a large collection of identically
prepared systems is found by taking the average over the fluctuating
potential as explained above. Formally and up to second order we
have
\begin{equation}\label{P2_Xe22}
\m{\rho_T}=\m{\rho_0 + \Ko_1(T)\rho_0+\rho_0
\Ko_1^{\dag}(T)+\Ko_2(T)\rho_0+\Ko_1(T)\rho_0 \Ko_1^{\dag}(T)+\rho_0
\Ko_2^{\dag}(T)}.
\end{equation}
The averaged density matrix $\m{\rho_T}$ will describe the average
evolution of the system including the effect of dephasing.

It is straightforward but lengthy to show that, up to second order
in the Dyson's expansion, the kinetic and potential parts of the
hamiltonian give independent, additive contributions to the average
evolution of the density matrix, i.e. $\m{\rho_T}=[\rho_T]_0 +
\m{[\rho_T]_1},$ where
\begin{eqnarray*}
[\rho_T]_0:=\rho_0 + [\Ko_1(T)]_0\rho_0+\rho_0
[\Ko_1(T)]_0^{\dag}\nonumber\\
+[\Ko_2(T)]_0\rho_0+[\Ko_1(T)]_0\rho_0 [\Ko_1(T)]_0^{\dag}+\rho_0
[\Ko_2(T)]_0^{\dag},
\end{eqnarray*}
\begin{eqnarray}\label{P1_P2_finale1}
\m{[\rho_T]_1}:=\left\langle\rho_0 + [\Ko_1(T)]_1\rho_0+\rho_0
[\Ko_1(T)]_1^{\dag}\right.\nonumber\\
\left.+[\Ko_2(T)]_1\rho_0+[\Ko_1(T)]_1\rho_0
[\Ko_1(T)]_1^{\dag}+\rho_0 [\Ko_2(T)]_1^{\dag}\right\rangle.
\end{eqnarray}
Here the kinetic propagators $[\Ko_1(T)]_0$ and $[\Ko_2(T)]_0$
depend solely on $\Ho^0$, while the potential propagators
$[\Ko_1(T)]_1$ and $[\Ko_2(T)]_1$ depend only on $\Ho^1(t)$. For a
proof of this statement we refer to Appendix \ref{ap2_1}. In the
next section we estimate the dephasing by calculating the term
$\m{[\rho_T]_1}$ alone.

\section{The conformal field and its correlation
properties}\label{P1_cfcp}

We now set the statistical properties of the conformal field $A$.
This is assumed to represent a real, stochastic process having a
zero mean. We further assume it to be isotropic. In Appendix
\ref{ap1} we review a series of important results concerning
stochastic processes, in particular in relation to real stochastic
signals satisfying the wave equation. The main quantity
characterizing the process is the power spectral density
$S(\omega)$. In the case of an isotropic bath of random radiation,
field averages such as $\ms{A^2}$, $\ms{\abs{\nabla A}^2}$ and
$\ms{(\pd_t A)^2}$ can be found in terms of $S(\omega)$, e.g.
\begin{equation*}
\m{A^2} = \frac1{(2\pi)^3} \int \d^3 k\, S(k),
\end{equation*}
where $kc = \omega$. In Appendix \ref{ap1} we show how the conformal
field can be resolved into components traveling along all possible
space directions according to
\begin{equation*}
A(\v{x},t)=\int
\d\hat{\v{k}}\,A_{\hat{\v{k}}}(\hat{\v{k}}\cdot\v{x}/c-t),
\end{equation*}
where $\d\hat{\v{k}}$ indicates the elementary solid angle. The
capacity of the fluctuations to maintain correlation is encoded in
the autocorrelation function $C(\tau)$. In the same appendix we
prove a generalization of the usual Wiener-Khintchine (WK) theorem,
valid for the case of a spacetime dependent process satisfying the
wave equation, and linking the autocorrelation function to the
Fourier transform of the power spectral density according to:
\begin{equation}\label{P1_WK}
C(\tau) = \frac{1}{(2\pi c)^3}\int\!\d \omega\, \omega^2
S(\omega)\cos(\omega\tau).
\end{equation}
This allows to prove that wave components traveling along
independent space directions are uncorrelated, i.e.
\begin{equation}\label{P1_P2_cfc}
\m{ A_{\hat{\v{k}}}(t)\, A_{\hat{\v{k}}'}(t+\tau) }=
\delta(\kvs,\kvs')\,C(\tau).
\end{equation}
The field mean squared amplitude is related to the correlation
function according to $\m{ A^2}=4\pi C_0$, as derived in Appendix
\ref{ap1}, and where $C_0:=C(0)$.

Isotropy implies that all directional components have the same
amplitude $A_0$. This is found introducing the \emph{normalized
correlation function} $R(\tau)$ through
\begin{equation*}
    R(\tau) := \frac{C(\tau)}{C_0},
\end{equation*}
so that $R(0)=1$. Equation \eqref{P1_P2_cfc} can now be re-written
as $\m{ A_{\hat{\v{k}}}(t)\, A_{\hat{\v{k}}'}(t+\tau)}=
\delta(\kvs,\kvs')\,C_0\,R(\tau)$ so that, introducing the
\emph{normalized directional components}, $ f_{\kvs}(t) :=
A_{\hat{\v{k}}}(t) / \sqrt{C_0} $ we have
\begin{equation*}
\m{ f_{\hat{\v{k}}}(t)\, f_{\hat{\v{k}}'}(t+\tau) }=
\delta(\kvs,\kvs')\,R(\tau).
\end{equation*}
We now define the constant $ A_0 := \sqrt{C_0} $ which is connected
to the \emph{squared amplitude per solid angle} according to $A_0^2
= C_0 = \m{A^2} / 4\pi. $ The directional components are given by
$A_{\kvs}(t)=A_0\,f_{\kvs}(t)$ and the general conformal field can
finally be expressed as an elementary superposition of the kind
\begin{equation}\label{P1_e6}
A(\xv,t)=A_0\int\! \d\kvs \,f_{\kvs}(\hat{\v{k}}\cdot\v{x}/c-t).
\end{equation}

\subsection{Summary of the correlation
properties of the conformal fluctuations}\label{P1_P2_prop}

The main statistical properties of the directional stochastic waves
$f_{\kvs}$ are summarized by
\begin{equation}\label{P1_e15}
\m{ f_{\kvs}(t) }=0,
\end{equation}
\begin{equation}\label{P1_e17}
\m{ f_{\kvs}(t)f_{\kvs'}(t') }=\delta(\kvs,\kvs')R(t-t'),
\end{equation}
i.e. each component has \emph{zero mean} and fluctuations traveling
along different space directions are perfectly \emph{uncorrelated}.
These two properties imply that odd products of directional
components have also a zero mean, i.e.
\begin{equation}\label{P1_e20}
\m{ f_{\kvs_1}(t_1)f_{\kvs_2}(t_2)f_{\kvs_3}(t_3) }=0.
\end{equation}
In the following dephasing calculation we will need to evaluate
means involving products of four directional components. To this
purpose we need to introduce the \emph{second order correlation
function} $R''(t-t')$ according to
\begin{equation}\label{P1_e21}
\m{ [f_{\kvs}(t)]^2[f_{\kvs'}(t')]^2
}=1+\delta(\kvs,\kvs')[R''(t-t')-1].
\end{equation}
This definition is compatible with the fact that the mean is one
when components traveling in different direction are involved, i.e.
$\m{ [f_{\kvs}(t)]^2[f_{\kvs'}(t')]^2 } =1$ if $\kvs \neq \kvs'.$

\section{Dephasing calculation outline}\label{P1_sec1}

To calculate the dephasing suffered by the probing particle we must
evaluate the average of all the individual terms in equation
\eqref{P1_P2_finale1}. The relevant propagators are
\begin{equation*}
[\Ko_1(T)]_1:= -\frac{i}{\hbar}\int_{0}^{T}\!\!\!{\d}t'{\Ho^1}(t'),
\end{equation*}
\begin{equation*}
[\Ko_2(T)]_1 := -\frac{1}{\hbar^2} \int_{0}^{T}\!\!\!\d
t\int_{0}^{t}{\d}t'{\Ho^1}(t){\Ho^1}(t').
\end{equation*}
The interaction Hamiltonian is given by
\begin{equation*}
\Ho^1(t)=\int \d^3 x V(\xv,t)\ket{\xv}\bra{\xv},
\end{equation*}
where the potential energy is
\begin{equation*}
     V(\xv,t)=\cI Mc^2 A_0\int\! \d\kvs
    \,f_{\kvs}(t-\xv\cdot\kvs/c) + \cII M c^2A_0^2 \left[\int\! \d\kvs
    \,f_{\kvs}(t-\xv\cdot\kvs/c)\right]^2.
\end{equation*}

\subsection{First order terms of the Dyson expansion}

We evaluate the two first order terms in the Dyson expansion. For a
more compact notation, we do not show the argument of the
directional components $f_{\kvs}$. The contribution of the linear
part of the potential $\cI Mc^2 A$ vanishes trivially since
$\m{f_{\kvs}} = 0.$ The quadratic part gives:
\begin{equation*}
\m{\Ko_1(T)\rho_0} = -\frac{i \cII M
c^2A_0^2}{\hbar}\int_{0}^{T}\!\!{\d}t\! \int\!\! \d^3 x\,
\ket{\xv}\bra{\xv}  \m{ \left[\int\! \d\kvs \,f_{\kvs}\right]^2 }
\rho_0.
\end{equation*}
Using $A_{\hat{\v{k}}}(t) = \sqrt{C_0} f_{\kvs}(t)$ and $\m{A^2} =
4\pi C_0 \equiv 4\pi A_0^2 $ it is seen that the average yields
$4\pi.$ Since $\int\!\! \d^3 x\, \ket{\xv}\bra{\xv} = \I$ and
integrating over $T$ we find
\begin{equation}\label{P1_P2_1st}
\m{\Ko_1(T)\rho_0} = -\frac{4\pi \cII i M c^2A_0^2 T}{\hbar}\rho_0.
\end{equation}

The calculation of the other first order term proceeds in the same
way. Since $\Ko_1^{\dag}(T) = -\Ko_1(T)$, it yields the same result
as in \eqref{P1_P2_1st} but with the opposite sign (more in general,
all the odd terms in the Dyson expansion have an $i$ factor and also
yield a vanishing contribution). We thus see that \emph{at first
order in the Dyson expansion there is no net dephasing and
$\ms{\Ko_1(T)\rho_0 + \rho_0\Ko_1^{\dag}(T)} = 0.$}

\subsection{Second order terms of the Dyson expansion}

The second order calculation is more complicated. A fundamental
point is that the linear part of the potential does again give a
vanishing contribution. Dephasing will be shown to come as a purely
nonlinear effect due to the nonlinear potential term $\sim A^2$.

\subsubsection{(Non)-contribution of the linear part of the
potential}\label{cdu}

To have an idea of how things work we consider e.g. the average of
the term $\Ko_2 \rho_0$. This has the following structure:
\begin{equation*}
\m{\Ko_2 \rho_0} \sim \int {\d}t\int {\d}t' \int \d^3 y \Pr{\yv}\int
\d^3 y'\Pr{\yv'} \m{V(\yv,t)V(\yv',t')}\rho_0.
\end{equation*}
The interesting part is the average $\m{V(\yv,t)V(\yv',t')}$. This
is:
\begin{equation*}
\m{V(\yv,t)V(\yv',t')} \sim \m{A(\yv,t)A(\yv',t')} +
\m{A(\yv,t)A^2(\yv',t')} + \m{A^2(\yv,t)A^2(\yv',t')}.
\end{equation*}
The second term vanishes in virtue of property \eqref{P1_e20}. This
is seen using the directional decomposition \eqref{P1_e6} and
writing:
\begin{equation*}
\m{A(\yv,t)A^2(\yv',t')} = A_0^3 \int\! \d\kvs_1 \int\! \d\kvs_2
\int\! \d\kvs_3 \,\m{f_{\kvs_1}f_{\kvs_2}f_{\kvs_3}} = 0.
\end{equation*}
The first term derives from the linear part of the potential. It
results in the contribution:
\begin{align*}
&\m{A(\yv,t)A(\yv',t')} \Rightarrow \\
&\hspace{-1cm}\int_0^T\!\!\! {\d}t\!\int_0^t \!\!\!{\d}t'\!\! \int
\d^3 y \Pr{\yv}\!\int \d^3 y'\Pr{\yv'}\! \int\! \d\kvs\!\! \int\!
\d\kvs'\m{f_{\kvs}(t - \yv\cdot\kvs)f_{\kvs'}(t' -
\yv'\cdot\kvs')}\rho_0.
\end{align*}
For convenience of notation we set $c=1$ in the arguments of the
directional functions $f_{\kvs}$. Using equation \eqref{P1_e17} the
average yields the 2-point correlation function according to
$\delta(\kvs,\kvs')R(t-t' + \yv'\cdot\kvs' - \yv\cdot\kvs)$.
Integrating with respect to $\kvs'$ yields:
\begin{eqnarray*}
\m{A(\yv,t)A(\yv',t')} \Rightarrow \int_0^T\!\!\! {\d}t\!\int_0^t
\!\!\!{\d}t'\!\! \int \d^3 y \Pr{\yv}\!\int \d^3 y'\Pr{\yv'}\!
\int\! \d\kvs R[t-t' + \kvs\cdot(\yv' - \yv)] \rho_0.
\end{eqnarray*}
The corresponding matrix element is found by inserting $\bra{\xv}$
and $\ket{\xv'}$ respectively on the left and on the right. Using
$\bra{\xv}\kett{\yv} = \delta(\xv-\yv)$ and exploiting the
properties of the delta function we find
\begin{equation}\label{P1_cont1}
\m{A(\yv,t)A(\yv',t')} \Rightarrow K \times \int_0^T\!\!\!
{\d}t\!\int_0^t \!\!\!{\d}t'\!\! \int\! \d\kvs R(t-t'),
\end{equation}
where $K$ is a constant given by
\begin{equation*}
K = -\frac{\cI^2 A_0^2 M^2 c^4\rho_{\xv\xv'}(0)}{\hbar^2},
\end{equation*}
and where $\rho_{\xv\xv'}(0) := \bra{\xv}\rho_0 \ket{\xv'}$. The
similar terms coming from $\ms{\rho_0 \Ko_2^{\dag}}$ will contribute
in the same way as in \eqref{P1_cont1}, thus yielding an extra
factor 2. Finally, through a similar calculation it is found that
the terms $\sim \ms{A(\yv,t)A(\yv',t')}$ coming from $\ms{\Ko_1
\rho_0 \Ko_1^{\dag}}$ contribute according to:
\begin{equation*}
\m{A(\yv,t)A(\yv',t')} \Rightarrow -K \times \int_0^T\!\!\!
{\d}t\!\int_0^T \!\!\!{\d}t'\!\! \int\! \d\kvs R[t-t' + \kvs\cdot
(\xv' - \xv)].
\end{equation*}
Bringing all together, the overall contribution deriving from the
linear part $\cI Mc^2 A$ of the effective potential is found to be
proportional to the expression:
\begin{equation}\label{P1_I}
    I:=  \int_0^T \! {\d}t\left\{
2\int_0^t \!\!\!{\d}t' R(t-t') - \int_0^T \!\!\!{\d}t' R[t-t' +
\kvs\cdot \Delta\xv]\right\},
\end{equation}
where $\Delta\xv := \xv' - \xv$. In Appendix \ref{ap2_3} we prove
that this vanishes provided $R(\tau)$ is an even function and the
drift time $T$ is much larger than the time needed by the
fluctuations to propagate through the distance $\abs{\Delta\xv}$,
i.e. if $T \gg \kvs\cdot\Delta\xv$, where $c=1$. This condition is
certainly satisfied in a typical interferometry experiment where the
drift time $T$ can be of the order of $\sim 1$ ms and $cT$ is indeed
much larger than the typical space separations $\abs{\Delta\xv}$
relevant to quantify the loss of contrast in the measured
interference pattern.

\emph{Thus we have here the important result that the linear part of
the potential doesn't induce in general any dephasing up to second
order in Dyson expansion}. In fact we show in the next section that
dephasing results \emph{purely} as an effect of the nonlinear
potential term $\cII M c^2 A^2.$

\subsubsection{Contribution of the nonlinear part of the potential}

This calculation requires estimating averages of the kind
$\m{A^2(\yv,t)A^2(\yv',t')}$, which will bring in the second order
correlation function $R''$ defined in \eqref{P1_e21}. This is
straightforward but algebraically lengthy. The full calculation is
reported in Appendix \ref{ap2_2}, where we show that proceeding in a
similar way as done above, exploiting the statistical properties
\eqref{P1_e15}-\eqref{P1_e21} and the already mentioned result $I =
0$ in relation to \eqref{P1_I}, then the general result for the
density matrix and valid up to second order in the Dyson expansion
follows as:
\begin{align}\label{P1_P2_mainB}\nonumber
&\rho_{\xv\xv'}(T)=\,\,\rho_{\xv\xv'}(0) -\frac{32\cII^2\pi^2 M^2
c^4 A_0^4 \rho_{\xv\xv'}(0)}{\hbar^2} \times \left[
\int_{0}^{T}\!\!\!\d
t \int_{0}^{T}\!\!\!{\d}t' R^2( t-t') \right.\\
&\hspace{1cm}\left. \hspace{-2cm}- \frac{1}{16\pi^2} \int\!
\d\kvs\int\! \d\Kvs\, \int_{0}^{T}\!\!\!\d t
\int_{0}^{T}\!\!\!{\d}t' R( t-t'-\kvs\cdot\Delta\xv/c) \times R(
t-t'-\Kvs\cdot\Delta\xv/c) \right].
\end{align}

Remarkably, \emph{the second order correlation function doesn't play
any role: the first order correlation function $R(\tau)$, and thus
the power spectral density $S(\omega)$, completely determines the
system evolution up to second order.} Equation \eqref{P1_P2_mainB}
implies that the diagonal elements of the density matrix are left
unchanged by time evolution. This is seen by setting $\Delta \xv=0$
which yields immediately $\rho_{\xv\xv}(T)=\rho_{\xv\xv}(0)$ for
every $T$.

\section{General density matrix evolution for large drift
times}\label{P1_lemma}

To verify that we have dephasing with an exponential decay of the
off diagonal elements we need further simplify the result
\eqref{P1_P2_mainB} by analyzing its behavior for appropriately
large evolution times. To this end we start from the following
identity
\begin{align}
\int_{0}^{T}\!\!\!\d t \int_{0}^{T}\!\!\!dt' g(t-t') &=
\frac{1}{2\pi} \int_{0}^{T}\!\!\!\d t \int_{0}^{T}\!\!\!dt'
\int_{-\infty}^{\infty}\!\!\!\d\omega\, \tilde{g}(\omega)e^{i\omega
(t-t')}
\nonumber\\
\nonumber\\
&= \frac{1}{2\pi}\int_{-\infty}^{\infty}\!\!\!\d\omega\,
\tilde{g}(\omega)\left[\frac{\sin(\omega T/2)}{\omega/2}\right]^2,
\label{P1_TTtt00}
\end{align}
where $\tilde{g}(\omega)$ denotes the Fourier transform of the
function $g(t)$. It is well known that the function in between
brackets can be used to define the Dirac delta function through
\begin{equation*}
\frac{1}{2\pi T}\left[\frac{\sin(\omega T/2)}{\omega/2}\right]^2
\xrightarrow[T\rightarrow\infty]{} \delta(\omega).
\end{equation*}
It is then clear that, for an appropriately large evolution time, we
have approximatively
\begin{equation}\label{lemmaI}
\int_{0}^{T}\!\!\!\d t \int_{0}^{T}\!\!\!dt' g(t-t') \approx T
\int_{-\infty}^{\infty}\!\!\!\d\omega\, \tilde{g}(\omega)
\delta(\omega) = \tilde{g}(0) T,
\end{equation}
where $\tilde{g}(0)$ is the Fourier transform of $g$ evaluated at
the frequency $\omega = 0$. To clarify what we mean by
`appropriately large evolution time', we re-write equation
\eqref{P1_TTtt00} as
\begin{equation*}
\frac{1}{2\pi}\int_{-\infty}^{\infty}\!\!\!\d\omega\,
\tilde{g}(\omega)\left[\frac{\sin(\omega T/2)}{\omega/2}\right]^2 =
\frac{T}{\pi}\int_{-\infty}^{\infty}\!\!\!\d
x\,\tilde{g}(2x/T)\left(\frac{\sin x}{x}\right)^2,
\end{equation*}
where we defined the dimensionless variable $x:=\omega T / 2$. We
now focus on the Fourier transform $\tilde{g}(\omega)$ and we
consider a frequency interval $[0,\Delta \omega]$  in which
$\tilde{g}(\omega)$ varies little and can thus be considered as
basically constant, in such a way that $\tilde{g}(\omega) \approx
\tilde{g}(0)$ for $\omega \in [0,\Delta \omega]$. Considering now
that
\begin{equation}
\frac{1}{\pi}\int_{-\infty}^{\infty}\!\!\!\d x\,\left(\frac{\sin
x}{x}\right)^2 = 1
\end{equation}
and
\begin{equation}
\frac{1}{\pi}\int_{-5}^{+5}\!\!\!\d x\,\left(\frac{\sin
x}{x}\right)^2 = 0.94
\end{equation}
we have the approximate result
\begin{equation*}
\frac{T}{\pi}\int_{-\infty}^{\infty}\!\!\!\d
x\,\tilde{g}(2x/T)\left(\frac{\sin x}{x}\right)^2 \approx
 T\,\tilde{g}(0)\, \frac{1}{\pi}\,\int_{-5}^{+5}\!\!\!\d
x\,\left(\frac{\sin x}{x}\right)^2 \approx T\,\tilde{g}(0),
\end{equation*}
provided that
\begin{equation}\label{Tmin}
    \frac{1}{T} \lesssim \Delta\omega,
\end{equation}
guaranteeing that the Fourier transform of the correlation function
remains practically constant across the relevant integration
interval. We have thus proven the approximate relation
\begin{equation}
\int_{0}^{T}\!\!\!\d t \int_{0}^{T}\!\!\!dt' g(t-t') \approx
\tilde{g}(0) T,\quad\text{for}\quad T \gtrsim (\Delta\omega)^{-1}.
\label{P1_TTtt01}
\end{equation}
This can now be used to simplify the equation \eqref{P1_P2_mainB}
governing the evolution of the density matrix. We remark that for
this result to hold the integrand function $g(t-t')$ need not be an
even function and the only requirement is that its Fourier transform
is slowly varying over the interval $[0,\Delta\omega]$.

\subsection{Correlation time and characteristic function}

Equation \eqref{P1_TTtt01} can now be used to evaluate the time
integrals appearing in \eqref{P1_P2_mainB}. This is done by
identifying in one case $g(t):=R^2(t)$ and in the other
$g_{\tau\tau'}(t):=R(t+\tau)R(t+\tau')$, where $\tau$ and $\tau'$
stand respectively for $-\kvs\cdot\Delta\xv/c$ and
$-\Kvs\cdot\Delta\xv/c$, and where the normalized correlation
function can be expressed, using the generalized WK theorem
\eqref{P1_WK}, as
\begin{equation}\label{tau}
R(\tau) \equiv \frac{C(\tau)}{C_0} = \frac{1}{C_0(2\pi
c)^3}\int_0^{\omega_c}\!\!\!\d \omega\, \omega^2
S(\omega)\cos(\omega\tau).
\end{equation}
Notice that the integration frequency has a cutoff at $\omega_c =
\omega_P / \lambda$, where the Planck frequency is $\omega_P := 2\pi
/ \Tp=1.166\times 10^{44}\mathrm{~s}^{-1}$. This is consistent with
the fact that below the scale $\ell = \lambda \Lp$ the approximation
of randomly fluctuating fields breaks down. In alternative this may
simply correspond to the fact that the probing particle is
insensitive to the short wavelengths as a result of its own finite
resolution scale $\lr.$

\subsubsection{Correlation time}

Application of \eqref{P1_TTtt01} to $g(t):=R^2(t)$ yields the result
\begin{equation}
\int_{0}^{T}\!\!\!\d t \int_{0}^{T}\!\!\!{\d}t' R^2(t-t') = \tau_{*}
T, \label{P1_I1c}
\end{equation}
where the \emph{correlation time} is defined as
\begin{equation}\label{P1_tts}
    \tau_{*} := \mathfrak{F}\left[R^2(t)\right](0) = \pi \frac{\int_0^{\omega_c}\d
\omega \,\omega^4\, S^2(\omega)}{\left[\int_0^{\omega_c}\d \omega
\,\omega^2\, S(\omega)\right]^2},
\end{equation}
$\mathfrak{F}$ denoting Fourier transform.

To see this we need to Fourier transform $R^2(t)$ and evaluate its
value for $\omega = 0.$ Explicitly we have:
\begin{align*}
\mathfrak{F}\left[R^2(t)\right](\omega) &=
\frac{1}{C_0^2}\int_{-\infty}^{\infty}\!\!\!\d t\, C(t)^2
e^{-i\omega t}
\nonumber\\
\nonumber\\
&\hspace{-1cm}=\frac{1}{C_0^2(2\pi c)^6}
\int_0^{\omega_c}\!\!\!\int_0^{\omega_c}\!\!\d \omega'\d \omega''
\int_{-\infty}^{\infty}\!\!\!\d t\, \omega'^2 \omega''^2
S(\omega')S( \omega'' ) e^{-i\omega t} \cos \omega' t \cos \omega''
t
\nonumber\\
\nonumber\\
&\hspace{-1cm}= \frac{1}{4C_0^2(2\pi c)^6}
\int_0^{\omega_c}\!\!\!\int_0^{\omega_c}\!\!\d \omega'\d \omega''
\int_{-\infty}^{\infty}\!\!\!\d t\, \omega'^2 \omega''^2
S(\omega')S( \omega'' ) e^{-i\omega t} (e^{i \omega' t}+e^{-i
\omega' t}) (e^{i \omega'' t}+e^{-i \omega'' t}).
\end{align*}
Multiplying the exponential functions and integrating with respect
to time we obtain
\begin{align*}
\mathfrak{F}\left[R^2(t)\right](\omega) &= \frac{1}{4C_0^2(2\pi
c)^6} \int_0^{\omega_c}\!\!\!\int_0^{\omega_c}\!\!\d \omega'\d
\omega'' \int_{-\infty}^{\infty}\!\!\!\d t\, \omega'^2 \omega''^2
S(\omega')S( \omega'' ) \left[e^{i (\omega'+\omega''-\omega) t}+e^{i
(\omega'-\omega''-\omega) t}\right.\nonumber\\
&\hspace{8.5cm}\left.+e^{i (-\omega'+\omega''-\omega) t}+e^{i
(-\omega'-\omega''-\omega) t}\right]
\nonumber\\
\nonumber\\
&= \frac{\pi}{2C_0^2(2\pi c)^6}
\int_0^{\omega_c}\!\!\!\int_0^{\omega_c}\!\!\d \omega'\d \omega''
\omega'^2 \omega''^2 S(\omega')S( \omega'' )
\left[\delta(\omega'+\omega''-\omega)+\delta(\omega'-\omega''-\omega)\right.
\nonumber \\
&\hspace{7.5cm}\left.+\delta(-\omega'+\omega''-\omega)+\delta(-\omega'-\omega''-\omega)\right].
\nonumber
\end{align*}
From the last expression we see that
$\mathfrak{F}\left[R^2\right](\omega)=0$ for $\abs{\omega} >
2\omega_c$. The value of the Fourier transform for $\omega = 0$
follows immediately from the properties of the $\delta$ function. We
have the sum of four integrals of which those containing
$\delta(\omega'+\omega'')$ and $\delta(-\omega'-\omega'')$ vanish
and we are left with the expression
\begin{equation*}
\mathfrak{F}\left[R^2\right](0) = \frac{\pi}{C_0^2(2\pi c)^6}
\int_0^{\omega_c}\!\!\!\int_0^{\omega_c}\!\!\d \omega'\d
\omega''\omega'^2 \omega''^2 S(\omega')S( \omega'' )
\delta(\omega'-\omega''),
\end{equation*}
where we have taken into account that then integrals containing
$\delta(\omega'-\omega'')$ and $\delta(-\omega'+\omega'')$ yield the
same contribution because of the exchange symmetry between the
integration variables $\omega'$ and $\omega''$. Carrying on the
remaining integration we have
\begin{equation}
\mathfrak{F}\left[R^2(t)\right](0) = \frac{\pi}{C_0^2(2\pi c)^6}
\int_0^{\omega_c}\d \omega \,\omega^4\, S^2(\omega). \label{G2F1}
\end{equation}
Using the explicit expression for the constant $C_0$ that is
obtained from \eqref{tau} for $\tau = 0$, we obtain the result
\eqref{P1_tts}, from which it is clear that
$\mathfrak{F}\left[R^2(t)\right](0)$ has indeed the dimension of a
time.

\subsubsection{Characteristic function}

We can also apply \eqref{P1_TTtt01} to
$g_{\tau\tau'}(t):=R(t+\tau)R(t+\tau')$, where $\tau$ and $\tau'$
stand respectively for $-\kvs\cdot\Delta\xv/c$ and
$-\Kvs\cdot\Delta\xv/c$. Writing
\begin{align*}
\mathfrak{F}\left[{g}_{\tau\tau'}(t)\right](\omega) &=
\frac{1}{C_0^2}\int_{-\infty}^{\infty}\d t C(t+\tau)C(t+\tau')
e^{-i\omega t}
\nonumber\\
\nonumber\\
&\hspace{-1cm}= \frac{1}{C_0^2(2\pi c)^6}
\int_0^{\omega_c}\!\!\!\int_0^{\omega_c}\!\!\d \omega'\d \omega''
\int_{-\infty}^{\infty}\d t\, \omega'^2 \omega''^2 S(\omega')S(
\omega'' ) e^{-i\omega t} \cos [\omega' (t+\tau)] \cos [\omega''
(t+\tau')],
\end{align*}
and re-expressing the cosine functions through the exponential
function, a calculation similar to the previous one shows that
$\mathfrak{F}\left[{g}_{\tau\tau'}\right](\omega)=0$ for
$\abs{\omega} > 2\omega_c$ and yields the Fourier transform for
$\omega = 0$ as
\begin{equation}\label{I2gex}
\mathfrak{F}[R(t+\tau)R(t+\tau')](0)=\frac{\pi}{C_0^2(2\pi c)^6}
\int_0^{\omega_c}\!\!\!\d \omega\,\omega^4 S^2(\omega) \cos [\omega
(\tau-\tau')].
\end{equation}
This expression reduces of course to \eqref{G2F1} for $\tau = \tau'
=0$ and it is thus convenient to re-write it as
\begin{equation}\label{I2gexb}
\mathfrak{F}[R(t+\tau)R(t+\tau')](0)=\tau_{*}\Gamma[\omega_c(\tau-\tau')],
\end{equation}
where we defined the \emph{characteristic function} $\Gamma$ as
\begin{equation}\label{P1_gamma}
\Gamma(\omega_c  t):=\frac{\int_0^{\omega_c}\d \omega\,\omega^4
S^2(\omega) \cos (\omega  t)}{\int_0^{\omega_c}\d \omega\,\omega^4
S^2(\omega)}.
\end{equation}
This is dimensionless and satisfies the following general
properties:
\begin{itemize}
  \item $\Gamma(\omega_c  t)=\Gamma(-\omega_c  t)$,
  \item
  $\Gamma(0)=1,$
  \item $
\Gamma(\omega_c  t)<1,\quad\mathrm{for}\quad t\neq 0, $
  \item $\Gamma(\omega_c  t) \rightarrow 0, \quad \mathrm{for} \quad t
\rightarrow \infty.$
\end{itemize}
Notice that both the correlation time $\tau_{*}$ and the
characteristic function $\Gamma$ solely depend on the fluctuations
power spectral density. We can now use equation \eqref{P1_TTtt01}
and have at once
\begin{equation}
\int_{0}^{T}\!\!\!\d t \int_{0}^{T}\!\!\!{\d}t'
R(t-t'+\tau)R(t-t'+\tau') = \tau_{*} \Gamma[\omega_c(\tau-\tau')]\,
T, \label{P1_I2c}
\end{equation}

The results \eqref{P1_I1c} and \eqref{P1_I2c} can now be used in
equation \eqref{P1_P2_mainB} to yield the neat result
\begin{equation}\label{P1_afe2BBIISS}
\rho_{\xv\xv'}(T) = \rho_{\xv\xv'}(0)\left[1 - \frac{32\cII^2\pi^2
M^2 c^4 A_0^4\tau_{*} T}{\hbar^2} \times F(\Delta\xv) \right],
\end{equation}
where
\begin{equation}\label{P1_fff}
F(\Delta\xv) := 1 - \frac{1}{16\pi^2} \int\! \d\kvs\! \int\!
\d\Kvs\, \Gamma[\omega_c(\Kvs-\kvs)\cdot\Delta\xv/c].
\end{equation}
This equation is important and represents one of the main results of
this chapter. It implies that \emph{dephasing due to conformal
fluctuations does indeed occur in general and independently of the
precise power spectrum characterizing the fluctuations}. Without the
need to evaluate the angular integrals, this follows from the
properties of the characteristic function $\Gamma$. The fact that
$\Gamma[\omega_c t]<1$ implies $0 \leq F(\Delta\xv) \leq 1$ with (i)
$F(\Delta\xv = 0) = 0$ and (ii) $F(\Delta\xv \rightarrow \infty) =
1$ as special limiting cases. As a consequence the diagonal elements
are unaffected while the off-diagonal elements decay exponentially
according to
\begin{equation*}
\dot{\rho}_{\xv\xv'}(0) := \frac{\rho_{\xv\xv'}(T) -
\rho_{\xv\xv'}(0)}{T} =  - \left[\frac{32\cII^2\pi^2 M^2 c^4
A_0^4\tau_{*}}{\hbar^2} \times F(\Delta\xv)\right]
\rho_{\xv\xv'}(0),
\end{equation*}
providing of course that $T$ is small enough so that the change in
the density matrix is small. Finally, if $\delta\rho :=
\rho_{\xv\xv'}(T) - \rho_{\xv\xv'}(0)$, we can define the
\emph{dephasing} as $\abs{\delta\rho/\rho_0}$. Thanks to the
property $F(\Delta\xv \rightarrow \infty) = 1$, this converges for
large spacial separations to the constant maximum value
\begin{equation}\label{P1_P2_3d}
\abs{\frac{\delta\rho}{\rho_0}} =  \frac{32\cII^2\pi^2 M^2 c^4
A_0^4\tau_{*}T}{\hbar^2}.
\end{equation}

This result based on the present three-dimensional analysis of the
conformal fluctuations can be compared to the analogue
one-dimensional result that PP found in \cite{powerpercival00}.
Using a gaussian correlation function from the outset they found
\begin{equation*}
\abs{\frac{\delta\rho}{\rho_0}}_{1D}= \frac{\sqrt{\pi} M^2 c^4 A_0^4
\tau_g\,T}{\sqrt{2}\hbar^2},
\end{equation*}
where $\tau_g$ stands for some characteristic correlation time of
the fluctuations. Identifying approximately $\tau_*\simeq\tau_g$, we
have $(32\cII^2\pi^2)/(\sqrt{\pi/2})\simeq 250$, assuming $\cII \sim
1$. \emph{Thus the present three-dimensional analysis is seen to
predict a dephasing two orders of magnitude larger than in the
idealized one-dimensional case.}

\subsection{A remark on the validity of the Dyson expansion}\label{P1_drift}

We have found that the change in the density matrix is given by:
\begin{equation*}
\abs{\frac{\delta\rho}{\rho_0}} \sim \left(\frac{M
c^2}{\hbar}\right)^2 \tau_{*} T \times A_0^4,
\end{equation*}
In order for the expansion scheme to be effective, the propagation
time $T$ must be short enough to guarantee that
$\abs{\delta\rho/\rho_0}$ is small. How short depends of course on
the statistical properties of the fluctuations, encoded in $\tau_*$,
and on the probing particle mass $M$. A fullerene $\text{C}_{70}$
molecule with ($M_{\text{C}_{70}} \approx 10^{-24}$ kg) gives $M c^2
/ \hbar \approx 10^{27}\mathrm{ s}^{-1}.$ Therefore the approach is
consistent only if the correlation time $\tau_{*}$, the flight time
$T$ and field squared amplitude $A_0^2$ are appropriately small. We
will come back on this issue in Section \ref{P1_longdrift}, where it
is shown that, in the case of vacuum fluctuations (introduced in the
next section), it is $\tau_{*} \sim \lambda \Tp$ and $A_0^2 \sim
1/\lambda^2.$ For a flight time $T \approx 1$ ms, typical of
interferometry experiments, this results in $\abs{\delta\rho/\rho_0}
\approx 10^6 / \lambda^{-3}$. For any reasonable value of $\lambda
\gtrsim 10^3$ the density matrix change is indeed small and the
Dyson expansion scheme well posed up to second order. In Appendix
\ref{ap2_4} we estimate the fourth order term in the expansion,
which will also yield a term proportional to $A_0^4$. It will be
shown that its contribution in fact vanishes under quite general
circumstances. This puts the result \eqref{P1_P2_3d} on an even
stronger basis.

\section{Explicit dephasing in the case of vacuum fluctuations}\label{P1_pcf}

The result \eqref{P1_afe2BBIISS} is quite general. The only
ingredients entering the analysis so far have been: (i) a spacetime
metric $g_{ab} = (1+A)^2 \eta_{ab}$ with $\abs{A} = O(\varepsilon
\ll 1)$, (ii) a randomly fluctuating conformal field $A$ satisfying
the wave equation $\pd^c\pd_c A = 0$, and (iii) isotropic
fluctuations characterized by an arbitrary power spectral density
$S(\omega)$. The dephasing then occurs as a result of the
nonlinearity in the effective potential $V = Mc^2 [\cI A + \cII
A^2]$.

A particularly interesting case, potentially related to the
possibility of detecting experimental signs of QG, is that in which
the fluctuations in $A$ are the manifestation, at the appropriate
semiclassical scale $\ell = \lambda \Lp$, of underlying \emph{vacuum
quantum fluctuations} close to the Planck scale. Strictly speaking
the presence of the probing particle perturbs the genuine quantum
vacuum state. For this reason it would be appropriate to talk of
\emph{effective vacuum}, i.e. up to the presence of the test
particle. By its nature, the present semiclassical analysis
\emph{cannot} take into account the backreaction of the system on
the environment. Therefore we simply assume that the modifications
on the vacuum state can be neglected as long as the probing particle
mass is not too large and the evolution time short. We thus model
the effective vacuum properties of the conformal field $A$ at the
semiclassical scale on the basis of the properties that real vacuum
is expected to possess at the same scale. It is a fact that vacuum
looks the same to all inertial observers far from gravitational
fields. In particular, its energy density content should be Lorentz
invariant. This can obtained through an appropriate choice of the
power spectrum $S(\omega)$.

\subsection{Isotropic power spectrum for vacuum conformal fluctuations}

According to the above discussion we expect the average properties
of $A$ above the scale $\ell$ to be Lorentz invariant. In
particular, the interesting quantities derived in Appendix \ref{ap1}
\begin{equation}\label{li1}
   \m{A^2} = \frac{1}{(2\pi)^3}\int \d^3 k\, S(k),
\end{equation}
\begin{equation}\label{li2}
   \m{\abs{\nabla A}^2} = \frac{1}{(2\pi)^3}\int \d^3 k\, k^2 S(k),
\end{equation}
\begin{equation}\label{li3}
   \m{(\pd_t A)^2} = \frac{1}{(2\pi)^3}\int \d^3 k\, \omega^2(k)
   S(k),
\end{equation}
should be invariant. As discussed in Appendix \ref{ap1}, for a
stationary, isotropic signal, the averages $\m{\cdot}$ can in fact
be carried out through suitable spacetime integrations over an
appropriate averaging scale $L \gg \ell$. In alternative they can be
expressed as in the above integrals depending on the power spectrum
and adopting a high energy cutoff set by $\kcut:=2\pi / (\lambda
L_P)$.

The problem of the Lorentz invariance of the above quantities has
been discussed in details by Boyer \cite{boyer1969} within his
random electrodynamics framework. He showed that the choice $S
\propto 1/\omega$ is unique in guaranteeing an energy spectrum
$\varrho(\omega) \propto \omega^3,$ also shown to be the only
possible choice for a Lorentz invariant energy spectrum of a
massless field. In the present case we take
\begin{equation}\label{P1_ps}
    S(k):= \frac{\hbar G}{2 c^2}\frac{1}{\omega(k)}.
\end{equation}
The combination of the constants $\hbar$, $G$ and $c$ gives the
correct dimensions for a power spectrum (i.e. $L^3$), while the
factor $1/2$ guarantees that the resulting energy density is
equivalent to that resulting from the superposition of zero-point
contributions $\hbar \omega / 2$ -see Appendix \eqref{ap3}-. In
particular this element makes the connection between the present
stochastic approach and quantum theory.

With this choice for the power spectrum the integrals \eqref{li2}
and \eqref{li3} are indeed Lorentz invariant as they are related to
the energy density of $A$. The same holds for the integral in
\eqref{li1} as $\d^3 k / \omega(k)$ is a Lorentz invariant measure
\cite{ryder85}. A final important point, which should not be
overlooked, is that these facts are true provided the cutoff $\kcut$
is given by the \emph{same number} for all inertial observers, as
also discussed in details by Boyer. In other words this means that
the critical length that sets the border line between the random
field approach and the full QG regime is supposed to be the same for
\emph{any} inertial observer. It represents some kind of structural
property of spacetime and \emph{not} an observer dependent property.
Accordingly it must not be transformed under Lorentz
transformations. It is important to note that this requirement will
naturally be satisfied later when we employ an effective cutoff set
by the particle Compton length.

\begin{figure}[!b]
\caption{\footnotesize{\textsl{Plot of $R^2(t - t')$. The
dimensionless variable $\sigma$ is basically $t - t'$ in units of
the correlation time $\tau_*$. It is seen that $\tau \approx \tau_*$
corresponds to the first of the secondary peaks.}}}\vspace{0.3cm}
\centerline{\psfig{figure=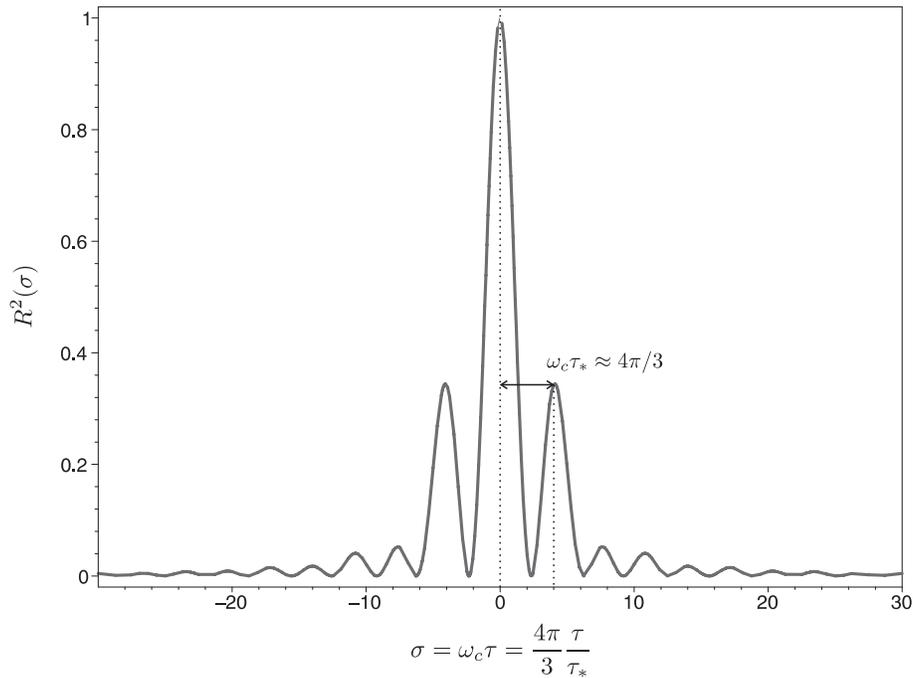,angle=0,width=12cm}}
\label{P1_fig2}
\end{figure}
Using \eqref{P1_ps} the normalized correlation function can be found
explicitly from the generalized WK theorem to be
\begin{equation}\label{P1_normcor}
R(\tau) := \frac{C(\tau)}{C_0} = 2\left[ \frac{\sin\omega_c
\tau}{\omega_c \tau} + \frac{\cos\omega_c \tau -1}{\omega_c^2
\tau^2} \right].
\end{equation}
The peak of the autocorrelation function is linked to the
fluctuations squared amplitude and gives explicitly:
\begin{equation}\label{P1_ap_C20b}
C_0 \equiv A_0^2 = \frac{1}{8\pi\lambda^2},
\end{equation}
implying $\m{A^2} = 1 / (2\lambda)^2$. The correlation time and
characteristic function follow from equations \eqref{P1_tts} and
\eqref{P1_gamma} as:
\begin{equation}\label{P1_P2_cti}
    \tau_* = \frac{2}{3}\lambda \Tp,
\end{equation}
and
\begin{equation}\label{P1_Gamma}
\Gamma(\sigma) =\frac{3\sin(\sigma)}{\sigma} +
\frac{6\cos(\sigma)}{\sigma^2} -\frac{6\sin(\sigma)}{\sigma^3},
\end{equation}
where $\sigma = \omega_c t$ is a dimensionless variable. The plot of
the squared normalized correlation function $R^2(t-t')$ is shown in
figure \ref{P1_fig2}: $t-t' = \tau_*$ corresponds to the first
secondary peak in the curve, where the correlation in the
fluctuations is reduced of $\sim 70\%$. This fully motivates the
choice of $\tau_*$ to represent the correlation time.

The explicit form of the characteristic function can be used in
\eqref{P1_fff} to evaluate the remaining angular integrals and find
the detailed expression for the density matrix evolution valid for
all $(\xv,\xv')$. Isotropy implies that the result must depend on
$\abs{\xv - \xv'}$ only. For convenience we can choose the reference
frame where a given $\Delta\xv$ lies along the $z$ axis. Then we can
define the dimensionless variable $\sigma$ through
$\omega_c\Delta\xv/c = \sigma\,\lvs,$ so that
\begin{equation}\label{sigma}
    \sigma=\frac{\omega_c\abs{\Delta\xv}}{c}=\frac{2\pi\abs{\Delta\xv}}{\ell}.
\end{equation}
With this choice of reference frame the unit vectors $\kvs$ and
$\Kvs$ have components
\begin{align}
\kvs &= \sin\vartheta\cos\varphi\,\ivs +
\sin\vartheta\sin\varphi\,\jvs + \cos\vartheta\,\lvs,\\
\Kvs &= \sin\vartheta'\cos\varphi'\,\ivs +
\sin\vartheta'\sin\varphi'\,\jvs + \cos\vartheta'\,\lvs
\end{align}
in such a way that
\begin{align}\label{kk}
\omega_c \, \kvs\cdot\Delta\xv/c&=\sigma\cos\vartheta,
\\
\omega_c \,\Kvs\cdot\Delta\xv/c&=\sigma\cos\vartheta'.
\end{align}
Now we have, for example, $\int\! \d\kvs = \int \! \d\vartheta
\sin\vartheta \d \varphi$ and since the integrand in \eqref{P1_fff}
depends only on $\vartheta$ and $\vartheta'$ the integrations over
$\varphi$ and $\varphi'$ are trivial and yield a factor $4\pi^2$.
Performing the change of variables
\begin{equation}
u := -\cos\vartheta,\quad u' := -\cos\vartheta',
\end{equation}
the integration is straightforward and yields the result:
\begin{equation}\label{P1_ffff}
F(\sigma) := 1-\frac{3}{2\sigma^2}\left(1 -
\frac{\sin\sigma\cos\sigma}{\sigma}\right).
\end{equation}

Substituting the results \eqref{P1_ap_C20b}, \eqref{P1_P2_cti},
\eqref{P1_Gamma} and \eqref{P1_ffff} into \eqref{P1_afe2BBIISS}
yields the explicit result for the dephasing, valid for vacuum
fluctuations described by
\begin{figure}[!t]
\caption{\footnotesize{\textsl{Plot of the function $F(\sigma)$ in
the range $\sigma = 0..10$, where $\sigma = 2\pi\abs{\xv-\xv'} /
(\lambda \Lp).$ The curve tends very rapidly to the limiting value 1
and for spacial separations $\abs{\xv-\xv'}$ which are slightly
larger than $\ell = \lambda\Lp$ the dephasing converges rapidly to
its maximum value.}}}\vspace{0.2cm}\label{P1_fig1}
\centerline{\psfig{figure=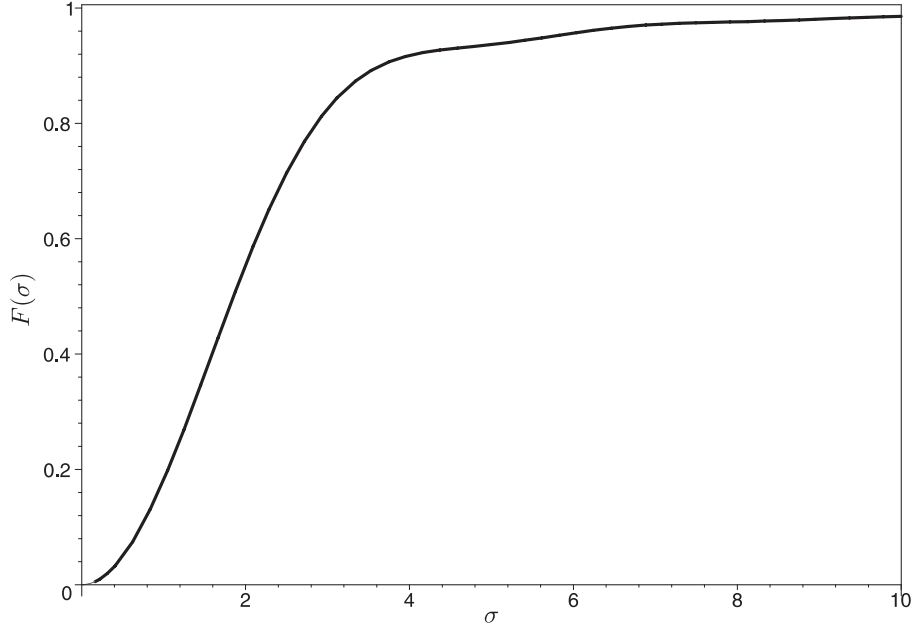,angle=0,width=12cm}}
\end{figure}
$S\propto 1/\omega$:
\begin{equation}\label{P1_fedf2b}
\left|\frac{\delta\rho_{\xv\xv'}}{\rho_0}\right|=
\frac{1}{3\lambda^3}\left(\frac{M}{M_P}\right)^2\left(\frac{T}{T_P}\right)\times
F\left( \frac{2\pi\abs{\xv-\xv'}}{\ell} \right),
\end{equation}
where we considered $\cII \sim 1$ and where
\begin{equation*}
\Mp := \frac{\hbar}{c^2\Tp} = \sqrt{\frac{\hbar c}{G}} = 2.176
\times 10^{-8} \mathrm{kg} = 1.310 \times 10^{19} \mathrm{amu}
\end{equation*}
is the Planck mass. The function $F$ is plotted in figure
\ref{P1_fig1}. It enjoys the properties $F(0) = 0$ and $F(\sigma)
\rightarrow 1$ for $\sigma \gg 1$, so that for $\abs{\xv-\xv'}
\gtrsim 10\ell$ the decoherence rate converges rapidly to its
maximum value.

\section{Discussion}\label{P1_disc}

\subsection{Probing the particle resolution scale and effective dephasing}\label{P1_pprs}

Equation \eqref{P1_fedf2b} gives the dephasing in the density matrix
of a quantum particle propagating in space under the only action of
a randomly fluctuating potential due to spacetime vacuum conformal
fluctuations. The fact that it predicts an exponential decay of the
off diagonal elements (which is the distinctive feature of quantum
decoherence) is interesting as a further confirmation that certain
effects involving quantum fluctuations can be mimicked by means of a
semi-classical treatment in the spirit of Boyer
\cite{boyer1969,boyer75}.

A significant feature of our dephasing formula is the quadratic
dependence on the probing particle mass $M$, which comes as a
consequence of the underlying non linearity. The coefficient
$1/(3\lambda^3)$ sets the overall strength of the effect. It is
proportional to $A_0^4$ and to the fluctuations correlation time
$\tau_*$: the more intense the fluctuations, the larger the
dephasing and the longer the various directional components stay
correlated, the higher their ability to induce dephasing. We have
found $\tau_* \approx \lambda \Tp,$ in such a way that the
correlation time directly depends on the spacetime intrinsic cutoff
parameter $\lambda$. According to this picture \emph{all} the
wavelength down to the cutoff $\ell = \lambda \Lp$ should be able to
affect the probing particle. However an atom or molecule is likely
to possess its own resolution scale $\lr$. Thus, whenever $\tau_* c
< \lr$, the ability of the fluctuations to affect the particle would
be reduced, as they would effectively average out. To characterize
this feature of the problem we write, in analogy to $\ell = \lambda
\Lp$,
\begin{equation*}
    \lr := \llr \Lp,
\end{equation*}
and use $\llr$ as a new, \emph{particle dependent}, cutoff
parameter. In general it is $\llr \geq \lambda$. The new effective
correlation time is now given by $\tau_* \approx \llr \Tp$. The
distance traveled by the fluctuations during a correlation time is
$L_* = c \tau_* \equiv 2\lr / 3$. Thus the effective
\emph{correlation distance} $L_*$ basically corresponds to the
particle resolution scale: short wavelengths that do not keep their
correlation up to the scale $\lr$ average out and cannot affect the
probing particle. The new, effective dephasing results by
substituting $\lambda$ with $\llr$ in \eqref{P1_fedf2b}:
\begin{equation*}
\left|\frac{\delta\rho_{\xv\xv'}}{\rho_0}\right|=
\frac{1}{3}\left(\frac{\Lp}{\lr}\right)^3\left(\frac{M}{M_P}\right)^2\left(\frac{T}{T_P}\right)\times
F\left( \frac{2\pi\abs{\xv-\xv'}}{\lr} \right).
\end{equation*}

\subsection{Validity of the long drift time regime}\label{P1_longdrift}

\begin{figure}[!t]
\caption{\footnotesize{\textsl{Fourier transform of $R^2(\sigma)$ as
a function of the frequency $\omega$ in units of the cutoff
frequency $\omega_c$.}}}\vspace{0.2cm}
\centerline{\psfig{figure=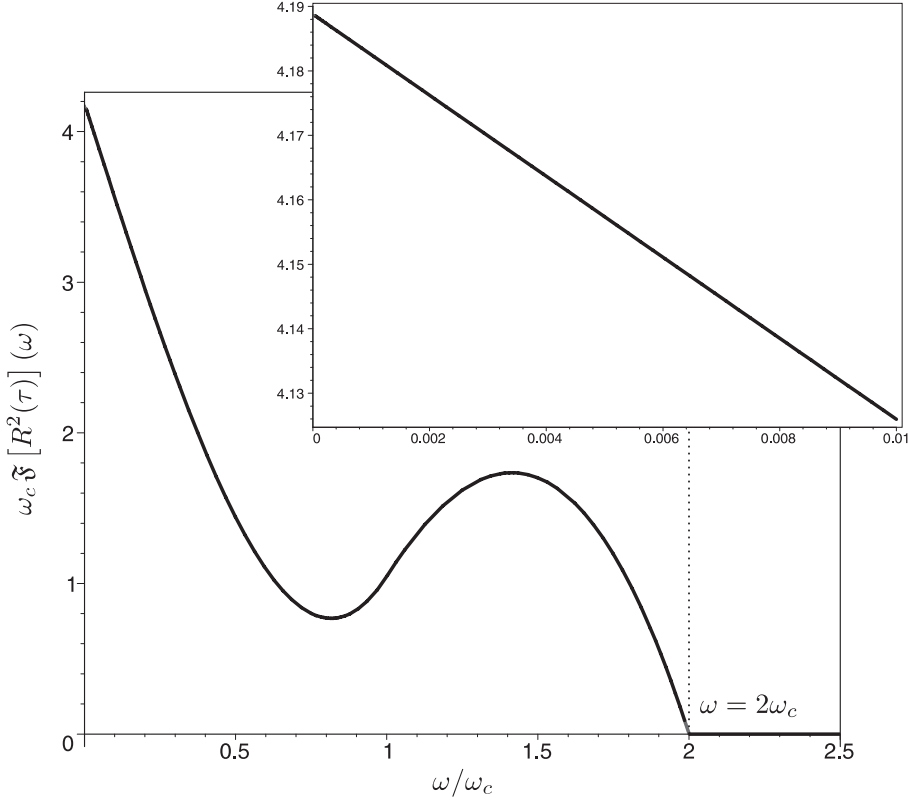,angle=0,width=12cm}}
\label{P1_fig3}
\end{figure}

As discussed in Section \ref{P1_lemma}, we recall that this result
holds for `long' drift times $T$, i.e. when $T \gtrsim (\Delta
\omega)^{-1}$, where $\Delta \omega$ is an appropriate frequency
range over which the Fourier transforms of $R^2(t)$ and
$R(t+\tau)R(t+\tau')$ vary little. We are now in the position to
make this precise and define clearly the limits of applicability of
the theory. To this end we consider the Fourier transform of
$R^2(\omega_c \tau)$:
\begin{equation*}
\mathfrak{F}[R^2(\omega_c \tau)](\omega) =
\frac{1}{\omega_c}\mathfrak{F}[R^2(\sigma)](\omega/\omega_c),
\end{equation*}
with $R(\sigma)$ given in \eqref{P1_normcor}. Its plot is displayed
in figure \ref{P1_fig3}.
The spectrum falls to 0 for $\omega \geq 2\omega_c$. The value of
the peak at $\omega = 0$ is precisely $4\pi / 3$, verifying that
$\tau_* \equiv \mathfrak{F}[R^2(\omega_c \tau)](0) = 4\pi /
(3\omega_c)$. The smaller box shows a zoom of the plot in the region
$\sigma \in [0, 1/100]$: the curve is slow varying in this range
since $\mathfrak{F}[R^2(\omega_c \tau)](0) = 4\pi /3 \approx 4.19$
and $\mathfrak{F}[R^2(\omega_c \tau)](1/100) \approx 4.13$.
Similarly it is possible to check that the Fourier transform of
$R(\sigma + \eta)R( \sigma + \eta')$, where the dimensionless
parameters $\eta$ and $\eta'$ depend on space direction and
locations as $\eta:= -\omega_c\kvs\cdot\Delta\xv/c$ and $\eta' :=
-\omega_c\Kvs\cdot\Delta\xv/c$, enjoys a similar property: for
\emph{every} choice of $\eta$ and $\eta'$, the resulting Fourier
transform is slow varying in the range $\sigma \in [0, 1/100]$.
Following this discussion we choose $\Delta\omega \approx [ 0,
\omega_c / 100]$. We can now quantify the concept of `long drift
time' by $T \gtrsim 100 / \omega_c$. From $\omega_c = 2\pi/(\llr
\Tp) = 4\pi/(3\tau_*)$ this yields the condition
\begin{equation*}
T \gtrsim 25 \, \tau_*.
\end{equation*}

\subsection{Some numerical estimates and outlook}\label{P1_sneo}

In summary we have studied the dephasing on a non-relativistic
quantum particle induced by a conformally modulated spacetime
$g_{ab} = (1+A)^2 \eta_{ab}$, where $A$ is a random scalar field
satisfying the wave equation. The important case of vacuum
fluctuations can be characterized by a suitable power spectrum $S
\propto 1/\omega$. If $\lr = \llr \Lp$ is the probing particle
resolution scale, the dephasing for $\abs{\xv - \xv'} \gg \lr$
converges rapidly to:
\begin{equation}\label{P1_dododo}
\left|\frac{\delta\rho}{\rho_0}\right|=
\frac{1}{3}\left(\frac{\Lp}{\lr}\right)^3\left(\frac{M}{M_P}\right)^2\left(\frac{T}{T_P}\right).
\end{equation}
The effective correlation time of the fluctuations is given by
$\tau_* \approx \llr \Tp$. The above result holds for `long' drift
times satisfying
\begin{equation*}
T \gtrsim (10 - 10^2) \llr \Tp.
\end{equation*}

\begin{figure}[!b]
\caption{\footnotesize{\textsl{Adimensional cutoff parameter as a
function of the dephasing for a $\text{C}_{70}$ molecule with a
drift time of $10^{-3}$ s.}}}\vspace{0.2cm}
\centerline{\psfig{figure=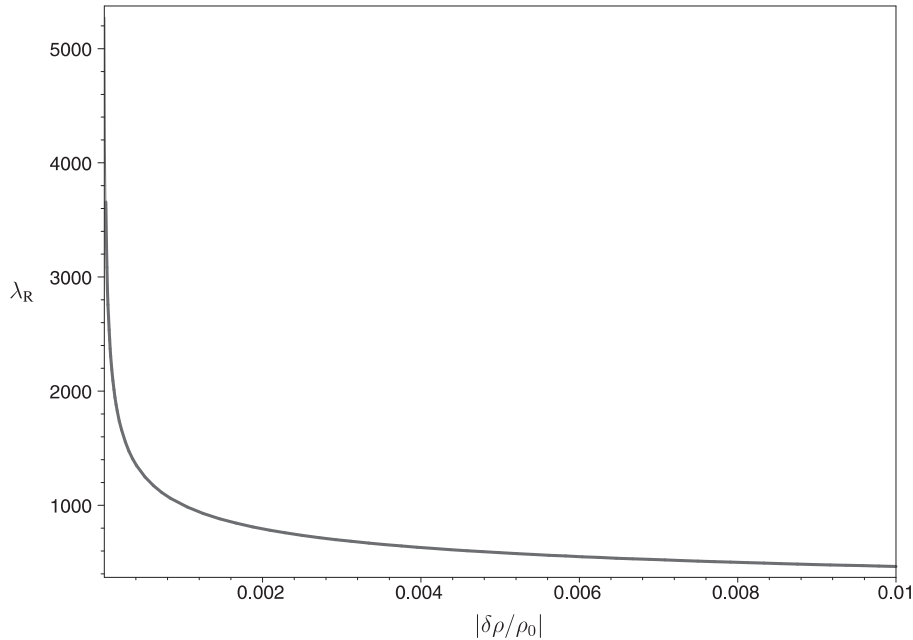,angle=0,width=12cm}}
\label{P1_fig4}
\end{figure}

To conclude we want to give some numerical estimates of the
dephasing that could be expected in a typical matter wave
interferometry experiment, e.g. like those described in
\cite{hornberger04}, where fullerene molecules have been employed
with drift times of the order of $T = : \tex \approx 10^{-3}$ s.
Consider e.g. a $\text{C}_{70}$ molecule with $M = \mc70 \approx
1.24 \times 10^{-24}$ kg. In comparison to the Planck units we have:
\begin{equation*}
\tex \approx 10^{40}\Tp,\quad \mc70 \approx 10^{-17}\Mp.
\end{equation*}
Thus, it is clear that the most critical factor controlling the
strength of the effect is set by the probing particle mass, together
with the effective resolution cutoff scale. Using these data in
\eqref{P1_dododo} we can estimate:
\begin{equation*}
\left|\frac{\delta\rho}{\rho_0}\right| \approx \frac{10^6}{\llr^3}
\quad \Leftrightarrow \quad \llr \approx \left(
\frac{10^6}{\abs{\delta\rho/\rho_0}}
    \right)^{\frac{1}{3}}.
\end{equation*}
\emph{This could be used to estimate $\llr$ if we were able to
identify within an experiment a residual amount of dephasing that
cannot be explained by other standard mechanisms} (e.g.
environmental decoherence, internal degrees of freedom). Figure
\ref{P1_fig4} plots
$\llr$ against $\abs{\delta\rho/\rho_0}$: a dephasing due to
conformal fluctuations in the range $1\% - 0.1\%$ would imply a
resolution parameter in the range $\llr \approx 10^3 - 10^4.$ This
would represent a lower bound on $\llr$, as interferometry
experiments will get more and more precise in measuring and modeling
environmental decoherence. Estimating the present typical
uncertainty of typical interferometry experiments as $\abs{\delta
\rho / \rho_0} \approx 0.01\%$ we get
\begin{equation*}
    \llr \gtrsim 10^3,\quad\mathrm{for}\quad \text{C}_{70}.
\end{equation*}
We remark that such order of magnitudes estimates are consistent
with a small change in the density matrix and the second order Dyson
expansion approach.

A value for $\llr$ as small as $10^3$ would probably approach the
intrinsic spacetime structural limit set by $\lambda$, i.e. $\ell=
\lambda \Lp$. It is interesting to ask what amount of dephasing our
model predicts, \emph{independently} of experimental data. To this
end we need to prescribe theoretically the particle resolution scale
$\lr$. Though no obvious choice exists, an interesting possibility
would be to set it equal to the particle \emph{Compton length}
\cite{goklu08,aloisio06}, i.e.
\begin{equation*}
\lr = \frac{h}{Mc}.
\end{equation*}
This choice is obviously Lorentz invariant and also motivated by the
fact that the Compton length represents a fundamental uncertainty in
the position of a nonrelativistic quantum particle. Indeed, by the
Heisenberg uncertainty principle, $\Delta x \approx h/Mc$ would
imply $\Delta p \gtrsim Mc,$ implying an uncertainty in the energy
of the same order of the rest mass $Mc^2$. In such a situation QFT
would become relevant. Alternatively it can also be argued that
wavelengths shorter than $h / Mc$ would have enough energy to create
a particle of mass $M$ from the vacuum. With this choice, equation
\eqref{P1_dododo} becomes
\begin{equation}\label{P1_dododo2}
\left|\frac{\delta\rho}{\rho_0}\right|=
\frac{1}{24\pi^3}\left(\frac{M}{M_P}\right)^5\left(\frac{T}{T_P}\right).
\end{equation}
This can be used to estimate the amount of dephasing induced by
vacuum conformal fluctuations.

In the case of $\text{C}_{70}$ the Compton wavelength is $ \approx
10^{-18}\mathrm{ m}\approx 10^{18}\Lp,$ corresponding to $\llr
\approx 10^{18}$. For a propagation time of $\approx 1$ ms this
gives a dephasing
\begin{equation*}
\left|\frac{\delta\rho}{\rho_0}\right|(\mc70,1\mathrm{ ms}) \approx
10^{-44},
\end{equation*}
which would be negligible and far beyond the possibility of
experimental detection. Thus, \emph{in order to achieve dephasing
within the current experimental accuracy, much heavier quantum
particles are needed}. In atomic mass units $\text{C}_{70}$ has a
mass $\mc70 \approx 10^3$ amu. Equation \eqref{P1_dododo2} applied
to a particle with mass $M \approx 10^{11}$ amu and with a drift
time $T \approx 100$ ms gives the estimate:
\begin{equation*}
\left|\frac{\delta\rho}{\rho_0}\right|(10^{11}\mathrm{
amu},100\mathrm{ ms}) \approx 10^{-2}.
\end{equation*}
A drift time of $\sim 100$ ms could possibly be achieved in a space
based experiment. On the other hand, the need of a quantum particle
as heavy as $10^{11}$ amu poses an extraordinary challenge. A
possibility would be to employ quantum entangles states. This is
already being considered in the literature, e.g. in
\cite{everitt08}, where entangled atomic states are studied and
suggested as a possible improved probe for future detection of
spacetime induced dephasing.

In relation to the issue of a possible future experimental detection
of dephasing due to vacuum effects, it is important to note that
even the vacuum fluctuations of the EM field can influence the
fringe visibility of a neutral particle \emph{if} this has a
permanent electric or magnetic dipole moment \cite{mazzitelli03}.
The amount of decoherence due to this effect, e.g. in a typical
interference experiment, is expected to depend quadratically on the
dipole moment of the particle. Fullerene molecules such as
$\text{C}_{70}$ (or even the spherically shaped $\text{C}_{60}$)
have symmetric charges distributions and possess no permanent
dipoles \cite{khudiakov96}. In this case no extra effect would be
expected. Drugged versions of fullerene where, e.g., $\text{Na}$ or
$\text{Li}$ atoms are combined with $\text{C}_{60}$ to form heavier
molecules such as $\text{Na}_{18}\text{C}_{60}$ or
$\text{Li}_{10}\text{C}_{60}$ can possess a permanent electric
dipole moment between $\sim$ 10 and $\sim$ 20 Debyes, depending on
the number of drugging atoms \cite{rabilloud07}. While in this case
some decoherence due to EM vacuum effects would be theoretically
expected, the mass of such molecules is still far too low for any
gravitational effect due to conformal fluctuations to be detectable.
Only in the case of more complex and heavier quantum systems, e.g.
gold clusters or entangled states, with a permanent dipole the two
effects would theoretically both contribute to the overall
dephasing. However the dephasing due to a permanent dipole moment is
in general predicted to be typically of the order of the square of
the typical dipole length in units of the total length of the
trajectory \cite{mazzitelli03}. In this sense, as the mass of the
quantum probe is increased, we expect the gravitationally induced
dephasing to represent the dominant effect.

The last important point that needs verification is that the
condition \eqref{P1_C2} given earlier at the beginning of this
chapter is indeed verified: that was required in order for the
change in momentum due to the fluctuations to be smaller than the
laboratory particle momentum $\pl = M\vl$. It read: $ \varepsilon^2
\sim \m{A^2} \ll (\vl / c)^2$. The field effective mean quadratic
amplitude interacting with the particle is given by $\m{A^2} \sim
{\llr}^{-2}$. Thus we have the condition:
\begin{equation*}
    \frac{1}{\llr} \equiv \frac{\Lp}{\lr} \ll \frac{\vl}{c}.
\end{equation*}
By using the expression for the Planck length and with $\lr$ given
by the particle Compton length, this yields a condition on the
particle mass $M$:
\begin{equation*}
\frac{M}{\Mp} \ll 2\pi\frac{\vl}{c}.
\end{equation*}
For typical laboratory velocities $\vl / c \approx 10^{-6}$ and,
since $\Mp \approx 10^{19}$ amu, this condition is met for particle
masses up to $M \approx 10^{13}$ amu, including the case of
$\text{C}_{70}$ molecules or the heavier entangled quantum states
discussed above. This limit would be reduced for slower particles.

We conclude by remarking that the theory described until now is
quite general, in the sense that as a starting input it only needs a
conformally modulated metric and a scalar field satisfying the wave
equation. Of course, it is important to identify in concrete which
theories of gravity can actually yield such a scenario. This problem
is central to this thesis and is be the object of the next three
chapters: in Chapter \ref{ch2} a general framework for the study of
fluctuating fields close to the Planck scale is introduced and the
resulting framework applied to standard GR in Chapter \ref{ch3}; in
Chapter \ref{ch4} we consider more general scenarios involving
scalar-tensor theories of gravity.

\newpage
\thispagestyle{empty}


\singlespacing
\chapter{The nonlinear random gravity framework}\label{ch2}

\begin{quote}
\begin{small}
In this chapter we introduce the nonlinear random gravity framework.
Our goal is modeling some aspects of the low energy physics related
to the spacetime metric and matter fields vacuum fluctuations. The
approach extends conceptually Boyer's random electrodynamics to a
theory of random geometry but has a somehow richer structure due to
the nonlinearity originally inherent to GR. Essentially, Isaacson's
perturbative approach over a curved background is applied to the
problem of spacetime fluctuations by introducing randomly
fluctuating solutions to the expansion equations. The fluctuations
are supposed to mimic some vacuum effects related to zero point
energy at an appropriate low energy scale. Isaacson's original
approach is generalized in order to seek for solutions in which the
physical metric may exhibit conformal fluctuations. The technique
must be extended to second order nonlinearity in order to take into
account the backreaction energy due to zero point GWs, as well as
matter fields and conformal fluctuations. This reveals interesting
connections to the well known cosmological constant problem. Finally
the framework is applied to standard GR as a first attempt to
identify a theory of gravity which may be compatible with the
presence of conformal fluctuations. To this end, the relevant first
and second order equations describing GWs and the vacuum
backreaction on the spacetime metric within GR are derived.
\end{small}
\end{quote}
\singlespacing

\section{Seeking a physical basis for the conformal fluctuations}\label{nme}

We denote the spacetime physical metric by $g_{ab}.$ This couples as
usual to matter fields and determines the geodesics of small
classical test particles. Conformal fluctuations can formally be
included by writing the physical metric as:
\begin{equation}\label{phm}
g_{ab} = \Omega^2 \gamma_{ab},
\end{equation}
where $\O$ is a conformal factor and $\gamma_{ab}$ will be referred
to as the \emph{\cfmp}. Denoting all matter fields collectively by
the symbol $\psi$ and imposing \ee
\begin{equation}\label{eeg}
G_{ab}[g] = 8\pi T_{ab}[\psi]
\end{equation}
on the physical metric $g_{ab}$, yields a corresponding equation for
the \cfmp. This reads \cite{wald84}:
\begin{equation}\label{eega1}
    G_{ab}[\gamma] = 8\pi T_{ab}[\psi] + \Sigma_{ab}^{1}[\Omega] +
    \Sigma_{ab}^{2}[\Omega],
\end{equation}
where we defined
\begin{equation}\label{sig11}
\Sigma_{ab}^{1}[\Omega] := 2\cd_a \cd_b \ln\Omega - 2 \gamma_{ab}
\cd^c\cd_c\, \ln\Omega,
\end{equation}
\begin{equation}\label{sig21}
\Sigma_{ab}^{2}[\Omega] := - \left( 2 \cd_a \ln\Omega \cd_b
\ln\Omega + \gamma_{ab} \cd^c \ln\Omega \cd_c \ln\Omega \right),
\end{equation}
and where $\cd_a$ denotes here the covariant derivative of the \cfm
$\gamma_{ab}$. By definition \ee is satisfied if and only if
equation \eqref{eega1} is satisfied and if the physical and
conformal metric are related by \eqref{phm}.

The correct point of view to have regarding the problem of whether
the physical metric \emph{truly has} a conformal modulation is the
following: we must think of the conformal metric $\gamma_{ab}$ as
having some concrete prescribed structure that \emph{does not}
depend upon $\O$, which implies that \eqref{eega1} simply represents
an equation constraining the conformal factor. In this sense
equation \eqref{phm} could be viewed as an \emph{ansatz} solution
for \eep. The problem of having a conformally modulated metric
within standard GR thus effectively reduces to that of finding a
suitable conformal metric such that the corresponding constraint
equation for $\O$ can be satisfied. If the solution for $\O$ is
compatible with a conformal factor of the form $\O = 1 + A$ with $A$
being a randomly fluctuating field then we would have a concrete
framework providing a basis for the results derived in Chapter
\ref{ch1}.

\subsection{The simplest scenario: a conformally modulated universe}

The simplest (and most dramatically restrictive!) choice to
constraint the \cfm is
\begin{equation*}
    \gamma_{ab} := \eta_{ab}.
\end{equation*}
This is PP choice in \cite{powerpercival00}. In this case we have
$\cd_a = \pd_a$, $G_{ab}[\eta] = 0$ and the constraint equation for
$\O$ reads
\begin{equation}\label{bal}
8\pi T_{ab}[\psi] + \Sigma_{ab}^{1}[\Omega] +
\Sigma_{ab}^{2}[\Omega] = 0.
\end{equation}
Moreover, in the simplified PP model all matter fields were
neglected by simply putting $T_{ab}[\psi] = 0$. Such an hypothetical
universe containing only conformal fluctuations should then satisfy
the equation:
\begin{equation*}
2\left(\pd_a \pd_b \ln\Omega - \eta_{ab} \pd^c\pd_c\,
\ln\Omega\right) - \left( 2 \pd_a \ln\Omega \pd_b \ln\Omega +
\eta_{ab} \pd^c \ln\Omega \pd_c \ln\Omega \right) = 0.
\end{equation*}
In the interesting case of conformal fluctuations we can write $\O =
1 + \delta\O$, where $\delta\O$ is a small fluctuating modulation.
Then by expanding $\ln\O \approx 1 + \delta\O - \delta\O^2/2$ we
would have:
\begin{equation*}
2\left(\pd_a \pd_b \delta\O - \eta_{ab} \pd^c\pd_c\, \delta\O\right)
-\left( \pd_a \pd_b \delta\O^2 - \eta_{ab} \pd^c\pd_c\, \delta\O^2
\right) - \left( 2 \pd_a \delta\O \pd_b \delta\O + \eta_{ab} \pd^c
\delta\O \pd_c \delta\O \right) = 0.
\end{equation*}

To linear order, the trace of the equation
\begin{equation}\label{lo}
\pd_a \pd_b \delta\O - \eta_{ab} \pd^c\pd_c\, \delta\O = 0
\end{equation}
implies that the perturbation $\delta\O$ should satisfy the wave
equation, i.e.
\begin{equation}\label{dowe}
\Box\delta \O = 0,
\end{equation}
where $\Box:=\pd^c\pd_c$. This is of course what PP found. However
their method employed the Hilbert action and completely overlooked
upon the fact that, beyond \eqref{dowe} being satisfied, \ee then
implies the very restrictive constraint $\pd_a \pd_b \delta\O = 0$,
as it follows from \eqref{lo} once the wave equation is also taken
on board. This constraint is too restrictive and it is hard to see
how it could be compatible with a fluctuating $\delta\O.$ Beyond
this PP model suffers from a further important problem: indeed the
second order constraint equation $-\left( \pd_a \pd_b \delta\O^2 -
\eta_{ab} \pd^c\pd_c\, \delta\O^2 \right) - \left( 2 \pd_a \delta\O
\pd_b \delta\O + \eta_{ab} \pd^c \delta\O \pd_c \delta\O \right) =
0$ is quadratic and could not be satisfied except in the trivial
case of $\delta \O = 0$. In conclusion a metric of the form $g_{ab}
= (1+\delta\O)^2\eta_{ab}$ seems to be incompatible with \eep. If we
insist in wishing to include conformal fluctuations in the formalism
we must seek a physical metric with a more general structure.

\subsection{The general basis for the model}

We will try to overcome the above difficulty by improving the model
upon three points:
\begin{enumerate}
  \item allow for the presence of GWs; this will correspond to a
  larger freedom in choosing the structure of $\gamma_{ab}$,
  \item account for the presence of matter fields through
  $T_{ab}[\psi]$;
  this is physically relevant as, when studying vacuum, matter
  fields contribution should not be neglected,
  \item in agreement with the nonlinear structure of GR, allow for a
  nonlinear correction to the conformal factor by writing $\O = 1 + A +
  B$, where $\abs{A} = O(\varepsilon \ll 1)$ and $\abs{B} = O(\varepsilon^2)$, with $\varepsilon \ll 1$.
\end{enumerate}
To implement these points we need to develop a framework allowing to
model some of the properties inherent to vacuum in a coherent way.
In the next sections we discuss the main features of the
\emph{random gravity framework}. This was first introduced in
\cite{wango8rg}, where the main goal was to show that the inclusion
of conformal fluctuations could offer a mechanism for a solution of
the cosmological constant problem \cite{Carroll2001}. This idea
could seem at first sight plausible if we look at the quadratic part
of the effective stress-energy tensor \eqref{sig21} due to the
conformal factor: this has a negative definite kinetic term and one
might expect that it could serve as a compensating factor to reduce
or cancel the large and positive vacuum energy density due to the
matter fields and GWs. Even in the case $\gamma_{ab} = \eta_{ab}$,
the constraint equation \eqref{bal} seems to motivate this idea.
This is also hinted by the canonical analysis of general relativity
\cite{DeWitt1967}. In particular the analysis performed by Wang in
\cite{wang05a,wang05b} led to arguing in \cite{wang2006} that
hypothetical quanta related to the conformal fluctuations could
offer a way to compensate for the vacuum energy associated to GWs.
This also was the result in \cite{wango8rg}. Unfortunately, as we
will show at the end of this chapter, that conclusion was invalid
due to the fact that the nonlinear backreaction affect on the
spacetime metric due to the conformal fluctuations was not accounted
for properly.

Before going into details we need to discuss in the next section
some points related to vacuum and the cosmological constant problem.

\section{Vacuum and some related problems}

The nature and properties of vacuum are not fully understood yet
\cite{Unruh1976, Ford1975, Ford1993, Haisch1994, Straumann1999,
Antoniadis2007, Milonni1994, Saunders2002}. A reasonably clear
definition of vacuum exists within QFT for quantum fields on a flat,
non-dynamical, Minkowski background. A satisfying theory of quantum
fields propagating on an assigned curved background geometry can
also be given \cite{wald94}. In this case the notion of vacuum state
is much more subtle, essentially because a generally curved geometry
lacks the notion of a global inertial observer, with respect to whom
vacuum is usually defined when no gravitational fields are present.
As it is well known, even in standard QFT, the ground state vacuum
energy receives a contribution from the zero point energies of an
infinite amount of harmonic oscillators. This results in a formally
infinite amount of vacuum zero point energy unless some high energy
cutoff is applied. Even with a reasonable cutoff, it has been
suggested that the resulting large energy density would lead either
to a rapid expansion of the universe as discussed by Weinberg
\cite{Weinberg1989} or to a drastic collapse of spacetime as noticed
by Pauli \cite{Straumann1999, Antoniadis2007}. Since there is no
obvious sign of such a catastrophic breakdown in the spacetime that
we can practically \emph{observe}, this rises the issue of the
\emph{reality} of the zero point energy. In this respect it is
important to note that vacuum fluctuations of the electromagnetic
field have long received support from experiments. These include the
Casimir effect \cite{Casimir1948, Casimir1948b, Mohideen1998}, Lamb
shift of atomic energy levels \cite{Welton1948, Beiersdorfer2005}
and spontaneous emission from atoms \cite{Louisell1973,
Gea-Banacloche1988, Wodkiewicz1988}. The fundamental
fluctuation-dissipation theorem \cite{Callen1951} provides a further
basis for the  vacuum fluctuations, manifested as superconductive
current noise in the Josephson junctions \cite{Josephson1962,
Koch1980, Koch1982}. Even so, elements of doubt still surround the
physical reality of the quantum vacuum. Some argue that e.g. the
Casimir effect could be explained without the vacuum energy
\cite{Jaffe2005}. Laboratory measurements of vacuum fluctuations can
at best detect differences in the vacuum energy by modifying their
boundary conditions \cite{Doran2006}.

Beyond these considerations, a further important issue related to
vacuum energy is that of its backreaction upon the spacetime
geometry; i.e. whether vacuum gravitates. It is generally believed
that if vacuum energy is real, then it must gravitate in accordance
with GR, and that only through the resulting gravity the net vacuum
energy could be determined. Recently, it has been shown that the
part of vacuum energy responsible for the Casimir effect does indeed
gravitate \cite{Fulling2007}. However, the gravitational consequence
of the total vacuum energy remains controversial \cite{Beck2005,
Jetzer2005, Jetzer2006, Mahajan2006}. Because of the observed
Lorentz invariance of vacuum at low energies, after any appropriate
regularization, the vacuum energy is expected to couple to \ee via
an effective cosmological constant only through a pure trace term of
the kind $-\Lambda g_{ab}$. There is a common perception that an
exceedingly large cosmological constant would built up from all
ground states up to the Planck energy density scale
\cite{Weinberg1989}. While a sizable cosmological constant could
have driven the cosmic inflation in the early universe, observations
indicate its present value to be just under the critical density
value \cite{Mottola1986,Carroll2001}. This discrepancy between the
\emph{observed} value of the cosmological constant and the
contribution that one would expect from QFT constitutes part of what
is known in the literature as the cosmological constant problem
\cite{rugh02}.

Despite the early suggested link between the vacuum energy of
elementary particles and the cosmological constant by Zel'dovich
\cite{Zeldovich1967}, a detailed mechanism is still lacking. This
motivates the ``dark energy'' models such as the quintessence fields
and their extensions \cite{Weinberg1989,Carroll2001}. Leaving aside
the original mysterious cancelation of the huge vacuum energy, these
models indeed offer arguably the most popular current approach to
the ``cosmological constant problem''. Nevertheless, efforts to
account for the observed cosmic acceleration using cosmological
perturbation back-reaction without resorting to dark energy are also
being made. The need for a better understanding of nonlinear metric
perturbations \cite{Kolb2006} and QG effects \cite{Li2008} is
clearly highlighted by the recent progress in this direction.

\section{Nonlinear random gravity}\label{rga}

Our point of view here is that spacetime metric fluctuations are
\emph{also} in fact an integral part of the quantum vacuum. Thus it
could be plausible that an attempt to disentangle these difficulties
should take them on board. However, a fundamental difficulty to
studying these effects is the absence of a consistent quantum theory
of gravity with an appropriate classical limit. Recently, progress
has been made using the stochastic gravity approach
\cite{CamposVerdaguer1996, MartinVerdaguer2000,
HuRouraVerdaguer2004, Hu2004}. This generalizes the previous
semiclassical approach where gravity is coupled to the expectation
value of quantum matter fields stress energy tensor. The stochastic
gravity approach attempts to take on board the \emph{fluctuations}
of the stress energy tensor around its average. By means of suitable
statistical considerations, the features of the quantum fluctuations
are `translated' into a suitable fluctuating classical, effective
tensor that acts back as a source on the spacetime metric, thus
inducing a stochastic metric perturbation. The approach is very
interesting and it has been applied to the interesting case of the
perturbations induced by an arbitrarily coupled scalar field upon
Minkowski geometry \cite{MartinVerdaguer2000}. This study has shown
that Minkowski spacetime is stable upon perturbations. Moreover the
induced metric perturbations have been found to be correlated only
very close to the Planck scale, in such a way that matter fields
tend in fact to suppress the short-scale perturbations for larger
scales. Though undoubtedly a very sound theory, the stochastic
gravity approach totally ignores the issue of zero point energy. In
fact this is done by regularizing from the outset the matter fields
stress energy tensor in such a way that, in the vacuum state, it is
$\ms{T_{ab}[\psi]} = 0.$ Various techniques exist to do this on a
curved background, as it is discussed in detail by Wald
\cite{wald94}. Without going here into these complexities it is
enough to say that, in the case of a flat background, these
correspond to the usual practise of normal ordering, by virtue of
which all zero point energies are in fact ignored. It follows that
if these are believed to be real and a source of gravity, then the
stochastic gravity approach is not suitable in its present form.
Another important consideration regards the fact that the stochastic
gravity approach accounts for the backreaction on the spacetime
geometry due to matter fields \emph{only}. To our knowledge,
possible backreaction effects due to the self energy of GWs have not
been investigated yet.

In order to account for zero point energy \emph{and} GWs, our
approach is somehow minimal: we seek to analyze \emph{some} aspects
related to vacuum at an appropriate energy regime, in between the
Planck scale and those characterizing the observed classical world;
we also wish to incorporate conformal fluctuations explicitly and to
account for the backreaction due to GWs zero point energy. The
purpose of this approach is thus twofold: (i) investigating whether
conformal fluctuations of the physical metric can provide a vacuum
energy balance mechanism that may help in relation to the riddle of
the cosmological constant problem; (ii) provide a coherent framework
that may provide a sound physical basis for the material presented
in Chapter \ref{ch1}.

\subsection{Stochastic geometry}

It is generally agreed that, independently of the true underlying QG
theory, spacetime structure at very small scales should `lose' its
usual classical appearance \cite{giulini03}. This is true in
particular within the two main approaches to QG mentioned in the
introduction: String Theory and LQG. In the first case the quantum
behavior of spacetime is inherent in the presence within the theory
of a particular excitation describing the graviton, while one of the
main results of the loop quantization program is the discrete,
quantized nature of the geometry (lengths, areas, volumes) close to
the Planck scale. This is characterized by the fundamental units,
introduced by Planck back in 1899:
\begin{align*}
& \Lp := \sqrt{\frac{\hbar G}{c^3}} \approx 1.62 \times
10^{-35}\text{ m},\\
& \Tp := \frac{\Lp}{c} \approx 5.40 \times 10^{-44} \text{ s},\\
& \Mp := \frac{\hbar}{\Lp c} \approx 2.17 \times 10^{-8} \text{ kg}
\approx 1.22 \times 10^{19} \text{ GeV}.
\end{align*}
The successful QG theory will have to provide a coherent way to
describe how the classical spacetime structure to be described in
terms of a smooth Lorentz manifold can emerge from the microphysics
at the Planck scale. Even though it is not known exactly at which
scale the `turbulent' quantum nature of spacetime gives in to yield
a classical structure, it is fair to state that the assumption of a
smooth regular manifold holds down to, at least, the subnuclear
scale, i.e. $\approx 10^{-15}$ m. In fact it is likely to hold down
to much shorter scales as it is implied e.g. by the fact that when
studying gluon-quarks states within QCD, gravity effects can still
be ignored. The natural realm of QG effects lies much further away
at the bottom and it is still directly inaccessible: even the LHC,
with its expected center of mass collision energy of $\sim 10$ TeV,
is still not even remotely close to the Planck energy of $10^{19}$
GeV.

It is commonly believed that, \emph{between} the Planck scale and
what we could call the minimal classical scale, there must exist an
intermediate regime in which spacetime begins to acquire a classical
character, though still inheriting features from the underlying
quantum regime \cite{powerpercival00,wang2006}. The usual assumption
is that these features are manifest as a stochastic behavior of the
spacetime geometry. This will also be our point of view. In a
nutshell this sums up to the expectation that some of the properties
of quantum spacetime can be modeled, at an appropriate low energy
scale, by employing a suitable stochastic approach.

That classical stochastic fields can be used to model and reproduce
various quantum effects related to the electromagnetic field is well
known, e.g. from the work of Boyer
\cite{boyer1969,boyer75,Boyer1975b,Boyer1980}. In his theory of
Random Electrodynamics (RED) he showed that a large variety of
quantum results can be obtained through the classical Maxwell
equation supplemented by suitable \emph{random boundary conditions}.
This had already been pointed out by Welton \cite{Welton1948}. These
can be applied to obtain a classical yet fluctuating solution of
Maxwell equations, whose statistical properties can be chosen in
order to match those of the true underlying zero point quantum field
in such a way to correspond to a Lorentz invariant spectrum.
Specifically, it was shown in \cite{Boyer1975b} that QED and RED
yield equivalent results for the $N$-points correlation functions of
free EM fields and also, e.g., for some (but not all!) expectation
values of an oscillator at zero temperature. Beyond Boyer's work,
this kind of classical stochastic behavior has also been advocated
by other authors in relation to vacuum related phenomena
\cite{Cavalleri1981, IbisonHaisch1996}. York championed a novel
gravitational analogue in which black hole entropy and radiance are
derived from quasinormal mode metric fluctuations \cite{York1983}.
These fluctuations are prescribed by a classical Vaidya geometry
satisfying the \ee with amplitudes set to the quantum zero point
level. An improved quantum treatment of the problem using path
integral is recently provided in~\cite{York2005}.

Even though we maintain the view that that quantum theory \emph{has}
a deeper motivation and truth than a classical theory supplemented
by boundary conditions, the work of Boyer shows that, \emph{in
certain regimes}, one may use the alternative classical theory to
\emph{mimic} some quantum related behaviors of the system under
study. Following these considerations we wish to apply this idea to
the interesting problem of spacetime close to the Planck scale. The
assumption we make is that this approach should allow us to model
some low energy effects due to vacuum metric fluctuations and assess
their effect on classical spacetime and other systems at a larger
scale.

In this respect a key ingredient we require from the framework is
that it allows to implement the passage from the stochastic scale to
the classical scale: in other words that it offers a way to recover
the smooth classical spacetime. The general idea is to use a
perturbation approach and employ suitably defined averaging
procedures. Effectively we shall generalize the classic framework
due to Isaacson \cite{isaacson1968I,isaacson1968II} for a theory of
metric perturbations over a curved background, including explicitly
the conformal fluctuations and applying it to the specific case of
vacuum. It was shown by Isaacson that a spacetime averaging
procedure allows indeed to treat coherently the backreaction affect
due to the \emph{local} energy content of high frequency GWs, whose
effective stress energy tensor then resembles that of a massless
spin-2 radiation fluid. As Isaacson work is central to the material
developed below, it is reviewed in some details in Appendix
\ref{ap4}.

\subsection{Definition of the relevant scales}

We consider the microscopic structure of spacetime at a scale $\l :=
\lambda L_P$, where $L_{P} \approx 10^{-35}$ m is the Planck scale
and where the dimensionless parameter satisfies $\lambda \gtrsim 1$.
As discussed in Chapter \ref{ch1}, it sets the benchmark between the
full QG domain and a semiclassical domain, in which spacetime
properties still inherit traces of the underlying QG physics, though
being expected to be treatable by semiclassical means.

Even in empty spacetime the vacuum energy of matter fields is still
a source of gravity. Therefore we consider \ee for the physical
metric $g_{ab}$
\begin{equation}\label{P1_e1B}
G_{ab}[g] = 8\pi T_{ab}[\psi,g],
\end{equation}
where ${T_{ab}}$ will be a model stress-energy tensor describing the
overall vacuum energy contributions coming from \emph{all} matter
fields, collectively denoted by $\psi$. The metric $g_{ab}$ and the
matter fields $\psi$ are considered to be randomly fluctuating at
the scale $\l$. Accordingly we shall refer to $\l$ as to the
\emph{random scale} and interpret it as the typical scale above
which quantum vacuum properties can approximately be described by
means of classical stochastic fields. Below $\l$ and closer to the
Planck scale a full theory of quantum spacetime is required.

We are considering here vacuum metric fluctuations in an otherwise
empty universe. These can in practice be assigned as random
perturbations about some background metric $\gb_{ab}$ which we will
assume to be set up in an autoconsistent way by the backreaction
effect due to vacuum energy. In the following calculations we work
\emph{locally} and with respect to a physical inertial laboratory
frame, whose typical scale $\lab$ is much larger than the random
scale, yet small enough for the background metric to be smooth and
slow varying in an appropriate coordinate system. We then expect the
vacuum metric to be described by an expansion of the kind
\begin{equation}\label{P1_mf}
    g_{ab} = \gb_{ab} + g_{ab}^{(1)} + g_{ab}^{(2)}
    + \ldots,
\end{equation}
where the $g_{ab}^{(n)}$ indicate small fluctuating terms. Here and
henceforth we follow the standard notation where a superscript
$^{(n)}$ denotes an $n$-th order perturbation with reference made to
some small parameter $\varepsilon \ll 1$. In the subsequent
expansion of field equations, the matter fields $\psi$ will be
treated as first order quantities. The classical equation
\eqref{P1_e1B} will be analyzed explicitly for the first and second
order metric perturbations, and suitable Boyer's type fluctuating
boundary conditions will be imposed. Derived physical quantities,
e.g. the Einstein tensor, are fluctuating as a result of the
fluctuating metric.

Of course vacuum is Lorentz invariant when viewed at some
appropriate macroscopic classical scale, in which case we usually
speak of empty space. As a result, the only way it can possibly
contribute to the Einstein equation is through an effective
cosmological constant term. Lorentz invariance also implies that
vacuum statistical properties must be the same regardless of space
position and direction, i.e. \emph{homogeneity} and \emph{isotropy}
hold in a statistical sense. In order to include these properties
into our formalism we want to recover classical and smooth
quantities from the fluctuating fields. This can be obtained as a
result of an averaging process. To this end we follow the spacetime
averaging procedure described in \cite{isaacson1968II, ADM1961} and
reviewed in Appendix \ref{ap4}, so that fluctuating tensors average
to tensors. This process involves a spacetime averaging over regions
whose typical dimensions are large in comparison to the fluctuations
typical wavelengths but smaller than the scale over which the
background geometry changes significantly. Accordingly we introduce
an averaging \emph{classical scale} $\L$ such that $\l \ll \L \ll
\lab.$

While keeping in mind that we are here only considering `apparently'
empty spacetime, a final comment about the involved physical scales
is in order. The following hierarchy holds, with $L_P \lesssim \l
\ll \L \ll \lab$. The precise characterization of the classical
scale $\L$ is that of the smallest scale at which classical, locally
Lorentz invariant spacetime starts to emerge as a result of the
averaging process. Though much larger than $\l$, the classical scale
$\L$ is still expected to be very small in comparison to the
laboratory scale (Table \ref{P1_scale}). The limits of the presently
suggested theory are thus clearly set: (i) below the random scale
$\l$ the random fields approximation breaks down and a full QG
theory would be needed; (ii) spacetime starts to be smooth and
classical when viewed at the classical scale $\L$.

\begin{table}[!h]
\begin{center}
\begin{tabular}{lcccc}
\hline \hline \noalign{\smallskip}  & $L_P$ & $\l$ & $\L$ & $\lab$
\\
Scale & Planck & Random\hspace{0.2cm} & Classical\hspace{0.3cm} & Laboratory\hspace{0.3cm} \\
\hline\vspace{-0.1cm} Order of & & & &
\\\vspace{-0.1cm} magn. (m)\hspace{0.2cm} & $\approx10^{-35}$ & $\gtrsim 10^{-35}$\hspace{0.2cm} &
$\gg 10^{-35}$ & $\sim 1$
\\ \\\hline\vspace{-0.1cm} Physical & Quantum\hspace{0.2cm} & Random &
Classical & Classical
\\ domain & gravity & gravity &
gravity & gravity\\
\hline Background & None & $\gb_{ab}$ & $\gb_{ab} + \text{\small{small corrections}}$ & $\gb_{ab} + \text{\small{small corrections}}$ \\
\hline\hline
\end{tabular}\caption{\footnotesize{\textsl{A guide to relevant physical scales.}}\label{P1_scale}}
\end{center}
\end{table}

\subsection{Matter fields stress energy tensor}\label{mfset}

In order to have a more precise idea of how the random gravity
framework can be applied it is useful to have a closer look at the
matter fields stress energy tensor $T_{ab}[\psi]$. This carries
contributions from all sectors of the Standard Model. At this stage
we adopt a \emph{minimal approach} by neglecting the
non-gravitational interactions between various matter fields
components. Then $T_{ab} = \sum_{j}{T}^{(j)}_{ab}$, where the index
$j$ runs over \emph{all} matter fields and where ${T}^{(j)}_{ab}$
represents the stress energy tensor describing the $j$-th matter
field as if it was free.

The detailed microscopic expression of the generic component
${T}^{(j)}_{ab}$ will depend \emph{quadratically} upon the
corresponding matter field, as well as on the random metric $g_{ab}
= \gb_{ab} + \sum_n g_{ab}^{(n)}$. As a result the stress energy
tensor at the random scale $\l$ is a also a stochastic quantity. The
dependence on the $g_{ab}^{(n)}$ would account for the coupling
between gravity fluctuations and matter fields. However, as long as
we work up to second order, only the smooth background $\gb_{ab}$
will appear. We will argue in Section \ref{cbg} that the background
can be considered Minkowski to a good approximation. In this sense
it is like having a collection of free fluctuating fields on a
Minkowski background and the effect of gravity upon the stress
energy tensor would only appear as a third order effect.

The corresponding energy density contribution can be defined at the
classical scale $\L$ in a statistical sense through a spacetime
averaging procedure. Provided the high frequency components are cut
off at the random scale $\l = \lambda \Lp$, the average will be well
defined and finite. Then the quantity $\m{T_{ab}}$ will represent a
macroscopic stress-energy tensor at the classical scale $\L$. An
idea of how this can be done is given in Appendix \ref{ap3}, where
for simplicity we model each of the matter fields independent
degrees of freedom by a Klein-Gordon scalar field. Using the generic
notation $\phi$ for such a degree of freedom, the microscopic
structure of the stress energy tensor is then given by the familiar
expression
\begin{equation}
\TKG[\phi]=\phi_a\phi_b-\frac12\eta_{ab}
\left(\phi^c\phi_c+\frac{m^2c^2}{\hbar^2}\phi^2\right).
\end{equation}
We show in Appendix \ref{ap1} that, under the assumption that $\phi$
is a zero mean, stationary stochastic process, spacetime averages
like e.g. $\m{(\pd_{\mu}\phi)^2}$ can be expressed equivalently as
an integral of the power spectral density $S(\omega)$ with an high
energy cutoff set at the random scale $\l$. Then we have e.g.
\begin{equation}
\ms{(\pd_t\phi)^{\;2}}= \frac{1}{(2\pi)^3}\int d^3 k\,\omega_k^2
\Sphi(k).
\end{equation}
The average of the stress energy tensor is found to be given by
\begin{equation}
\m{\TKG}=\m{\phi_a\phi_b},
\end{equation}
and can be calculated if a power spectrum is assigned. In particular
the energy density is given by
\begin{equation}\label{ed}
\rho=\m{T_{00}}=\frac1{c^2}\ms{(\pd_t\phi)^{\;2}}
=\frac{1}{(2\pi)^3}\int d^3 k\,\frac{\omega_k^2}{c^2} \Sphi(k).
\end{equation}

The key point to describe zero point vacuum energy is that, in the
same way as it was done by Boyer, we can link the random field
approach to the quantum properties of vacuum by requiring that the
energy density \eqref{ed} matches the zero point energy for a scalar
quantum field in its ground state, i.e. we impose
\begin{equation}
\rho:=\frac{1}{(2\pi)^3}\int d^3 k\,\frac{\hbar\omega_k}{2},
\end{equation}
where in the r.h.s. the contributions from all normal modes of
$\phi$ are included. As found by Boyer for the EM field, this
implies that the power spectrum must have the following simple form:
\begin{equation}\label{vsd}
\Sphi(k)=\frac{\hbar c^2}{2\omega_k}.
\end{equation}
Because of homogeneity and isotropy holding at the classical scale
and assuming the stochastic properties of the fluctuations in
different spacetime directions to be uncorrelated, the averaged
matter fields stress energy tensor will take the perfect fluid form
\begin{equation}\label{P1_set2}
    \m{T_{ab}} = \left(
               \begin{array}{cccc}
                 \rho & 0 & 0 & 0 \\
                 0 & p & 0 & 0 \\
                 0& 0 & p & 0 \\
                 0 & 0 & 0& p \\
               \end{array}
             \right)
\end{equation}
where $\rho := \m{{T}_{00}}$ is the energy density and $ p :=
\frac13 \m{{T}^i{}_i} $, $i=1,2,3$ is the pressure. The dominant
energy condition, i.e. $\rho \geq 0$ and $\rho \geq p$, is normally
thought to be valid for all known reasonable forms of matter
\cite{Hawking&Ellis1973}, at least as long as the adiabatic speed of
sound $dp / d\rho$ is less than the speed of light. This is true for
massless fields since, in this case, $\rho = 3p$. By assuming the
vacuum spectral density \eqref{vsd}, we show in Appendix \ref{ap3}
that the stronger condition $\rho > 3p$ is satisfied by massive
fields in their vacuum state, at least in the ideal case in which
interactions can be neglected and the field masses are much smaller
than the Planck mass.

To assess $\rho$ and $P$ we use the results \eqref{phirho6} and
\eqref{phiP6} derived in Appendix \ref{ap3} and providing an
estimate of the zero point energy $\rho_{\phi}$ and pressure
$p_{\phi}$ for one scalar field $\phi$ of mass $m$:
\begin{equation}
\rho_{\phi} = \frac{\pi^2 \rho_P}{\lambda^4} +
    \frac{\rho_P}{4\lambda^2}\left(\frac{m}{M_P}\right)^2
\end{equation}
and
\begin{equation}
p_{\phi} = \frac{\pi^2 \rho_P}{3\lambda^4} -
    \frac{\rho_P}{12\lambda^2}\left(\frac{m}{M_P}\right)^2.
\end{equation}
Describing each independent component of the Standard Model matter
fields as a simple scalar field and neglecting fields interactions
we can obtain an estimate of the zero point energy and pressure by
simply summing up the contributions from all individual components.
Indicating with $N_i$ the number of independent components of the
particular field having mass $m_i$ we thus obtain:
\begin{equation}
\rho = \frac{N \pi^2 \rho_P}{\lambda^4} +
    \frac{N
    \rho_P}{4\lambda^2}\left(\frac{M}{M_P}\right)^2,
\end{equation}
where $N := \sum_{i}N_i$ is the total number of independent
components for all matter fields and where we defined
\begin{equation}
M^2:=\frac{\sum_{i} N_i m_i^2}{N},
\end{equation}
i.e. the weighted average of the squared masses of all matter
fields. Notice that for indices $i$ corresponding to massless fields
it is $m_i = 0$. The pressure is found to be
\begin{equation}
p = \frac{N \pi^2 \rho_P}{3\lambda^4} -
    \frac{N
    \rho_P}{12\lambda^2}\left(\frac{M}{M_P}\right)^2.
\end{equation}
These expressions are valid when $M \ll \Mp,$ which is certainly the
case within the standard model since it is expected $M \approx 10^2$
GeV.

To separate the contributions of the massless and massive fields it
is useful to perform a trace decomposition in \eqref{P1_set2}. We
have
\begin{align}
    \m{\T[\psi]} =& \left(
      \begin{array}{cccc}
        \rhom & 0 & 0 & 0 \\
        0 & \frac{\rhom}{3} & 0 & 0 \\
        0 & 0 & \frac{\rhom}{3} & 0 \\
        0 & 0 & 0 & \frac{\rhom}{3} \\
      \end{array}
    \right)
- \rho_M \eta_{ab}
\end{align}
where
\begin{align}
  &\rhom := \frac{3}{4}\left(\rho + p\right) =
  \frac{N \pi^2 \rho_P}{\lambda^4} +
    \rho_M,\label{P1_aaa1}
\end{align}
is the energy density of the traceless part and where the trace part
is defined in terms of
\begin{align}
  &\rho_M := \frac{1}{4}(\rho-3p)
  =\frac{N
    \rho_P}{8\lambda^2}\left(\frac{M}{M_P}\right)^2\label{P1_aaa2}.
\end{align}
We remark that the trace part corresponds to a cosmological constant
term, where the cosmological constant would be defined by
\begin{equation}
\LM := \frac{8\pi G}{c^4}\rho_M.
\end{equation}
Defining the Planck cosmological constant as
\begin{equation}
\LP := \frac{8\pi G}{c^4}\rho_P
\end{equation}
we find
\begin{equation}
\LM = \frac{N}{8\lambda^2}\left(\frac{M}{M_P}\right)^2\LP.
\end{equation}
Thus we see that only the massive fields are expected to contribute
to the cosmological constant through their zero point energy and the
trace part of the matter stress energy tensor. On the other hand
this also presents a traceless contribution $\rhom$ which is even
more important than $\rho_M$. Indeed
\begin{equation}
    \frac{\rho_M}{\rhom} \propto \left(\frac{M}{\Mp/\lambda}\right)^2 \ll 1,
\end{equation}
where the inequality holds within the Standard Model, with $M
\approx 10^2$ GeV, and for any realistic value of the cutoff
parameter $\lambda$.

Summarizing, the effect of vacuum zero point energy is two fold: (1)
the averaged stress energy tensor $\ms{T_{ab}[\psi]}$ would
contribute to the overall background geometry curvature while (2)
the fluctuations defined by $T_{ab}[\psi] - \ms{T_{ab}[\psi]}$ would
further induce extra metric fluctuations. We notice that it is this
second element only that is accounted for, though via a different
technique, in the stochastic gravity approach. We will discuss some
issues related the stress energy tensor average and the cosmological
constant in Section \ref{cbg}.

The next step is to include GWs backreaction and the conformal
fluctuations into the analysis. To this end it is natural to perform
an expansion of \eep, as we start to illustrate in the next section.

\section{Characterizing vacuum at the random scale}

In this section we start applying the nonlinear random gravity
framework to standard GR while trying to incorporate conformal
fluctuations. To this end we want to seek for a solution of \ee
$G_{ab}[g] = 8\pi T_{ab}[\psi]$ at the random scale where the vacuum
spacetime physical metric $g_{ab}$ takes the form
\begin{equation}\label{phm2}
g_{ab} = \Omega^2 \gamma_{ab},
\end{equation}
where $\gamma_{ab}$ is a conformal metric and $\O$ a conformal
factor. As already discussed above this could make sense if we
succeed in imposing some structure on the conformal metric
$\gamma_{ab}$ in such a way that it is \emph{not} depending on $\O$.

A preliminary formal structure can be imposed on the conformal
metric by splitting it into a smooth, slow varying background
$\gmb_{ab}$ with superimposed fast varying perturbations according
to
\begin{equation}\label{gaexpn}
\gamma_{ab} = \gmb_{ab} + \xi_{ab} + \gamma^{(2)}_{ab} +
O(\varepsilon^3),
\end{equation}
where $\varepsilon \ll 1$ is a small dimensionless parameter. The
perturbations and background typical variations scales are supposed
to be set respectively by the random scale $\l = \lambda \Lp$ and by
some macroscopic scale $L \gg \L$, where $\L$ is the classical scale
(see Table \ref{P1_scale}). As to their typical magnitudes we
assume:
\begin{equation*}
|\gmb_{ab}| = O(1),\quad\quad|\xi_{ab}| = O(\varepsilon),\quad\quad
|\gamma^{(2)}_{ab}| = O(\varepsilon^2).
\end{equation*}
The conformal fluctuations are formally inserted by expressing the
conformal factor as
\begin{equation}\label{1AB}
    \O = 1 + A + B + O(\varepsilon^3),
\end{equation}
where
\begin{equation*}
\abs{A} = O(\varepsilon),\quad\quad |B| = O(\varepsilon^2).
\end{equation*}
In particular, $A$ will be referred to as the \emph{conformal field}
and will be later on characterized as a zero mean and stationary
stochastic process, i.e.
\begin{equation*}
\m{A} = 0,\quad\quad\m{\pd_a A} = 0,\quad\quad \m{A^2}
=\text{const.}
\end{equation*}

Imposing \ee $G_{ab}[g] = 8\pi T_{ab}[\psi]$ yields:
\begin{equation}\label{nhpv}
    G_{ab}[\gamma] = 8\pi T_{ab}[\psi] + \Sigma_{ab}^{1}[\Omega] +
    \Sigma_{ab}^{2}[\Omega],
\end{equation}
where
\begin{equation}
\Sigma_{ab}^{1}[\Omega] := 2\cd_a \cd_b \ln\Omega - 2 \gamma_{ab}
\cd^c\cd_c\, \ln\Omega,
\end{equation}
\begin{equation}
\Sigma_{ab}^{2}[\Omega] := - \left( 2 \cd_a \ln\Omega \cd_b
\ln\Omega + \gamma_{ab} \cd^c \ln\Omega \cd_c \ln\Omega \right).
\end{equation}
As explained in Section \ref{nme} this can be viewed as an equation
constraining the conformal factor if the conformal metric can
somehow be specified. In our attempt to implement this concretely we
will tackle equation \eqref{nhpv} perturbatively, basically
extending Isaacson's framework for small perturbations over a curved
background \cite{isaacson1968I,isaacson1968II} to allow for the
presence of conformal fluctuations.

If we succeed, the intuitive picture of vacuum spacetime that would
emerge is that of a smooth manifold presenting, at the random scale
$\l$, metric perturbations due to random stretchings of the scale
\emph{and} GWs. In addition there will be higher order metric
perturbations due to all sorts of backreaction effects. A key point
is that such a scenario should be \emph{autoconsistent}: i.e. we are
supposing that the background geometry is precisely set up by the
contributions due to \emph{all} vacuum components. We remark thus
how, in this sense, Isaacson's framework provides an ideal basis for
our model. Since all these perturbations are small, the physical
metric is expected to be also expandable as
\begin{equation}\label{gexpn}
g_{ab} = \gb_{ab} + h_{ab} + g^{(2)}_{ab} + O(\varepsilon^3),
\end{equation}
with background, linear and higher order perturbations orders of
magnitude characterized by
\begin{equation*}
|\gb_{ab}| = O(1),\quad\quad|h_{ab}| = O(\varepsilon),\quad\quad
|g^{(2)}_{ab}| = O(\varepsilon^2).
\end{equation*}

Before embarking into the expansion scheme we now establish some
algebraic constraints relating the physical and conformal metric
perturbations that will be useful later on. By putting together
equations \eqref{phm2}, \eqref{gaexpn} and \eqref{1AB}, retaining
terms up to second order and comparing with \eqref{gexpn} we obtain:
\begin{align}\label{constrNEW}
&\gb_{ab} = \gmb_{ab},\nonumber\\
\nonumber\\
&h_{ab} = \xi_{ab} + 2A \,\gb_{ab},\\
\nonumber\\
&g^{(2)}_{ab} = \gamma^{(2)}_{ab} + 2A \, \xi_{ab} + 2B\gb_{ab} +
A^2 \,\gb_{ab}\nonumber.
\end{align}
In particular the relation $\gb_{ab} = \gmb_{ab}$ shows that we just
need introduce a single background geometry. In the following we
will use the notation $\gb_{ab}$.

\section{Setting up the expansion scheme}

We assume that all first order perturbations $h_{ab}$, $\xi_{ab}$
and $A$ are fast varying with typical variation scale $\ell =
\lambda \Lp$ set by the random scale. Moreover the smooth slow
varying background $\gb$ is characterized by the typical variation
scale $L \gg \L$. By the observed local properties of vacuum
spacetime it is in fact fair to assume that $L \gtrsim \lab$. In
this sense, using the laboratory macroscopic scale as a reference,
we will consider from now on $L = O(1)$.

The matter fields in their vacuum state are indicated collectively
by $\psi$, also assumed to be a first order quantity varying on the
typical scale $\l$, i.e. $\abs{\psi} = O(\varepsilon)$. The stress
energy tensor $T_{ab}$ is has a quadratic structure $T \sim \pd \psi
\pd \psi$, which yields the order of magnitude estimate:
\begin{equation}\label{upi2}
T[\psi] = O(\varepsilon^2 / \ell^2).
\end{equation}

As it shown in Appendix \ref{ap4}, the effective backreaction energy
due to GWs will also be of order $(\varepsilon/\l)^2$. On the other
hand the typical background curvature will be of order $(\pd \gb)^2
\sim 1/L^2$. As it will be shown shortly, even the $A$ dependent
terms formally contribute to the backreaction energy with terms of
order $(\varepsilon/\l)^2$. Then the autoconsistent framework, in
which by definition all of the background curvature is produced by
vacuum components, is characterized by the condition:
\begin{equation*}
\varepsilon \sim \frac{\ell}{L}.
\end{equation*}
In particular, using $L = O(1)$, this gives the estimate
\begin{equation}
T_{ab}[\psi] = O(1).
\end{equation}

\subsection{Einstein tensor expansion}

In expanding equation \eqref{nhpv} we start by the Einstein tensor
term $G_{ab}[\gamma]$. More details of how this is done are given in
Appendix \ref{ap4}. Expanding the Ricci tensor $R_{ab}[\gamma]$
gives:
\begin{equation}
R_{ab}[\gb + \xi + \gamma^{(2)}] = R_{ab}^{(0)}[\gb] +
R_{ab}^{(1)}[\xi] + R_{ab}^{(1)}[\gamma^{(2)}] + R_{ab}^{(2)}[\xi] +
R_{ab}^{(3+)},
\end{equation}
where $R_{ab}^{(0)}[\gb]$ is the Ricci tensor of the background
metric and where the linear operator $R_{ab}^{(1)}[\,\cdot\,]$ and
nonlinear operator $R_{ab}^{(2)}[\,\cdot\,]$ are given explicitly by
equations \eqref{ord1} and \eqref{ord2}. The term $R_{ab}^{(3+)}$
represents higher order contributions. As done in Appendix
\ref{ap4}, by using the information $O(\varepsilon) = O(\ell)$, we
have the order of magnitude estimates:
\begin{equation}
    R^{(0)}_{ab}[\gb]\sim {\gb}^{-1}\pd^2\gb\sim 1/L^2 = O(1),
\end{equation}
\begin{equation}
    R^{(1)}_{ab}[\xi]\sim {\gb}^{-1}\pd^2\xi\sim \varepsilon/\ell^2 =
    O(1/\varepsilon),
\end{equation}
\begin{equation}
    R^{(1)}_{ab}[\gamma^{(2)}]\sim {\gb}^{-1}\pd^2\gamma^{(2)}\sim \varepsilon^2/\ell^2 =
    O(1),
\end{equation}
\begin{equation}
    R^{(2)}_{ab}[\xi]\sim \xi{\gb}^{-2}\pd^2\xi\sim \varepsilon^2/\ell^2 =
    O(1),
\end{equation}
together with $R_{ab}^{(3+)} = O(\varepsilon)$.

The expansion of the Einstein tensor follows as
\begin{equation}\label{etexp}
G_{ab}[\gamma] = \underbrace{G^{(0)}_{ab}[\gb]}_{O(1)} +
\underbrace{G^{(1)}_{ab}[\xi]}_{O(1/\varepsilon)} +
\underbrace{G^{(1)}_{ab}[\gamma^{(2)}] + G^{(2)}_{ab}[\xi] +
G^{(\xi\xi)}_{ab}[\xi]}_{O(1)} +
\underbrace{G_{ab}^{(3+)}}_{O(\varepsilon)},
\end{equation}
where
\begin{equation}
    G^{(0)}_{ab}[\gb]:= R^{(0)}_{ab}[\gb] -
    \frac{1}{2}\gb_{ab}R^{(0)}[\gb],
\end{equation}
is the background metric Einstein tensor. Moreover the \emph{linear
tensor} $G^{(1)}_{ab}$ is defined as
\begin{equation}\label{linop}
    G^{(1)}_{ab}[\,\cdot\,]:= R^{(1)}_{ab}[\,\cdot\,] -
    \frac{1}{2}\gb_{ab}R^{(1)}[\,\cdot\,],
\end{equation}
and the \emph{quadratic tensors} $G^{(2)}_{ab}$ and
$G^{(\xi\xi)}_{ab}$ by
\begin{equation}\label{linop2}
    G^{(2)}_{ab}[\,\cdot\,]:= R^{(2)}_{ab}[\,\cdot\,] -
    \frac{1}{2}\gb_{ab}R^{(2)}[\,\cdot\,],
\end{equation}
\begin{equation}\label{linop3}
    G^{(\xi\xi)}_{ab}[\xi]:= \frac{1}{2}\gb_{ab}\xi^{cd}R^{(1)}_{cd}[\xi] -
    \frac{1}{2}{\xi}_{ab}R^{(1)}[\xi].
\end{equation}

\subsection{Terms dependent on the conformal factor}

To implement the expansion scheme we must also expand the terms that
depend on the conformal factor. These are given in \eqref{sig11} and
\eqref{sig21} as
\begin{equation}\label{s1}
\Sigma_{ab}^{1}[\Omega] = 2\cd_a \cd_b \ln\Omega - 2 \gamma_{ab}
\cd^c\cd_c\, \ln\Omega,
\end{equation}
\begin{equation}\label{s2}
\Sigma_{ab}^{2}[\Omega] = - \left( 2 \cd_a \ln\Omega \cd_b \ln\Omega
+ \gamma_{ab} \cd^c \ln\Omega \cd_c \ln\Omega \right).
\end{equation}
The covariant derivative of the conformal metric appears. When the
conformal metric is expressed as in \eqref{gaexpn} its Christoffel
symbol can be expanded up to linear order in the perturbation as
\cite{flanagan2005}:
\begin{align}\label{christo}
{\Gamma^a}_{bc} &= \frac{1}{2}\gamma^{ad}\left( \pd_c \gamma_{db} +
\pd_b
\gamma_{dc} - \pd_d \gamma_{bc} \right)\nonumber\\
&=\frac{1}{2}\left(\gbu^{ad} - \xi^{ad}\right)\left( \pd_c \gb_{db}
+ \pd_c \xi_{db} + \pd_b \gb_{dc} + \pd_b \xi_{dc} - \pd_d \gb_{bc}
-
\pd_d \xi_{bc} \right)+O(\varepsilon^2)\nonumber\\
&= {\Csb^a}_{bc} + \delta{\Gamma^a}_{bc}[\xi] + O(\varepsilon^2),
\end{align}
where ${\Csb^a}_{bc}$ is the Christoffel symbol of the background
metric and the first order correction $\delta{\Gamma^a}_{bc}[\xi]$
is given by
\begin{equation}\label{dch}
\delta{\Gamma^a}_{bc} := \frac{1}{2}\gbu^{ad}\left( \cdb_c \xi_{db}
+ \cdb_b \xi_{dc} - \cdb_d \xi_{bc} \right),
\end{equation}
$\cdb_a$ being the covariant derivative operator associated to the
background metric. If $L = O(1)$ is the typical variation scale of
the background geometry we have the orders of magnitudes:
\begin{equation*}
    {\Csb^a}_{bc} = O(1), \quad\quad \delta{\Gamma^a}_{bc}[\xi] =
    O(\varepsilon/\ell).
\end{equation*}

The covariant derivatives in \eqref{s1} and \eqref{s2} act upon
$\ln\Omega$, which must also be expanded up to second order. Using
$\Omega = 1 + A + B$ gives
\begin{equation}
\ln\Omega \equiv \ln(1 + A + B) = A + B - \frac{A^2}{2} +
O(\varepsilon^3).
\end{equation}
Since we are dealing with scalar functions we have e.g.
\begin{equation}
\cd_a \ln\Omega = \pd_a A + \pd_a B - \frac{1}{2}\pd_a A^2 +
O(\varepsilon^3/\ell),
\end{equation}
where $\ell$ is the typical variation scale of the fluctuations.

The term $\Sigma_{ab}^{2}[\Omega]$ being already quadratic in
$\ln\O$ gives a single contribution quadratic in $A$ which is found
immediately to be
\begin{equation}\label{sig22n}
\Sigma_{ab}^{2(2)}[A] := - \left( 2 \pd_a A \pd_b A +
\gb_{ab}\gbu^{cd}\pd_c A \pd_d A \right) = O(\varepsilon^2/\ell^2).
\end{equation}
Here the suffix $^{(2)}$ stands for `\emph{the second order terms
of}'. This notation will be used from now on, e.g. the suffix $^{(n
)}$ will indicate `\emph{the $n$-th order terms of}' for some given
quantity whose expansion we are considering.

This is useful for the term $\Sigma_{ab}^{1}[\Omega]$ since it will
yield both a linear and a quadratic contribution, which we will
denote by $\Sigma_{ab}^{1(1)}[\Omega]$ and
$\Sigma_{ab}^{1(2)}[\Omega]$. To find these we have to consider
\begin{equation*}
\cd_a \cd_b \ln\Omega \equiv \cd_a [\pd_b A + \pd_a B -
\frac{1}{2}\pd_b A^2 + O(\varepsilon^3/\ell)].
\end{equation*}
This gives:
\begin{align*}
\cd_a \cd_b \ln\Omega &= \cd_a \pd_b A + \cd_a \pd_b B -
\frac{1}{2}\cd_a \pd_b A^2
+ O(\varepsilon^3/\ell^2)\\
&= \left(\pd_a \pd_b A - {\Gamma^c}_{ab}\pd_c A \right) +
\left(\pd_a \pd_b B - {\Gamma^c}_{ab}\pd_c B \right)  -
\frac{1}{2}\left( \pd_a \pd_b A^2  - {\Gamma^c}_{ab} \pd_c A^2
\right) + O(\varepsilon^3/\ell^2).
\end{align*}
Using equation \eqref{christo} we have
\begin{align}\label{cco}
\cd_a \cd_b \ln\Omega=&\left[\pd_a \pd_b A - \left({\Csb^c}_{ab} +
\delta{\Gamma^c}_{ab}[\xi]\right)\pd_c A \right] + \left[\pd_a \pd_b
B - \left({\Csb^c}_{ab} +
\delta{\Gamma^c}_{ab}[\xi]\right)\pd_c B \right]\nonumber\\
&\hspace{1cm}- \frac{1}{2}\left[ \pd_a \pd_b A^2  -
\left({\Csb^c}_{ab} + \delta{\Gamma^c}_{ab}[\xi]\right) \pd_c A^2
\right] + O(\varepsilon^3/\ell^2).
\end{align}
The typical order of these terms are:
\begin{equation}\label{oorr1}
\pd\pd A = O(\varepsilon / \ell^2),\quad \Csb \pd A = O(\varepsilon
/ \ell), \quad \delta\Gamma \pd A = O(\varepsilon^2 / \ell^2),
\end{equation}
\begin{equation}\label{oorr1b}
\pd\pd B = O(\varepsilon^2 / \ell^2),\quad \Csb \pd B =
O(\varepsilon^2 / \ell), \quad \delta\Gamma \pd B = O(\varepsilon^3
/ \ell^2),
\end{equation}
\begin{equation}\label{oorr2}
\pd\pd A^2 = O(\varepsilon^2 / \ell^2),\quad \Csb \pd A^2 =
O(\varepsilon^2 / \ell), \quad \delta\Gamma \pd A^2 =
O(\varepsilon^3 / \ell^2).
\end{equation}
Using \eqref{cco}, $\gamma^{cd} = {\gb}^{cd} - \xi^{cd} +
O(\varepsilon^2)$, and collecting equal order terms according to the
above estimates, the term \eqref{s1} is easily found to give the
linear contribution
\begin{equation}\label{sigB11}
\Sigma_{ab}^{1(1)}[A] := 2\pd_a\pd_b A - 2\gb_{ab}
{\gb}^{cd}\pd_c\pd_d A = O(1/\varepsilon)
\end{equation}
and the nonlinear contribution
\begin{equation}\label{sigB12}
\Sigma_{ab}^{1(2)}[A,B] := \Sigma_{ab}^{1(1)}[B] -\left(\pd_a\pd_b
A^2 - \gb_{ab} {\gb}^{cd}\pd_c\pd_d A^2\right) + S_{ab}[A,\xi] =
O(1),
\end{equation}
where $S_{ab}[A,\xi]$ is defined by
\begin{align}\label{S}
S_{ab}[A,\xi]:=&-2{\Csb^c}_{ab}\pd_c A +
2\gb_{ab}\gbu^{cd}{\Csb^e}_{cd}\pd_e A\nonumber\\
&-2\delta{\Gamma^c}_{ab}[\xi]\pd_c A + 2\gb_{ab}\gbu^{cd}
\delta{\Gamma^e}_{cd}[\xi]\pd_e A\nonumber\\
&+2\gb_{ab}\xi^{cd}\pd_c\pd_d A - 2\xi_{ab} \gbu^{cd}\pd_c\pd_d A =
O(1).
\end{align}

\subsection{The expansion vacuum equations on a curved background}

We have found that the random scale \ee for the conformal metric up
to second order in the perturbations has the structure:
\begin{equation}
\underbrace{G^{(0)}_{ab}[\gb]}_{O(1)} +
\underbrace{G^{(1)}_{ab}[\xi]}_{O(1/\varepsilon)} +
\underbrace{G^{(1)}_{ab}[\gamma^{(2)}] + G^{(2)}_{ab}[\xi] +
G^{(\xi\xi)}_{ab}[\xi]}_{O(1)} = \underbrace{8\pi
T_{ab}^{(2)}[\psi]}_{O(1)} +
\underbrace{\Sigma_{ab}^{1(1)}[A]}_{O(1/\varepsilon)} +
\underbrace{\Sigma_{ab}^{1(2)}[A,B] + \Sigma_{ab}^{2(2)}[A]}_{O(1)},
\end{equation}
where we neglected terms of order $O(\varepsilon).$

We can finally write down the expansion equations. To highest order
we can collect all terms of order $O(1/\varepsilon)$ -equations
\eqref{etexp}, \eqref{sigB11}- to get
\begin{equation}\label{EQ1}
G^{(1)}_{ab}[\xi] = \Sigma_{ab}^{1(1)}[A],
\end{equation}
governing the propagation of the linear first order perturbation
$\xi_{ab}$ upon the curved background $\gb_{ab}$. Note that this
enters the equation in the r.h.s. \emph{and} through the linear
operator $G^{(1)}_{ab}$ defined in \eqref{linop}.

Similarly, we can collect all terms of order $O(1)$ -equations
\eqref{upi2}, \eqref{etexp}, \eqref{sig22n}, \eqref{sigB12}- to get
the nonlinear backreaction equation for the background geometry as:
\begin{align}\label{EQ2}
G^{(0)}_{ab}[\gb] + G^{(1)}_{ab}[\gamma^{(2)}] = & \;8\pi
T_{ab}^{(2)}[\psi] - G^{(2)}_{ab}[\xi] - G^{(\xi\xi)}_{ab}[\xi] -
\left( 2 \pd_a A \pd_b A + \gb_{ab}{\gb}^{cd}\pd_c A \pd_d A
\right)\nonumber\\
& +\Sigma_{ab}^{1(1)}[B] -\left(\pd_a\pd_b A^2 - \gb_{ab}
{\gb}^{cd}\pd_c\pd_d A^2\right) + S_{ab}[A,\xi],
\end{align}
where $S_{ab}[A,\xi]$ is given in \eqref{S}.

These equations must be solved simultaneously and generalize
Isaacson's equations to the case where conformal fluctuations
described by $\Omega = 1 + A + B$ are included.

\section{Considerations on the cosmological constant problem}\label{cbg}

The above equations are very general and are based upon an arbitrary
smooth background geometry. By taking the average in \eqref{EQ2} we
obtain the equation for the background as:
\begin{align}\label{EQ2av}
G^{(0)}_{ab}[\gb] = \m{8\pi T_{ab}^{(2)}[\psi] - G^{(2)}_{ab}[\xi] -
G^{(\xi\xi)}_{ab}[\xi] - \left( 2 \pd_a A \pd_b A +
\gb_{ab}{\gb}^{cd}\pd_c A \pd_d A \right) + S_{ab}[A,\xi]},
\end{align}
where we have used $\ms{G^{(0)}} = G^{(0)}$, $\ms{\gamma^{(2)}} =
0$, $\ms{A^2} = \text{const}$, $\m{B}=0$ and the fact that the
averaging procedure commutes with linear differential operators such
as $G^{(1)}_{ab}$ \cite{flanagan2005}. This equation establishes the
overall, large scale effect of zero point energy on the background
spacetime geometry. It is a well known fact that observed vacuum,
which we can alternatively call apparently empty space, is locally
Lorentz invariant. This implies that it must be:
\begin{equation*}
\m{8\pi T_{ab}^{(2)}[\psi] - G^{(2)}_{ab}[\xi] -
G^{(\xi\xi)}_{ab}[\xi] - \left( 2 \pd_a A \pd_b A +
\gb_{ab}{\gb}^{cd}\pd_c A \pd_d A \right) + S_{ab}[A,\xi]} \equiv -
\LRG \gb_{ab},
\end{equation*}
where $\LRG$ denotes an effective cosmological constant induced at
the classical scale by the combined effect of matter fields, GWs
and, possibly, conformal fluctuations. The corresponding energy
density due to vacuum behavior at the random scale would be
\begin{equation*}
    \rRG:=\frac{c^4 \LRG}{8\pi G},
\end{equation*}
where for clarity we have re-inserted the constants $G$ and $c$.

Now, if we denote by $\rV$ the hypothetical \emph{true} theoretical
amount of vacuum energy due to matter fields and QG effects, we
reasonably expect $\rRG \lesssim \rV$. This is because the random
gravity framework provides a mean to model zero point energies down
to the random scale $\l$ but is not expected to model vacuum
phenomena that happen closer to the Planck scale or other
contributions to vacuum energy such as those due to phase
transitions in the electro-weak sector or non linear effects typical
of QCD. It is well known that theoretical estimates of $\rV$ made
within QFT yield a result which is of many order of magnitudes
larger than what we expect from observations. This discrepancy
between theoretical expectations and observations represents one of
the main aspects of the so called \emph{cosmological constant
problem}. In his brilliant review Carroll \cite{Carroll2001} points
out how, taking into account all possible contributions to vacuum
energy (scalar fields, ZPE, QCD and EW effects), one finds in fact
the planckian value $\rV \approx \rho_P$, where $\rho_P \approx
10^{19}\text{ GeV}/L_P^3$ is the Planck energy density. On the other
hand the current estimated amount of observed vacuum energy in the
present epoch universe is \cite{Carroll2001}
\begin{equation}\label{P1_occ}
    \rho^{\text{\scriptsize{obs}}}_{\text{\scriptsize{vacuum}}}
    \approx 5\times 10^{-11}\text{ J m}^{-3} \approx 10^{-125}
    \rho_P.
\end{equation}
This value is deduced from the observed large scale properties of
the universe, including the approximatively flat geometry, as
suggested by CMB anisotropies \cite{boomerang}, and the present,
small, cosmic acceleration indicated by type Ia supernovae
observations \cite{Ia_supernovae}. Defining the Planck cosmological
constant as
\begin{equation}
\LP := \frac{8\pi G}{c^4}\rho_P,
\end{equation}
then \eqref{P1_occ} implies the following observational constraint
on the present epoch cosmological constant
\begin{equation}\label{P1_CCP}
0 \leq \Lobs \lesssim 10^{-125} \LP.
\end{equation}
The upper bound is extremely small, especially in comparison to the
QFT theoretical expectations usually reported in the literature.
Because of this dramatic discrepancy it is commonly believed that
either vacuum energy makes a case on its own in that it does not
gravitate or, more likely, that there must exist some sort of
balancing mechanism. Such a mechanism is to date unknown. Some
possibilities include supersymmetry, extra scalar fields, dark
energy models such as quintessence \cite{Carroll2001} or new
symmetry principles such as invariance of the gravitational
lagrangian under shift by a constant \cite{Padmanabhan2008}.

\subsection{Choice of the background geometry}

It had been one of the original goal of the present work, as
discussed in \cite{wango8rg}, to verify whether conformal
fluctuations could offer a mean to obtain such a balancing
mechanism. That this cannot be the case will be shown in the
forthcoming sections. For the moment we also recall how the
effective averaged stress energy tensor due to matter fields alone
in their vacuum state is predicted by the random gravity framework
to be
\begin{align}\label{cp}
    \m{\T[\psi]} =& \left(
      \begin{array}{cccc}
        \rhom & 0 & 0 & 0 \\
        0 & \frac{\rhom}{3} & 0 & 0 \\
        0 & 0 & \frac{\rhom}{3} & 0 \\
        0 & 0 & 0 & \frac{\rhom}{3} \\
      \end{array}
    \right)
- \rho_M \eta_{ab}.
\end{align}
The contribution to the effective cosmological constant coming from
the standard model matter fields corresponds to the energy density
$\rho_M = \frac{N \rho_P}{8\lambda^2}\left(\frac{M}{M_P}\right)^2$,
as given in equation \eqref{P1_aaa2}. In comparison to the usual
asserted planckian value, this is reduced by a factor $(M/M_P)^2
\approx 10^{-34}$. The suggested link between vacuum related
cosmological constant and mass is interesting but we will not
further investigate this issue in the present work. Rather we notice
another interesting point, somehow rarely mentioned in the
literature, but remarked e.g. by Padmanabhan in
\cite{Padmanabhan2008}. As equation \eqref{cp} shows explicitly,
only matter fields whose stress energy tensor has a non vanishing
trace are expected to contribute to the cosmological constant. On
the other hand massless fields such as the EM field and in general
all other forms of matter with a \emph{traceless} component in their
stress energy tensor should contribute to vacuum through a term
which, beyond attaining a large planckian value, is \emph{not}
locally Lorentz invariant. Since such a contribution is obviously
not observed the vacuum energy problem appears to be enriched:
beyond a mechanism for canceling the contribution to the
cosmological constant one appears to need some mechanism to also
balance this traceless contribution.

These problems are all very interesting. However providing a deep
analysis goes beyond the scope of the present thesis. The approach
we will maintain at this point is a pragmatic one. Since what we are
really interested in here is whether we can find a solution of \ee
which presents conformal fluctuations at the random scale, we will
simply assume that the theoretical framework can in principle be
made general enough to conform to the observational bounds. Thus we
assume:
\begin{equation*}
    \rRG \lesssim \rho^{\text{\scriptsize{obs}}}_{\text{\scriptsize{vacuum}}}
    \approx 5\times 10^{-11}\text{ J m}^{-3}.
\end{equation*}
This corresponds to the assumption that the hypothetical missing
ingredient that can (and hopefully will) provide the energy balance
mechanism \emph{is} actually thought to be included within the
collection of matter fields $\psi$. An interesting alternative which
we will investigate later on is whether the conformal fluctuations
$A$ can actually represent the missing ingredient.

To conclude this discussion, we have quite a strong bound on the
magnitude of the averaged stress energy tensor that we expect to
obtain from the random scale physics. The limiting value is so close
to zero that many authors believe it is natural that the actual bulk
contribution of vacuum to the large scale energy density is actually
\emph{exactly} zero. In this respect, the other and non trivial
aspect of the cosmological constant problem is to find a mechanism
that yields an almost, but not quite vanishing, value of the
cosmological constant. The problem is all the more relevant since in
terms of the critical cosmological energy density, the cosmological
constant contribution is roughly 70\%. Leaving this aspect of the
problem aside our point of view here will be the simplest: i.e. the
missing theoretical ingredient to the vacuum energy can provide an
\emph{exact} balance at every macroscopic scale. In any case we can
reasonably expect that, at the laboratory scale $\lab$, we should
have
\begin{equation}\label{EQ2CON}
\m{8\pi T_{ab}^{(2)}[\psi] - G^{(2)}_{ab}[\xi] -
G^{(\xi\xi)}_{ab}[\xi] - \left( 2 \pd_a A \pd_b A +
\gb_{ab}{\gb}^{cd}\pd_c A \pd_d A \right) + S_{ab}[A,\xi]} = 0.
\end{equation}
As a result, a natural choice for the background geometry that
neglects secular evolution on long time scales and appears to be a
good approximation in line with observations is a flat Minkowski
background geometry, i.e. we put
\begin{equation}
\gb_{ab} := \eta_{ab}.
\end{equation}

\subsection{Simplified form of the expansion equations}

With this choice of the background geometry, the perturbation
equations \eqref{EQ1} and \eqref{EQ2} simplify a lot. Indeed we now
have
\begin{equation}
{\Csb^a}_{bc} = 0,\quad \gbu^{ab}=\eta^{ab},\quad \cdb_a =
\pd_a,\quad G^{(0)}_{ab}[\eta] = 0,
\end{equation}
and we can use $\eta_{ab}$ to rise and lower indices without
ambiguity. Equations \eqref{dch} and \eqref{S} now read
respectively:
\begin{equation}\label{dgxi}
\delta{\Gamma^a}_{bc}[\xi] = \frac{1}{2}\eta^{ad}\left( \pd_c
\xi_{db} + \pd_b \xi_{dc} - \pd_d \xi_{bc} \right),
\end{equation}
and
\begin{align}\label{SxiA}
S_{ab}[A,\xi]=-2\delta{\Gamma^c}_{ab}[\xi]\pd_c A +
2\eta_{ab}\eta^{cd} \delta{\Gamma^e}_{cd}[\xi]\pd_e A +
2\eta_{ab}\xi^{cd}\pd_c\pd_d A - 2\xi_{ab}\pd^c\pd_c A.
\end{align}
We can now re-write the first and second order equations as
\begin{equation}\label{EQ1b}
G^{(1)}_{ab}[\xi] = 2\pd_a\pd_b A - 2\eta_{ab}\pd^c\pd_c A,
\end{equation}
and
\begin{align}\label{EQ2b}
G^{(1)}_{ab}[\gamma^{(2)}] = & \;8\pi T_{ab}^{(2)}[\psi] -
G^{(2)}_{ab}[\xi] - G^{(\xi\xi)}_{ab}[\xi] - \left( 2 \pd_a A \pd_b
A + \eta_{ab}\pd^c A \pd_c A
\right)\nonumber\\
& +\left(2\pd_a\pd_b B - 2\eta_{ab}\pd^c\pd_c
B\right)-\left(\pd_a\pd_b A^2 - \eta_{ab}\pd^c\pd_c A^2\right) +
S_{ab}[A,\xi].
\end{align}
These are to be considered together with \eqref{dgxi} and
\eqref{SxiA} and are compatible with the condition \eqref{EQ2CON}.
Notice that $T_{ab}^{(2)}[\psi]$ is quadratic in $\psi$ and contains
otherwise just the Minkowski tensor $\eta_{ab}$. The same is true
for the expressions \eqref{linop}, \eqref{linop2}, \eqref{linop3}
defining $G^{(1)}_{ab}$, $G^{(2)}_{ab}$ and $G^{(\xi\xi)}_{ab}$,
which now all contain $\eta_{ab}$. The algebraic constraints
\eqref{constrNEW} now read
\begin{align}\label{con}
&h_{ab} = \xi_{ab} + 2A \,\eta_{ab},\nonumber\\
\\
&g^{(2)}_{ab} = \gamma^{(2)}_{ab} + 2A \, \xi_{ab} + 2B\eta_{ab} +
A^2 \,\eta_{ab}\nonumber.
\end{align}

Equations \eqref{EQ1b} and \eqref{EQ2b} are the starting point of
the next chapter, where we can finally provide the answer to the
question of whether conformal fluctuations can play any role within
standard GR.

\newpage
\thispagestyle{empty}


\singlespacing
\chapter{GR and conformal fluctuations}\label{ch3}

\begin{quote}
\begin{small}
In this chapter we study the perturbation equations derived in the
previous chapter following the application of the random gravity
framework to standard GR. An attempt is made to build explicit
solutions in which the physical metric presents true conformal
fluctuations. A secondary but equally important goal is to assess
whether conformal fluctuations can offer a balancing mechanism that
may cancel part of the large vacuum energy related to matter fields
and GWs, as it was recently claimed in \cite{wango8rg}, and
previously in \cite{wang2006}. We show that a technical imprecision
affected the conclusion in \cite{wango8rg}: in particular, the
nonlinear backreaction effect due to the conformal fluctuations was
not treated properly. When this is fixed for, the correct second
order equations show that no vacuum energy balance in fact occurs.
What is more, we show quite in general that conformal fluctuations
cannot have any physical effect within the standard GR framework:
the attempt to build a vacuum solution where the physical metric has
a true conformal modulation leads to unphysical conditions for the
conformal field. If GR is the correct theory of gravitation down to
at least the random scale, the main conclusion is that dephasing of
a quantum particle due to gravity related conformal fluctuations is
not expected to occur.
\end{small}
\end{quote}
\singlespacing

\section{Analysis of the linear equation and GWs}\label{sec1}

We now proceed to study the perturbation equations derived in the
previous chapter. We start the analysis from the linear equation
\eqref{EQ1b}
\begin{equation}\label{eqlin}
G^{(1)}_{ab}[\xi] = 2\pd_a\pd_b A - 2\eta_{ab}\pd^c\pd_c A,
\end{equation}
which will allow us to introduce GWs. The linear operator
$G^{(1)}_{ab}$ is given by
\begin{equation}\label{g1def3}
    G^{(1)}_{ab}[\xi]= R^{(1)}_{ab}[\xi] -
    \frac{1}{2}\eta_{ab}R^{(1)}[\xi],
\end{equation}
where the linear part of the Ricci tensor is
\begin{equation}\label{R123}
R_{ab}^{(1)}[\xi] = \frac{1}{2}\pd^c\pd_b {\xi}_{ac} +
\frac{1}{2}\pd^c \pd_a {\xi}_{bc} -\frac{1}{2} \pd^c\pd_c {\xi}_{ab}
- \frac{1}{2}\pd_a \pd_b {\xi}.
\end{equation}

Equation \eqref{eqlin} can be re-written in a very simple form. By
using equations \eqref{g1def3} and \eqref{R123} it is
straightforward to verify that
\begin{equation}\label{G1aeta}
G^{(1)}_{ab}[-2A\eta] = 2\pd_a \pd_b A - 2 \eta_{ab} \pd^c\pd_c\,A.
\end{equation}
From \eqref{con} we know that $\xi_{ab} + 2A\eta_{ab} = h_{ab}$,
i.e. the first order perturbation of the physical metric $g_{ab}$.
Since $G^{(1)}_{ab}$ is a linear operator, equation \eqref{eqlin}
can be written in an equivalent way as
\begin{equation}\label{lin3}
G^{(1)}_{ab}[\xi + 2A\eta] \equiv G^{(1)}_{ab}[h] = 0.
\end{equation}
This is precisely the linear equation which would result if the
expansion scheme had been applied directly to \ee expressed in terms
of $g_{ab}$. At first order we thus have two equivalent equations:
\begin{equation}
G^{(1)}_{ab}[h] = 0 \quad \Leftrightarrow \quad G^{(1)}_{ab}[\xi] =
2\pd_a \pd_b A - 2 \eta_{ab} \pd^c\pd_c\,A,
\end{equation}
and one algebraic constraint
\begin{equation}\label{con1}
h_{ab} = \xi_{ab} + 2A\eta_{ab},
\end{equation}
linking the physical metric and conformal metric linear
perturbations $h_{ab}$ and $\xi_{ab}$.

The key point\label{comment} now is to \emph{decide} how to
represent the GWs. We introduce the notation $\hgw_{ab}$ to indicate
GWs, i.e. a metric perturbation satisfying the linear equation
$G^{(1)}_{ab}[\hgw] = 0.$ The above equations seem to suggest the
obvious choice $\hgw_{ab} := h_{ab}$. This choice was also made in
\cite{wango8rg} and we will explore its consequences below. We point
out however that one could also, at least formally, set $\hgw_{ab}
:= \xi_{ab}$ by \emph{imposing} $G^{(1)}_{ab}[\xi] = 0.$ This would
provide an example of what we mean by forcing a structure on the
conformal metric. It would imply the \emph{constraint} $\pd_a \pd_b
A - \eta_{ab} \pd^c\pd_c\,A = 0$ on the conformal field. This choice
would lead to an alternative formalism in which the conformal
4-metric encodes GWs and it will be explored further in
\ref{choice2}.

We now proceed in studying the scenario in which the physical metric
linear perturbation $h_{ab}$ describes GWs. In practise this is
achieved by \emph{choosing a solution} for the homogeneous linear
equation $G^{(1)}_{ab}[h] = 0$, to which we apply fluctuating
boundary conditions as explained by Boyer in the case of the zero
point classical fluctuations of the EM field \cite{boyer75}. It is
also convenient to put the perturbation in the \texttt{TT} gauge, as
this will slightly simplify some formulas later. Then vacuum
fluctuations due to GWs are represented by:
\begin{equation}
\hgw_{ab} := h_{ab}.
\end{equation}
From now on the notation $\hgw_{ab}$ indicates a concrete
fluctuating, \texttt{TT} gauge, solution of the homogeneous
linearized \eep. Concretely and coherently with the properties of
vacuum as described in the previous chapter, $\hgw_{ab}$ would have
to be given by a superposition of random phase waves traveling in
all space directions and having suitable statistical properties that
should be yielding a homogeneous, isotropic and Lorentz invariant
character to all averaged quantities we may derive from $\hgw_{ab}$.
In the following this is understood but we won't need to make these
properties formally explicit.

The conformal metric linear perturbation now follows from the
general constraint \eqref{con1} as
\begin{equation}
\xi_{ab} = \hgw_{ab} - 2A\eta_{ab},
\end{equation}
in such a way that the equation
\begin{equation}
G^{(1)}_{ab}[\xi] = -G^{(1)}_{ab}[2A\eta_{ab}] \equiv 2\pd_a \pd_b A
- 2 \eta_{ab} \pd^c\pd_c\,A
\end{equation}
is automatically satisfied. Notice that, at this stage, $A$ is still
completely unconstrained and could be anything at all. We can
exploit this freedom by \emph{demanding} that the conformal
fluctuations satisfy a simple homogeneous wave equation
\begin{equation}
\pd^c \pd_c A = 0.
\end{equation}
This equation should be solved next employing Boyer's random
boundary conditions. As a result, hereafter, the fluctuations in $A$
must be considered as being concretely assigned. As a reminder of
this fact, in the following we will use the notation $ \ar$, where
the little tilde indicates an explicit \emph{fluctuating} solution.
The conformal metric linear perturbation is now completely
determined and given explicitly by
\begin{equation}\label{xi}
\xi_{ab} = \hgw_{ab} - 2\ar\eta_{ab}.
\end{equation}
By construction this satisfies
\begin{equation}
G^{(1)}_{ab}[\xi] = 2\pd_a \pd_b \ar.
\end{equation}
Contracting the identity $\pd_a (\ar \pd_b \ar) = \pd_a \ar \pd_b
\ar + \ar \pd_a \pd_b \ar$ and using the wave equation we get
\begin{equation}
\pd^c (\ar \pd_c \ar) = \pd^c \ar \pd_c \ar,
\end{equation}
implying the important condition
\begin{equation}
\m{\pd^c \ar \pd_c \ar} = 0.
\end{equation}
Since the solutions $\hgw_{ab}$ and $\ar$ are deduced in totally
independent ways, they are statistically uncorrelated, i.e.
\begin{equation}\label{noncor}
\m{\hgw_{ab,\ldots} \ar_{,\ldots}} = 0,
\end{equation}
where $,\ldots$ denote arbitrary derivatives.

\section{Analysis of the second order equation}\label{ch2ansecordew}

The second order equation is central to the random gravity framework
as it describes how matter fields $\psi$, GWs given by $\hgw_{ab}$
and conformal fluctuations $\ar$ build up the background spacetime
curvature through their energy content. As discussed in the previous
chapter, it is our assumption that \emph{if} all the ingredients
entering the matter fields symbol $\psi$ were known and explicitly
included, then it would be possible to `see' concretely how the
overall backreaction energy at the classical scale due to zero point
energy is basically vanishing, in agreement with observations.
Thanks to this assumption it was possible to select a flat Minkowski
background. Our two main objectives at this point are:
\begin{enumerate}
  \item verifying whether it is possible to find a vacuum solution
  where the physical metric $g_{ab}$ depends nonlinearly on the
  conformal fluctuations, e.g. through a term like $\ar^2 \eta_{ab}$
  which, according to the analysis in chapter \ref{ch1} would induce
  dephasing;
  \item in relation to the issue of zero point vacuum energy balance,
  verifying whether the conformal fluctuations $\ar$ themselves can
  actually provide the (or one) ingredient for a total (or partial)
  balancing mechanism.
\end{enumerate}

The second order equation for the nonlinear conformal 4-metric
perturbation $\gamma^{(2)}_{ab}$ with a source term depending on
$\psi$, $\ar$ and $\hgw_{ab}$ is given in \eqref{EQ2b} as:
\begin{align}\label{nonlin1}
G^{(1)}_{ab}[\gamma^{(2)}] = & \;8\pi T_{ab}^{(2)}[\psi] -
G^{(2)}_{ab}[\xi] - G^{(\xi\xi)}_{ab}[\xi] - \left( 2 \pd_a \ar
\pd_b \ar + \eta_{ab}\pd^c \ar \pd_c \ar
\right)\nonumber\\
& +\left(2\pd_a\pd_b B - 2\eta_{ab}\pd^c\pd_c
B\right)-\left(\pd_a\pd_b \ar^2 - \eta_{ab}\pd^c\pd_c \ar^2\right) +
S_{ab}[\ar,\xi],
\end{align}
where $S_{ab}[A,\xi]$ is defined in \eqref{SxiA}. The quadratic part
of the Einstein tensor is defined by
\begin{equation}\label{G2def3}
    G^{(2)}_{ab}[\,\cdot\,]:= R^{(2)}_{ab}[\,\cdot\,] -
    \frac{1}{2}\eta_{ab}R^{(2)}[\,\cdot\,],
\end{equation}
with
\begin{align}\label{G2n3}
R_{ab}^{(2)}[\xi]:=&\,\frac{1}{2}\,\xi^{cd}\pd_a\pd_b {\xi}_{cd} -
\xi^{cd}\pd_c \pd_{(a}{\xi}_{b)d} + \frac{1}{4}\, (\pd_a
{\xi}_{cd})\pd_b \xi^{cd} + (\pd^d
{\xi^c}_b)\pd_{[d}{\xi}_{c]a} + \frac{1}{2}\, \pd_d (\xi^{dc}\pd_c {\xi}_{ab})\nonumber\\
& - \frac{1}{4}\, (\pd^c \xi)\pd_c {\xi}_{ab} - (\pd_d \xi^{cd} -
\frac{1}{2}\, \pd^c \xi)\pd_{(a}{\xi}_{b)c}.
\end{align}
Substituting the first order solution \eqref{xi} into
\eqref{nonlin1} we have
\begin{align}\label{nonlin12}
G^{(1)}_{ab}[\gamma^{(2)}] = & \;8\pi T_{ab}^{(2)}[\psi] -
G^{(2)}_{ab}[\hgw - 2\ar\eta] - G^{(\xi\xi)}_{ab}[\hgw - 2\ar\eta] -
\left( 2 \pd_a \ar \pd_b \ar + \eta_{ab}\pd^c \ar \pd_c \ar
\right)\nonumber\\
& G^{(1)}_{ab}[-2B\eta + \ar^2\eta] + S_{ab}[\ar,\hgw - 2\ar\eta],
\end{align}
where we have used \eqref{G1aeta} to re-express the terms depending
on $B$ and $A^2$ more compactly. Symbolically, the r.h.s. will have
the following structure:
\begin{equation}
\text{r.h.s. }\sim (\pd \ar)(\pd \ar) + (\pd\hgw)(\pd\hgw) +
(\pd\psi)^2 + (\pd \ar \pd\hgw).
\end{equation}
The average of the cross terms $(\pd \ar \pd\hgw)$ will vanish
thanks to property \eqref{noncor}. On the other hand the three
quadratic contributions will describe, in order, the backreaction
energy due to conformal fluctuations, GWs and matter fields
(possibly including the relevant unknown exotic components that
provide the energy balance). To proceed we must thus evaluate each
term
in equation \eqref{nonlin12} explicitly.\\
\\
\textbf{(i) Analysis of the term} $- G^{(\xi\xi)}_{ab}[\hgw - 2\ar\eta]$:\\
\\
From the general expression \eqref{linop3} with $\gb_{ab} =
\eta_{ab}$ we have
\begin{equation*}
-G^{(\xi\xi)}_{ab}[\hgw - 2\ar\eta]=\frac{1}{2}\left(\hgw_{ab} -
2\ar\eta_{ab}\right)R^{(1)}[\xi] -
\frac{1}{2}\eta_{ab}\left({\hgw}^{cd} -
2\ar\eta^{cd}\right)R^{(1)}_{cd}[\xi].
\end{equation*}
Using the fact that $R^{(1)}_{ab}[\,\cdot\,]$ is a linear operator
we have
\begin{equation*}
R^{(1)}_{ab}[\xi] = R^{(1)}_{ab}[\hgw] - R^{(1)}_{ab}[2\ar\eta] = -
R^{(1)}_{ab}[2\ar\eta]
\end{equation*}
since $\hgw_{ab}$ satisfies the linear GWs equation, equivalent to
$R^{(1)}_{ab}[\hgw] = 0.$ Exploiting this we find
\begin{align*}
-G^{(\xi\xi)}_{ab}[\xi] &=\frac{1}{2}\left(2\ar\eta_{ab} -
\hgw_{ab}\right)R^{(1)}[2\ar\eta] +
\frac{1}{2}\eta_{ab}\left({\hgw}^{cd} -
2\ar\eta^{cd}\right)R^{(1)}_{cd}[2\ar\eta],\\
&=2\left(\ar\eta_{ab}R^{(1)}[\ar\eta] -
\ar\eta_{ab}\eta^{cd}R^{(1)}_{cd}[\ar\eta]\right) + \left(
\eta_{ab}{\hgw}^{cd}R^{(1)}_{cd}[\ar\eta] - \hgw_{ab}
R^{(1)}[\ar\eta] \right).
\end{align*}
The first bracketed term is zero since
$\eta^{cd}R^{(1)}_{cd}[\ar\eta] = R^{(1)}[\ar\eta]$ and we see that
only a cross term of the kind $\pd \ar \pd\hgw$ survives. Because of
the statistical property $\m{\hgw_{ab,\ldots} \ar_{,\ldots}} = 0$,
the average of such a term will vanish, in such way that it cannot
affect the large scale vacuum energy balance. However its
fluctuations will induce extra fluctuations on the spacetime
geometry and it is thus useful for later purposes to make it
explicit. We can now use the expression \eqref{R123} to find, in any
gauge:
\begin{equation}\label{R1aeta}
R_{ab}^{(1)}[\ar\eta] = -\pd_a \pd_b \ar - \frac{1}{2}\eta_{ab}
\pd^c\pd_c \ar,
\end{equation}
implying
\begin{equation}\label{R1aetaT}
R^{(1)}[\ar\eta] = - 3 \pd^c\pd_c \ar.
\end{equation}
Using these we get the general result
\begin{equation}\label{xixi}
-G^{(\xi\xi)}_{ab}[\xi] = 3 \hgw_{ab} \pd^c\pd_c \ar -
\eta_{ab}{\hgw}^{cd}\pd_{c}\pd_{d} \ar - \frac{1}{2} \hgw \eta_{ab}
\pd^c\pd_c \ar.
\end{equation}
We exploit now the fact that the GWs perturbation $\hgw_{ab}$ is in
the \texttt{TT} gauge, i.e. its components in the laboratory
reference frame satisfy:
\begin{equation}\label{TT}
\hgw:=\eta^{\mu\nu}\hgw_{\mu\nu} = 0; \quad\quad \hgw_{0\mu} = 0;
\quad\quad \pd^\mu \hgw_{\mu\nu} = 0.
\end{equation}
The expression \eqref{xixi} now simplifies to
\begin{equation}\label{xixi2}
-G^{(\xi\xi)}_{\aa\bb}[\xi] = 3 \hgw_{\aa\bb} \pd^\cc\pd_\cc \ar -
\eta_{\aa\bb}{\hgw}^{\cc\dd}\pd_{\cc}\pd_{\dd} \ar.
\end{equation}
Using the wave equation for $\ar$ this reduces to
\begin{equation}\label{xixi3 }
-G^{(\xi\xi)}_{ab}[\xi] = -\eta_{ab}{\hgw}^{cd}\pd_{c}\pd_{d} \ar,
\end{equation}
which is valid in an arbitrary gauge.\\
\\
\textbf{(ii) Analysis of the term} $S_{ab}[\ar,\hgw - 2\ar\eta]$:\\
\\
From equation \eqref{SxiA} we have explicitly
\begin{align}\label{nnn}
S_{ab} & = -\eta^{cd}\left( \pd_a \xi_{bd} + \pd_b \xi_{ad} - \pd_d
\xi_{ab} \right)\pd_cA + \eta_{ab}\eta^{dc}\eta^{ef}\left( \pd_d
\xi_{cf} +
\pd_c \xi_{df} - \pd_f \xi_{dc} \right)\pd_eA\nonumber\\
&\hspace{.4cm}-2\xi_{ab}\pd^c\pd_cA + 2\eta_{ab}\xi^{dc}\pd_d
\pd_cA.
\end{align}
Using $\xi_{ab} = \hgw_{ab} - 2\ar\eta_{ab}$ to eliminate
$\xi_{ab}$, we get a series of terms involving cross products
$\hgw_{ab}\ar$, whose corresponding average will vanish, plus terms
quadratic in $\ar$. Explicitly and in an arbitrary gauge we find:
\begin{align}\label{sigma12}
S_{ab}[\ar,\hgw] = &\,4\pd_a \ar\pd_b \ar +
2\eta_{ab}\pd^cA\pd_cA\nonumber\\
&- \left[ \pd^dA \left( \pd_a \hgw_{bd} + \pd_b \hgw_{ad} -\pd_d
\hgw_{ab} \right) -  \eta_{ab} \pd^fA \left( \pd^c\hgw_{cf} + \pd^d
\hgw_{df} - \pd_f \hgw \right)\right.\nonumber\\
& \hspace{0.5cm}\left.+ 2\hgw_{ab}\pd^c\pd_cA - 2\eta_{ab}
h^{\text{\scriptsize{GW}}de}\pd_d\pd_e \ar \right].
\end{align}
Exploiting the \texttt{TT} gauge properties we have
\begin{align}\label{sigma12cross}
S_{\aa\bb}[\ar,\hgw] = &\,4\pd_\aa \ar\pd_\bb \ar +
2\eta_{\aa\bb}\pd^\cc \ar\pd_\cc \ar\nonumber\\
&- \left[ \pd^\dd \ar \left( \pd_\aa \hgw_{\bb\dd} + \pd_\bb
\hgw_{\aa\dd} -\pd_\dd \hgw_{\aa\bb} \right) +
2\hgw_{\aa\bb}\pd^\cc\pd_\cc \ar - 2\eta_{\aa\bb}
h^{\text{\scriptsize{GW}}\dd\eee}\pd_\dd\pd_\eee \ar \right].
\end{align}
Using the wave equation for $\ar$ this simplifies to
\begin{align}\label{sigma12crossw}
S_{\aa\bb}[\ar,\hgw] = &\,4\pd_\aa \ar\pd_\bb \ar +
2\eta_{\aa\bb}\pd^\cc \ar\pd_\cc \ar\nonumber\\
&- \left[ \pd^\dd \ar \left( \pd_\aa \hgw_{\bb\dd} + \pd_\bb
\hgw_{\aa\dd} -\pd_\dd \hgw_{\aa\bb} \right) - 2\eta_{\aa\bb}
h^{\text{\scriptsize{GW}}\dd\eee}\pd_\dd\pd_\eee \ar \right].
\end{align}
\\
\\
\textbf{(iii) Analysis of the term} $-G^{(2)}_{ab}[\hgw - 2\ar\eta]$:\\
\\
The analysis of this term is relatively straightforward but
particularly lengthy due to the nonlinear structure of the operator
$G^{(2)}_{ab}[\,\cdot\,]$. This is where the error in
\cite{wango8rg} was made.

More explicitly, the structure of this nonlinear term is:
\begin{equation}\label{G2cross}
-G^{(2)}_{ab}[\hgw - 2\ar\eta] = -G^{(2)}_{ab}[2\ar\eta]
-G^{(2)}_{ab}[\hgw] + [\pd \ar \pd \hgw]_{G^{(2)}},
\end{equation}
where the symbolic notation $[\pd \ar \pd \hgw]_{G^{(2)}}$ stands
for the collection of all the cross term involving $\ar$ and
$\hgw_{ab}$ that will result and will be analyzed in details in
Section \ref{equi}. This term has a zero average and, again, it
cannot affect the net amount of large scale vacuum energy. We recall
now that $\hgw_{ab}$ satisfies the linearized \ee
$G^{(1)}_{ab}[\hgw] = 0.$ Then, Isaacson's general results
summarized in Appendix \ref{ap4} guarantee that the averaged
quantity $\langle -G^{(2)}_{ab}[\hgw] \rangle$ is given by equations
\eqref{Tgw} and \eqref{finally2} as
\begin{equation}\label{finally22}
\langle -G^{(2)}_{ab}[\hgw] \rangle = \m{ \frac{1}{4} \cd_a
\bar{h}_{cd} \cd_b \bar{h}^{cd} - \frac{1}{8} \cd_a \bar{h}\cd_b
\bar{h} - \frac{1}{2}\cd_d \bar{h}^{cd} \cd_{(a}\bar{h}_{b)c} }.
\end{equation}
Isaacson showed that this is (1) \emph{gauge invariant}, (2)
\emph{positive definite} and, for $\hgw_{ab}$ given by an arbitrary
superposition of plane waves, it is also seen to be (3)
\emph{traceless}. Because of these reasons it can represent the GWs
stress energy tensor. When considering the averaged second order
equation this contributes, together with the other standard matter
fields in $\psi$, to build up a positive, large, zero point vacuum
energy.

The mistake in \cite{wango8rg} came about as follows: there, the
expression \eqref{finally22} was used to evaluate the averaged
quantity $\langle -G^{(2)}_{ab}[2\ar\eta] \rangle$. The result was
$-6\m{\pd_a \ar \pd_b \ar}$. When this is added to the other
contributions from all other terms quadratic in $\ar$, the net
result was that a \emph{negative definite}, traceless, effective
tensor quadratic in $\ar$ given by $-\langle 4\pd_a \pd_b \ar -
\eta_{ab} \pd^c \pd_c \ar\rangle$ appeared on the r.h.s. of averaged
second order \eep. By suitably adjusting the statistical amplitude
of the fluctuations in $\ar$ this was then used to cancel out the
positive vacuum energy contribution coming from GWs plus the
massless matter fields in $\psi.$ Unfortunately the flaw in the
above reasoning is that, in fact, the expression \eqref{finally22}
\emph{cannot} be used to evaluate $\langle -G^{(2)}_{ab}[2\ar\eta]
\rangle$. As explained in detail in Appendix \ref{ap4} at page
\pageref{rules}, the reason for this is that \eqref{finally22}
\emph{holds only when the general operator $G^{(2)}_{ab}[\,\cdot\,]$
is acting on a perturbation satisfying itself the linearized \eep}.
However, in the problem we are studying, the perturbation upon which
$G^{(2)}_{ab}$ is acting on is $2\ar\eta$. This does \emph{not}
satisfy the linearized equation and it is instead
\begin{equation}\label{general}
G^{(1)}_{ab}[-2\ar\eta] = 2\pd_a \pd_b \ar - 2 \eta_{ab}
\pd^c\pd_c\,\ar = 2\pd_a \pd_b \ar \neq 0.
\end{equation}

From these considerations it follow that we must calculate
$-G^{(2)}_{ab}[2\ar\eta]$ \emph{exactly} and \emph{before} any
average is done, using the full expression of the nonlinear operator
$G^{(2)}_{ab}$. To do this we estimate first
$R^{(2)}_{ab}[2\ar\eta]$ starting from equation \eqref{G2n3} and
find:
\begin{align*}
R_{ab}^{(2)}[2\ar\eta] = 4\ar \pd_a \pd_b \ar + 6 \pd_a \ar \pd_b
\ar + 2\ar\eta_{ab} \pd^c\pd_c \ar.
\end{align*}
After finding the trace $R^{(2)}[2\ar\eta] = 12\ar\pd^c\pd_cA +
6\pd^cA\pd_c \ar$ we get
\begin{align}\label{G2aeta0}
G_{ab}^{(2)}[2\ar\eta] &\equiv R_{ab}^{(2)}[2\ar\eta]
-\frac{1}{2}\eta_{ab} R^{(2)}[2\ar\eta] \nonumber\\
&= 4\ar \pd_a \pd_b \ar + 6 \pd_a \ar \pd_b \ar - 4 \ar\eta_{ab}
\pd^c\pd_c \ar - 3\eta_{ab} \pd^cA\pd_c \ar.
\end{align}
This can be conveniently written as the sum of two contributions,
one of which is just a total derivative and will have a zero
average. Using the identity $\pd_a (\ar \pd_b \ar) = \pd_a \ar \pd_b
\ar + \ar \pd_a \pd_b \ar$ and its contracted form we finally find
\begin{align}\label{G2aeta}
-G_{ab}^{(2)}[2\ar\eta] = - 2 \pd_a \ar \pd_b \ar - \eta_{ab}
\pd^cA\pd_c \ar - 4 \left[ \pd_a (\ar \pd_b \ar) - \eta_{ab} \pd^c
(\ar \pd_c \ar) \right],
\end{align}
where the terms in brackets have a vanishing average. Using the wave
equation for $\ar$ then $\pd^c (\ar \pd_c \ar) = \pd^c \ar \pd_c
\ar$ and it is
\begin{align}\label{G2aeta2}
-G_{ab}^{(2)}[2\ar\eta] = - 2 \pd_a \ar \pd_b \ar +3 \eta_{ab}
\pd^cA\pd_c \ar - 4 \pd_a (\ar \pd_b \ar).
\end{align}

\section{Averaged second order equation and zero point energy balance}

We are finally in the position to present the second order equation
\eqref{nonlin12} in a more explicit form. This will allow to address
the question of whether conformal fluctuations can function as a
balancing agent for the large amount of zero point energy. By
collecting together the results \eqref{xixi}, \eqref{sigma12},
\eqref{G2cross} and \eqref{G2aeta} we find
\begin{align}\label{secordF1}
G^{(1)}_{ab}[\gamma^{(2)}] = 8\pi& \left( T^{(2)}_{ab}[\psi]
-\frac{1}{8\pi} G^{(2)}_{ab}[\hgw] \right)  + 4 \left [ \eta_{ab}
\pd^c (\ar \pd_c \ar) - \pd_a (\ar \pd_b \ar) \right ]\nonumber\\
& + G^{(1)}_{ab}[-2B\eta + \ar^2\eta]  + \left[ \text{terms } (\pd
\ar\pd\hgw ) \right],
\end{align}
which, as long as we do not consider the explicit structure of the
cross terms, is valid in any gauge. Taking the spacetime average
these vanish thanks to the property $\m{\hgw_{ab,\ldots}
\ar_{,\ldots}} = 0$. All the other terms containing $A$ or $B$ are
total derivatives. Their average also vanishes since all
fluctuations are assumed to be stationary. Thus the averaged,
classical scale equation is:
\begin{equation}
G^{(1)}_{ab}[\m{\gamma^{(2)}}] = 8\pi \left(\m{ T^{(2)}_{ab}[\psi]}
+ \Tgw_{ab}[\hgw] \right),
\end{equation}
where, as defined in Appendix \ref{ap4}, the GWs stress energy
tensor is
\begin{equation}
\Tgw_{ab} := -\frac{1}{8\pi}\m{G^{(2)}_{ab}[\hgw]}.
\end{equation}
This is traceless and contributes together with other matter fields
to build up vacuum energy at the classical scale. \emph{This results
shows that conformal fluctuations cannot provide, within GR, a
mechanism to obtain vacuum energy cancelation}. This would seem to
imply that either zero point energy is not real and it does not
gravitate or else that there should be a insofar undiscovered
hypothetical ingredient for cancelation. In this case, this would be
thought to be included within the collection of matter fields
$\psi$.

\section{Equivalence of the physical and conformal
metric formalisms}\label{equi}

We now come to the important matter of building an explicit
\emph{microscopic} solution for the spacetime physical metric. The
formalism developed so far was based on the idea of choosing an
ansatz metric $g_{ab} = [\O(A)]^2 \gamma_{ab}$ \emph{first}, on
which \ee is imposed; \emph{secondly} the perturbation scheme is
applied using $\gamma_{ab} = \eta_{ab} + \xi_{ab}
+\gamma^{(2)}_{ab}$. We will refer to this procedure as to the
\emph{Conformal Metric Scheme}, i.e. \emph{impose ansatz solution
first-expand second}. An alternative procedure would be to expand
formally \ee in terms of the metric $g_{ab} = \eta_{ab} + h_{ab}
+g^{(2)}_{ab}$ first and after impose the ansatz solution directly
on the linear and second order perturbations $h_{ab}$ and
$g^{(2)}_{ab}$ by using the expressions \eqref{con}. We will refer
to this procedure as to the \emph{Physical Metric Scheme}, i.e.
\emph{expand first-impose ansatz solution second}. The analysis in
Section \ref{sec1} shows that these two procedures are totally
equivalent at first order. This is implied by the result
\begin{equation}
G^{(1)}_{ab}[h] = 0 \quad \Leftrightarrow \quad G^{(1)}_{ab}[\xi] =
2\pd_a \pd_b A - 2 \eta_{ab} \pd^c\pd_c\,A,
\end{equation}
that holds together with the linear algebraic constraint
\begin{equation}
h_{ab} = \xi_{ab} + 2A\eta_{ab}.
\end{equation}

An important question now is whether the two procedures are in fact
equivalent \emph{even to second order}. We stress that this is not a
priori obvious. Indeed, when carrying out the expansion scheme, it
may well be that the strong nonlinearity inherent to \ee could lead
in the two cases to physically inequivalent results. In fact this
belief had been central to the work in \cite{wango8rg}. Ironically,
this had been fueled by the fact that the already cited mistake
actually led to vacuum energy cancelation; a result which would have
indeed been physically inequivalent to what one finds within the
\emph{Physical Metric Scheme}!\footnote{In this footnote we leave
the \emph{pluralis maestatis} aside to highlight some details
regarding the historic development of this PhD work. Notwithstanding
the plausible hints, it was towards the end of my first attempt to
write up my PhD results that I came to look at the possibility that
the two expansion schemes are physically inequivalent as suspicious.
That was mainly because the identity of the two formalisms at linear
order was so obvious and clear that I found surprising it would not
hold at second order. I came to a point in which I believed there
must be some error in the treatment of the second order terms in
\cite{wango8rg}. This effectively allowed me to spot the mistake
described in the last part of section \ref{ch2ansecordew}. This then
quickly led to the results presented in the reminder of this
chapter.} Contrary to our early expectations, we now show that the
two procedures are in fact \emph{equivalent}, \emph{even to second
order}. The fact that conformal fluctuations cannot provide a
mechanism to balance vacuum energy at the large scale also depends
heavily on this equivalence.

This equivalence will simplify the structure of the second order
equation in a very severe way: at the end of this section we will
show that it is still possible, at least formally, to impose a
suitable microscopic condition on the second order conformal
fluctuations, in such a way that the physical metric \emph{seems} to
depend on $\ar^2$; however we will show how the constraint that must
be imposed on $\ar$ to achieve this is so strict that the solution
must be ruled out as unphysical. The conclusion will be that
standard GR does seem \emph{not to allow} for the possibility of a
conformally modulated metric that may in turn induce quantum
dephasing.

To see how this conclusion can be achieved let us see how the
physical metric scheme would work. We start from the physical metric
$g_{ab} = \eta_{ab} + h_{ab} + g^{(2)}_{ab}$ and expand \ee $
G_{ab}[g] = 8\pi T_{ab}[\psi] $ up to second order. This is a
standard procedure and results in the two equations:
\begin{equation}\label{ee1}
G^{(1)}_{ab}[h] = 0,
\end{equation}
\begin{equation}\label{ee2}
G^{(1)}_{ab}[g^{(2)}] = 8\pi \left( T^{(2)}_{ab}[\psi]
-\frac{1}{8\pi} G^{(2)}_{ab}[h] \right).
\end{equation}
Next we bring in the information related to the conformal
fluctuations, i.e. we use $g_{ab} = (1+A+B)^2\gamma_{ab}.$ This
translates in the algebraic constraints:
\begin{equation}\label{1stconst2}
h_{ab} = \xi_{ab} + 2A\eta_{ab},
\end{equation}
\begin{equation}\label{2ndconst2}
g^{(2)}_{ab} = \gamma^{(2)}_{ab} + 2 A \xi_{ab} + 2 B \eta_{ab} +
A^2\eta_{ab},
\end{equation}
which must now be substituted into equations \eqref{ee1} and
\eqref{ee2} starting from first order. We will thus get two
equations for the conformal 4-metric perturbations in the Physical
Metric Scheme.

In order to parallel the formalism developed above we need to solve,
with Boyer's boundary conditions, equation \eqref{ee1} first. This
simply gives
\begin{equation}\label{sol1a}
h_{ab} = \hgw_{ab}.
\end{equation}
Then the constraint \eqref{1stconst2} yields again
\begin{equation}\label{sol1b}
\xi_{ab} = \hgw_{ab} - 2A\eta_{ab}.
\end{equation}
This completes the first order analysis which yields of course the
same results as we found above with the Conformal Metric Scheme.

To find the second order equation for $\gamma^{(2)}$ we substitute
the first order solution \eqref{sol1a}-\eqref{sol1b} and the
constraint \eqref{2ndconst2} into \eqref{ee2} to find:
\begin{equation*}
G^{(1)}_{ab}[\gamma^{(2)} + 2A \hgw - 3A^2\eta + 2B\eta] = 8\pi
\left( T^{(2)}_{ab}[\psi] -\frac{1}{8\pi} G^{(2)}_{ab}[\hgw]
\right).
\end{equation*}
This is better rearranged as
\begin{equation}\label{secordF2}
G^{(1)}_{ab}[\gamma^{(2)}] = 8\pi \left( T^{(2)}_{ab}[\psi]
-\frac{1}{8\pi} G^{(2)}_{ab}[\hgw] \right) + G^{(1)}_{ab}[2A^2\eta]
- G^{(1)}_{ab}[2A \hgw] - G^{(1)}_{ab}[2B\eta - A^2\eta].
\end{equation}
By comparing with the previous result \eqref{secordF1} we notice
that the two equations have at least the same structure. The two
equations (i.e. the two alternative schemes) are truly equivalent if
it is true that
\begin{equation}\label{v1}
G^{(1)}_{ab}[2A^2\eta] \equiv 4 \left [ \eta_{ab} \pd^c (A \pd_c A)
- \pd_a (A \pd_b A) \right ]
\end{equation}
and
\begin{equation}\label{v2}
- G^{(1)}_{ab}[2A \hgw] \equiv \left[ \text{terms } (\pd A\pd\hgw )
\right].
\end{equation}
We now show that these are in fact two identities.\\
\\
\textbf{Verification for the term} $G^{(1)}_{ab}[2A^2\eta]$:\\
\\
This is straightforward. We start by noting the identity
\begin{equation*}
4 \left [ \eta_{ab} \pd^c (A \pd_c A) - \pd_a (A \pd_b A) \right ] =
2 \eta_{ab} \pd^c \pd_c A^2 - 2 \pd_a \pd_b A^2.
\end{equation*}
Then, by simply looking at \eqref{general}, we see immediately that
\begin{equation}\label{yyy}
G^{(1)}_{ab}[2A^2\eta] = 2 \eta_{ab} \pd^c\pd_c\,A^2 - 2\pd_a \pd_b
A^2,
\end{equation}
so that \eqref{v1} \emph{is} verified in an arbitrary gauge.\\
\\
\textbf{Verification for the term} $- G^{(1)}_{ab}[2A \hgw]$:\\
\\
This is more tricky as there are a lot of cross terms to check. For
this reason it will simplify things a bit if we work with
$\hgw_{ab}$ in the \texttt{TT} gauge. We start with the r.h.s. of
\eqref{v2}, i.e. by putting together all the cross terms we found in
the previous section. To have shorter formulas we will use here the
notation $h$ instead of $\hgw$. Equations \eqref{xixi2},
\eqref{sigma12cross}, \eqref{G2cross} yield
\begin{align}\label{temp1}
\left[ \text{terms } (\pd A\pd\hgw ) \right] = &\left[3 h_{\aa\bb}
\pd^\cc\pd_\cc A - \eta_{\aa\bb}{h}^{\cc\dd}\pd_{\cc}\pd_{\dd} A
\right]\nonumber\\
&- \left[ \pd^\dd A \left( \pd_\aa h_{\bb\dd} + \pd_\bb h_{\aa\dd}
-\pd_\dd h_{\aa\bb} \right) + 2h_{\aa\bb}\pd^\cc\pd_\cc A -
2\eta_{\aa\bb} h^{\dd\eee}\pd_\dd\pd_\eee A \right]\nonumber\\
&+ [\pd A \pd h]_{G^{(2)}},
\end{align}
where we recall that $[\pd A \pd h]_{G^{(2)}}$ indicates all the
cross terms resulting from $-G^{(2)}_{ab}[\hgw - 2A\eta]$.

To find these we start from the expression \eqref{G2n3} for
$R_{ab}^{(2)}[\,\cdot\,]$. We act on $\hgw_{\aa\bb} -
2A\eta_{\aa\bb}$ and collect all the cross terms. By using the
\texttt{TT} gauge conditions, $\pd^\bb h_{\aa\bb} = 0,$
$\eta^{\aa\bb} h_{\aa\bb} = 0$ and the wave equation $\pd^\cc
\pd_\cc h_{\aa\bb} = 0$ we find:
\begin{align*}
R_{\aa\bb}^{(2)}[\hgw_{\aa\bb} - 2A\eta_{\aa\bb}]\Rightarrow&\,
\pd_\aa({h^\cc}_{\bb}\pd_\cc A) - \pd^\cc A \pd_\cc h_{\aa\bb} +
\pd^\dd A \pd_\bb h_{\dd\aa} - \pd^\cc A \pd_\aa h_{\bb\cc} -
\pd^\cc A \pd_\bb h_{\aa\cc}\\
&-\eta_{\aa\bb} h^{\dd\cc} \pd_\dd \pd_\cc A + {h^\cc}_\aa
\pd_\cc\pd_\bb A.
\end{align*}
In evaluating the trace, many terms vanish because of the
\texttt{TT} conditions. We have:
\begin{align*}
R^{(2)}[\hgw_{\aa\bb} - 2A\eta_{\aa\bb}]\Rightarrow&\,
\pd^\bb({h^\cc}_{\bb}\pd_\cc A) - 4 h^{\dd\cc} \pd_\dd \pd_\cc A +
h^{\cc\bb} \pd_\cc\pd_\bb A = -2 h^{\cc\dd}\pd_\dd\pd_\cc A.
\end{align*}
Finally collecting all the cross terms in $-G_{\aa\bb}^{(2)}[\hgw -
2A\eta] \equiv -R_{\aa\bb}^{(2)} + \frac{1}{2}\eta_{ab} R^{(2)}$
yields
\begin{align*}
[\pd A \pd h]_{G^{(2)}} = & - \pd_\aa({h^\cc}_{\bb}\pd_\cc A) +
\pd^\cc A \pd_\cc h_{\aa\bb} - \pd^\dd A \pd_\bb h_{\dd\aa} +
\pd^\cc A \pd_\aa h_{\bb\cc} +
\pd^\cc A \pd_\bb h_{\aa\cc}\\
& - {h^\cc}_\aa \pd_\cc\pd_\bb A.
\end{align*}
Going now back to equation \eqref{temp1} and putting all terms
together gives the neat result:
\begin{equation}\label{comp1}
\left[ \text{terms } (\pd A\pd\hgw ) \right] =
h_{\aa\bb}\pd^\cc\pd_\cc A + \eta_{\aa\bb}h^{\dd\cc}\pd_\dd\pd_\cc A
+ 2\pd_\cc h_{\aa\bb} \pd^\cc A - \pd_\aa (h_{\cc \bb} \pd^\cc A) -
\pd_\bb (h_{\cc\aa}\pd^\cc A),
\end{equation}
which should be compared to $G^{(1)}_{ab}[-2A h]$. This is very
straightforward. Using \eqref{g1def3} and \eqref{R123} and the
\texttt{TT} gauge conditions for $h_{ab}$ we have
\begin{equation}\label{comp2}
G^{(1)}_{\aa\bb}[-2A h] = h_{\aa\bb}\pd^\cc\pd_\cc A +
\eta_{\aa\bb}h^{\dd\cc}\pd_\dd\pd_\cc A + 2\pd_\cc h_{\aa\bb}
\pd^\cc A - \pd_\aa (h_{\cc \bb} \pd^\cc A) - \pd_\bb
(h_{\cc\aa}\pd^\cc A).
\end{equation}
We see that the r.h.s. is indeed the same as in \eqref{comp1}. We
have thus verified equation \eqref{v2}, at least in the \texttt{TT}
gauge, i.e. it is indeed true that
\begin{equation}
- G^{(1)}_{\aa\bb}[2A \hgw] = \left[ \text{terms } (\pd A\pd\hgw )
\right].
\end{equation}

To summarize, \emph{this proves that the Conformal and Physical
Metric Schemes are in fact equivalent procedures as they result in
the same set of equations}. Namely, \emph{in the case in which
$G^{(1)}_{ab}[h] = 0$ is solved first and $\hgw_{ab} = h_{ab}$}, we
have by either procedure:
\begin{equation}
G^{(1)}_{ab}[\xi] = -G^{(1)}_{ab}[2A\eta_{ab}] \quad \Leftrightarrow
\quad G^{(1)}_{ab}[\hgw] = 0,
\end{equation}
and
\begin{equation}\label{xxx}
G^{(1)}_{ab}[\gamma^{(2)}] = 8\pi \Teff_{ab}[\psi,\hgw] +
G^{(1)}_{ab}[3A^2\eta - 2B \eta - 2A \hgw] \quad \Leftrightarrow
\quad G^{(1)}_{ab}[g^{(2)}] = 8\pi \Teff_{ab}[\psi,\hgw],
\end{equation}
where
\begin{equation}
\Teff_{ab}[\psi,\hgw] := \left( T^{(2)}_{ab}[\psi] -\frac{1}{8\pi}
G^{(2)}_{ab}[\hgw] \right)
\end{equation}
and
\begin{equation}
\hgw_{ab} = \xi_{ab} + 2A\eta_{ab},
\end{equation}
\begin{equation}\label{umf}
g^{(2)}_{ab} = \gamma^{(2)}_{ab} + 2 A \xi_{ab} + 2 B \eta_{ab} +
A^2\eta_{ab}.
\end{equation}
In particular this implies that, when one tries to assess the
overall amount of vacuum energy at the classical scale by taking the
average of the second order equation(s), it is
\begin{equation}
G^{(1)}_{ab}[\m{\gamma^{(2)}}] = 8\pi \m{\Teff_{ab}[\psi,\hgw]}
\quad \Leftrightarrow \quad G^{(1)}_{ab}[\m{g^{(2)}}] = 8\pi
\m{\Teff_{ab}[\psi,\hgw]},
\end{equation}
showing again that, if implemented as done so far, conformal
fluctuations do not contribute to the net vacuum energy amount.
Either the averaged physical or the conformal metric can serve to
describe the large scale structure of vacuum spacetime. The two
averaged metrics can be related as $\langle\gamma^{(2)}_{ab}\rangle
= \langle g^{(2)}_{ab}\rangle + C \eta_{ab}$, where $C$ can be an
arbitrary constant. This is indeed compatible with \eqref{umf} which
shows that $C = 3\m{A^2}$.

\section{Are spontaneous conformal fluctuations compatible with GR?}

Although the two expansion schemes are formally equivalent and lead
to no vacuum energy balance mechanism, the question remains to be
addresses about which metric between $g_{ab}$ and $\gamma_{ab}$ has
a \emph{true} dependence on $A^2$ at the microscopic level. Indeed
all we know, in general, is that the constraint \eqref{umf} must
hold. Answering whether $g^{(2)}_{ab}$ or rather $\gamma^{(2)}_{ab}$
really depends on $A^2$ is an important matter as this affects the
theoretical occurrence of quantum dephasing. This is so because,
within the standard GR framework we are analyzing now, a test
particle couples to the physical metric $g_{ab}$. This must present
a genuine dependence on $A^2$ if the particle wave function is to
suffer dephasing. As we discussed for the first order equation, the
question now boils down to which equation is solved first in
\eqref{xxx}. If one solves first the equation on the right then the
physical metric will \emph{not} depend on $A$ up to second order. In
this case the conformal field $A$ would just represent a formal
device that has no physical effect at all. As anticipated earlier
the key lies now in trying to \emph{fix the second order structure
of the conformal metric} in such a way that it does \emph{not}
contain any $A^2$ dependence. This would in turn induce a constraint
on the conformal field $A$ and its nonlinear correction $B$. If this
constraint can be satisfied, then the algebraic constraint
\eqref{umf} would automatically imply that the physical metric would
present a true nonlinear modulation $A^2\eta_{ab}$. We turn to this
central issue in the next section.

\subsection{Attempt for a solution of the second order equation}\label{fixalpha}

In order to implement these ideas a more detailed analysis of the
second order equation for the conformal metric nonlinear
perturbation is required. This is in general composed of a smooth
component $\ms{\gamma^{(2)}_{ab}}$ and a randomly fluctuating
component $\Delta\gamma^{(2)}_{ab}$:
\begin{equation}
\gamma^{(2)}_{ab} = \ms{\gamma^{(2)}_{ab}} +
\Delta\gamma^{(2)}_{ab}.
\end{equation}
Thus the second order equation gives rise to two separate equations.
One determines the background geometry and can be found by averaging
\eqref{xxx}. It reads:
\begin{equation}\label{bckgr}
G^{(1)}_{ab}[\ms{\gamma^{(2)}}] = 8\pi \m{\Teff_{ab}[\psi,\hgw]}
\quad \Leftrightarrow \quad G^{(1)}_{ab}[\m{g^{(2)}}] = 8\pi\m{
\Teff_{ab}[\psi,\hgw]},
\end{equation}
from which one can calculate the smooth perturbation induced over
the flat Minkowski. The second equation determines how the
fluctuations in the effective stress energy tensor induce extra
metric perturbations. It can be found by subtracting equation
\eqref{bckgr} from \eqref{xxx}. It reads:
\begin{equation}
G^{(1)}_{ab}[\Delta\gamma^{(2)}] = 8\pi \left(\Teff_{ab}[\psi,\hgw]
- \m{\Teff_{ab}[\psi,\hgw]} \right) + G^{(1)}_{ab}[3A^2\eta - 2B
\eta - 2A \hgw],
\end{equation}
which is equivalent to
\begin{equation}
G^{(1)}_{ab}[\Delta g^{(2)}] = 8\pi \left(\Teff_{ab}[\psi,\hgw] -
\m{ \Teff_{ab}[\psi,\hgw]}\right).
\end{equation}
In both these equations it is
\begin{equation}
\Teff_{ab}[\psi,\hgw] = \left( T^{(2)}_{ab}[\psi] -\frac{1}{8\pi}
G^{(2)}_{ab}[\hgw] \right)
\end{equation}
and
\begin{equation}\label{vacuum}
\m{\Teff_{ab}[\psi,\hgw]} = \m{T^{(2)}_{ab}[\psi]} + \Tgw_{ab}.
\end{equation}
Moreover
\begin{equation}
\hgw_{ab} = \xi_{ab} + 2A\eta_{ab},
\end{equation}
\begin{equation}\label{umf2}
\ms{g^{(2)}_{ab}} + \Delta g^{(2)}_{ab} = \ms{\gamma^{(2)}_{ab}} +
\Delta\gamma^{(2)}_{ab} + 2A \hgw_{ab} + 2B \eta_{ab} -
3A^2\eta_{ab}.
\end{equation}

As discussed in Section \ref{cbg} we assume
\begin{equation}
\m{\Teff_{ab}[\psi,\hgw]} = 0,
\end{equation}
as implied by observations. The second order equations thus take the
form:
\begin{equation}\label{bckcor}
G^{(1)}_{ab}[\ms{\gamma^{(2)}}] = 0 \quad \Leftrightarrow \quad
G^{(1)}_{ab}[\m{g^{(2)}}] = 0,
\end{equation}
for the background correction, and
\begin{equation}\label{zzz}
G^{(1)}_{ab}[\Delta\gamma^{(2)}] = 8\pi\Teff_{ab}[\psi,\hgw] +
G^{(1)}_{ab}[3A^2\eta - 2B \eta - 2A \hgw]
\end{equation}
equivalent to
\begin{equation}
G^{(1)}_{ab}[\Delta g^{(2)}] = 8\pi\Teff_{ab}[\psi,\hgw],
\end{equation}
for the second order fluctuations. Equations \eqref{bckcor} for the
background correction imply that $\ms{\gamma^{(2)}}$ (and
$\m{g^{(2)}}$) could in general be given by any smooth GWs
perturbation. Since we are focusing here on the random scale
structure of vacuum only, we can make the simple choice
$\ms{\gamma^{(2)}} = \m{g^{(2)}} = 0$.

\subsection{Second order constraint on the conformal
field}\label{Esol}

We now proceed to build explicitly a solution in which the physical
metric seems to depend \emph{formally} on the conformal
fluctuations. To this end we start from equation \eqref{zzz} and
first re-write it as
\begin{equation}
G^{(1)}_{ab}[\Delta\gamma^{(2)} + 2A \hgw] =
8\pi\Teff_{ab}[\psi,\hgw] + G^{(1)}_{ab}[3A^2\eta - 2B \eta].
\end{equation}
The main idea now is that $\gamma^{(2)}$ would \emph{not} depend on
$A^2$ if we could somehow manage to impose the condition
\begin{equation}
G^{(1)}_{ab}[3A^2\eta - 2B \eta] = 0.
\end{equation}
From equation \eqref{G1aeta} this reads explicitly
\begin{equation}\label{ccoonn}
G^{(1)}_{ab}[3A^2\eta - 2B \eta] = \eta_{ab} \pd^c\pd_c\,(3A^2 - 2B)
-\pd_a \pd_b (3A^2 - 2B) = 0.
\end{equation}
Thus it is satisfied provided that
\begin{equation}\label{fuch}
\pd_a \pd_b B = \frac{3}{2}\pd_a \pd_b A^2.
\end{equation}
As it had to be expected, this shows that the hope to be able to
build a solution where the physical metric contains a true physical
dependence on the conformal fluctuations is doomed to failure.
Indeed we have two options to satisfy equation \eqref{fuch}. The
most obvious is to set
\begin{equation*}
    B = \frac{3}{2}A^2.
\end{equation*}
In this case however we see from equation \eqref{umf2} that the
physical metric wouldn't have in fact any dependence on either $A^2$
or $B$! Its fluctuating part would simply be given by
\begin{equation}
\Delta g^{(2)}_{ab} = \Delta\gamma^{(2)}_{ab} + 2A \hgw_{ab}.
\end{equation}

The term $\Delta\gamma^{(2)}_{ab}$ represents the fluctuations
induced upon the Minkowski background by the fluctuations in the
effective stress energy tensor. We notice that this is precisely the
same kind of problem that Verdaguer and collaborators studied in
\cite{MartinVerdaguer2000} using the stochastic gravity approach. As
we already mentioned at the beginning of Section \ref{rga} their
result is that the matter fields tend to suppress the induced metric
perturbations above the Planck scale. Too see that even our
classical random approach leads to a qualitatively similar
conclusion let us consider the fluctuations backreaction second
order equation:
\begin{equation}\label{gf}
G^{(1)}_{ab}[\Delta\gamma^{(2)} + 2A \hgw] =
8\pi\Teff_{ab}[\psi,\hgw].
\end{equation}
This could in principle be solved for $E_{ab} := \Delta\gamma^{(2)}
+ 2A \hgw$. Working in the Lorentz gauge this is a simple
inhomogeneous wave equation for the trace reversed perturbation
$\bar{E}_{ab}$
\begin{equation*}
\pd^c\pd_c \bar{E}_{ab} = -16\pi \Teff_{ab}[\psi,\hgw].
\end{equation*}
The general solution can be written in terms of the retarded Green's
function \cite{wald84}. In coordinates components it reads:
\begin{equation*}
\bar{E}_{\mu\nu}(x) =
4\int_{\Sigma}\frac{\Teff_{\mu\nu}(x')}{\abs{\xv - \xv'}} dS(x'),
\end{equation*}
where $\Sigma$ denotes the past light cone of the point $x$ and the
volume element on the light cone is $dS = r^2 dr d\Omega$. Without
going into the details that would not affect anyway the main
conclusion, it is clear that, being the source a zero average term,
the solution for $\bar{E}_{ab}$ will be slow varying and smoother
than the fields $\psi$ and $\hgw$ inducing the fluctuations in the
source. Thus is because performing the spacetime integral of the
zero average fast varying source will tend to smooth out and
suppress the fluctuations. For our present purposes we do not need
to calculate the induced fluctuations $E_{ab}$ in any detail, and we
simply need to know that these would be characterized by larger
variations scales and would have a more regular behavior than the
fluctuations in $\psi$ or $\hgw_{ab}$. We thus write the solution of
equation \eqref{gf} symbolically as
\begin{equation*}
\Delta\gamma^{(2)} = \widetilde{E_{ab}} - 2A \hgw,
\end{equation*}
where the notation $\widetilde{E_{ab}}$ indicates a slow varying,
long wavelength second order perturbation. The physical metric thus
would follow as
\begin{equation*}
    g_{ab} = \eta_{ab} + \hgw_{ab} + \widetilde{E_{ab}}.
\end{equation*}
This is the result that one expects from a standard perturbation
scheme applied to \ee without even trying to introduce any conformal
fluctuations: the Minkowski background here presents a linear
perturbation $\hgw_{ab}$ describing GWs and a slow varying
perturbation $\widetilde{E_{ab}}$ induced by the fluctuations in the
backreaction effective stress energy tensor. The conformal
fluctuations have disappear altogether and do not play any role.

An alternative possibility to satisfy the constraint \eqref{ccoonn}
would appear to require
\begin{equation*}
    B=0, \quad  \text{and} \quad \pd_a \pd_b A^2 = 0.
\end{equation*}
Proceeding exactly as before to deduce $\Delta\gamma^{(2)} = E_{ab}
- 2A \hgw$ we would now find that the physical metric is formally
given by
\begin{equation}\label{out1}
g_{ab} = \eta_{ab} + \hgw_{ab} - 3A^2\eta_{ab} + E_{ab},
\end{equation}
which appears to possess the desired formal dependence on $A^2$.
However, even though the solution we have built seems to be
mathematically feasible, it is based on the very severe constraint
$\pd_a \pd_b A^2 = 0$. This looks very restrictive and artificial
and would imply a very unphysical spacetime dependence for the
conformal field which, we recall, has also been assumed to satisfy
$\pd^c\pd_c A = 0.$

The conclusion is quite strong: \emph{all the elements suggest that
within the random gravity framework, a vacuum solution of \ee for
standard GR is not compatible with the presence of spontaneous
metric conformal fluctuations. Surely they do not lead to any vacuum
energy cancelation and the attempt to built a solution with a true
conformal modulation leads to formal inconsistences.}

\subsection{An alternative scenario? conformal metric and
GWs}\label{choice2}

The treatment given above was based on the fact that the physical
metric was assumed to represent GWs through its linear perturbation
$h_{ab}$. As anticipated earlier we now wish to explore whether the
alternative formalism in which GWs are described through the
conformal 4-metric linear perturbation $\xi_{ab}$ can be given any
sense. The idea that the conformal 3-metric may represent GWs in the
sense of carrying GR `true' degrees of freedom was proposed by York
\cite{york1971,york1972} in his study of the canonical formulation
of the theory and related initial value problem. More recently this
as also been advocated in \cite{wang05a}, where the usual
geometrodynamics approach has been extended in such a way that the
conformal metric and related conformal factor play a central role.
Moreover this is the approach in another paper by Wang and
co-workers \cite{wang2006}, where a metric of the form $g_{ab} =
(1+A)^2\gamma_{ab}$ was considered and where $\gamma_{ab}$ was
claimed to encode the GWs perturbation. In the same work the authors
claimed to find a vacuum energy balancing mechanism between a
negative contribution due to $A$ and the usual positive contribution
due to the GWs.

To reconsider this possibility in detail we start from the first
order equation \eqref{eqlin}
\begin{equation*}
G^{(1)}_{ab}[\xi] = 2\pd_a \pd_b A - 2 \eta_{ab} \pd^c\pd_c\,A.
\end{equation*}
The requirement $\hgw_{ab}:=\xi_{ab}$, i.e. $G^{(1)}_{ab}[\xi] = 0$,
implies the constraint equation on $A$:
\begin{equation}\label{constrI}
\pd_a \pd_b A = 0.
\end{equation}
By taking the trace we have
\begin{equation}\label{constrIb}
\pd^c \pd_c A = 0.
\end{equation}
Technically the next step would be to find the general solution of
the wave equation with Boyer's fluctuating boundary conditions and
then restrict its form appropriately so that \eqref{constrI} is
satisfied. Again we denote the result of this procedure by the
symbol $\ar.$

Next, after having solved the equation $G^{(1)}_{ab}[\xi] = 0$ in
the \texttt{TT} gauge, leading to $\xi_{ab} = \hgw_{ab}$, the
physical metric linear perturbation follows from the algebraic
constraint \eqref{con1} as
\begin{equation}
h_{ab} = \hgw_{ab} + 2\ar\eta_{ab}.
\end{equation}
We remark that $h_{ab}$ also satisfies the linearized \ee equation
$G^{(1)}_{ab}[h] = 0.$ However this comes in this case only as a
consequence of $\hgw_{ab}$ and $\ar$ being already \emph{fixed}
through their respective equations. In particular we can e.g. assume
that $\hgw_{ab}$ has been found in such a way as to satisfy the
\texttt{TT} gauge conditions. We then see that the physical metric
perturbation would present a \emph{randomly} fluctuating trace
\emph{induced} by $\ar$ and given by
\begin{equation}
h:=\eta^{ab}h_{ab} = 8\ar.
\end{equation}
Notice that one cannot think of performing a further gauge
transformation to put $h_{ab}$ in a \texttt{TT} gauge, thus
reabsorbing the effect of $\ar$ in the definition of a new
coordinate frame. We observe indeed that to do so one should
consider a change of coordinates $x^{\mu} \mapsto x^{\mu} + \delta
v^{\mu}$ whose generator $\delta v^{\mu}$ would have to be itself
fluctuating. This would correspond to pass from the macroscopic
observer frame to an effective microscopic fluctuating frame. We
simply assume that this cannot be done as long as we just wish to
describe physics from the point of view of a macroscopic observer.
As a consequence of the above construction we now have that the
conformal 4-metric linear perturbation and the conformal
fluctuations are independent and, as a consequence, statistically
uncorrelated, i.e. we have
\begin{equation}\label{uncorr2}
\m{\xi_{ab,\ldots} \ar_{,\ldots}} \equiv \m{\hgw_{ab,\ldots}
\ar_{,\ldots}} = 0.
\end{equation}

We now consider the second order equation \eqref{nonlin1}
\begin{align}
G^{(1)}_{ab}[\gamma^{(2)}] = & \;8\pi T_{ab}^{(2)}[\psi] -
G^{(2)}_{ab}[\xi] - G^{(\xi\xi)}_{ab}[\xi] - \left( 2 \pd_a \ar
\pd_b \ar + \eta_{ab}\pd^c \ar \pd_c \ar
\right)\nonumber\\
& +\left(2\pd_a\pd_b B - 2\eta_{ab}\pd^c\pd_c
B\right)-\left(\pd_a\pd_b \ar^2 - \eta_{ab}\pd^c\pd_c \ar^2\right) +
S_{ab}[\ar,\xi],
\end{align}
and we simply use $\xi_{ab} = \hgw_{ab}$. This leads to a much
simpler formalism than before. Indeed $G^{(1)}_{ab}[\hgw] = 0$
implies that it is simply $G^{(\xi\xi)}_{ab}[\hgw] = 0$ -see
equation \eqref{linop3}- and that $-G^{(2)}_{ab}[\hgw]$ satisfies
all of Isaacson's property so that it will represent, after average,
the GWs stress energy tensor. We have
\begin{align}\label{soe}
G^{(1)}_{ab}[\gamma^{(2)}] = & 8\pi \Teff_{ab} - \left( 2 \pd_a \ar
\pd_b \ar + \eta_{ab}\pd^c \ar \pd_c \ar
\right)\nonumber\\
& +\left(2\pd_a\pd_b B - 2\eta_{ab}\pd^c\pd_c
B\right)-\left(\pd_a\pd_b \ar^2 - \eta_{ab}\pd^c\pd_c \ar^2\right) +
S_{ab}[\ar,\hgw],
\end{align}
where we defined again
\begin{equation*}
\Teff_{ab} := \left(T^{(2)}_{ab}[\psi]
-\frac{1}{8\pi}G^{(2)}_{ab}[\hgw]\right).
\end{equation*}
If we take the average we obtain now
\begin{align}\label{ji}
G^{(1)}_{ab}[\m{\gamma^{(2)}}] =  8\pi\m{\Teff_{ab} -
\frac{1}{4\pi}\pd_a \ar \pd_b \ar},
\end{align}
where we have used $\m{\pd^c\ar\pd_c\ar} = 0$, as implied by
$\pd^c\pd_c\ar = 0,$ and $\m{\hgw_{ab,\ldots} \ar_{,\ldots}} = 0.$
At first sight it would seem that in this formalism the conformal
fluctuations could really serve as an explicit balancing mechanism
for the traceless part of the vacuum energy contained in
$\Teff_{ab}$. However this is \emph{not} the case. Indeed to arrive
at \eqref{ji} we had to impose the constraint $\pd_a\pd_b \ar = 0.$
In this case we see that the identity $\pd_a\pd_b \ar^2 = 2\left(
\pd_a\ar \pd_b\ar + \ar \pd_a \pd_b \ar \right)$ reads:
\begin{equation}\label{fw}
\pd_a\pd_b \ar^2 = 2 \pd_a\ar \pd_b\ar.
\end{equation}
Because the left side is a total derivative we have that, in fact,
it must be
\begin{equation*}
\m{\pd_a\ar \pd_b\ar} = 0.
\end{equation*}
We thus conclude that even if one attempts to describe GWs through
the conformal metric linear perturbation, conformal fluctuations do
\emph{not} lead to any explicit vacuum energy balance mechanism, not
even at a formal level.

In order to find an explicit microscopic second order solution we
can again exploit the fact established previously according to which
the conformal and the physical metric expansion schemes are
equivalent. Starting from $G^{(1)}_{ab}[g^{(2)}] = 8\pi \left(
T^{(2)}_{ab}[\psi] -\frac{1}{8\pi} G^{(2)}_{ab}[h] \right)$, it is
then straightforward to verify that the second order equation
\eqref{soe} can be written in an equivalent and more compact form as
\begin{equation}
G^{(1)}_{ab}[\gamma^{(2)}] = 8\pi \Teff_{ab} -
G^{(2)}_{ab}[2\ar\eta] - G^{(1)}_{ab}[2\ar\hgw + A^2\eta + 2B\eta] +
[\pd \ar \pd h]_{G^{(2)}},
\end{equation}
where $[\pd \ar \pd h]_{G^{(2)}}$ indicates the cross terms
resulting from $-G^{(2)}_{ab}[\hgw + 2\ar\eta]$. This can be
re-written as
\begin{equation}
G^{(1)}_{ab}[\gamma^{(2)} + 2\ar\hgw] = \left\{8\pi \Teff_{ab} -
G^{(2)}_{ab}[2\ar\eta] + [\pd \ar \pd h]_{G^{(2)}}\right\} -
G^{(1)}_{ab}[(A^2 + 2B)\eta],
\end{equation}
where the term in brackets has again a zero average. The situation
now is even worse than in the previous case. Indeed, in order to
obtain a microscopic solution in which $\gamma^{(2)}$ does not
depend alone on $\ar^2$ we would need to impose the formal
constraint $B=0$ and $\pd_a\pd_b \ar^2 = 0$. From \eqref{fw} it
would then follow
\begin{equation}
\pd_a\ar \pd_b\ar = 0,
\end{equation}
which, again, appears to be too restrictive and leading to an
unphysical result.

\section{Summary}

In this and the previous chapter we have provided a general
framework which allows to model some of the effects related to
vacuum down to the random scale $\l = \lambda \Lp$ by using
classical fluctuating fields. The main motivation was to provide a
concrete framework in which conformal fluctuations of the metric
could enter as a new physical actor and thus be seen as a cause of
dephasing through the nonlinear effect they would have on the
spacetime metric. This was attempted within standard GR by looking
for a solution of \ee in the form $g_{ab} = (1 + A + B)^2
\gamma_{ab}$, where $A$ and $B$ encode the conformal fluctuations.
The main idea was to fix somehow the structure of the conformal
metric $\gamma_{ab}$ in such a way to induce a true dependence of
the physical metric on the conformal field $A$. This can be formally
done by inserting the ansatz solution form for $g_{ab}$ into \eep,
thus obtaining a linear and second order constraint equations for
$A$ and $B$. The hope that such an approach could lead to physically
viable results was based upon the possibility that the two expansion
schemes of \ee based on the metric $g_{ab}$ or $\gamma_{ab}$ were in
fact inequivalent to second order due to the nontrivial nonlinear
features of the problem. However we have verified that the expansion
schemes are in fact equivalent both to first and second order. This
implies that the set of constraints that one would need to impose to
build a solution in which the physical metric contains an $A^2$
dependence are mathematically too restrictive and unphysical.
Formally, there are two main scenarios: one in which the physical
metric encodes the GWs and an alternative scenario in which one
seeks to describe GWs through the conformal metric linear
perturbation. We have shown that in both cases acceptable solutions
for the physical metric do not contain a true dependence on $A^2$.
Moreover we have shown that, contrary to previous claims, conformal
metric fluctuations within standard GR do not offer a viable way for
a vacuum energy balance mechanism. It is our main conclusion at this
point that, though an interesting possibility, the idea that
conformal fluctuations of the metric may represent a true physical
actor of the microscopic physics close to the Planck scale is not
tenable within GR. The next chapter is devoted to investigating
whether and how such conclusions may change within some modified
theories of gravity such e.g. scalar-tensor.

\newpage
\thispagestyle{empty}


\singlespacing
\chapter{Scalar-tensor theories and conformal fluctuations}\label{ch4}

\begin{quote}
\begin{small}
In this chapter we study conformal transformations from the deeper
point of view of local units transformations. In this case all
lengths, times and masses also transform along with the metric
tensor. As noted by Bekenstein and other authors before him this
implies that all the fields of the standard model have in fact a
conformally invariant dynamics. For this to be true the
transformation properties of rest masses are essential. Requiring
that gravity be also conformally invariant leads to Bekenstein
action for the metric tensor and a novel gauge field which couples
directly to rest masses. As pointed out by Bekenstein, this theory
of conformally invariant gravity is in fact physically equivalent to
GR, the equivalence being manifest in the particular conformal frame
where the mass field is constant. Reinterpreting Bekenstein theory
within the random gravity framework, the new gauge field is also
expected to possess spontaneous fluctuations at the random scale.
These are mathematically equivalent to conformal fluctuations and
this fact allows to re-interpret conformal fluctuations of the
metric as fluctuations of the mass field. In this sense the
framework of conformally invariant gravity offers a natural way to
introduce conformal fluctuations. Of course, this theory is still
essentially GR and these cannot have any physically observable
effect, in agreement with the results of the previous chapter.
However the renewed point of view makes clear how the action for
gravity should be modified in order for conformal fluctuations to
play a role. This leads us to consider scalar-tensor theories. After
reviewing some general facts, the analysis focuses on the special
class of theories due to Brans and Dicke. Application of the random
gravity scheme provides a valid vacuum solution where the physical
metric does possess a conformal modulation. This is induced by the
Brans-Dicke field spontaneous linear fluctuations, which naturally
satisfy a standard wave equation. This facts happen independently of
the specific value of the Brans-Dicke coupling parameter and provide
a physical basis for the dephasing of a quantum particle studied in
Chapter \ref{ch1}. Together with matter fields and GWs, the
Brans-Dicke field linear fluctuations consistently contribute to a
correction of the background geometry through a standard
Klein-Gordon stress-energy tensor whose coupling to gravity depends
linearly on the Brans-Dicke parameter. In particular, for large
negative values the theory would provide a ghost field which could
alleviate the usual problem of the large zero point energy.
\end{small}
\end{quote}
\singlespacing

\section{Conformally invariant physics and Bekenstein's theory}

We start the analysis by showing that, even if they do not seem to
lead to physically observable effects, conformal fluctuations within
GR can at least be introduced in a more satisfactory way through the
random fluctuations of a novel gauge scalar field. This is important
in that it sheds light on how one should properly interpret
conformal transformations (CT) of the metric. As already noted by
Dicke in his second paper on the Brans-Dicke theory \cite{dicke62},
these have to be seen as \emph{local change of physical units} in
such a way that all dimensional quantities, including lengths, times
\emph{and} masses must be transformed simultaneously. The failure to
do this leads to the usually asserted fact that masses break the
conformal invariance of physics. In fact, as it was already noted by
Bekenstein in his 1980 remarkable paper \cite{bekenstein1980}
\emph{Conformal invariance, microscopic physics, and the nature of
gravitation}, physics is indeed conformally invariant, in the sense
of being invariant under local changes of units. This fact has been
more recently acknowledged by a number of authors
\cite{flanagan04,faraoni06,Dabrowski2008}. Bekenstein's very lucid
treatment shows that the actions of \emph{all} matter fields in the
standard model of particles are conformally invariant in the sense
specified above. Moreover the requirement that gravity be itself
conformally invariant leads to the introduction of a new gauge field
that couples directly to, and in fact defines, particle masses.
Because of its importance for our results in relation to
scalar-tensor theories we will now review Bekenstein's theory in
some detail. This will allow us to show that the conformal
fluctuations introduced somehow `by hand' in the previous chapters
can be interpreted more elegantly as natural fluctuations in the
Bekenstein's gauge field.

\subsection{Conformal transformations as local transformation of physical units}

We start by considering a CT for the metric tensor
\begin{equation}\label{P2_ct}
\gc_{ab} = \O^2 g_{ab},
\end{equation}
where $\O^2$ is an arbitrary, positive and smooth dimensionless
spacetime function. In the following all the conformally transformed
quantities will be denoted with a bar.

To understand the role played by the mass in guaranteeing a
conformally invariant physics, it is instructive to consider the
example of a scalar field $\phbd$. It is well known that the
standard minimally coupled action is not conformally invariant. On
the other hand conformal invariance (CI) arises naturally if $\phbd$
couples to the metric scalar curvature. A straightforward
calculation shows that \cite{wald84}
\begin{equation}\label{P2_ce}
    \left( \gc^{ab}\bar{\cd}_a\bar{\cd}_b - \frac{1}{6}\bar{R}
    \right)(\O^{-1}\phbd) = \O^{-3}\left( {g}^{ab}{\cd}_a{\cd}_b - \frac{1}{6}{R}
    \right)\phbd,
\end{equation}
where all the barred quantities refer to the conformally transformed
metric $\gc_{ab}$. Defining the conformally transformed field
\begin{equation}
    \bar{\phbd}:=\O^{-1}\phbd,
\end{equation}
\eqref{P2_ce} says that the equation
\begin{equation}\label{P2_cise}
\left( {g}^{ab}{\cd}_a{\cd}_b - \frac{1}{6}{R}
    \right)\phbd = 0
\end{equation}
is conformally invariant.

We consider now a massive field, satisfying the equation $\left(
{g}^{ab}{\cd}_a{\cd}_b - \frac{1}{6}{R} \right)\phbd = m^2\phbd$.
Now equation \eqref{P2_ce} gives
\begin{equation}\label{P2_ce2}
    \left( \gc^{ab}\bar{\cd}_a\bar{\cd}_b - \frac{1}{6}\bar{R} -
    m^2\O^{-2}
    \right)\bar{\phbd} = \O^{-3}\left( {g}^{ab}{\cd}_a{\cd}_b -
    \frac{1}{6}{R} - m^2
    \right)\phbd.
\end{equation}
Because of the $\Omega$ dependent term in the left hand side it is
usually concluded that mass spoils CI. However, as discussed very
clearly by Bekenstein \cite{bekenstein1980}, this conclusion is a
misconception deriving from a common confusion between CT and scale
transformations. The latter require to transform the metric
according to \eqref{P2_ct} and to rescale the fields by powers of
$\O$ and are equivalent to an enlargement of the physical system
under study. On the other side, as originally discussed by Weyl
\cite{Weyl1952}, Dicke \cite{dicke62}, Hoyle and Narlikar
\cite{Hoyle1972}, CT should properly be interpreted as \emph{local
units transformations}. In this case \emph{all} lengths, time
intervals \emph{and} masses are stretched by a factor that depends
on the spacetime location. All lengths and durations are multiplied
by a factor $\Omega$ in such a way that velocities are unchanged, in
particular the speed of light. In order to deduce the transformation
law for the masses the key point here is to note that Compton
wavelengths make no exception and must also be transformed as any
other length: Bekenstein refers to them properly as
``\emph{fundamental metersticks in physics}''. As any other length,
they must transform under a CT like \eqref{P2_ct} as
\begin{equation}\label{P2_lt}
    \bar{\lambda}_c = \lambda_c \O.
\end{equation}
Considering that
\begin{equation}
    \lambda_c := \frac{h}{m c},
\end{equation}
where $c$ is the (conformally invariant) speed of light and $m$ the
rest mass of the particle under exam, and assuming that $h
\rightarrow h$ under a CT, then equation \eqref{P2_lt} implies the
following transformation law for the rest mass
\begin{equation}\label{P2_mt}
    \bar{m} =  m \Omega^{-1}.
\end{equation}
Including the transformation of rest mass, equation \eqref{P2_ce2}
now reads
\begin{equation}\label{P2_ce3}
    \left( \gc^{ab}\bar{\cd}_a\bar{\cd}_b - \frac{1}{6}\bar{R} -
    \bar{m}^2
    \right)\bar{\phbd} = \O^{-3}\left( {g}^{ab}{\cd}_a{\cd}_b -
    \frac{1}{6}{R} - m^2
    \right)\phbd,
\end{equation}
showing that the massive scalar field equation
\begin{equation}\label{P2_cise2}
\left( {g}^{ab}{\cd}_a{\cd}_b - \frac{1}{6}{R}
    \right)\phbd = m^2\phbd
\end{equation}
is indeed conformally invariant once the conformal transformation is
interpreted as a local change of all units.

\subsection{CI action for a massive scalar field}

The relativistic action that leads to equation \eqref{P2_cise2} is
\begin{equation}\label{P2_cia}
\S_{\phbd} = -\frac{1}{2}\int\left( \phbd_{,a}\phbd^{,a} +
\frac{1}{6}R\phbd^2 + m^2 \phbd^2\right)\rg d^4 x,
\end{equation}
where $g$ is the metric determinant and $_{,a}$ indicates coordinate
partial differentiation. We can verify that this is CI by
considering the transformation laws
\begin{align*}
    &g_{ab} \rightarrow  \gc_{ab} = \O^{2} g_{ab} \\
    \\
    &\rg \rightarrow \rgc = \rg\O^{4},\\
    \\
    &\phbd \rightarrow \bar{\phbd} = \phbd\, \O^{-1},\\
    \\
    &m \rightarrow \bar{m} = m\, \O^{-1}.
\end{align*}
The action in the conformally transformed frame is
\begin{equation}
\bar{\S}_{\bar{\phbd}} = -\frac{1}{2}\int\left(
\bar{\phbd}_{,a}\bar{\phbd}^{,a} + \frac{1}{6}\bar{R}\bar{\phbd}^2 +
\bar{m}^2 \bar{\phbd}^2\right)\rgc d^4 x.
\end{equation}
Using the above transformation laws one gets
\begin{equation}\label{P2_odt}
\rgc\bar{\phbd}_{,a}\bar{\phbd}^{,a} = \rg\phbd_{,a}\phbd^{,a} +
\rg\left( \O^{-2} \phbd^2 \O_{,a}g^{ab}\O_{,b} -
2\O^{-1}\phbd\,\phbd_{,a}g^{ab}\Omega_{,b} \right),
\end{equation}
\begin{equation}
\bar{m}^2 \bar{\phbd}^2 \rgc = m^2 \phbd^2 \rg.
\end{equation}
The scalar curvature term transforms as
\begin{equation}
    \rgc\frac{1}{6}\bar{R}\bar{\phbd}^2 = \rg\frac{1}{6}R\phbd^2 -
    \rg\phbd^2\O^{-1}\B_{g} \O,
\end{equation}
where $\B_g := g^{ab}\pd_a\pd_b$. Performing an integration by parts
on the second term and neglecting the vanishing boundary term gives
\begin{align*}
-\int d^4 x\rg\phbd^2\O^{-1}\B_{g} \O &= \int d^4 x\rg
\left(\phbd^2\O^{-1}\right)_{,a}g^{ab}\O_{,b}\\
&= \int d^4 x\rg \left( 2\O^{-1}\phbd\,\phbd_{,a}g^{ab}\Omega_{,b} -
\O^{-2} \phbd^2 \O_{,a}g^{ab}\O_{,b} \right).
\end{align*}
This cancels the $\O$ dependent terms in \eqref{P2_odt} so that $
\bar{\S}_{\bar{\phbd}} = \S_{\phbd}$ as required.

\subsection{Conformal invariance of Standard Model fields}

In fact \emph{all} the physical fields of the Standard Model are
conformally invariant once the transformation laws \eqref{P2_lt},
\eqref{P2_mt} are adopted together with \eqref{P2_ct}
\cite{bekenstein1980,Dabrowski2008}. Bekenstein shows this
explicitly for the actions of abelian and non-abelian gauge fields,
massive vector field, Dirac field, scalar field, Higgs field,
including all the relevant coupling terms. In particular, a physical
field with dimensions $L^{a}T^{b}M^{c}$ must transform under a CT
with $\O^{a+b-c}$. The fact that all the microscopic physics of the
standard model is CI is remarkable. To quote Bekenstein's own words
\begin{quote}
\emph{The extension of units transformations to the local level
(CT's), and the requirement of CI of physical laws then parallel the
promotion of gauge and internal invariances to the local level by
the introduction of gauge fields.}
\end{quote}
In this sense the conformal transformation $g_{ab} \rightarrow \O^2
g_{ab}$ is like a gauge transformation. As explained in the next
section, the requirement of local invariance under this kind of
gauge transformation implies the existence of a new gauge field,
deeply related to mass.

\subsection{Bekenstein's gauge field and GR as a natural consequence
of CI}\label{P2_rtm}

As by the usual practise we use the term \emph{conformal frame} (CF)
to indicate an arbitrary system of units, \emph{locally defined at
each point in the spacetime manifold}. If physics is to be CI, an
important consequence is that the rest mass $m_i$ of any given
particle field must be itself a spacetime field. This is because, in
an arbitrary CF, the rest mass of a given physical field will vary
over spacetime. If one insists on the fact that there is no
preferred system of units and that the laws of physics must be the
same within any arbitrary CF, then there is no other choice but to
admit that the rest mass corresponds to a physical field. A priori
there are as many mass fields as there are different types of
fundamental rest masses. However, if the ratios of all possible kind
of rest masses are strictly constant, i.e. $m_i / m_j =
\text{cont}$, it follows that there must exist \emph{only one} mass
field. This must be a scalar and defines fields masses through
\begin{equation}\label{P2_mf}
m_i := \frac{\ggg_{i} \hbar}{c} \mf,
\end{equation}
where $\ggg_i$ is a dimensionless mass coupling constant for the
physical field labeled by $i$. It follows that $\mf$ has the
dimensions of $L^{-1}$ and must transform under a $CT$ according to
\begin{equation}
\mf \rightarrow \bar{\mf} = \mf\O^{-1}.
\end{equation}
We will call $\mf$ the \emph{Bekenstein mass field}. It plays the
role of gauge field associated with CTs.

As all the other gauge fields in physics, even the mass field $\mf$
should have a dynamics. Bekenstein identifies the proper action
$\Sb$ by requiring the following properties:
\begin{enumerate}
  \item $\Sb$ is covariant;
  \item $\Sb$ depends only on $\mf$, $g_{ab}$ and their derivatives;
  \item the dynamical equation for $\mf$ can contain up to second
  derivatives;
  \item $\Sb$ is CI.
\end{enumerate}
In particular the last requirement is important and it is put
forward by Bekenstein mainly for esthetic reasons, in analogy with
the fact that all the other standard model fields are known to have
a CI dynamics. It corresponds to the idea that \emph{the whole of
physics should be CI}. The most general action satisfying these
requirements comes in the form \cite{bekenstein1980}
\begin{equation}\label{P2_mfaaa}
\Sb = \pm\frac{1}{2}\hbar c \int\left( \mf_{,a}\mf^{,a} +
\frac{1}{6}R\mf^2 + \lb\, \mf^4\right)\rg d^4 x,
\end{equation}
where $\lb$ is a dimensionless parameter, $d^4 x = c\,dt\, d^3 x$
and $\hbar c$ guarantees that $\Sb$ has the dimensions of a
relativistic action. Apart for the $\pm$ sign, this action has the
same form as that in \eqref{P2_cia} with the quartic
self-interaction term playing the role of the mass term.

It is important to note that the requirement of CI (with the metric
transformation $g_{ab} \rightarrow \O^2 g_{ab}$) automatically
implies that one \emph{must} consider a general curved geometry
which then enters the action through the scalar curvature $R$. This
fact has a fundamental consequence. Indeed, by the principle of CI
one can always perform a CT to a conformal frame where the mass
field is represented by a constant spacetime function, i.e.
$\bar{\mf} = \mf_0$. In this CF particle masses are also constant
and the action reads
\begin{equation}\label{PS_EH}
\bar{\S}_{\mf_0} =  \pm\kappa\int\left(\bar{R} -
2\Lambda_0\right)\rgc d^4 x,
\end{equation}
where we defined
\begin{equation}\label{P2_GC}
    \kappa:=\frac{\hbar c \, \mf^2_0}{12},
\end{equation}
and
\begin{equation}
    \Lambda_0:=-3 \lb \mf_0^2.
\end{equation}
The masses of the physical fields are constant and given by
\begin{equation}\label{P2_m}
m_i = \frac{\ggg_i\hbar}{c}\mf_0.
\end{equation}
Selecting the plus sign, one recognizes in \eqref{PS_EH} the usual
Hilbert Einstein action in empty space with a \emph{bare
cosmological constant} term. Performing the functional derivative
with respect to the inverse metric yields indeed
\begin{equation}
    \frac{\delta\bar{\S}_{\mf_0}}{\delta\gc^{ab}} = \bar{R}_{ab} - \frac{1}{2}\gc_{ab}\bar{R} + \Lambda_0
    \gc_{ab} = 0.
\end{equation}
Since conventionally $\kappa = c^4 / (16 \pi G)$ the gravitational
constant $G$ follows from \eqref{P2_GC} as
\begin{equation}\label{P2_GG}
    G = \frac{3c^3}{4\pi\hbar \mf_0^2}.
\end{equation}

The variability of $G$ due to its dependence upon the arbitrary
constant $\mf_0$ simply reflects the variation in its numerical
value in different \emph{global} systems of units. To check whether
there is any intrinsic variation that could lead to physically
observable effects one has to look at the dimensionless
gravitational strength, defined as the ration between the
gravitational and electric forces that two protons exert on each
other. If
\begin{equation}
    F_g = \frac{G m_p^2}{d^2},\hspace{1cm} F_e =
    \frac{1}{4\pi\varepsilon_0}\frac{e^2}{d^2},
\end{equation}
this is given by
\begin{equation}
    \frac{F_g}{F_e} = \frac{G m^2_p}{e^2 / 4\pi\varepsilon_0},
\end{equation}
where $m_p$ is the proton mass. On the other hand
\begin{equation}
\frac{e^2}{4\pi\varepsilon_0} = \alpha \hbar c,
\end{equation}
where $\alpha \approx 1 / 137$ is the fine structure constant. The
strength of gravity is accordingly defined as
\begin{equation}
    \gamma:=\frac{G m_p^2}{\hbar c}.
\end{equation}
A real, intrinsic variation of the strength of gravity would
manifest itself as a variation of $\gamma$. Within the present CI
theoretical framework and in the CF introduced above with $\mf =
\mf_0$, equations \eqref{P2_m} and \eqref{P2_GG} yield
\begin{equation}
    \gamma = \frac{3c^3}{4\pi\hbar \mf_0^2}\times g_p^2
    \mf_0^2\left(\frac{\hbar}{c}\right)^2
    \times \frac{1}{\hbar c} \equiv \frac{3 g_p^2}{4\pi}.
\end{equation}
Since the mass coupling constants are assumed to be constant it
follows that $\gamma$ is constant \emph{in every} CF. Thus, in the
CF where particle masses are constant, the action
\begin{equation}\label{P2_fca}
\Sb = \frac{1}{2}\hbar c \int\left( \mf_{,a}\mf^{,a} +
\frac{1}{6}R\mf^2 + \lb\, \mf^4\right)\rg d^4 x,
\end{equation}
is manifestly equivalent to the standard Hilbert action of GR with
no matter sources. However, since this action is CI, the physical
predictions of the theory will be the same in \emph{every} CF, in
such a way that $\Sb$ should be regarded, in general, as \emph{the}
action for gravity. It is remarkable that the principle of CI
together with that of General Covariance, according to which the
laws of physics must be invariant under general coordinate
transformations \emph{and} local change of units, basically
naturally imply that gravity be described by GR.

\subsubsection{Bare cosmological constant}

The term $\lb \mf^4$ plays in \eqref{P2_fca} the role of mass term.
The mass of the mass field would then follow as
\begin{equation}
    \mu := \frac{\sqrt{\lb}\hbar}{c}\mf.
\end{equation}
This formula is consistent with the general equation \eqref{P2_mf}
holding for all the other massive fields so that $\sqrt{\lb}$ plays
the role of a self-coupling constant for the mass field $\mf$. It
follows that a bare cosmological constant emerges in the theory if
and only if the mass field suffers a self-interaction. The simplest
possibility is that with no self-interaction and $\lb = \Lambda_0 =
\mu = 0.$ In this case the action reads
\begin{equation}\label{P2_fca2}
\Sb = \frac{1}{2}\hbar c \int\left( \mf_{,a}\mf^{,a} +
\frac{1}{6}R\mf^2\right)\rg d^4 x,
\end{equation}
and, as far as $\mf$ is concerned, it is equivalent to the action
for a massless scalar field. The corresponding dynamical equation
for $\mf$ is of course
\begin{equation}\label{P2_dem}
    \frac{\delta \Sb}{\delta \mf} = \left( \B - \frac{1}{6}R \right)\mf =
    0.
\end{equation}

\subsection{Gravitational equations in an arbitrary conformal frame}\label{P2_sec}

In an arbitrary CF, and with $\lb = 0$ the field equations for the
metric follow from
\begin{equation}
    \frac{\delta \Sb}{\delta g^{ab}} = 0.
\end{equation}
As we will see more in detail in the next section, \emph{as long as
the coupling to matter is neglected}, the action $\Sb$ is formally
equivalent to that of gravity non minimally coupled to a scalar
field $\phbd$. Capozziello \emph{et al}. \cite{Capozziello1997}
consider a general action of the form
\begin{equation}\label{P2_cpz}
\S = \int\left( \frac{1}{2}\phbd_{,a}\phbd^{,a} + F(\phbd)R -
V(\phbd)\right)\rg d^4 x,
\end{equation}
where, for simplicity of notation we set $\hbar = c = 1$. This
arises naturally within scalar-tensor theories of gravity
\cite{wagoner70} and is also closely related to the action for
Brans-Dicke theories \cite{fujii04}. Upon functional variation with
respect to $g^{ab}$, this yields the equation
\begin{equation}\label{capo}
    F(\phbd)G_{ab} = -\frac{1}{2}\Xi_{ab} - g_{ab}\B F(\phbd) +
    F(\phbd)_{;ab},
\end{equation}
where $_{;a}$ indicates the covariant derivative and where the
tensor $\Xi_{ab}$ is defined as
\begin{equation}
    \Xi_{ab} := \phbd_{,a}\phbd_{,b} -
    \frac{1}{2}g_{ab}\phbd_{,a}g^{ab}\phbd_{,b} + g_{ab} V(\phbd).
\end{equation}

The mass field action \eqref{P2_fca2} formally corresponds to
\eqref{P2_cpz} with the choice
\begin{equation}\label{ffs}
    F(\phbd) = \frac{\phbd^2}{12},\hspace{1cm}V(\phbd)=0.
\end{equation}
It follows that, in an arbitrary CF, the dynamical equation for the
spacetime metric corresponding to $\Sb$ takes the form
\begin{equation}
    G_{ab} = \frac{12}{\mf^2}\left[ F(\mf)_{;ab} - g_{ab}\B F(\mf)
    - \frac{1}{2} \left( \mf_{,a}\mf_{,b} -
    \frac{1}{2}g_{ab}\mf_{,c}\mf^{,c} \right)  \right].
\end{equation}
Considering that
\begin{equation}
F(\mf)_{;ab} = \left[\pd_a \left( \frac{\mf^2}{12}\right)
\right]_{;b} = \frac{1}{6}\left(  \mf\, \mf_{;ab} + \mf_{,a}\mf_{,b}
\right)
\end{equation}
and
\begin{equation}
\B F(\mf) = g^{ab}F(\mf)_{;ab} = \frac{1}{6}\left(  \mf\, \B\mf
+\mf_{,c}\mf^{,c} \right)
\end{equation}
we obtain
\begin{equation}\label{P2_de1}
    G_{ab} = 2\left( \frac{\mf_{;ab}}{\mf} - g_{ab}\frac{\B \mf}{\mf}
    \right) - \left( 4\frac{\mf_{,a}\mf_{,b}}{\mf^2} - g_{ab} \frac{\mf_{,c}\mf^{,c}}{\mf^2}
    \right).
\end{equation}
Defining now a smooth scalar field ${\a}$ through
\begin{equation}\label{P2_ia}
{\a} = \ln\frac{\mf}{\mf_0},
\end{equation}
where $\mf_0$ is an arbitrary constant with dimensions of $L^{-1}$
yields
\begin{equation}\label{P2_psg}
    \mf =: e^{{\a}} \mf_0.
\end{equation}
The following relations hold
\begin{align}
&\frac{\mf_{,a}}{\mf} = {\a}_{,a},\\
\nonumber\\
&\frac{\mf_{;ab}}{\mf} = {\a}_{;ab} + {\a}_{,a}{\a}_{,b},\\
\nonumber\\
&\frac{\B\mf}{\mf} = \B{\a} + {\a}_{,c}{\a}^{,c}.
\end{align}
The field equation expressed in terms of ${\a}$ reads
\begin{equation}\label{P2_de}
    G_{ab} = 2\left({\a}_{;ab} - g_{ab} \B {\a} \right)
    - 2 \left(  {\a}_{,a}{\a}_{,b} + \frac{1}{2}g_{ab} {\a}_{,c}{\a}^{,c}
    \right).
\end{equation}
We can re-write this as
\begin{align}\label{cie}
G_{ab}[g] =
&\,\,2\left[{\left(\ln\frac{\mf}{\mf_0}\right)}_{\!\!;ab} - g_{ab}
\B \left(\ln\frac{\mf}{\mf_0}\right) \right]\nonumber\\
&    - 2 \left[
{\left(\ln\frac{\mf}{\mf_0}\right)}_{\!\!,a}{\left(\ln\frac{\mf}{\mf_0}\right)}_{\!\!,b}
+ \frac{1}{2}g_{ab}
{\left(\ln\frac{\mf}{\mf_0}\right)}_{\!\!,c}{\left(\ln\frac{\mf}{\mf_0}\right)}^{\!\!,c}
\right],
\end{align}
which, apart for the matter fields not being included yet, is
formally equivalent to equation \eqref{eega1} with $\mf / \mf_0$
playing the role of $\O$.

We close this section by pointing out that the dynamical equation
\eqref{P2_dem} for $\mf$ that one derives by varying the action
$\Sb$ with respect to $\mf$ is automatically satisfied if the \ee is
satisfied, for an \emph{arbitrary spacetime function} $\mf$. We can
easily verify this by taking the trace of equation \eqref{P2_de1}.
Since $g^{ab}G_{ab} = -R$ we have
\begin{equation}
    -R \mf = 2\B \mf -8\B \mf - 4\frac{\mf_{,c}\mf^{,c}}{\mf} +
    4\frac{\mf_{,c}\mf^{,c}}{\mf},
\end{equation}
which yields $(\B - R/6)\mf = 0.$

\subsubsection{From the Jordan frame to the Einstein
frame}\label{JtoE}

To reveal clearly the connection to our treatment of conformal
fluctuations presented in the previous chapters we need to analyze
more closely the properties of the explicit CT that allows to
connect an arbitrary CF to the CF with $\mf = \mf_0 = \text{const}.$
In this section we shall refer to the arbitrary CF where $\mf$ is a
non-constant function of the spacetime coordinates non minimally
coupled to gravity as to the \emph{Jordan frame}. The CF in which
$\mf = \mf_0 = \text{const}$ will be referred to as the
\emph{Einstein frame}. We remark that, within the present framework
based on the CI of physics, it is clear a priori that \emph{these
two frames are equally good in providing a description of any given
physical system}. In particular they must lead to the same physical
predictions.

Imagine to select any Jordan frame where $\mf$ is not constant, in
such a way that the action \eqref{P2_fca2} yields the dynamical
equation \eqref{cie} for the Jordan frame metric $g_{ab}$. We then
perform a CT to a new CF, i.e. we transform the metric and the mass
field as
\begin{align}
&g_{ab} \rightarrow \gc_{ab} = \O^2 g_{ab},\\
\nonumber\\
&\mf \rightarrow \bar{\mf} = \mf \O^{-1}.
\end{align}
We now require that this CT allows to pass to an Einstein frame,
i.e. a frame where the mass field and all particle masses are
constant. Since it was introduced arbitrarily, there is no loss of
generality in requiring that, in the transformed frame, one has
\begin{equation}
    \bar{\mf} \equiv \mf_0.
\end{equation}
Comparing to equation \eqref{P2_psg} we see that it must be
\begin{equation}
    \O = e^{\a} \equiv \frac{\mf}{\mf_0}.
\end{equation}
The action in the transformed frame (Einstein frame) takes of course
the usual Hilbert action form and yields the \ee in empty space for
the metric $\gc_{ab}$ as discussed in Section \ref{P2_rtm}. In the
Einstein frame particle masses are constant and the mass field
doesn't appear explicitly in the equations of motion. Thus it
appears that the transformation $\gc_{ab} = e^{2\a} g_{ab}$ is
simply that CT that connects the particular Jordan frame with metric
$g_{ab}$ and spacetime dependent mass field $\mf(x) = e^{\a(x)}
\mf_0$ to the Einstein frame with metric $\gc_{ab}$ and constant
mass field $\bar{\mf} = \mf_0$.

We thus see how, within this new perspective of a conformally
invariant theory of gravity, conformal fluctuations would formally
be equivalent to the spontaneous fluctuations in the mass field
$\mf.$

\subsubsection{From the Einstein frame to the Jordan frame}

Conversely one can start from the action expressed in an arbitrary
Einstein frame with metric $\gc_{ab}$ and constant mass function
$\bar{\mf} = \mf_0 $, namely
\begin{equation}\label{P2_aef}
    \bar{\S}_{\mf_0} = \int
\frac{\mf_0^2}{12} \bar{R} \rgc d^4 x.
\end{equation}
Perform now the CT
\begin{align}\label{P2_ppp}
&\gc_{ab} \rightarrow g_{ab} = \omega^2 \gc_{ab},\\
\nonumber\\
&\mf_0 \rightarrow \mf = \mf_0 \omega^{-1},
\end{align}
where $\omega$ is an arbitrary and non-constant function of the
spacetime coordinates. To find the action in the new CF (Jordan
frame with variable mass function) we simply have to express the
`old' quantities $(\mf_0, \bar{R}, \sqrt{-\gc})$ in terms of the
`new' related quantities $(\mf, R, \sqrt{-g})$ and substitute
directly in \eqref{P2_aef}. We have
\begin{equation}\label{P2_eq}
    \mf_0 = \omega\,\mf,
\end{equation}
\begin{equation}
    \sqrt{-\gc} = \omega^{-4}\sqrt{-g}.
\end{equation}
The corresponding equation for $\bar{R}$ is \cite{Dabrowski2008}
\begin{equation}
 \bar{R} = \omega^2 R + 6 \omega \B \omega - 12
 \omega_{,c}\omega^{,c}.
\end{equation}
The action \eqref{P2_aef} now yields
\begin{equation}
\bar{\S}_{\mf_0} = \int \frac{\mf^2}{12} R \rg d^4 x + \int
\mf^2\left( \frac{1}{2}\frac{\B\omega}{\omega} -
\frac{\omega_{,c}}{\omega}\frac{\omega^{,c}}{\omega} \right) \rg d^4
x =: \S_1 + \S_2,
\end{equation}
where, in the second term $\S_2$, indices are raised by the new
inverse metric $g^{ab}$. Using equation \eqref{P2_eq} this term can
be re-written in terms of $\mf$ only as
\begin{equation}
\S_2 = \int \rg d^4 x \left[ \frac{1}{2}\mf^3 \B(\mf^{-1}) - \mf^4
(\mf^{-1})_{,c}(\mf^{-1})^{,c} \right].
\end{equation}
The terms containing the derivatives of the mass field yield
\begin{equation}
- \mf^4 (\mf^{-1})_{,c}(\mf^{-1})^{,c} = -\mf_{,c}\mf^{,c}
\end{equation}
\begin{equation}
\frac{1}{2}\mf^3 \B(\mf^{-1}) = \mf_{,c}\mf^{,c} -
\frac{1}{2}\mf\B\mf,
\end{equation}
so that
\begin{equation}
\S_2 = -\frac{1}{2}\int \rg d^4 x \mf\B\mf = \frac{1}{2}\int \rg d^4
x \mf_{,c}\mf^{,c}
\end{equation}
after an integration by parts. Finally, the action in the new Jordan
frame takes the desired form
\begin{equation}
\bar{\S}_{\mf_0} = \int \left(\frac{\mf^2}{12} R + \frac{1}{2}
\mf_{,c}\mf^{,c}\right) \rg d^4 x \equiv \Sb.
\end{equation}

\subsection{Inclusion of matter fields}

We have shown that the action $\Sb$ provides a CI framework for GR
at the expense of introducing a mass field $\mf$ that plays the role
of a new gauge field in the theory. This come as a consequence of
the fact that CT should properly be interpreted as local units
transformations that one must carry out \emph{simultaneously} on the
the metric $g_{ab}$, the mass field $\mf$ and all the other matter
fields that one wishes to include in the formalism. We stress again
the fact that, since the mass field couples to particle masses, its
transformation law implies that particle masses themselves transform
under CT. In particular, in an arbitrary Jordan frame, particle
masses will be varying from point to point in spacetime in
proportion to the variations in $\mf$. The usual Einstein frame
corresponds to the particular situation in which the mass field, and
thus particle masses, are constant.

Moreover we have shown how fluctuations in the mass field could
equivalently be interpreted as conformal fluctuations of the metric.
In this sense, if one accepts that gravity is conformally invariant
and by applying the random gravity framework to study vacuum
properties at the random scale, it would follow that conformal
fluctuations are indeed expected to be there after all! However the
question of whether these can actually affect test particles is
doomed to yield a negative answer. This depends entirely on how one
prescribes now the coupling of matter to gravity.

To see this let us notice that, as mentioned above, in the original
work by Bekenstein it is shown that \emph{all} the fields of the
particle Standard Model have a CI action once one transforms their
masses accordingly. In fact particle masses are directly coupled to
the mass field $\mf$ through $m_i = (\ggg_{i} \hbar / c) \mf$. It
follows that, in the presence of matter fields, the total
\emph{conformally invariant} action one should consider in an
arbitrary Jordan frame is
\begin{equation}\label{P2_tot}
    \Sjf := \Sb[g_{ab},\mf] +
    \S^{(m)}[\psi,g_{ab},\mf],
\end{equation}
where $\S^{(m)}$ is the action of matter fields, described
collectively by $\psi$. We stress here the important point that, in
an arbitrary Jordan frame, only the massive matter fields included
in $\S^{(m)}$ get a dependence on the mass field $\mf$ through their
rest masses. Apart for this fact, $\psi$ couples otherwise to the
metric $g_{ab}$ in the usual way.

The equation resulting from the above action follows upon functional
differentiation with respect to $g^{ab}$. In particular, matter
fields stress energy tensor will be defined as usual as
\begin{equation}
T_{ab}^{(m)}[\psi,g,\mf] := -\frac{2}{\rg}\frac{\delta
\S^{(m)}[\psi,g_{ab},\mf]}{\delta g^{ab}}.
\end{equation}
We notice from equations \eqref{capo} and \eqref{ffs} that the
variation of $\Sb[g_{ab},\mf]$ yields on the l.h.s. the Einstein
tensor multiplied by $\mf^2 / 12$. It follows that the conformal
invariant gravitational equation in an arbitrary conformal frame
take the form \cite{Dabrowski2008}:
\begin{equation}\label{BELLA}
    G_{ab}[g] = 2\left({\a}_{;ab} - g_{ab} \B {\a} \right)
    - 2 \left(  {\a}_{,a}{\a}_{,b} + \frac{1}{2}g_{ab} {\a}_{,c}{\a}^{,c}
    \right) + \frac{6}{\mf^2} T_{ab}^{(m)}[\psi,g,\mf],
\end{equation}
where
\begin{equation}
{\a} = \ln\frac{\mf}{\mf_0}.
\end{equation}

We thus see that the whole formalism of the previous chapters is
recovered if one inserts fluctuations in the mass field by
prescribing $\a = 1 + A + B$. The extra term $6 / \mf^2$ that
couples to $T_{ab}^{(m)}$ defines the gravitational constant and
contains fluctuations that would appear as higher than second order
corrections. Thus, up to second order, equation \eqref{BELLA} is
formally equivalent to equation \eqref{eega1}. Now however $g_{ab}$
indicates the physical metric which necessarily couples to the
intrinsic fluctuations in the mass field. In this sense our vacuum
solution would imply that $g_{ab}$ does indeed depend upon the
fluctuations in $\mf$. Unfortunately a massive test particle would
still be \emph{not} affected by such fluctuations. The reason is
that if it is true that the metric depends on $\a$, so does the
particle mass, which also fluctuates. This happens in such a way
that there cannot be any observable effect. This must be so exactly
because the theory is conformally invariant. As shown above one can
always transform away from the Jordan frame to an Einstein frame in
which the new $\bar{\mf}$ is constant. The corresponding Einstein
frame action would be
\begin{equation}
    \Sef := \bar{\S}_{\mf_0}[\gc_{ab}] +
    \bar{\S}^{(m)}[\bar{\psi},\gc_{ab}],
\end{equation}
where rest masses are exactly constant and the dependence on the
constant $\mf_0$ is simply absorbed in the particular \emph{global}
units one employs for the gravitational constant $G$. In this frame
the field equation is manifestly \ee with a source represented by
the stress energy tensor of matter coupled to the metric $\gc_{ab}$.
Indeed, when $\mf = \mf_0 = \text{const},$ equation \eqref{BELLA}
above reduces to the standard \ee with gravitational constant
\begin{equation}
    G = \frac{3}{4\pi\mf_0^2},
\end{equation}
in agreement with what we had defined in equation \eqref{P2_GG} with
$c = \hbar = 1$. The physical prediction in this gauge is that a
massive quantum particle would not suffer dephasing. \emph{By
conformal invariance this must then also be the result in the
general Jordan frame where the mass field is fluctuating}.

Our previous result that conformal fluctuations are not effective
within GR is thus confirmed. However the present deeper point view
also provides hints of how the situation could change. Indeed it is
clear that an action of the kind $\S = \Sb[g_{ab},\mf] +
\S^{(m)}[\psi,g_{ab}]$, where matter fields do \emph{not} couple
directly to $\mf$ would indeed lead to a different physical result.
Of course such a theory would exit the realm of standard GR: the
above action would have to be re-interpreted as a special case of a
theory of gravity coupled to an external field. In fact we will see
in the next section that such an action actually corresponds to a
Brans-Dicke theory with coupling parameter $\omega = - 3/2.$ Even
though such a value seems to be ruled out by observations within the
solar system \cite{fujii04}, this suggests considering scalar-tensor
theories of gravity more in detail and verifying what our random
gravity framework can then predict in relation to the problem of
dephasing of a quantum particle.

\section{Scalar-Tensor Theories of gravity}

Scalar tensor theories refer to alternative theories of gravity
where the gravitational field is described by a rank 2 symmetric
tensor $g_{ab}$ and one spin 0 scalar field $\phbd$. The first
examples of such theories were considered by Jordan \cite{jordan59}
and by Brans and Dicke \cite{brans61} in the attempt to incorporate
Mach principle within GR and find a general framework which could
accommodate for a varying gravitational `constant' $G$ whose value
was related to the local value of the scalar field $\phbd$, in turn
related to the energy-matter distribution of the universe. The
original class of Brans-Dicke theories was extended in the 70s by
Wagoner \cite{wagoner70}, who considered a more general action for
gravity and a scalar field. Such theories go under the name of
scalar-tensor. Since their original introduction, they have been
constrained by many observational tests, mainly within the solar
system. From the theoretical point of view it is known that gravity
description \emph{must} include a rank 2 symmetric tensor. However
there are not valid reasons to exclude the theoretical possibility
for an extra scalar field. More recently, more general theories with
nonlinear terms in $R$ included in the Einstein Hilbert action have
also been considered \cite{Capozziello1997}, in particular in
relation to their cosmological consequences. Beyond the original
motivation that had inspired Brans and Dicke's work, an extra
universal scalar field could have an important role in relation to
dark energy, the cosmological constant and inflation. Moreover
String Theory in its low energy limit also leads to the prediction
of a scalar field (dilaton) whose dynamics is governed by an action
in all respect similar to Brans-Dickes's \cite{brans05,fujii04}.

Wagoner action for a scalar-tensor theory is \cite{faraoni06}:
\begin{equation}\label{scten}
    \S_{W}=\int \d^4 x \rg \left[ \frac{f(\phbd)}{2}R - \frac{\omega(\phbd)}{2}g^{ab}\cd_a\phbd\cd_b\phbd - V(\phbd)
    \right] +  \int \d^4 x \rg \alpha_m \L^{(m)}[g,\psi],
\end{equation}
where $\L^{(m)}[g,\psi]$ is the matter fields lagrangian and $\phbd$
is the \emph{Brans-Dicke scalar field}. The functions $f, \omega$
and $V$ are arbitrary. Matter fields are only coupled to the metric
$g_{ab}$ and their stress-energy tensor is defined as usual as
\begin{equation*}
    T_{ab}^{(m)}:=\frac{-2}{\rg}\frac{\delta\S^{(m)}}{\delta
    g^{ab}}.
\end{equation*}
It is possible to show that this is conserved in the usual sense:
\begin{equation}
\cd^a T^{(m)}_{ab} = 0,
\end{equation}
which also implies that point-like test particles follow the
geodesic of the physical metric $g_{ab}$, so that the WEP is
satisfied. Experimental constraints on $\omega$, deduced from
measurements of the frequency shift of radio signals to and from the
Cassini-Huygens spacecraft, suggest \cite{bertotti03,shapiro04}:
\begin{equation*}
    \abs{\omega(\phbdo)} > 40000,
\end{equation*}
where $\phbdo$ represents the present value of the scalar field.
Such a form of the action defines the theory in the so called
\emph{Jordan Frame} (JF) in terms of the gravitational variables
$(g_{ab},\phbd)$. In the Jordan frame the scalar field is
non-minimally coupled to the metric and matter fields suffer gravity
through direct interaction with the metric $g_{ab}$ \emph{only}. On
the other hand the scalar field contributes, together with other
form of matter, to set up the physical metric. Such a theory is
characterized by a varying gravitational constant defined by
\begin{equation*}
    G_{\text{\scriptsize{eff}}}:= \frac{1}{8\pi f(\phbd)}.
\end{equation*}

It is well known that the theory can be equivalently formulated in
such a way that the action goes over to the usual Hilbert-Einstein
with a standard, minimally coupled scalar field. This can be
achieved by means of the following CT
\cite{wagoner70,flanagan04,faraoni06}:
\begin{equation}
\gc_{ab}=\O^2 g_{ab},\quad\text{with}\quad \O = \sqrt{f(\phbd)},
\end{equation}
\begin{equation}
\phic= \int
\frac{d\phbd}{f(\phbd)}\sqrt{f(\phbd)+\frac{3}{2}\left(\frac{df}{d\phbd}\right)^2}
\end{equation}
with allows to pass to the \emph{Einstein Frame} (EF), in terms of
the variables $(\gc_{ab},\phic)$. The action then takes the form:
\begin{equation*}
    \S_{W}=\int \d^4 x \rgc \left[ \frac{\Rc}{12\pi} - \frac{1}{2}\gc^{ab}\cdc_a\phic\cdc_b\phic - \Vc(\phic)
     + (G\phbd)^{-2} \L^{(m)}[\gc,\psi] \right],
\end{equation*}
where
\begin{equation*}
    \Vc := \frac{V}{\O^2}.
\end{equation*}
In this frame the action resembles the standard HE action with a
canonically coupled scalar field $\phic$. However this theory is
\emph{not} GR since the matter field lagrangian is now coupled to
$\phic$ through the coupling constant $[G\phbd(\phic)]^{-2}$. In
particular test particles do \emph{not} follow the geodesic of the
metric $\gc_{ab}$.

As already shown by Dicke \cite{dicke62} and as discussed in the
previous section in relation to Bekenstein's theory, the two frames
are physically equivalent \cite{faraoni06,flanagan04} provided that
the units of mass, length, time, and quantities derived therefrom
scale with appropriate powers of the conformal factor $\O$ in the
Einstein frame. In this sense it is still true that \emph{physics is
invariant under choice of the units}. In this point of view the
Einstein Frame contains a different system of units at each
spacetime location and the symmetry group of classical physics is
enlarged to include conformal transformations with the associated
rescaling of units. These must transform according to:
\begin{equation*}
    d\bar{t} = \O dt,\quad d\bar{x}^i = \O dx^i,\quad
    \bar{m}=\O^{-1}m .
\end{equation*}
The lucid discussion of Flanagan \cite{flanagan04} and Faraoni
\cite{faraoni06} make this point clear while providing many examples
of calculations in the literature that employ different conformal
frames and actually do yield the same physical prediction because
the scaling of units under CTs was properly taken into account.

Summarizing one has: in the \emph{Jordan frame} $h$, $c$, particle
masses $m$ and all coupling constant of physics are constant, while
the gravitational `constant' $G$ varies and test particles follows
metric geodesic; in the \emph{Einstein frame} $G$, $h$ and $c$ are
constant while particle masses, fields couplings and units of
length, time and mass vary with spacetime location.

\section{Brans-Dicke theory}

The original Brans-Dicke theory \cite{brans61} is a special case of
scalar-tensor theory and is recovered with the choice:
\begin{equation*}
    f(\phbd) = 2\phbd; \quad \omega(\phbd) =
    \frac{2\omega}{\phbd},\quad\!\!\omega=\text{const};
    \quad V(\phbd) = 0.
\end{equation*}
Equation \eqref{scten} then yields the Brans-Dicke action:
\begin{equation}
    \S_{BD}=\int \d^4 x \rg \left[ \phbd R - \frac{\omega}{\phbd}g^{ab}\cd_a\phbd\cd_b\phbd
    \right] + \int \d^4 x \rg \alpha_m\L^{(m)}[g,\psi].
\end{equation}
By applying the usual variational principle with respect to the
metric this yields the dynamical equation \cite{brans05}:
\begin{equation}\label{pme}
G_{ab}[g] = \frac{1}{2\phbd}T^{(m)}_{ab} + \frac{1}{\phbd}\left(
\cd_a\cd_b\phbd - g_{ab}\cd^c\cd_c\phbd \right) +
\frac{\omega}{\phbd^2}\left( \cd_a\phbd \cd_b \phbd -
\frac{1}{2}g_{ab}\cd^c \phbd \cd_c \phbd \right)
\end{equation}
for the spacetime metric $g_{ab}$. By varying with respect to the
scalar field yields instead:
\begin{equation}\label{sfe}
\cd^c\cd_c \phbd = \frac{1}{4\omega + 6} T^{(m)},
\end{equation}
where $T^{(m)}$ denotes the trace of the matter fields stress energy
tensor. In this theory the gravitational constant $G$ is related to
$1/\phbd$ and in general varies from point to point. Its local
values is determined through $\phbd$ in response to the
matter-energy distribution throughout the universe, in agreement
with Mach's principle.

The Brans Dicke action can be written in an equivalent way that will
be useful for us by defining \cite{fujii04}:
\begin{equation*}
\epsilon:= \text{Sign}(\omega),\quad \xi:= \frac{\epsilon}{4\omega}
> 0,\quad \phbd := \frac{1}{2}\xi \ph^2.
\end{equation*}
Then we have:
\begin{equation}\label{BD2}
    \S_{BD}=\int \d^4 x \rg \left[ \frac{1}{2}\xi\ph^2 R - \frac{\epsilon}{2}g^{ab}\cd_a\ph\cd_b\ph
    \right] + \int \d^4 x \rg \alpha_m \L^{(m)}[g,\psi].
\end{equation}
This reveals that, as long as matter fields are neglected,
Bekenstein's action \eqref{P2_fca2} corresponds to a Brans Dicke
model with $\epsilon = -1$ and $\xi = 1/6$, i.e. $\omega = -3/2.$

This form of the action is also useful to prove the equation for the
scalar field \eqref{sfe}. Starting from equation \eqref{pme} for the
metric and taking the trace we have:
\begin{equation}\label{equ1}
    \Box \phbd - \frac{1}{3}R\phbd +
    \frac{\omega}{3}\frac{\cd^c\phbd\cd_c\phbd}{\phbd} =
    \frac{T^{(m)}}{6}.
\end{equation}
On the other hand, varying the Brans-Dicke action in its form
\eqref{BD2} with respect to the field $\ph$ yields\footnote{This is
obviously so since it is already known that that action must yield
the conformally invariant Klein-Gordon equation when $\xi = 1/6$ and
$\epsilon = -1$.}
\begin{equation*}
\epsilon \Box \ph + \xi \ph R = 0.
\end{equation*}
Multiplying by $\ph$ and using $\phbd = \xi \ph^2 / 2$ gives
\begin{equation}\label{equ2}
    \epsilon \ph \Box \ph + 2\phbd R = 0.
\end{equation}
It is also
\begin{equation}\label{equ3}
    \frac{\xi}{2}\Box\ph^2 \equiv \Box \phbd = \xi\left( \ph\Box\ph + \cd^c \ph \cd_c \ph
    \right)
\end{equation}
and
\begin{equation}\label{equ4}
\frac{\cd^c\phbd\cd_c\phbd}{\phbd} = 2\xi\cd^c\ph \cd_c\ph.
\end{equation}
Using \eqref{equ2} and \eqref{equ4} into \eqref{equ3} gives:
\begin{equation}
\frac{\cd^c\phbd\cd_c\phbd}{\phbd} = 2\Box \phbd + 4\phbd
R\xi\epsilon^{-1} = 2\Box \phbd + \frac{\phbd R}{\omega}.
\end{equation}
Substituting into equation \eqref{equ1} gives
\begin{equation}
    \Box \phbd - \frac{1}{3}R\phbd +
    \frac{\omega}{3}\left(2\Box \phbd + \frac{\phbd R}{\omega}\right) =
    \frac{T^{(m)}}{6}.
\end{equation}
Thus we see that the curvature terms simplify and we obtain
\begin{equation}
    \Box \phbd = \frac{T^{(m)}}{6+4\omega},
\end{equation}
which is indeed equation \eqref{sfe}.

\section{Random gravity framework and Brans-Dicke theory}

We now apply our random gravity framework to the problem of vacuum
fluctuations within the special class of scalar-tensor theories
given by the Brans-Dicke action. We work in the Jordan frame, in
such a way that the scalar field is non-minimally coupled and test
particles follow geodesics of the physical metric $g_{ab}$. The
relevant dynamical equation for the spacetime geometry is obtained
upon variation of the Brans-Dicke action as given by Fujii
\cite{fujii03}
\begin{equation}\label{BD22}
    \delta\S_{BD}=\delta \int \d^4 x \rg \left[ \frac{1}{2}\xi\ph^2 R - \frac{\epsilon}{2}g^{ab}\cd_a\ph\cd_b\ph
    +\alpha_m \L^{(m)}[g,\psi]\right] = 0.
\end{equation}
Apart for the constant $\epsilon = \pm 1$ this corresponds to the
action we already considered in \eqref{P2_cpz}. It is then
straightforward to verify that the modified \ee results in:
\begin{equation}\label{fineq}
G_{ab}[g]=\frac{1}{\xi\ph^2}T^{(m)}_{ab}[g,\psi]+\frac{1}{\ph^2}\left(\cd_a\cd_b\ph^2
- g_{ab}\cd^c\cd_c \ph^2\right)+\frac{\epsilon}{\xi\ph^2}\left(
\cd_a\ph\cd_b\ph - \frac{1}{2}g_{ab}\cd^c\ph\cd_c\ph \right),
\end{equation}
where we recall that the field $\ph$ is related to the Brans-Dicke
field $\phbd$ by
\begin{equation}
\phbd = \frac{1}{2}\xi\ph^2
\end{equation}
and where the matter fields stress energy tensor is defined by
\begin{equation}
    T_{ab}^{(m)}:=\frac{-2}{\rg}\frac{\delta\S^{(m)}}{\delta
    g^{ab}}.
\end{equation}
It is immediate to verify that \eqref{fineq} is indeed equivalent to
\eqref{pme}. We also remark that in the action \eqref{BD22} we
didn't include a potential term $V(\phbd)$, which could appear in
more general scalar-tensor theories having implications for the
cosmological constant.

We now wish to apply the random gravity framework to equation
\eqref{fineq} and study its implications at the level of the random
scale $\l = \lambda \Lp.$ The situation is now different from what
we had considered in the previous chapters. Indeed the scalar field
$\ph$ (or equivalently $\phbd$) now affects the spacetime metric
together with the other matter fields. It is thus expected that, in
the problem of vacuum at the random scale, the physical metric will
depend on the scalar field. A test particle or any physical field
couple indeed to the metric but are nonetheless influenced
indirectly by the scalar field since this does influence the metric.
That this must be the case can also be inferred by the fact that the
physical predictions in the Jordan frame we are using must be
equivalent to those that one would find in the Einstein frame after
a local units transformation (CT): in the Einstein frame the scalar
field minimally couples to gravity \emph{and} also couples directly
to all matter fields; as a result of this a test particle would
deviate from geodesic motion through the effect of an effective
force induced by the scalar field \cite{faraoni06}.

Ad discussed in Chapter \ref{ch2} we expect some effects related to
the quantum fluctuations of the various fields in their vacuum state
to be mimicked at the random scale by classical but randomly
fluctuating fields. To see how these fluctuations manifest
themselves we start by re-writing equation \eqref{fineq} in a more
convenient way. Using $\cd_a\cd_b \ph^2 = 2(\ph \cd_a\cd_b\ph +
\cd_a\ph\cd_b\ph)$ the term containing $\ph^2$ becomes:
\begin{equation}\label{pass}
\frac{1}{\ph^2}\left(\cd_a\cd_b\ph^2 - g_{ab}\cd^c\cd_c \ph^2\right)
= 2\left(\frac{\ph_{;ab}}{\ph}-g_{ab}\frac{\Box\ph}{\ph}\right)+
2\left(\frac{\ph_{,a}\ph_{,b}}{\ph^2}-g_{ab}\frac{\ph^{,c}\ph_{,c}}{\ph^2}\right),
\end{equation}
where $\Box = \cd^c\cd_c$. We define now
\begin{equation*}
    \a:=\ln\frac{\ph}{\ph_0},
\end{equation*}
where $\ph_0$ is an arbitrary constant. The Brans-Dicke field
$\phbd$ is then represented as
\begin{equation}\label{bdf}
\phbd = \phbdo e^{2\a},\quad \text{with} \quad \phbdo:=
\frac{1}{2}\xi\ph_0^2,
\end{equation}
in such a way that the field $\a$ can later be employed to describe
the vacuum random fluctuations at the random scale. We have
\begin{equation*}
    \frac{\ph_{,a}}{\ph}=\a_{,a}\quad\text{and}\quad\frac{\ph_{;ab}}{\ph}=\a_{;ab}+\a_{,a}\a_{,b}.
\end{equation*}
Then equation \eqref{pass} becomes
\begin{equation*}
\frac{1}{\ph^2}\left(\cd_a\cd_b\ph^2 - g_{ab}\cd^c\cd_c \ph^2\right)
= 2\left( \a_{;ab} - g_{ab}\Box\a \right) + 4\left( \a_a\a_b -
g_{ab} \a^c\a_c \right).
\end{equation*}
In terms of $\a$ the modified \ee now reads:
\begin{equation}\label{fineq2}
G_{ab}[g]=2\left( \a_{;ab} - g_{ab}\Box\a \right) +
\frac{e^{-2\a}}{2\phbdo}T^{(m)}_{ab}[g,\psi] + \left(\frac{4\xi +
\epsilon}{\xi}\right)\left[\a_{,a}\a_{,b} -
\frac{1}{2}\left(\frac{8\xi + \epsilon}{4\xi +
\epsilon}\right)\a^{,c}\a_{,c}\right]
\end{equation}
which, as anticipated, correctly reduces to \eqref{BELLA} for
$\epsilon=-1$ and $\xi=1/6$. The constant $\phbdo$ is related to the
value of the gravitational constant in the particular units that one
wishes to employ.

\subsection{Brans-Dicke field fluctuations at the random scale}

Equation \eqref{fineq2} provides the basis to study vacuum at the
random scale. We will employ the same technique that we used in our
study of conformal fluctuations within GR and we will perform an
expansion scheme up to second order. The fluctuations in the
Brans-Dicke scalar field will be encoded in $\a$, which we assume to
be a small quantity.

As a first step we consider the wave equation that holds for the
Brans-Dicke field:
\begin{equation}\label{we}
\Box \phbd = \frac{1}{4\omega + 6} T^{(m)}.
\end{equation}
Expanding the exponential to second order gives
\begin{equation*}
    \phbd = \phbdo e^{2\a} = \phbdo\left[1 + 2\a + 2\a^2
    +O(\varepsilon^3) \right].
\end{equation*}
Equation \eqref{we} now reads in terms of $\a$:
\begin{equation*}
\Box \phbdo\left[1 + 2\a + 2\a^2
    +O(\varepsilon^3) \right] = \frac{1}{4\omega + 6} T^{(m)},
\end{equation*}
which can conveniently be rearranged as
\begin{equation*}
\Box \a = \frac{1}{2\phbdo(4\omega + 6)} T^{(m)} - \Box \a^2,
\end{equation*}
and where we neglected third order corrections. Using now $\Box\a^2
= 2(\a\Box\a + \a^{,c}\a_{,c})$ we have
\begin{equation*}
(1 + 2\a) \Box \a   = \frac{1}{2\phbdo(4\omega + 6)} T^{(m)} +
2\a^{,c}\a_{,c}.
\end{equation*}
Multiplying by $(1+2\a)^{-1}$ and retaining only terms up to second
order we find
\begin{equation}
\Box \a  = \frac{1}{2\phbdo(4\omega + 6)} T^{(m)} + 2\a^{,c}\a_{,c},
\end{equation}
showing that the fluctuations field $\a$ satisfies a wave equation
with a source that depends on the trace $T^{(m)}$ of the matter
fields stress energy tensor and on an auto-interaction term
$2\a^{,c}\a_{,c}$.

It is in the spirit of the random gravity framework that the
Brans-Dicke field should have spontaneous vacuum fluctuations at the
random scale due to its supposed quantum nature. These do not depend
on other matter fields or energy sources. Accordingly we express the
fluctuations field $\a$ as
\begin{equation}
\a = A + B.
\end{equation}
Here $\abs{A} = O(\varepsilon)$ is a first order field satisfying
the homogeneous wave equation, while $\abs{B}=O(\varepsilon^2)$ is a
nonlinear correction that depends on the sources:
\begin{equation}\label{AB}
    \Box A = 0\quad \text{and} \quad \Box{B} = \frac{1}{2\phbdo(4\omega + 6)} T^{(m)} +
    2A^{,c}A_{,c}.
\end{equation}
The first of these equation can be solved with Boyer random boundary
conditions to yield a fluctuating field which we will denote by
$\ar.$ Thus the Brans-Dicke field at the random scale and up to
second order is given by
\begin{equation}
\phbd = \phbdo e^{2\a} = \phbdo\left[1 + 2\ar + 2B + 2\ar^2 +
O(\varepsilon^3)\right].
\end{equation}

\section{Expansion equations and vacuum solution}

We now proceed in expanding the equation
\begin{equation}\label{hs}
G_{ab}[g]=2\left( \a_{;ab} - g_{ab}\Box\a \right) +
\frac{e^{-2\a}}{2\phbdo}T^{(m)}_{ab}[g,\psi] + \left(\frac{4\xi +
\epsilon}{\xi}\right)\left[\a_{,a}\a_{,b} -
\frac{1}{2}\left(\frac{8\xi + \epsilon}{4\xi +
\epsilon}\right)\a^{,c}\a_{,c}\right],
\end{equation}
taking into account that $\a = \ar + B$. This equation establishes
how the spacetime metric is affected at the random scale by the
fluctuations in the Brans-Dicke field $\phbd$ and in all other
matter fields $\psi$. In general we expect that $g_{ab}$ will indeed
depend upon $\ar$ nonlinearly. We thus write down an \emph{ansatz}
solution of the kind
\begin{align}
g_{ab} &= \eta_{ab} + h_{ab} + g^{(2)}_{ab} +
O(\varepsilon^3)\nonumber\\
&= \eta_{ab} + \underbrace{\left(\xi_{ab} + \ki \ar
\eta_{ab}\right)}_{h = O(\varepsilon)} + \underbrace{\left( \kii B
\eta_{ab} + \kiii\ar^2\eta_{ab} + \kiiii \ar \xi_{ab} +
\gamma^{(2)}_{ab} \right)}_{g^{(2)} = O(\varepsilon^2)} +\,
O(\varepsilon^3).
\end{align}
Here the constants $\ki,$ $\kii$, $\kiii$, $\kiiii$ and the metric
perturbations $\xi_{ab}$ and $\gamma^{(2)}_{ab}$ are to be
determined by imposing \eqref{hs}. The background has been chosen
equal to $\eta_{ab}$ in virtue of the discussion in Section
\ref{cbg}.

\subsection{First order solution}

Using the same notation of Chapter \ref{ch2} we define
$\Sigma_{ab}^1[\a]:=2\left( \a_{;ab} - g_{ab}\Box\a \right)$. This
will be giving rise to both first order and second order terms which
we denote by $\Sigma_{ab}^{1(1)}$ and $\Sigma_{ab}^{1(2)}$. Then the
first order expansion equation that follow from \eqref{hs} reads:
\begin{equation}\label{hs1}
G^{(1)}_{ab}[h] = \Sigma_{ab}^{1(1)}[\ar],
\end{equation}
where
\begin{equation*}
\Sigma_{ab}^{1(1)}[\ar] := 2\left( \pd_a\pd_b\ar -
g_{ab}\pd^c\pd_c\ar \right).
\end{equation*}
We know from equation \eqref{G1aeta} that $2\pd_a \pd_b \ar - 2
\eta_{ab} \pd^c\pd_c\,\ar = G^{(1)}_{ab}[-2\ar\eta]$. Then, using
$h_{ab} = \xi_{ab} + \ki \ar \eta_{ab}$ we have
\begin{equation}\label{hs12}
G^{(1)}_{ab}[\xi + (\ki + 2) \ar \eta] = 0.
\end{equation}
This can be easily satisfied by:
\begin{align}\label{solI}
\ki = -2,\quad \text{and} \quad G^{(1)}_{ab}[\xi] = 0.
\end{align}
In the \texttt{TT} gauge we thus have a wave equation for $\xi$. As
done previously this can be solved with fluctuating boundary
conditions and we have the solution
\begin{equation}
\xi_{ab} = \hgw_{ab}
\end{equation}
representing vacuum fluctuations of GWs at the random scale. To
first order the spacetime physical metric is thus:
\begin{align}
g_{ab} = \eta_{ab} + \left(\hgw_{ab} - 2 \ar \eta_{ab}\right) +
O(\varepsilon^2),
\end{align}
with
\begin{equation}\label{fos}
h_{ab} = \hgw_{ab} - 2 \ar \eta_{ab}.
\end{equation}
Moreover the GWs and Brans-Dicke field linear fluctuations are
totally uncorrelated, implying
\begin{equation}
    \m{\hgw_{,\ldots}\ar_{,\ldots}} = 0.
\end{equation}

\subsection{Second order equation}

The second order equation is very similar to what we already studied
in Chapter \ref{ch3}. Expanding the Einstein tensor and selecting
the second order terms coming from $\Sigma_{ab}^1[\a=A+B]$ gives:
\begin{equation}\label{hs2}
G^{(1)}_{ab}[g^{(2)}] = - G^{(2)}_{ab}[h] - G^{(hh)}_{ab}[h] +
\Sigma_{ab}^{1(1)}[B] + \Sigma_{ab}^{1(2)}[\ar,h] +
\Sigma_{ab}^{2(2)}[\ar] + \frac{1}{2\phbdo}T^{(m)(2)}_{ab}[\psi],
\end{equation}
where
\begin{equation}
\Sigma_{ab}^{2(2)}[\ar] := \left(\frac{4\xi +
\epsilon}{\xi}\right)\left[\ar_{,a}\ar_{,b} -
\frac{1}{2}\left(\frac{8\xi + \epsilon}{4\xi +
\epsilon}\right)\ar^{,c}\ar_{,c}\right]
\end{equation}
and where $T^{(m)(2)}_{ab}[\psi]$ is quadratic in the matter fields
$\psi$ and contains otherwise the Minkowski tensor; note that the
factor $\exp(-2\ar)$ does not add extra contributions to second
order.

To proceed we simply have to use the first order solution $h_{ab} =
\hgw_{ab} - 2 \ar \eta_{ab}$ and make the various terms in the
r.h.s. explicit. These are the same we already found in Chapter
\ref{ch3}. Equations \eqref{xixi2}, \eqref{G2cross} and
\eqref{G2aeta} tell us immediately:
\begin{equation}\label{d1}
-G^{(hh)}_{ab}[\hgw - 2\ar\eta] = 3 \hgw_{ab} \pd^c\pd_c \ar -
\eta_{ab}{\hgw}^{cd}\pd_{c}\pd_{d} \ar,
\end{equation}
and
\begin{align}\label{d2}
-G^{(2)}_{ab}[\hgw - 2\ar\eta] = &-G^{(2)}_{ab}[\hgw] - \left(2
\pd_a \ar \pd_b \ar + \eta_{ab}
\pd^c\ar\pd_c \ar\right)\nonumber\\
& - 4 \left[ \pd_a (\ar \pd_b \ar) - \eta_{ab} \pd^c (\ar \pd_c \ar)
\right] + [\pd \ar \pd \hgw]_{G^{(2)}},
\end{align}
where the first equation holds for $\hgw$ in the \texttt{TT} gauge
and where $[\pd \ar \pd \hgw]_{G^{(2)}}$ indicates all the cross
terms deriving from $G^{(2)}$.

The terms deriving from $\Sigma_{ab}^{1(2)}[\ar,h]$ can be found by
keeping just the second order terms in the expression
\begin{equation*}
\Sigma_{ab}^{1} = 2\cd_a \pd_b \ar - 2 g_{ab} \cd^c\pd_c\, \ar.
\end{equation*}
Using $\cd_a \pd_b \ar = \pd_a \pd_b \ar - {\Gamma^{c}}_{ab}[g]
\pd_c \ar$ and the linearized connection
\begin{equation}
    {{\Gamma^{c}}_{ab}}^{(1)} = \frac{1}{2}\eta^{cd}
    \left(\pd_a h_{bd} + \pd_b h_{ad} - \pd_d h_{ab}\right),
\end{equation}
we have
\begin{align}
\Sigma_{ab}^{1(2)} & = -\eta^{cd}\left( \pd_a h_{bd} + \pd_b h_{ad}
- \pd_d h_{ab} \right)\pd_c\ar + \eta_{ab}\eta^{dc}\eta^{ef}\left(
\pd_d h_{cf} +
\pd_c h_{df} - \pd_f h_{dc} \right)\pd_e\ar\nonumber\\
&\hspace{.4cm}-2h_{ab}\pd^c\pd_c\ar + 2\eta_{ab}h^{dc}\pd_d
\pd_c\ar,
\end{align}
where the last two term arise because $g_{ab}\cd^c = g_{ab}g^{dc}
\cd_d \approx (\eta_{ab} + h_{ab})(\eta^{dc} - h^{dc})\cd_d.$ Using
$h_{ab} = \hgw_{ab} - 2\ar\eta_{ab}$ to eliminate $h_{ab}$, we get a
series of terms involving cross products $\hgw_{ab}\ar$ plus terms
quadratic in $\ar$. Explicitly and in an arbitrary gauge:
\begin{align}
\Sigma_{ab}^{1(2)}[\ar,\hgw] = &\,4\pd_a\ar\pd_b\ar +
2\eta_{ab}\pd^c\ar\pd_c\ar\nonumber\\
&- \left[ \pd^d\ar \left( \pd_a \hgw_{bd} + \pd_b \hgw_{ad} -\pd_d
\hgw_{ab} \right) -  \eta_{ab} \pd^f\ar \left( \pd^c\hgw_{cf} +
\pd^d
\hgw_{df} - \pd_f \hgw \right)\right.\nonumber\\
& \hspace{0.5cm}\left.+ 2\hgw_{ab}\pd^c\pd_c\ar - 2\eta_{ab}
h^{\text{\scriptsize{GW}}de}\pd_d\pd_e \ar \right].
\end{align}
Specializing to the \texttt{TT} gauge we have
\begin{align}\label{sigma12crossBBB}
\Sigma_{ab}^{1(2)}[\ar,\hgw] = &\,4\pd_a\ar\pd_b\ar +
2\eta_{ab}\pd^c\ar\pd_c\ar\nonumber\\
&- \left[ \pd^d\ar \left( \pd_a \hgw_{bd} + \pd_b \hgw_{ad} -\pd_d
\hgw_{ab} \right) + 2\hgw_{ab}\pd^c\pd_c\ar - 2\eta_{ab}
h^{\text{\scriptsize{GW}}de}\pd_d\pd_e \ar \right],
\end{align}
and we see that this coincides with the result \eqref{sigma12cross}
for $S_{ab}[\ar,\hgw]$ in Chapter \ref{ch3}.

By putting together the results \eqref{d1} \eqref{d2} and
\eqref{sigma12crossBBB} we get
\begin{align}
&\hspace{1cm}\Sigma_{ab}^{1(2)}[\ar,\hgw] - G^{(hh)}_{ab}[\hgw,\ar]
- G^{(2)}_{ab}[\hgw,\ar] = \nonumber\\
&=-G^{(2)}_{ab}[\hgw] + \left(2 \pd_a \ar \pd_b \ar + \eta_{ab}
\pd^c\ar\pd_c \ar\right) + 4 \left[\eta_{ab} \pd^c (\ar \pd_c \ar)
-\pd_a (\ar \pd_b \ar) \right] + \left[ \text{terms } (\pd
\ar\pd\hgw ) \right],
\end{align}
where, from the results \eqref{v1} and \eqref{v2} derived in Chapter
\ref{ch3}, the last two groups of terms are simply:
\begin{equation}
4 \left [ \eta_{ab} \pd^c (\ar \pd_c \ar) - \pd_a (\ar \pd_b \ar)
\right ] = G^{(1)}_{ab}[2\ar^2\eta]
\end{equation}
and
\begin{equation}
\left[ \text{terms } (\pd \ar\pd\hgw ) \right]=- G^{(1)}_{ab}[2\ar
\hgw].
\end{equation}

Collecting together all these results, the second order equation
\eqref{hs2} reads:
\begin{align}\label{af}
G^{(1)}_{ab}[g^{(2)}] = &\,G^{(1)}_{ab}[2\ar^2\eta] -
G^{(1)}_{ab}[2\ar
\hgw] - G^{(1)}_{ab}[2B\eta] +\nonumber\\
&+\frac{1}{2\phbdo}T^{(m)(2)}_{ab}[\psi] -G^{(2)}_{ab}[\hgw]
+\left(6 + \frac{\epsilon}{\xi}\right)\left( \pd_a\ar\pd_b\ar -
\frac{1}{2}\eta_{ab}\pd^c\ar\pd_c\ar \right)
\end{align}
where we used $\Sigma_{ab}^{1(1)}[B] = 2\left( \pd_a\pd_b B -
g_{ab}\pd^c\pd_c B\right) \equiv G^{(1)}_{ab}[-2B\eta].$

\section{Second order equation solution}

To solve equation \eqref{af} we consider the ansatz for the second
order metric perturbation:
\begin{equation*}
g^{(2)}_{ab} = \kii B \eta_{ab} + \kiii\ar^2\eta_{ab} + \kiiii \ar
\hgw_{ab} + \gamma^{(2)}_{ab}.
\end{equation*}
Then the second order equation can be rearranged as
\begin{align}
&G^{(1)}_{ab}[(\kii+2) B \eta + (\kiii-2) \ar^2\eta + (\kiiii+2)\ar
\hgw +\gamma^{(2)}] = \nonumber\\
&\hspace{1cm}=\frac{1}{2\phbdo}T^{(m)(2)}_{ab}[\psi]
-G^{(2)}_{ab}[\hgw] +\left(6 + \frac{\epsilon}{\xi}\right)\left(
\pd_a\ar\pd_b\ar - \frac{1}{2}\eta_{ab}\pd^c\ar\pd_c\ar \right).
\end{align}
We see immediately that this can be satisfied with the choice
\begin{equation*}
    \kii = -2,\quad\text{and}\quad\kiii=2, \quad \kiiii= -2.
\end{equation*}
In this case the physical metric up to second order is given by:
\begin{align}
g_{ab} = \eta_{ab} + \hgw_{ab} -2 \ar \eta_{ab} + 2\ar^2\eta_{ab} -2
\ar \hgw_{ab} + \gamma^{(2)}_{ab} -2 B \eta_{ab},
\end{align}
where the second order metric correction satisfies the equation:
\begin{align}
&G^{(1)}_{ab}[\gamma^{(2)}] =\frac{1}{2\phbdo}\Teff[\psi,\hgw,\ar].
\end{align}
The effective stress energy tensor representing the effect of vacuum
at the random scale is defined by
\begin{equation}\label{teff}
\Teff[\psi,\hgw,\ar]:= T^{(m)(2)}_{ab}[\psi] -2\phbdo
G^{(2)}_{ab}[\hgw] +2\phbdo\left(6 +
\frac{\epsilon}{\xi}\right)\left( \pd_a\ar\pd_b\ar -
\frac{1}{2}\eta_{ab}\pd^c\ar\pd_c\ar \right).
\end{equation}
We see that, remarkably, the vacuum spontaneous fluctuations at the
random scale of Brans-Dicke field, described by $\ar$, contribute
through the usual Klein-Gordon stress energy tensor for a massless
field and with coupling constant
\begin{equation*}
    \cbd := 2\phbdo\left(6 +
\frac{\epsilon}{\xi}\right) = 8\phbdo\left(\frac{3}{2} +
\omega\right).
\end{equation*}
This is consistent with the fact that the equation $\pd^c\pd_c \ar =
0$ holds.

Following the discussion of Section \ref{cbg} we will assume that
\begin{equation}
\m{\Teff[\psi,\hgw,\ar]} = 0
\end{equation}
at the classical scale. This implies that the yet unknown ingredient
that can achieve the vacuum energy balance and bring it down to the
observed almost vanishing value is included within the collection of
matter fields $\psi$. As a result, considering that $\gamma^{(2)}$
is not expected to have its own zero point spontaneous fluctuation,
it will simply represent an extra second order induced perturbation
in response of the microscopic behavior of all the sources. As
noticed in Section \ref{Esol} $\gamma_{ab}$ can be found from the
usual retarded Green's function and is expected to be slow varying
in comparison to $\ar,$ $\hgw$ and $\psi$. The same kind of argument
holds for the Brans-Dicke field fluctuations second order correction
$B$. From \eqref{AB} this satisfies
\begin{equation}
\Box{B} = \frac{1}{2\phbdo(4\omega + 6)} T^{(m)} + 2\ar^{,c}\ar_{,c}
\equiv \frac{T^{(m)}}{\cbd} + 2\ar^{,c}\ar_{,c}.
\end{equation}
The source term has again a zero average. Indeed
$\ms{\ar^{,c}\ar_{,c}} = 0$ because $\pd^c\pd_c \ar = 0$. On the
other hand we know from the discussion in Section \ref{mfset} that
the average of the trace of matter fields stress energy tensor is
related to the effective cosmological constant. This again will be
basically vanishing if the unknown ingredient for vacuum energy
balance is included into $\psi$. As a result $B$ will also be a slow
varying perturbation in comparison to $\ar,$ $\hgw$ and $\psi$.

To indicate specific, slow varying solutions for $\gamma_{ab}$ and
$B$ we employ the symbol $\widetilde{\,\,}$. Thus the explicit
solution for the spacetime physical metric within a Brans-Dicke
model, and quite independently of the details of the model as set by
$\xi$ (or $\omega$), is given by:
\begin{equation}
g_{ab} = \eta_{ab} -2 \ar \eta_{ab} + 2\ar^2\eta_{ab} + \hgw_{ab} -2
\ar \hgw_{ab} + \widetilde{\gamma^{(2)}_{ab}} -\widetilde{2 B
\eta_{ab}},
\end{equation}
where $\pd^c\pd_c \ar = 0$ and $\pd^c\pd_c \hgw_{ab} = 0$ for the
GWs perturbation described by $\hgw_{ab}$ in the \texttt{TT} gauge.
Moreover, noticing that $\exp(-2\ar) = 1 - 2\ar + 2\ar^2 +
O(\varepsilon^3)$ we can re-write the result more compactly as:
\begin{equation}
g_{ab} = \exp(-2\ar)\eta_{ab} + \hgw_{ab} -2 \ar \hgw_{ab} +
\widetilde{\gamma^{(2)}_{ab}} -\widetilde{2 B \eta_{ab}} +
O(\varepsilon^3),
\end{equation}
where it is understood the the exponential is to be considered only
up to second order. The emerging physical picture is that of a
spacetime which presents a conformal modulation around a Minkowski
background induced by the Brans-Dicke field linear perturbation;
superimposed to this there is a totally independent and uncorrelated
perturbation due to GWs. Extra perturbations include a cross term
due to some form of interaction between $\ar$ and $\hgw$ plus second
order slow varying corrections that depend on the average behavior
of the vacuum source defined in term of the matter fields $\psi$,
$\hgw$ and $\ar$.

\section{Discussion and outlook}

The main important conclusion is that the random gravity framework
applied to Brans-Dicke theory predicts that the microscopic
structure of the spacetime metric should present a conformal
modulation in response to the linear part of the fluctuations in the
Brans-Dicke field $\phbd = \phbdo \exp[2(A + B)]$, where
$\abs{A}=O(\epsilon)$ and $\pd^c\pd_c A = 0.$ The vacuum solution
representing spacetime geometry has the structure $g_{ab} =
\exp(-2\ar)\eta_{ab} + [\hgw\text{terms}]$. The extra terms, also
including second order slow varying corrections, depend on
$\hgw_{ab}$, describing gravitational waves fluctuations, are
uncorrelated to the terms in $A$. As far as the dephasing problem
studied in Chapter \ref{ch1} is concerned this is important because,
in a first approximation, we can assume that the conformal term
$\exp(-2\ar)\eta_{ab}$ and the extra terms affect a test particle
independently. The prediction is thus that, beyond a possible
dephasing effect due to interaction with GWs, a quantum particle
should suffer extra dephasing due to the conformal modulation of the
metric. We stress that, as proven in Chapter \ref{ch3}, this is
\emph{not} expected to occur within standard GR. Thus it is a
precise prediction of the Brans-Dicke theory (and possibly of more
general scalar-tensor theories) that dephasing of a quantum particle
due to spacetime conformal fluctuations should occur.

The treatment has now gone all the way back to where we started: in
Chapter \ref{ch1} we proved quite in general that a quantum particle
interacting with a conformally modulated metric
$g_{ab}=\O^2\eta_{ab}$ where $\O^2 = 1 + 2A + A^2$ and with $A$
satisfying the wave equation should suffer dephasing. In the case of
vacuum fluctuations at the random scale this is quantified by
equation \eqref{P1_dododo}. The key feature behind this result is
the fact that the conformal fluctuation induce an effective
nonlinear newtonian potential given in equation \eqref{P1_pot} as $V
:= \left(\cI A + \cII A^2\right)Mc^2.$

We now see how the Brans-Dicke theory can provide such a physical
scenario, \emph{independently} of the value of the Brans-Dicke
parameter $\omega$ specifying the model. In particular the wave
equation for $A$ is a direct consequence of the wave equation
governing the dynamics of the Brans-Dicke field $\phbd$. The fact
that the conformal factor is now mathematically given by $\O =
\exp(-A)$ rather then $1+A$ as we considered in Chapter \ref{ch1}
doesn't affect in any major way the conclusion that leads to
dephasing. Indeed the explicit form of the Klein-Gordon equation for
a minimally coupled massive Klein-Gordon $\phi$ still reads, like in
\eqref{P1_KG1},
\begin{equation}
\left( -\frac{1}{c^2}\frac{\pd^2}{\pd t^2} + \nabla^2 \right) \phi =
\frac{\Omega^2 M^2 c^2}{\hbar^2} \phi - 2 \pd_a (\ln \Omega) \pd^a
\phi.
\end{equation}
The squared conformal factor $\O^2 = \exp(-2A)$ still leads to a
quadratic potential. We remark that the situation is now actually
simper than before. Indeed the extra term containing $\ln \O$ now
simply contributes to the effective potential through an extra
linear term in $A$. The analysis in Chapter \ref{ch1} showed in
general that the linear part of the potential does not contribute to
the dephasing at all. Following the derivation of the potential in
Chapter \ref{ch1} and using $\O = \exp(-A)$ we can now conclude
that, within the Brans-Dicke framework, the effective potential that
can induce dephasing is
\begin{equation}
V = A^2 M c^2.
\end{equation}
The maximum dephasing is thus given, as in equation
\eqref{P1_fedf2b}, by
\begin{equation}
\left|\frac{\delta\rho_{\xv\xv'}}{\rho_0}\right|=
\frac{1}{3\lambda^3}\left(\frac{M}{M_P}\right)^2\left(\frac{T}{T_P}\right).
\end{equation}

To conclude we observe that the structure of the effective stress
energy tensor in equation \eqref{teff}, including the backreaction
due to the conformal fluctuations, seems to suggest that a possible
zero point energy balance mechanism could be obtained for a suitably
negative and large value of the Brans-Dicke coupling parameter
$\omega$, in which case the Brans-Dicke field would be equivalent to
a ghost. This of course would require fine tuning, but indicates
nonetheless that further research in this direction, possibly
exploring scenarios involving more general scalar tensor theories
than Brans-Dicke and modeling the matter fields contribution to
vacuum energy more realistically, will be worth pursuing.

\newpage
\thispagestyle{empty}


\appendix
\singlespacing
\chapter{Stochastic scalar waves and generalized Wiener-Khintchine
theorem}\label{ap1}

\begin{quote}
\begin{small}
In this appendix we work out some general results used in Chapter
\ref{ch1} and related to scalar stochastic waves. We focus on the
case of interest of a scalar, real stochastic field satisfying the
wave equation in three-dimensional space. The main result is the
generalization to a spacetime defined random process of the
Wiener-Khintchine theorem, relating the autocorrelation function to
its power spectral density $S(\omega)$. We also derive general
relations allowing to estimate a variety of interest statistical
quantities, such as the field mean squared amplitude, through
suitable integrals involving $S(\omega)$.
\end{small}
\end{quote}
\singlespacing

\section{General solution to the wave equation}

Let $\phi$ be a scalar field defined on spacetime. Working in units
with $c=1$, the solution to the wave equation $(\nabla^2 -
\partial_t^2)\phi(\v{x},t)=0$ can be written as
\begin{equation}\label{ap_Afe}
\phi(\v{x},t) = \frac1{(2\pi)^3} \int d^3 k \tilde{\phi}(\v{k},t)
e^{i \v{k}\cdot\v{x}},
\end{equation}
where
\begin{equation}\label{ap_Awsolk}
\tilde{\phi}(\v{k},t) = \int d^3 x \phi(\v{x},t) e^{-i
\v{k}\cdot\v{x}}
\end{equation}
and where $\phi \in L^2(\mathbb{R}^3)$. The Fourier coefficients
take the general form
\begin{equation}\label{ap_Atilfk}
\tilde{\phi}(\v{k},t) = a(\v{k}) e^{-i kt} + b(\v{k}) e^{i kt}
\end{equation}
for some \emph{complex functions} $a(\v{k})$ and $b(\v{k})$ and
where $k := \abs{\v{k}}$. These can be obtained by Fourier
transforming $\phi(\v{x},0)$ and $\phi_t(\v{x},0):=\pd_t
\phi(\v{x},0) $. It is indeed straightforward to show that
\begin{align}\label{ap_Aab}
  a(\v{k}) &= \frac{1}{2}\int d^3 x \left[\phi(\v{x},0)+\frac{i}{k}\phi_t(\v{x},0)\right]e^{-i
\v{k}\cdot\v{x}},\\
  b(\v{k}) &= \frac{1}{2}\int d^3 x \left[\phi(\v{x},0)-\frac{i}{k}\phi_t(\v{x},0)\right]e^{-i
\v{k}\cdot\v{x}}\label{ap_Aab1}.
\end{align}
The general solution can thus be written as
\begin{equation}\label{ap_Asol}
    \phi(\v{x},t) = \frac1{(2\pi)^3} \int d^3 k \left[a(\v{k})e^{i
(\v{k}\cdot\v{x}-kt)}+b(\v{k})e^{i (\v{k}\cdot\v{x}+kt)}\right].
\end{equation}

\subsection{Real waves}

It is readily shown that $\phi$ is real, that is $\phi=\phi^*$, if
the following condition is satisfied
\begin{equation}\label{ap_Acr}
    a(\v{k})=b^*(-\v{k}).
\end{equation}
It follows from equations \eqref{ap_Aab} and \eqref{ap_Aab1} that,
if $\phi(\v{x},0)$ and $\phi_t(\v{x},0)$ are real, then
$\phi(\v{x},t)$ is real for any $t$. In this case the solution takes
the form
\begin{equation}\label{ap_Asr}
    \phi(\v{x},t) = \frac1{(2\pi)^3} \int d^3 k \left[a(\v{k})e^{i
(\v{k}\cdot\v{x}-kt)}+a^*(\v{-k})e^{i
(\v{k}\cdot\v{x}+kt)}\right]\,\,\,\,\,\in \mathbb{R}.
\end{equation}
Taking the real part it is easy to show that
\begin{equation}\label{ap_Asrr}
    \phi(\v{x},t)\equiv\mbox{Re}\phi(\v{x},t)=\frac1{(2\pi)^3} \int d^3 k
    \left\{[2\,\mbox{Re}\,a(\v{k})]\cos(\v{k}\cdot\v{x}-kt)+[2\,\mbox{Im}\,a(\v{-k})]\sin(\v{k}\cdot\v{x}+kt)\right\}.
\end{equation}

\subsection{Decomposition in components traveling along different
directions}

Considering the general solution \eqref{ap_Asol}, swapping $\v{k}$
to $-\v{k}$ in the second term and splitting the integration over
$d^3 k$ into the \emph{solid angle} part $\d\hat{\v{k}}$ in the $2$
dimensional space and the \emph{magnitude} part $k$ we have
\begin{equation}
    \phi(\v{x},t) = \frac1{(2\pi)^3} \int \d\hat{\v{k}}\,\int_0^{\infty}\!\!\!\d k\, k^{2}
    \left[a(\v{k})e^{ik
(\hat{\v{k}}\cdot\v{x}-t)}+b(-\v{k})e^{-ik
(\hat{\v{k}}\cdot\v{x}-t)}\right].
\end{equation}
Using spherical coordinates $\v{k}(\vartheta,\varphi,k)$ in momentum
space and employing the notation $a_{\hat{\v{k}}}(k):=
a(\vartheta,\varphi;k)$ and $b_{-\hat{\v{k}}}(k):=
b(\pi-\vartheta,\varphi+\pi;k),$ we define the \emph{directional
wave} along $\hat{\v{k}}(\vartheta,\varphi)$ as
\begin{equation}\label{ap_Adw1}
    \phi_{\hat{\v{k}}}(\hat{\v{k}}\cdot\v{x}/c-t):=\frac1{(2\pi)^3}\int_0^{\infty}\!\!\!\d k\, k^{2}
    \left[a_{\hat{\v{k}}}(k)e^{ikc
(\hat{\v{k}}\cdot\v{x}/c-t)}+b_{-\hat{\v{k}}}(k)e^{-ikc(\hat{\v{k}}\cdot\v{x}/c-t)}\right],
\end{equation}
where for clarity we have restored the speed of light. The general
solution can thus be written according to the directional
decomposition
\begin{equation}\label{ap_Aerdd}
\phi(\v{x},t) = \int
\d\hat{\v{k}}\,\phi_{\hat{\v{k}}}(\hat{\v{k}}\cdot\v{x}/c-t).
\end{equation}

\section{Stochastic waves}\label{ap_Asw}

The scalar field satisfies the wave equation but has a stochastic
nature. We are now going to extend the general results of the
previous section to this interesting case. The advantage of the
directional decomposition \eqref{ap_Aerdd} is that the wave
traveling in 3D space can be seen as the superposition of signals
propagating in all possible directions. Once a given direction is
chosen we are practically dealing with a function of just one time
variable $\kvs\cdot \xv/c - t$. In this section we thus consider
some elementary properties of stochastic signals of one single
variable $t$.

\subsection{Average and correlation properties of the fluctuations}

Consider a generic stochastic process $\phi(t)$. The precise outcome
of the variable $\phi$ at time $t$ is unpredictable. Nevertheless
its properties can be well defined in a statistical sense. At time
$t$ we can ideally think of an \emph{ensemble} of values for
$\phi(t)$, distributed according to a probability density
$p_t(\phi)$ in general depending on time and such that
$p_t(\phi)d\phi$ gives the probability that the actual value
$\phi(t)$ at time $t$ is found to be between $\phi$ and
$\phi+d\phi$. The \emph{average} value of $\phi$ at time $t$ is
defined as
\begin{equation}\label{2e7}
\m{\phi(t)}:=\int p_t(\phi)\phi d\phi.
\end{equation}
The \emph{variance} is defined as
\begin{equation}\label{2e8}
\m{ [\phi(t)]^2 } := \int p_t(\phi)\phi^2 d\phi.
\end{equation}
Equations \eqref{2e7} and \eqref{2e8} are not enough to completely
describe a stochastic field in a statistical sense. \emph{It is also
important to know whether and how the value of $\phi$ at time $t'>t$
is related to the actual value $\phi$ had at time $t$.} In principle
the probability of having the particular outcomes $\phi(t)$
\emph{and} $\phi(t')$ is governed by a \emph{joint distribution
function} $p_{t t'}(\phi,\phi')$. Then the quantity $p_{t
t'}(\phi,\phi')d\phi\,d\phi'$ gives the probability for $\phi$ to
have values lying between $\phi$ and $\phi+d\phi$ at time $t$
\emph{and} between $\phi'$ and $\phi'+d\phi'$ at time $t'$. The
\emph{first order autocorrelation function} of $\phi(t)$ can then be
defined as
\begin{equation}\label{2e9}
R(t-t'):=\m{ \phi(t)\phi(t') }\equiv\int p_{t t'}(\phi,\phi') \phi
\phi' d\phi d\phi'.
\end{equation}
If the probability of having the value $\phi(t')$ is completely
independent from the previous outcome $\phi(t)$ then
\begin{equation}\label{2e10}
p_{t t'}(\phi,\phi')=p_t(\phi)p_{t'}(\phi')
\end{equation}
and the stochastic process described by $\phi(t)$ is said to be
perfectly \emph{uncorrelated}. In this case
\begin{equation}\label{2e11}
\m{\phi(t)\phi(t')}=\m{ \phi(t) }\m{ \phi(t') }.
\end{equation}
Higher order autocorrelation functions can also be defined. The
\emph{second order correlation function} is given by
\begin{equation}\label{2e12}
R''(t-t'):=\m{ [\phi(t)]^2[\phi(t')]^2 } \equiv\int p_{t
t'}(\phi,\phi') \phi^2 \phi'^2 d\phi d\phi'.
\end{equation}
Finally, if two different stochastic processes $\phi_1(t)$ and
$\phi_2(t)$ are given, their \emph{correlation function} is defined
as
\begin{equation}\label{2e13}
g(t-t')=\m{ \phi_1(t)\phi_2(t') }\equiv\int q_{t t'}(\phi_1,\phi_2)
\phi_1 \phi_2 d\phi_1 d\phi_2,
\end{equation}
where $q_{t t'}(\phi_1,\phi_2)$ is the joint probability
distribution for the two stochastic processes. The processes are
perfectly uncorrelated if
\begin{equation}\label{2e14}
\m{ \phi_1(t)\phi_2(t') }=\m{ \phi_1(t) }\m{ \phi_2(t') }.
\end{equation}
All the mean quantities defined in this section through the
underlying distribution functions are called \emph{ensemble or
statistical averages}.

\subsection{Characterization of the stochastic waves}

We now go back to the scalar wave equation but we consider the
situation in which $\phi$ is \emph{stochastic wave}. For our
purposes, the interesting statistical properties of the process are
specified once the average value at different spacetime points and
the autocorrelation properties are known. As explained above it is
impossible to predict the value that the field will have at the
location $\v{x}$ at time $t$. Nonetheless it is conceptually
possible to consider many measurements of the values that the
stochastic wave takes at nearby spacetime points. This would ideally
gives an \emph{a posteriori} knowledge of a particular
\emph{realization} of the process. We indicate this with
$\phi(\v{x},t)$, representing a concrete \emph{sample function} of
the underlying stochastic process. Since we are interested in the
case in which this models the conformal fluctuations of vacuum
spacetime, we will assume the stochastic process to be
\emph{stationary}, i.e. all mean properties do not depend on time.
Moreover, because of Lorentz invariance, all the mean properties
cannot depend on the space location either. In the non-relativistic
formalism employed in Chapter \ref{ch1} the space and time continua
are split: in the following we assume the stochastic process
underlying the conformal fluctuations to be \emph{ergodic} with
respect to time and 3 dimensional space. By this we mean that
ensemble averages are assumed to be equal to time or space averages
taken on any given sample function representing the process. The
time and space averages are thought to be obtained through integrals
of the kind $\left(\int d^3 x \ldots \right)/\L^3$ and $\left(\int d
t \ldots \right)/T$, where $T:=\L/c$. The scale $\L$ is small from a
macroscopic point of view but still supposed to contain many
wavelengths. It can be identified with the classical scale defined
in chapter \ref{ch1} (see table \ref{P1_scale}).

Our main goal is to derive a generalization of the Wiener-Khintchine
theorem linking the power spectrum to the first order
autocorrelation function. This will enable us to evaluate
\emph{expectation values} of the field at different spacetime points
like $\m{ \phi(\v{x}_1,t_1)\phi(\v{x}_2,t_2)}$. We proceed by steps,
reviewing the standard theorem holding for a stochastic functions of
three variables first, and generalizing it to the case of a
stochastic wave propagating in $4D$ spacetime. We assume that a
particular sample function representing the stochastic process can
be written in analogy to \eqref{ap_Aerdd} according to
\begin{equation}\label{ap_AerddB}
\phi(\v{x},t) = \int
\d\hat{\v{k}}\,\phi_{\hat{\v{k}}}(\hat{\v{k}}\cdot\v{x}/c-t),
\end{equation}
in such a way that the wave equation is satisfied. Some care must be
taken in using the Fourier expansions relations of the previous
section. Indeed, for a given $t$, $\phi$ \emph{is not} in general a
square integrable function belonging to $L^2(\mathbb{R}^3)$. To
circumvent this problem, given a sample function of the process and
for any given $t$, we define
\begin{equation}\label{ap_AwsolkB}
\tilde{\phi}^L(\v{k},t) := \int_{\mathcal{D}_L} d^3 x \phi(\v{x},t)
e^{-i \v{k}\cdot\v{x}},
\end{equation}
where $\mathcal{D}_L$ is a $3D$ cubic domain of side $L$. Then, the
function
\begin{equation}\label{ap_AfeB} \phi^L(\v{x},t) :=
\frac1{(2\pi)^3} \int d^3 k \tilde{\phi}^L(\v{k},t) e^{i
\v{k}\cdot\v{x}},
\end{equation}
satisfies the wave equation provided that
\begin{equation}\label{ap_AtilfkB}
\tilde{\phi}^L(\v{k},t) = a^L(\v{k}) e^{-i kt} + b^L(\v{k}) e^{i
kt}.
\end{equation}
These expressions will be used later while taking the limit with $L
\rightarrow \infty$.

\section{Wiener-Khintchine theorem for a stochastic process}

\subsection{Standard $3D$ case}\label{ap_nd}

Consider a complex stationary stochastic function $\phi(\v{x})$ in
$3$-dimensions. Its Fourier transform over the compact domain
$\mathcal{D}_L$ of volume $L^3$ is given by
\begin{equation}\label{ap_Antf}
\tilde{\phi}^L(\v{k}) := \int_{\mathcal{D}_L} d^3 x \phi(\v{x})
e^{-i\v{k}\cdot\v{x}}.
\end{equation}
The mean power spectral density over this interval is defined by
\begin{equation}\label{ap_AnSL}
S^L(\v{k}) := \frac{1}{L^3}\m{ \tilde{\phi}^L(\v{k})^*
\tilde{\phi}^L(\v{k}) }.
\end{equation}
In the limit $L\rightarrow\infty$ we have
\begin{equation}\label{ap_AnSk}
S(\v{k}) := \lim_{L\rightarrow\infty} S^L(\v{k}).
\end{equation}
The autocorrelation function of $\phi(\v{x})$ for any two values
$\v{x}_1$ and $\v{x}_2=\v{x}_1+\vg{\xi}$ is given by
\begin{equation}\label{ap_AnRDel}
C(\vg{\xi})=\m{ \phi(\v{x}_1)^*\phi(\v{x}_2) }
\end{equation}
satisfying
\begin{equation}\label{ap_AnRR}
 C(-\vg{\xi}) = C(\vg{\xi})^*.
\end{equation}
Consider now
\begin{align*}
\m{ \tilde{\phi}^L(\v{k})^* \tilde{\phi}^L(\v{k}) } &=
\int_{\mathcal{D}_L} d^3 x_1 \int_{\mathcal{D}_L} d^3 x_2 \m{
\phi(\v{x}_1)^*\phi(\v{x}_2)} e^{-i \v{k}\cdot(\v{x}_2-\v{x}_1)}
\\
&= \int_{\mathcal{D}_L} d^3 x_1 \int_{\mathcal{D}_L} \d^3 \vg{\xi}\,
C(\vg{\xi}) e^{-i \v{k}\cdot \vg{\xi}},
\end{align*}
where \eqref{ap_Antf} and \eqref{ap_AnRDel} have been used. Since
$\int_{\mathcal{D}_L} d^3 x_1=L^3$, we have
\begin{equation}
\m{ \tilde{\phi}^L(\v{k})^* \tilde{\phi}^L(\v{k}) } = L^3 \int \d^3
\vg{\xi}\, C(\vg{\xi}) e^{-i \v{k}\cdot\vg{\xi}}.
\end{equation}
In the limit $L\rightarrow\infty$, we see that
\begin{equation}\label{ap_AnSkk}
S(\v{k}) = \int \d^3 \vg{\xi}\, C(\vg{\xi}) e^{-i
\v{k}\cdot\vg{\xi}}.
\end{equation}
That is, the power spectral density $S(\v{k})$ is the Fourier
transform of the autocorrelation function $C(\vg{\xi})$. Conversely,
\begin{equation}\label{ap_AnSS}
C(\vg{\xi}) = \frac1{(2\pi)^3} \int d^3 k S(\v{k}) e^{i
\v{k}\cdot\vg{\xi}}.
\end{equation}
These two equations express the well known WK theorem.

\subsubsection{Real functions}

If $\phi(\v{x})$ is real it follows from \eqref{ap_AnRDel} that the
correlation function $C(\vg{\xi})$ is also real, and hence
\eqref{ap_AnRR} becomes
\begin{equation}\label{ap_AnRRR}
C(-\vg{\xi}) = C(\vg{\xi}).
\end{equation}
Inserting this into \eqref{ap_AnSkk}, we get
\begin{eqnarray}\label{ap_AnSSS}
S(\v{k}) = \int \d^3\vg{\xi}\, C(\vg{\xi}) \cos(\v{k}\cdot\vg{\xi}).
\end{eqnarray}
This implies that the spectral density $S(k)$ is even
\begin{equation}\label{ap_AnSeven}
S(-\v{k}) = S(\v{k}).
\end{equation}
Therefore, \eqref{ap_AnSS} becomes
\begin{eqnarray}\label{ap_AnSR}
C(\vg{\xi}) = \frac1{(2\pi)^3} \int d^3 k S(\v{k})
\cos(\v{k}\cdot\vg{\xi}).
\end{eqnarray}

\section{Generalized Wiener-Khintchine Theorem for stochastic
waves} \label{ap_AappC}

Consider now a complex, ergodic time-dependent stochastic function
$\phi(\v{x},t)$ in an $3$-dimensional space with time $t$,
satisfying the wave equation $(\partial_t^2 -
\nabla^2)\phi(\v{x},t)=0$. We now assume that the autocorrelation
function of $\phi(\v{x},t)$ for any two events $(\v{x}_1,t_1)$ and
$(\v{x}_2,t_2)=(\v{x}_1+\vg{\xi},t_1+\tau)$ is a function of
$\vg{\xi}$ and $\tau$ given by
\begin{equation}\label{ap_ATnRDel}
C(\vg{\xi},\tau)=\m{ \phi(\v{x}_1,t_1)^*\phi(\v{x}_2,t_2) },
\end{equation}
and having the property
\begin{equation}\label{ap_ATnRR}
C(-\vg{\xi},-\tau) = C(\vg{\xi},\tau)^*.
\end{equation}
For any fixed choice of $\v{x}_1$ and $t_1$, it follows from the
definition \eqref{ap_ATnRDel} that $C(\vg{\xi},\tau)$ also satisfies
the wave equation, i.e.
\begin{equation}\label{ap_ARweq}
\left(
\partial_\tau^2 - \nabla_{\vg{\xi}}^2
\right)C(\vg{\xi},\tau)=0.
\end{equation}
Assuming that, for any $\tau$, $C(\vg{\xi},\tau)$ belongs to
$L^2(\mathbb{R}^3)$ and using equation \eqref{ap_Asol} we have,
\begin{equation}\label{ap_ARwsol}
C(\vg{\xi},\tau) = \frac1{(2\pi)^3} \int d^3 k \left[ \a(\v{k})
e^{-i k\tau} + \b(\v{k}) e^{i k\tau} \right]\, e^{i
\v{k}\cdot\vg{\xi}}.
\end{equation}
The correlation function satisfies
\begin{align}\label{ap_ARwsol4}
C(-\vg{\xi},-\tau) &= \frac1{(2\pi)^3} \int d^3 k \left[ \a(\v{k})
e^{i k\tau} + \b(\v{k}) e^{-i k\tau} \right]\, e^{-i
\v{k}\cdot\vg{\xi}},
\\\label{ap_ARwsol5}
C(\vg{\xi},\tau)^* &= \frac1{(2\pi)^3} \int d^3 k \left[ \a(\v{k})^*
e^{i k\tau} + \b(\v{k})^* e^{-i k\tau} \right]\, e^{-i
\v{k}\cdot\vg{\xi}}.
\end{align}
From \eqref{ap_ATnRR}, \eqref{ap_ARwsol4}, \eqref{ap_ARwsol5} we see
that
\begin{equation}\label{ap_AAB}
\a(\v{k})^* = \a(\v{k}),\quad \b(\v{k})^* = \b(\v{k})
\end{equation}
i.e. both $\a(\v{k})$ and  $\b(\v{k})$ are real. Considering that
the process is stationary the power spectrum cannot depend on time.
Setting $\tau=0$, we have that \eqref{ap_ARwsol} correctly reduces
to \eqref{ap_AnSS} if
\begin{equation}\label{ap_AredS}
S(\v{k})= \a({\v{k}}) + \b({\v{k}}).
\end{equation}

\subsection{Evaluation of $\a(\v{k})$ and $\b(\v{k})$}

To determine $\a(\v{k})$ and $\b(\v{k})$ let us consider the
stochastic process for some fixed time $t_0$. From equations
\eqref{ap_AfeB} and \eqref{ap_AtilfkB} we have
\begin{equation}\label{ap_AfeC}
\phi^L(\v{x},t_0) = \frac1{(2\pi)^3} \int d^3 k
\tilde{\phi}^L(\v{k},t_0) e^{i \v{k}\cdot\v{x}},
\end{equation}
with
\begin{equation}\label{ap_AtilfkC}
\tilde{\phi}^L(\v{k},t_0) = a^L(\v{k}) e^{-i kt_0} + b^L(\v{k}) e^{i
kt_0}.
\end{equation}
We are thus dealing with a $3$ dimensional stochastic process and we
can use the results of Section \ref{ap_nd}. Exploiting the fact that
the stochastic process is stationary we define mean power spectral
density as
\begin{equation}
S(\kv) := \lim_{L,T \rightarrow \infty} \frac{1}{T}\int_0^T \d
    t_0\,\frac{1}{L^3}\m{\tilde{\phi}^L(\v{k},t_0)^*
\tilde{\phi}^L(\v{k},t_0)}.
\end{equation}
Substituting equations \eqref{ap_AfeC} and \eqref{ap_AtilfkC} we get
\begin{align}\label{ap_SsS}
    S(\kv) &= \lim_{L,T \rightarrow \infty} \frac{1}{T}\int_0^T \d
    t_0\,\frac{1}{L^3}\m{[ a^L(\v{k})^* e^{i kt_0} + b^L(\v{k})^* e^{-i
kt_0} ][ a^L(\v{k}) e^{-i kt_0} + b^L(\v{k}) e^{i kt_0}] }\nonumber\\
\nonumber\\
& = \lim_{L,T \rightarrow \infty} \frac{1}{T}\int_0^T \d
    t_0\,\frac{1}{L^3}\m{\abs{a^L(\v{k})}^2 + \abs{b^L(\v{k})}^2 +
    2\mbox{Re}[a^L(\v{k})b^L(\v{k})^* e^{-2i
kt_0}]}\nonumber\\
\nonumber\\
& = \lim_{L \rightarrow \infty}\frac{1}{L^3}\m{\abs{a^L(\v{k})}^2 +
\abs{b^L(\v{k})}^2}.
\end{align}
Comparing \eqref{ap_SsS} with \eqref{ap_AredS} we have
\begin{eqnarray}
\a(\v{k}) = \lim_{L\rightarrow\infty}\frac{1}{L^3}\m{\abs{a^L(\v{k})}^2} \label{ap_AABeqs} \\
\b(\v{k}) =
\lim_{L\rightarrow\infty}\frac{1}{L^3}\m{\abs{b^L(\v{k})}^2}
\label{ap_AABeqsb}
\end{eqnarray}
and
\begin{equation}\label{ap_Scr}
    S(\v{k})= \lim_{L\rightarrow\infty}\frac{1}{L^3}\m{ \abs{a^L(\v{k})}^2 + \abs{b^L(\v{k})}^2
    }.
\end{equation}

\subsection{Real functions}

If $\phi(\v{x},t)$ is real we know from \eqref{ap_Acr} that
$a^L(\v{k})^*=b^L(-\v{k})$ and equation \eqref{ap_ARwsol} becomes
\begin{equation}\label{ap_ARaabb1}
C(\vg{\xi},\tau) = \frac1{(2\pi)^3}
\lim_{L\rightarrow\infty}\frac{1}{L^3}\m{ \int d^3 k [
\abs{a^L(\v{k})}^2 e^{-i k\tau} + \abs{a^L(-\v{k})}^2 e^{i k\tau}
]\, e^{i \v{k}\cdot\vg{\xi}}}
\end{equation}
and
\begin{equation}\label{ap_Scr2}
    S(\v{k})= \lim_{L\rightarrow\infty}\frac{1}{L^3}\m{ \abs{a^L(\v{k})}^2 + \abs{a^L(-\v{k})}^2
    },
\end{equation}
implying that $S(\v{k})$ \emph{is even}, i.e. $S(\v{k})=S(-\v{k})$.
Moreover, from \eqref{ap_ARaabb1} we see that $\a^*(\v{k}) =
\b(-\v{k})$, implying that $C(\vg{\xi},\tau)$ \emph{is real}.
Swapping $\v{k}$ to $-\v{k}$ in the second integral in
\eqref{ap_ARaabb1} we have
\begin{align*}
C(\vg{\xi},\tau) &=
\frac1{(2\pi)^3}\lim_{L\rightarrow\infty}\frac{1}{L^3}\m{  \int d^3
k\, \abs{a^L(\v{k})}^2 e^{i(\v{k}\cdot\vg{\xi}-
k\tau)}+\abs{a^L(\v{k})}^2
e^{-i(\v{k}\cdot\vg{\xi}- k\tau)}} \\
\\
&=\frac1{(2\pi)^3}\lim_{L\rightarrow\infty}\frac{1}{L^3}\m{  \int
d^3 k\, 2\abs{a^L(\v{k})}^2 \cos(\v{k}\cdot\vg{\xi}- k\tau)}.
\end{align*}
This reduces to \eqref{ap_AnSR} for $\tau=0$ if
\begin{equation}\label{ap_AredS1b}
S(\v{k})= \lim_{L\rightarrow\infty}\frac{2}{L^3}\m{
\abs{a^L(\v{k})}^2}.
\end{equation}

To summarize the main result of this section, we have found that the
generalized Wiener-Khintchine theorem for a real, stationary
stochastic scalar wave takes the form
\begin{equation}\label{ap_Agwkt1}
C(\vg{\xi},\tau)=\frac1{(2\pi)^3} \int d^3 k\, S(\v{k})
\cos(\v{k}\cdot\vg{\xi}- kc\tau),
\end{equation}
where we have restored the speed of light and where the mean power
spectrum is defined in \eqref{ap_AredS1b}.

\section{Correlation properties of wave components in different
directions}

The results that we have come to establish allow to show that wave
components traveling in different directions are uncorrelated. We
evaluate $\m{ \phi^*(\v{x}_1,t_1)\phi(\v{x}_2,t_2) }$ using equation
\eqref{ap_AerddB} and we have
\begin{equation}\label{ap_Adc1}
   C(\vg{\xi},\tau)=\m{
\phi^*(\v{x},t)\phi(\v{x}+\vg{\xi},t+\tau)} =\int
\d\hat{\v{k}}\,\int \d\hat{\v{k}}'\,\m{
\phi^{*}_{\hat{\v{k}}}(\hat{\v{k}}\cdot\v{x}/c-t)\,
\phi_{\hat{\v{k}}'}([\hat{\v{k}}'\cdot\v{x}/c-t]+[\hat{\v{k}}'\cdot\vg{\xi}/c-\tau])
}.
\end{equation}
Using the Wiener-Khintchine theorem in its form as given by
\eqref{ap_ARwsol}, restoring the speed of light, swapping $\v{k}$ to
$-\v{k}$ in the second integral and splitting the $d^3 k$ integral
in its angular and magnitude parts we can write as well
\begin{align*}\label{ap_Adc2}
    C(\vg{\xi},\tau) &= \frac1{(2\pi)^3} \int d^3 k [ \a(\v{k}) e^{-i
kc\tau} + \b(\v{k}) e^{i kc\tau} ]\, e^{i
\v{k}\cdot\vg{\xi}}\\\nonumber
\\\nonumber
&= \frac1{(2\pi)^3} \int d^3 k [ \a(\v{k}) e^{i(\v{k}\cdot\vg{\xi}-
kc\tau)} + \b(-\v{k}) e^{-i(\v{k}\cdot\vg{\xi}- kc\tau)}
]\\\nonumber
\\\nonumber
&= \frac1{(2\pi)^3} \int \d\hat{\v{k}}\,\int_0^{\infty}\!\!\!\d k\,
k^{2}[\a(\v{k}) e^{ikc(\hat{\v{k}}\cdot\vg{\xi}/c- \tau)} +
\b(-\v{k}) e^{-ikc(\hat{\v{k}}\cdot\vg{\xi}/c- \tau)}].\nonumber
\end{align*}
Therefore
\begin{equation}\label{ap_Adircor}
    C(\vg{\xi},\tau)=\int
    \d\hat{\v{k}}\,C_{\hat{\v{k}}}(\hat{\v{k}}\cdot\vg{\xi}/c-\tau),
\end{equation}
where we defined the correlation function in the direction
$\hat{\v{k}}$ as
\begin{equation}\label{ap_Acordir}
    C_{\hat{\v{k}}}(\hat{\v{k}}\cdot\vg{\xi}/c-\tau):=\frac1{(2\pi)^3}\int_0^{\infty}\!\!\!\d k\, k^{2}
 [\a_{\hat{\v{k}}}(k) e^{ikc(\hat{\v{k}}\cdot\vg{\xi}/c-\tau)} + \b_{-\hat{\v{k}}}(k)
 e^{-ikc(\hat{\v{k}}\cdot\vg{\xi}/c-\tau)}],
\end{equation}
with $\a_{\hat{\v{k}}}(k) := \a(\v{k})$ and $\b_{-\hat{\v{k}}}(k) :=
\b(-\v{k})$. The two equations \eqref{ap_Adc1} and
\eqref{ap_Adircor} must be equivalent. This implies at once the
following equation
\begin{equation}\label{ap_Afff}
    \m{ \phi^{*}_{\hat{\v{k}}}(\hat{\v{k}}\cdot\v{x}/c-t)\,
\phi_{\hat{\v{k}}'}([\hat{\v{k}}'\cdot\v{x}/c-t]+[\hat{\v{k}}'\cdot\vg{\xi}/c-\tau])
}=
\delta(\kvs,\kvs')\,C_{\hat{\v{k}}}(\hat{\v{k}}\cdot\vg{\xi}/c-\tau)
\end{equation}
or, equivalently, since $\hat{\v{k}}\cdot\v{x}/c$ has the dimensions
of a time,
\begin{equation}\label{ap_Afff2}
     \m{ \phi^{*}_{\hat{\v{k}}}(t)\,
\phi_{\hat{\v{k}}'}(t+\tau) }=
\delta(\kvs,\kvs')\,C_{\hat{\v{k}}}(\tau).
\end{equation}
We thus see that the fluctuating field can be resolved into
components along different directions represented by completely
uncorrelated functions of \emph{just one time variable}.

\subsection{Real functions}\label{ap_As1}

If $\phi$ is real equation \eqref{ap_Afff2} translates to
\begin{equation}\label{ap_Afff2r}
   \m{ \phi_{\hat{\v{k}}}(t)\,
\phi_{\hat{\v{k}}'}(t+\tau) }=
\delta(\kvs,\kvs')\,C_{\hat{\v{k}}}(\tau),
\end{equation}
with
\begin{equation}\label{ap_Adsp}
    C_{\hat{\v{k}}}(\tau)=\frac1{(2\pi)^3}\int_0^{\infty}\!\!\!\d k\,
    k^{2}S(\v{k})\cos(kc\tau),
\end{equation}
as deduced from equation \eqref{ap_Agwkt1}.

\section{Isotropic power spectrum and field averages}\label{A04s}

In general wave components traveling in different directions can
have different autocorrelation properties. However, in the case
relevant to us and connected to vacuum conformal fluctuations, the
isotropy of space implies that the spectral density is also
isotropic, i.e. $S(\v{k})\equiv S(k)$. In this case we can introduce
a single \emph{isotropic correlation function} defined as
\begin{equation}\label{ap_Adsp2}
    C(\tau) := \frac{1}{(2\pi)^3}\int_0^{\infty}\!\!\!\d k\,
    k^{2}S(k)\cos(kc\tau)
\end{equation}
and \eqref{ap_Afff2r} simply becomes
\begin{equation}\label{ap_Afff2rb}
\m{ \phi_{\hat{\v{k}}}(t)\, \phi_{\hat{\v{k}}'}(t+\tau) }=
\delta(\kvs,\kvs')\,C(\tau).
\end{equation}
We conclude this section by deducing a useful formula for the mean
squared amplitude. Using equations \eqref{ap_Afff2rb} together with
\eqref{ap_Aerdd}, this follows as
\begin{equation}
    \m{ \phi^2}=\m{\int\d\kvs \phi_{\kvs}(\xv,t)\int\d\kvs'
    \phi_{\kvs'}(\xv,t)} =C_0\int\d\kvs
    \int\d\kvs'\delta(\kvs,\kvs').
\end{equation}
Thus we have
\begin{equation}\label{ap_mswa}
    \m{ \phi^2}=4\pi C_0,
\end{equation}
where ${C_0}$, the peak of the autocorrelation function, follows
from \eqref{ap_Adsp2} as
\begin{equation}\label{ap_Adsp2b}
    {C_0} := \frac1{(2\pi)^3}\int_0^{\infty}\!\!\!\d k\,
    k^{2}S(k).
\end{equation}

The power spectrum appears to be the most fundamental quantity
related to the stochastic waves. Indeed, once $S(k)$ in known, the
correlation properties, as well as all other sort of averages of the
fluctuating field can be calculated. For instance, equations
\eqref{ap_mswa} and \ref{ap_Adsp2b} can be combined to yield
\begin{equation}
\m{ \phi^2} = \frac1{(2\pi)^3} \int d^3 k\, S(k). \label{ap_AM01}
\end{equation}
An analogue result holds more in general for a complex and non
isotropic stochastic process.

It is also useful to deduce two expressions for the averaged time
and space derivatives of the field. Exploiting the fact that the
stochastic process is ergodic and stationary we can write (for
arbitrary complex and non isotropic signal)
\begin{equation}
\m{ \partial_t \phi^*\partial_t \phi }:=
\lim_{L,T\rightarrow\infty}\frac{1}{L^3} \int d^3 x
\,\frac{1}{T}\int_0^T \d t \,\partial_t \phi^L(\xv,t)^*\partial_t
\phi^L(\xv,t).
\end{equation}
Using equation \eqref{ap_AfeB} the right hand side gives
\begin{equation}
\text{r.h.s.} =
\frac1{(2\pi)^{6}}\lim_{L,T\rightarrow\infty}\frac{1}{L^3} \int d^3
x \,\frac{1}{T}\int_0^T \d t \int d^3 k_1 \int d^3 k_2\,
\m{\partial_t \tilde{\phi}^L(\kv_1,t)^* \partial_t
\tilde{\phi}^L(\kv_2,t)} e^{i \xv \cdot (\kv_2 - \kv_1)},
\end{equation}
where the statistical average $\m{\,\,}$ has been inserted to deal
with the stochasticity in the Fourier coefficients
$\tilde{\phi}^L(\kv_1,t)$. Integrating over the space variable and
using the properties of the $\delta$ function this gives
\begin{align}
\text{r.h.s.} &=
\frac1{(2\pi)^3}\lim_{L,T\rightarrow\infty}\frac{1}{L^3}
\,\frac{1}{T}\int_0^T \d t \int d^3 k_1 \int d^3 k_2\,\m{
\partial_t \tilde{\phi}^L(\kv_1,t)^* \partial_t \tilde{\phi}^L(\kv_2,t)} \delta(\kv_2-\kv_1)
\nonumber\\
\\
&= \frac1{(2\pi)^3}\lim_{L,T\rightarrow\infty}\frac{1}{L^3}
\,\frac{1}{T}\int_0^T \d t \int d^3 k\, \m{\partial_t
\tilde{\phi}^L(\kv,t)^* \partial_t \tilde{\phi}^L(\kv,t)}.
\end{align}
Finally using equation \eqref{ap_AtilfkB} to express the Fourier
coefficients we have
\begin{equation}
\text{r.h.s.} =
\frac1{(2\pi)^3}\lim_{L,T\rightarrow\infty}\frac{1}{L^3}
\,\frac{1}{T}\int_0^T \d t  \int d^3 k\, \m{\partial_t[a^L(\kv)^*
e^{i k t}+b^L(\kv)^*e^{-i kt}] \partial_t[a^L(\kv) e^{-i
kt}+b^L(\kv) e^{i kt}]}.
\end{equation}
Performing the derivatives and the time integration we finally get
the result
\begin{equation}
\m{ \partial_t \phi^*\partial_t \phi } = \frac1{(2\pi)^3} \int d^3
k\, k^2 S(\kv). \label{ap_AM01.5}
\end{equation}
A perfectly analogue calculation allows to find
\begin{equation}
\m{ \nabla \phi^* \cdot\nabla \phi } = \frac1{(2\pi)^3} \int d^3 k\,
k^2 S(\kv). \label{ap_AM02}
\end{equation}

\section{Treatment of the term $T_4$ in the effective \schr
equation}\label{ap2XXX}

In this last section we sketch a semi-qualitative argument by which
we can infer the behavior of the extra term $T_4$ appearing in the
effective \schr equation \eqref{P1_eqeq}.

This term reads $T_4 = -i\hbar\left( \dot{A} - A\dot{A} \right)\psi
- i\hbar \vl \left( A_{,x} - A A_{,x} \right)\psi,$ as derived in
Section \ref{P1_lvle}. Let us write the conformal field $A$, in the
isotropic and real case, as:
\begin{equation}\label{sig1T4}
A \approx \int \d\hat{\v{k}}\,\int_0^{2\pi/\lr}\!\!\!\d k\, k^2
a(k)e^{ik (\hat{\v{k}}\cdot\v{x}-ct)},
\end{equation}
where the upper cutoff is set by the particle resolution scale
$\lr.$ The power spectrum is basically proportional to the square of
the Fourier component $a(k)$. For $S(k) \sim 1 / k$ we have $a(k)
\sim k^{-1/2}$, so that the effective coefficient appearing in the
expansion is $k^2 a(k) \sim k^{3/2}$. Thus the short wavelengths
close to the cutoff give the most important contribution. For this
reason we can approximate the field as
\begin{equation}\label{sig2T4}
A \approx \int \d\hat{\v{k}}\,\int_0^{2\pi/\lr}\!\!\!\d k\, k^{3/2}
\Delta(k - k_A)e^{ik (\hat{\v{k}}\cdot\v{x}-ct)},
\end{equation}
where the function $\Delta(k - k_A)$ is peaked around a typical wave
number $k_A$. This can in principle be selected in such a way that
the average properties of \eqref{sig1T4} are equivalent to those of
\eqref{sig2T4}. Effectively we get:
\begin{equation}\label{sig3}
A \approx \int \d\hat{\v{k}} {k_A}^{3/2} e^{ik_A
(\hat{\v{k}}\cdot\v{x}-ct)},
\end{equation}
so that the conformal field is approximated as a fast varying and
isotropic random signal characterized by a \emph{single typical
wavelength} $\lambda_A \equiv 2\pi / k_A$. In relation to the
fluctuations ability to affect the particle, this will be close to
the particle resolution scale, i.e. we put $\lambda_A \equiv \kappa
\lr$, with $\kappa \gtrsim 1$. From \eqref{sig3} we now have:
\begin{equation*}
\dot{A} \approx -i k_A c A = -\frac{2\pi i}{\kappa\lr}cA = - \frac{i
Mc^2}{\kappa \hbar}A,
\end{equation*}
where we used $\lr = h/Mc$. The space derivatives yield:
\begin{equation*}
A_{,x} \approx  i{k_A}^{5/2} \int \d\hat{\v{k}} \hat{k}_x e^{ik_A
(\hat{\v{k}}\cdot\v{x}-ct)} = 0.
\end{equation*}
Using these two relations in \eqref{P1_t4} yields the result
\eqref{P1_t4bis}.

\newpage
\thispagestyle{empty}


\singlespacing
\chapter{Technical derivations related to Chapter
\ref{ch1}}\label{ap2}

\begin{quote}
\begin{small}
This Appendix reports some technical derivations related to Chapter
\ref{ch1}. In the first part \ref{ap2_1} we show that the kinetic
and potential parts of the Hamiltonian give separate, additive
contributions to the Dyson expansion and can be considered
separately. In \ref{ap2_2} we report the details of the second order
calculation leading to the most general expression
\eqref{P1_P2_mainB} for the density matrix evolution. In \ref{ap2_3}
we prove an approximate integral identity which allows to achieve an
important simplification of the final result. Finally, in
\ref{ap2_4} we briefly outline the analysis of the fourth order term
in the Dyson expansion and show how this is expected to yield a
vanishing contribution, at least in the case of vacuum fluctuations.
\end{small}
\end{quote}
\singlespacing

\section{Separability of the kinetic and potential part in the Dyson
expansion}\label{ap2_1}

The potential energy of a particle of mass $M$ due to its coupling
to the conformal fluctuations is $V(\xv,t)=Mc^2 (\cI A(\xv,t) + \cII
A^2)$. Using the spectral theorem for hermitian operators the
corresponding abstract operator representing the interaction part of
the hamiltonian can be written as\footnote{Indeed, using the
$\delta$ function properties, it is easily verified that
$\Ho^1(t)\ket{\xv}=V(\xv,t)\ket{\xv}$. This implies
$\Ho^1_{\xv\xv'}(t)\equiv\bra{\xv}\Ho^1(t)\ket{\xv'}=V(\xv,t)\delta(\xv-\xv')$.}
\begin{equation}\label{P2_Xe26}
\Ho^1(t)=\int d^3 x V(\xv,t)\Pio(\xv),
\end{equation}
where $\Pio(\xv)=\ket{\xv}\bra{\xv}$ is the \emph{projection
operator} on the space spanned by the position operator eigenstate
$\ket{\xv}$.

We now proceed to show that, up to second order in the Dyson's
expansion, the kinetic and conformal part of the hamiltonian give
additive contributions to the average evolution of the density
matrix. This fact provides an important simplification and the
kinetic part is simply responsible for the free evolution while the
dephasing effect solely depends upon the fluctuating part of the
Hamiltonian. Since $\Ho(t)=\Ho^0+\Ho^1(t)$, the linear terms of
equation \eqref{P2_Xe22} containing $\Ko_1(T)$ and $\Ko_1^{\dag}(T)$
do not pose any problem. Turning to the second order terms we have
\begin{align}\label{P2_TT1}
&\m{ \Ko_2(T)\rho_0 }
=\nonumber\\
&\hspace{1cm}-\frac{1}{\hbar^2}\,\m{\int_{0}^{T}\!\!\!\d t
\int_{0}^{t}\!\!\d t' \left[\int\! {d^3 y}\,
V(\v{y},t)\Pio(\v{y})+\Ho^0\right]\times\left[ \int\! {d^3 y'}
V(\v{y}',t')\Pio(\v{y}')+\Ho^0\right]\rho_0}.
\end{align}
This expression can be simplified using the standard properties of
the projectors operators. Going to the position representation and
exploiting the fact that
\begin{equation}
\bra{\v{x}}\left[\int\! {d^3 y}\,
V(\v{y},t)\ket{\v{y}}\bra{\v{y}}\right] = \int\! {d^3 y}\,
V(\v{y},t)\delta(\v{x}-\v{y})\bra{\v{y}} = \bra{\v{x}}V(\v{x},t),
\end{equation}
the matrix elements $[\m{ \Ko_2(T)\rho_0 }]_{\xv\xv'} :=
\bra{\v{x}}\m{ \Ko_2(T)\rho_0 }\ket{\v{x}'}$ read
\begin{align}\label{P2_TT2}
&[\m{ \Ko_2(T)\rho_0 }]_{\xv\xv'} =\nonumber\\
&\hspace{1cm}-\frac{1}{\hbar^2}\,\m{ \int_{0}^{T}\!\!\!\d t
\int_{0}^{t}\!\!\d t' \left[
\bra{\xv}V(\v{\xv},t)+\bra{\xv}\Ho^0\right]\times\left[ \int\! {d^3
y'}
V(\v{y}',t')\Pio(\v{y}')+\Ho^0\right]\rho_0\ket{\xv'}}\nonumber\\
\nonumber\\
&=-\frac{1}{\hbar^2}\!\! \int_{0}^{T}\!\!\!\d t \int_{0}^{t}\!\!\d
t'
\left\{\,\m{ V(\v{\xv},t)V(\v{\xv},t')}\rho_{\xv\xv'}(0)+\bra{\xv}(\Ho^0)^2\rho_0\ket{\xv'}\right.\nonumber\\
\nonumber\\
&\hspace{1cm}+\left.\m{
V(\v{\xv},t)}\bra{\xv}\Ho^0\rho_0\ket{\xv'}+\bra{\xv}\Ho^0\!\!\!\int\!
{d^3 y'}\m{ V(\v{y}',t')}\Pio(\v{y}')\rho_0\ket{\xv'}\,\right\},
\end{align}
where $\rho_{\xv\xv'}(0)=\bra{\xv}\rho_0\ket{\xv'}$. Since the
statistical properties of the conformal fluctuations $A$ are
stationary and spatially homogeneous and the potential $V$ is
quadratic, its average will take on a finite value, say $\m{
V(\v{\xv},t)} =: V_0$. As we will discuss more in details later on,
the precise value depends on the spectral properties of the
conformal fluctuations, as well as on the cutoff parameter that sets
the boundaries of the random scale. We have
\begin{align}
[\m{ \Ko_2(T)\rho_0 }]_{\xv\xv'} &=-\frac{1}{\hbar^2}\!\!
\int_{0}^{T}\!\!\!\d t \int_{0}^{t}\!\!\d t'
\left\{\,\m{ V(\v{\xv},t)V(\v{\xv},t')}\rho_{\xv\xv'}(0)+\bra{\xv}(\Ho^0)^2\rho_0\ket{\xv'}\right.\nonumber\\
\nonumber\\
&\hspace{1cm}+\left.V_0\bra{\xv}\Ho^0\rho_0\ket{\xv'}+V_0\bra{\xv}\Ho^0
\left[\int\! {d^3 y'}\Pio(\v{y}')\right]\rho_0\ket{\xv'}\,\right\}\nonumber\\
\nonumber\\
&\hspace{-2cm}=-\frac{1}{\hbar^2}\!\! \int_{0}^{T}\!\!\!\d t
\int_{0}^{t}\!\!\d t' \{\,\m{
V(\v{\xv},t)V(\v{\xv},t')}\rho_{\xv\xv'}(0)+\bra{\xv}(\Ho^0)^2\rho_0\ket{\xv'}+2V_0\bra{\xv}\Ho^0\rho_0\ket{\xv'}\},
\end{align}
where in the last line we used the fact that the position
eigenstates $\ket{\v{x}}$ form a complete set, as expressed by the
projectors identity $\int\! {d^3 y'}\Pio(\v{y}') = \I$. Performing
the time integrals in the last term and going back to the abstract
operators notation we finally have
\begin{equation}\label{P2_TT4}
\m{ \Ko_2(T)\rho_0} = [\Ko_2(T)]_0\rho_0 + \m{
\left[\Ko_2(T)\right]_1\rho_0} -\frac{V_0 T^2}{\hbar^2}\Ho^0\rho_0,
\end{equation}
where we defined the second order \emph{kinetic and potential
propagator} as
\begin{equation}\label{P2_NN5}
 [\Ko_2(T)]_0 := -\frac{1}{\hbar^2}
\int_{0}^{T}\!\!\!\d t\int_{0}^{t}dt'\Ho^0\,\Ho^0 =
-\frac{T^2}{2\hbar^2}(\Ho^0)^2,
\end{equation}
\begin{equation}\label{P2_NN6}
[\Ko_2(T)]_1 := -\frac{1}{\hbar^2} \int_{0}^{T}\!\!\!\d
t\int_{0}^{t}dt'{H^1}(t){H^1}(t').
\end{equation}
Of course, in equation \eqref{P2_TT4}, the statistical average
appears only in the term depending upon the potential propagator
$[\Ko_2(T)]_1 $.

The calculation of the term $\m{ \rho_0 \Ko_2^{\dag}(T)}$ can be
carried on in the same way since $\Ko_2(T)=\Ko_2^{\dag}(T)$ and
yields the final result
\begin{equation}\label{P2_TT7}
\m{ \rho_0 \Ko_2^{\dag}(T)} = \rho_0[\Ko_2(T)]_0^{\dag} + \m{
\rho_0\left[\Ko_2(T)\right]_1^{\dag}} -\frac{V_0
T^2}{\hbar^2}\rho_0\Ho^0.
\end{equation}

The last term $\m{ \Ko_1(T)\rho_0 \Ko_1(T)^{\dag}}$ must be
considered separately
\begin{align}\label{P2_TT8}
&\m{ \Ko_1(T)\rho_0 \Ko_1(T)^{\dag}}
=\nonumber\\
&\hspace{0.5cm}\frac{1}{\hbar^2}\,\m{ \int_{0}^{T}\!\!\!\d t
\left[\int\! {d^3 y}\,
V(\v{y},t)\Pio(\v{y})+\Ho^0\right]\times\rho_0\times\int_{0}^{T}\!\!\!\d
t\left[ \int\! {d^3 y'} V(\v{y}',t')\Pio(\v{y}')+\Ho^0\right]}.
\end{align}
Taking the matrix elements we have
\begin{align}\label{P2_TT9}
    [\m{ \Ko_1(T)\rho_0 \Ko_1(T)^{\dag}}]_{\xv\xv'}&=\frac{1}{\hbar^2}\,\m{
\int_{0}^{T}\!\!\!\d t\int_{0}^{T}\!\!\!\d t'\bra{\xv}[V(\xv,t)+\Ho^0]\,\rho_0\,[V(\xv',t')+\Ho^0]\ket{\xv'}}\nonumber\\
\nonumber\\
&=\frac{1}{\hbar^2}\!\! \int_{0}^{T}\!\!\!\d t\int_{0}^{T}\!\!\!\d
t'\,\{\,\m{
V(\xv,t)V(\xv',t')}\rho_{\xv\xv'}(0)+\bra{\xv}\Ho^0\rho_0\Ho^0\ket{\xv'}\nonumber\\
\nonumber\\
&\hspace{1.5cm}+\m{
V(\xv,t)}\bra{\xv}\rho_0\Ho^0\ket{\xv'}+\bra{\xv}\Ho^0\rho_0\ket{\xv'}\m{
V(\xv',t')}\,\}.
\end{align}
Performing the time integrals in the last term and going back to the
abstract operator notation yields
\begin{align}\label{P2_TT11}
     \m{ \Ko_1(T)\rho_0 \Ko_1(T)^{\dag}} =& [\Ko_1(T)]_0\rho_0
     [\Ko_1(T)]_0^{\dag} + \m{ [\Ko_1(T)]_1\rho_0
     [\Ko_1(T)]_1^{\dag}} +\frac{V_0 T^2}{\hbar^2}\rho_0\Ho^0+\frac{V_0 T^2}{\hbar^2}\Ho^0\rho_0,
\end{align}
where we defined the first order kinetic and potential propagator
\begin{equation}\label{P2_TT13}
[\Ko_1(T)]_0:=
-\frac{i}{\hbar}\int_{0}^{T}{\Ho^0}dt'=-\frac{i}{\hbar}\Ho^0 T,
\end{equation}
and
\begin{equation}\label{P2_TT12}
[\Ko_1(T)]_1:= -\frac{i}{\hbar}\int_{0}^{T}{H^1}(t')dt'.
\end{equation}
Substituting now equations \eqref{P2_TT4}, \eqref{P2_TT7} and
\eqref{P2_TT11} into equation \eqref{P2_Xe22} we see that the cross
terms containing the operators $\rho_0\Ho^0$ and $\Ho^0\rho_0$
cancel out and the average evolution for the density matrix follows
as
\begin{equation}\label{P2_finale}
\rho_T=[\rho_T]_0 + [\rho_T]_1,
\end{equation}
where
\begin{equation}\label{P2_finale2}
[\rho_T]_0:=\rho_0 + [\Ko_1(T)]_0\rho_0+\rho_0
[\Ko_1(T)]_0^{\dag}+[\Ko_2(T)]_0\rho_0+[\Ko_1(T)]_0\rho_0
[\Ko_1(T)]_0^{\dag}+\rho_0 [\Ko_2(T)]_0^{\dag},
\end{equation}
\begin{equation}\label{P2_finale1}
[\rho_T]_1:=\m{\rho_0 + [\Ko_1(T)]_1\rho_0+\rho_0
[\Ko_1(T)]_1^{\dag}+[\Ko_2(T)]_1\rho_0+[\Ko_1(T)]_1\rho_0
[\Ko_1(T)]_1^{\dag}+\rho_0 [\Ko_2(T)]_1^{\dag}}.
\end{equation}
We thus see that, at least up to second order in the Dyson
expansion, the kinetic and the potential parts contributions to the
density matric evolution can be calculated separately. We also
remark that this result is independent of the precise form of the
operator $\Ho^0$. In our present case though we have
$\Ho^0=\hat{p}^2/2M$ and equation \eqref{P2_finale2} simply
describes free evolution. Equation \eqref{P2_finale1} on the other
hand will describe dephasing.

\section{Nonlinear part of the
potential and density matrix evolution}\label{ap2_2}

In this section we prove the result \eqref{P1_P2_mainB} for the
general evolution of the density matrix. In Chapter \ref{ch1} we
showed that only the nonlinear part of the potential, i.e. $\cII
Mc^2 A^2$, gives a non vanishing contribution, coming from the
second order terms in the Dyson expansion \eqref{P2_finale1}. This
means we need to evaluate explicitly the terms $\Ko_2(T)\rho_0$,
$\Ko_1(T)\rho_0 \Ko_1^{\dag}(T)$ and $\rho_0 \Ko_2^{\dag}(T)$, where
the propagators are given in \eqref{P2_NN6} and \eqref{P2_TT12} (for
simplicity of notation here we do not indicate $[\,\,]_1$ for the
potential propagators).

Let us start to evaluate $\ms{ \Ko_2(T)\rho_0}$ for $\cII Mc^2 A^2$.
We have explicitly
\begin{align}\label{P1_P2_t1}
\m{ \Ko_2(T)\rho_0} = -&\frac{{\cII^2}M^2 c^4 A_0^4}{\hbar^2}
\int_{0}^{T}\!\!\!\d t \int d^3 y \ket{\yv}\bra{\yv}
\int_{0}^{t}dt'\int d^3 y' \ket{\yv'}\bra{\yv'}\times \nonumber\\
\times&\m{ \left[\int\! \d\kvs \,f_{\kvs}(t-\yv\cdot\kvs)\right]^2
\times \left[\int\! \d\kvs' \,f_{\kvs'}(t'-\yv'\cdot\kvs')\right]^2
}\rho_0.
\end{align}
To carry on the mean we write
\begin{equation}
\left[\int\! \d\kvs \,f_{\kvs}(t-\yv\cdot\kvs)\right]^2 =
\int\!\!\!\int\! \d\kvs \,\d\Kvs \,f_{\kvs}(t-\yv\cdot\kvs)
\,f_{\Kvs}(t-\yv\cdot\Kvs),
\end{equation}
in such a way that
\begin{align}\label{P2_4i}
&\m{ \left[\int\! \d\kvs \,f_{\kvs}(t-\yv\cdot\kvs)\right]^2 \times
\left[\int\! \d\kvs' \,f_{\kvs'}(t'-\yv'\cdot\kvs')\right]^2 }
=\\\nonumber
\\\nonumber
&=\int\!\!\!\int\!\!\!\int\!\!\!\int\! \d\kvs \,\d\Kvs\, \d\kvs'
\,\d\Kvs' \m{ f_{\kvs}(t-\yv\cdot\kvs)
\,f_{\Kvs}(t-\yv\cdot\Kvs)\,f_{\kvs'}(t'-\yv'\cdot\kvs')
\,f_{\Kvs'}(t'-\yv'\cdot\Kvs') }.
\end{align}
The four angular integration variables $\kvs, \Kvs, \kvs', \Kvs'$
are independent and care must be used in evaluating the multiple
mean while applying the statistical properties listed in Section
\ref{P1_P2_prop}. While performing the integral, the four angular
variables will take on all possible values and the explicit form of
the averaged integrand will depend upon these. In most of cases, for
example if $\kvs \neq (\Kvs, \kvs' \text{and } \Kvs')$, the
integrand simply vanishes since it involves at least one
uncorrelated directional component. For a correct evaluation, the
quadruple integral must be spit into a sum of various integrals
where the integrand is properly expressed in order to take into
account all possible reciprocal values of angular variables. Below
we list all the possible combinations that can occur. For a more
compact notation we use $\tau_{\kvs} := t-\v{y}\cdot\kvs$,
$\tau'_{\kvs'} := t'-\v{y}'\cdot\kvs'$ and $\tau_{\Kvs} :=
t-\v{y}\cdot\Kvs$ and $\tau'_{\Kvs'} := t'-\v{y}'\cdot\Kvs'$.
Keeping in mind that the directional components characterized by
$\kvs$ and $\Kvs$ depend on the unprimed coordinates $t, \yv$ while
those characterized by $\kvs'$ and $\Kvs'$ depend on the primed
coordinates $t', \yv'$, there are three main sets of possibilities:
\begin{description}
  \item[S1.] If $\kvs = \Kvs$ \emph{and} $\kvs' = \Kvs'$ then we need
  the second order correlation function defined in \eqref{P1_e21} as we are dealing
  with the squares of two directional components at the same point. At
  the same time, if $\kvs \neq \Kvs$ \emph{or} $\kvs' \neq \Kvs'$
  then the mean yields zero thanks to the statistical properties
  \eqref{P1_e15} and \eqref{P1_e20}. The occurrence of all these possibilities
  is summarized by expressing the integrand as
\begin{equation}\label{P1_e21B}
\delta(\kvs,\Kvs)\delta(\kvs',\Kvs')\m{
[f_{\kvs}(\tau)]^2[f_{\kvs'}(\tau')]^2
}=\delta(\kvs,\Kvs)\delta(\kvs',\Kvs')\left\{1+\delta(\kvs,\kvs')
\left[R''(\tau_{\kvs}-\tau'_{\kvs'})-1\right]\right\}.
\end{equation}
  \item[S2.] If $\kvs = \kvs'$ \emph{and} $\Kvs = \Kvs'$ then we are
  dealing with means of products involving a given directional component at
  different points. In this case we need the autocorrelation function and
  the integrand can be expressed with the help of \eqref{P1_e17} as
\begin{equation}\label{P1_e17B}
\m{ f_{\kvs}(\tau)f_{\kvs'}(\tau') }\m{
f_{\Kvs}(\tau)f_{\Kvs'}(\tau')
}=\delta(\kvs,\kvs')\delta(\Kvs,\Kvs')R(\tau_{\kvs}-\tau'_{\kvs'})
R(\tau_{\Kvs}-\tau'_{\Kvs'})\left[1 - \delta(\kvs,\Kvs)\right].
\end{equation}
  \item[S3.] If $\kvs = \Kvs'$ \emph{and} $\Kvs = \kvs'$ then we have
  in a similar way
\begin{equation}\label{P1_e17C}
\m{ f_{\kvs}(\tau)f_{\Kvs'}(\tau') }\m{
f_{\Kvs}(\tau)f_{\kvs'}(\tau')
}=\delta(\kvs,\Kvs')\delta(\Kvs,\kvs')R(\tau_{\kvs}-\tau'_{\Kvs'})
R(\tau_{\Kvs}-\tau'_{\kvs'})\left[1 - \delta(\kvs,\Kvs)\right].
\end{equation}
\end{description}
We remark that the cases S2 and S3 are indeed complementary to S1,
since they occur when $\kvs \neq \Kvs$ \emph{and} $\kvs' \neq \Kvs'$
(in which case S1 yields zero) while giving a non vanishing
contribution. We also notice that the extra delta function factor
$\left[1 - \delta(\kvs,\Kvs)\right]$ in \eqref{P1_e17B} and
\eqref{P1_e17C} has been introduced in order to have a vanishing
integrand when $\kvs = \Kvs = \kvs' = \Kvs'$. Indeed this situation
involving only one directional component must be properly expressed
with the second order correlation function and, as such, it is
already included in S1.

Given these rules the multiple integral in \eqref{P2_4i} must be
correctly split into the sum of three integrals
\begin{small}
\begin{align*}\label{P1_P2_4iB}\nonumber
&\hspace{-0.2cm}\int\!\!\!\int\!\!\!\int\!\!\!\int\! \d\kvs
\,\d\Kvs\, \d\kvs' \,\d\Kvs' \m{ f_{\kvs}(t-\yv\cdot\kvs)
\,f_{\Kvs}(t-\yv\cdot\Kvs)\,f_{\kvs'}(t'-\yv'\cdot\kvs')
\,f_{\Kvs'}(t'-\yv'\cdot\Kvs') }=\\\nonumber
\\\nonumber
&\hspace{-0.1cm}=\int\!\!\!\int\!\!\!\int\!\!\!\int\! \d\kvs
\,\d\Kvs\, \d\kvs' \,\d\Kvs'\,\,
\delta(\kvs,\Kvs)\delta(\kvs',\Kvs')\left\{1+\delta(\kvs,\kvs')\left[R''(t-t'-\v{y}\cdot\kvs
+\v{y}'\cdot\kvs')-1\right]\right\}\\\nonumber
\\\nonumber
&+\!\!\int\!\!\!\int\!\!\!\int\!\!\!\int\! \d\kvs \,\d\Kvs\, \d\kvs'
\,\d\Kvs' \delta(\kvs,\kvs')\delta(\Kvs,\Kvs')R( t-t'-\v{y}\cdot\kvs
+\v{y}'\cdot\kvs' ) R(t-t'-\v{y}\cdot\Kvs + \v{y}'\cdot\Kvs'
)\left[1 - \delta(\kvs,\Kvs)\right]\\\nonumber
\\\nonumber
&+\!\!\int\!\!\!\int\!\!\!\int\!\!\!\int\! \d\kvs \,\d\Kvs\, \d\kvs'
\,\d\Kvs' \delta(\kvs,\Kvs')\delta(\Kvs,\kvs')R( t-t'-\v{y}\cdot\kvs
+\v{y}'\cdot\Kvs' ) R(t-t'-\v{y}\cdot\Kvs + \v{y}'\cdot\kvs'
)\left[1 - \delta(\kvs,\Kvs)\right]\\
\\
&\hspace{-0.2cm}=: I_1 + I_2 + I_3,
\end{align*}
\end{small}
where we also re-expressed explicitly the arguments of all the
correlation functions and with $I_1, I_2$ and $I_3$ defining
respectively the three integrals corresponding to the possibilities
S1, S2 and S3. These integral must be evaluated one by one. Since
the integrand in $I_1$ doesn't depend upon $\Kvs$ or $\Kvs'$ we use
the fact that $\int\d\Kvs\delta(\kvs,\Kvs) = 1$ and
$\int\d\Kvs'\delta(\kvs',\Kvs') = 1$ to get
\begin{equation*}
I_1 = \int\!\!\!\int\! \d\kvs \,\d\kvs' \,\,
\left\{1+\delta(\kvs,\kvs')\left[R''(t-t'-\v{y}\cdot\kvs
+\v{y}'\cdot\kvs')-1\right]\right\}.
\end{equation*}
We can now integrate with respect to $\kvs'$ and we get three terms.
Since $\int \d \kvs' = 4\pi$ we find
\begin{equation}\label{P1_P2_i1}
I_1 = \int\! \d\kvs \left\{ 4\pi + R''[t-t'-\kvs\cdot(\v{y}-\v{y}')]
- 1 \right\}.
\end{equation}
To evaluate $I_2$ we start by integrating over $\kvs'$ and $\Kvs'$
to obtain
\begin{equation*}
I_2 = \!\!\int\!\!\!\int\! \d\kvs \,\d\Kvs\, R[
t-t'-\kvs\cdot(\v{y}-\v{y}')] \times R[
t-t'-\Kvs\cdot(\v{y}-\v{y}')]\left[1 - \delta(\kvs,\Kvs)\right].
\end{equation*}
Carrying on the integration with respect to $\Kvs$ this yields two
contributions
\begin{equation*}
I_2 = \!\!\int\! \d\kvs\left\{\int\! \d\Kvs\,  R[
t-t'-\kvs\cdot(\v{y}-\v{y}')]\times R[ t-t'-\Kvs\cdot(\v{y}-\v{y}')]
- R^2[ t-t'-\kvs\cdot(\v{y}-\v{y}')]\right\}.
\end{equation*}
The integral $I_3$ gives exactly the same result and we can write
\begin{align}\label{P1_P2_i2i3}
I_2 + I_3=& \nonumber\\
&\hspace{-1cm}2\!\!\int\! \d\kvs\left\{\int\! \d\Kvs\,  R[
t-t'-\kvs\cdot(\v{y}-\v{y}')] \times R[
t-t'-\Kvs\cdot(\v{y}-\v{y}')] - R^2[
t-t'-\kvs\cdot(\v{y}-\v{y}')]\right\}.
\end{align}

To further simplify equation \eqref{P1_P2_t1} and perform the space
integration with respect to $\yv$ and $\yv'$, it is useful to go to
the position representation and evaluate $\ms{
\Ko_2(T)\rho_0}_{\xv\xv'} :=
\bra{\xv}\ms{\Ko_2(T)\rho_0}\ket{\xv'}$. The density matrix $\rho_0$
operates on the ket $\ket{\xv'}$ while the projection operators act
from the right on the bra $\bra{\xv}$ producing delta functions. By
investigating the structure of the right hand side in
\eqref{P1_P2_t1} it is easily seen that the effect of integrating
with respect to $\yv$ first and $\yv'$ next is to transform all the
$\yv$ and $\yv'$ comparing in the arguments of the correlation
functions into $\xv$, in such a way that all the factors containing
the difference $\yv - \yv'$ disappear. Therefore the following
results hold
\begin{align*}
&\bra{\xv}\left[\int d^3 y \ket{\yv}\bra{\yv}\int d^3 y'
\ket{\yv'}\bra{\yv'}\,\, I_1\right] = \bra{\xv} \int\! \d\kvs
\left\{ 4\pi - 1 + R''(t-t')\right\},\\
\\
&\bra{\xv}\left[\int d^3 y \ket{\yv}\bra{\yv}\int d^3 y'
\ket{\yv'}\bra{\yv'}\,\, (I_2+I_3)\right] = \bra{\xv} \!\!\int\!
\d\kvs 2(4\pi - 1)R^2( t-t').
\end{align*}
We can finally plug these back into equation \eqref{P1_P2_t1} which
gives, in the position representation
\begin{equation}\label{P1_2ord1}
\m{ \Ko_2(T)\rho_0}_{\xv\xv'}\!\!\! = -\frac{{\cII^2}M^2 c^4 A_0^4
\rho_{\xv\xv'}(0)}{\hbar^2} \int_{0}^{T}\!\!\!\d t \int_{0}^{t}dt'
\!\!\int\! \d\kvs \left\{ 4\pi - 1 + R''(t-t') + 2(4\pi - 1)R^2(
t-t')\right\},
\end{equation}
where $\rho_{\xv\xv'}(0):=\bra{\xv}\rho_0\ket{\xv'}$. Since
$\Ko_2(T)=\Ko_2(T)^{\dag}$, the evaluation of the second order term
$\ms{ \rho_0 \Ko_2(T)^{\dag} }$ would proceed in exactly the same
way to yield
\begin{equation}\label{P1_2ord1B}
\m{ \rho_0 \Ko_2(T)^{\dag}}_{\xv\xv'} = \m{
\Ko_2(T)\rho_0}_{\xv\xv'}.
\end{equation}

Finally we must consider the remaining second order term $\m{
\Ko_1(T)\rho_0 \Ko_1(T)^{\dag}}$. This reads explicitly
\begin{scriptsize}
\begin{align*}
\m{ \Ko_1(T)\rho_0 \Ko_1(T)^{\dag} } = \frac{{\cII^2}M^2 c^4
A_0^4}{\hbar^2} \m{ \int_{0}^{T}\!\!\!\d t \int_{0}^{T}dt' \int d^3
y \ket{\yv}\bra{\yv}  \left[\int\! \d\kvs
\,f_{\kvs}(t-\yv\cdot\kvs)\right]^2 \times \rho_0 \times \int d^3 y'
\ket{\yv'}\bra{\yv'} \left[\int\! \d\kvs'
\,f_{\kvs'}(t'-\yv'\cdot\kvs')\right]^2 }
\end{align*}
\end{scriptsize}
and it should be noticed that the both time integrals run from 0 to
$T$. The main difference when compared to \eqref{P1_P2_t1} is that
the density matrix is now `squeezed' in between space and
directional integrals defining the potential. For this reason it is
convenient to go to the position representation and carry out the
integrations with respect to $\yv$ and $\yv'$ and evaluate the
statistical average afterwards. This gives
\begin{scriptsize}
\begin{align*}
\m{ \Ko_1(T)\rho_0 \Ko_1(T)^{\dag} }_{\xv\xv'} \!\!\!=
\frac{{\cII^2}M^2 c^4 A_0^4\rho_{\xv\xv'}(0)}{\hbar^2}
\int_{0}^{T}\!\!\!\d t \int_{0}^{T}\!\!\!dt' \m{\left[\int\! \d\kvs
\,f_{\kvs}(t-\xv\cdot\kvs)\right]^2\!\! \times\! \left[\int\!
\d\kvs' \,f_{\kvs'}(t'-\xv'\cdot\kvs')\right]^2 }.
\end{align*}
\end{scriptsize}
The statistical average has the same structure as in \eqref{P2_4i}
and the result can be read off equations \eqref{P1_P2_i1} and
\eqref{P1_P2_i2i3} by simply replacing $\yv$ and $\yv'$ with $\xv$
and $\xv'$. By1 defining
\begin{equation}
    \Delta\xv := \xv - \xv'
\end{equation}
we have
\begin{align}\label{P1_P2_sot}
\m{ \Ko_1(T)\rho_0 \Ko_1(T)^{\dag} }_{\xv\xv'} \!\!\!=
&\frac{{\cII^2}M^2 c^4 A_0^4\rho_{\xv\xv'}(0)}{\hbar^2}
\int_{0}^{T}\!\!\!\d t \int_{0}^{T}\!\!\!dt' \int\! \d\kvs \left\{
4\pi - 1 + R''(t-t'-\kvs\cdot\Delta\xv)\right.\\\nonumber
\\\nonumber
&\left.\hspace{-1cm}+ \,2\!\int\! \d\Kvs\,  R(
t-t'-\kvs\cdot\Delta\xv) \times R( t-t'-\Kvs\cdot\Delta\xv) - 2R^2(
t-t'-\kvs\cdot\Delta\xv) \right\}.
\end{align}

Bringing together the result \eqref{P1_2ord1}, \eqref{P1_2ord1B} and
\eqref{P1_P2_sot} into the genera Dyson expansion expression, we get
a formula for the evolved density matrix at time $t$:
\begin{align}\label{P1_azz}
\rho_{\xv\xv'}(T)=&\,\,\rho_{\xv\xv'}(0) -\frac{{\cII^2}M^2 c^4
A_0^4
\rho_{\xv\xv'}(0)}{\hbar^2} \int\! \d\kvs \nonumber\\
\nonumber\\
&\left\{ 2\!\!\int_{0}^{T}\!\!\!\d t \int_{0}^{t}dt'  \left[ (4\pi -
1) + R''(t-t') -2R^2( t-t') + 8\pi R^2( t-t') \right]
\right.\nonumber\\
\nonumber\\
&- \int_{0}^{T}\!\!\!\d t \int_{0}^{T}\!\!\!dt' \left[  (4\pi - 1) +
R''(t-t'-\kvs\cdot\Delta\xv) - 2R^2(
t-t'-\kvs\cdot\Delta\xv)\right.\nonumber\\
\nonumber\\
&\hspace{2.3cm}\left.+ \left.\,2\!\int\! \d\Kvs\,  R(
t-t'-\kvs\cdot\Delta\xv) \times R( t-t'-\Kvs\cdot\Delta\xv)\right]
\right\}.
\end{align}
This seemingly complicated expression, involving multiple time
integrals and directional integrals of the first and second order
correlation functions can still be simplified significantly using
the result proven in Section \ref{ap2_3}:
\begin{equation}\label{P1_canc}
2\int_{0}^{T}\!\!\!\d t \!\!\int_{0}^{t}\!\!\d t'
f(t-t')-\int_{0}^{T}\!\!\!\d t\!\!\int_{0}^{T}\!\!\!\d
t'f(t-t'-\kvs\cdot\Delta \xv) = 0,
\end{equation}
holding for an even function $f$ and for $T \gg \kvs\cdot\Delta\xv$.

Since the first and second order correlation functions are even
functions of their argument we see that the first three time
integrals in \eqref{P1_azz} involving the quantities $(4\pi -1)$,
$R''(t-t')$ and $-2R^2(t-t')$ are precisely in the form
\eqref{P1_canc}. Therefore their contribution vanishes and the
general formula for the density matrix evolution finally reduces to
\begin{align}\nonumber
\rho_{\xv\xv'}(T)=\,\,\rho_{\xv\xv'}(0) -&\frac{{2\cII^2}M^2 c^4
A_0^4 \rho_{\xv\xv'}(0)}{\hbar^2} \int\! \d\kvs \int_{0}^{T}\!\!\!\d
t \int_{0}^{T}\!\!\!dt' \\\nonumber
\\
&\left\{ 4\pi R^2( t-t')-  \int\! \d\Kvs\, R(
t-t'-\kvs\cdot\Delta\xv/c) \times R( t-t'-\Kvs\cdot\Delta\xv/c)
\right\}.
\end{align}
This can be written in a slightly more compact form as
\begin{eqnarray}\nonumber
\rho_{\xv\xv'}(T)=\,\,\rho_{\xv\xv'}(0) -\frac{32\cII^2\pi^2 M^2 c^4
A_0^4 \rho_{\xv\xv'}(0)}{\hbar^2} \times \left[ \int_{0}^{T}\!\!\!\d
t \int_{0}^{T}\!\!\!{\d}t' R^2( t-t') \right.\\\nonumber
\\
\left. \hspace{-2cm}- \frac{1}{16\pi^2} \int\! \d\kvs\int\! \d\Kvs\,
\int_{0}^{T}\!\!\!\d t \int_{0}^{T}\!\!\!{\d}t' R(
t-t'-\kvs\cdot\Delta\xv/c) \times R( t-t'-\Kvs\cdot\Delta\xv/c)
\right],
\end{eqnarray}
which is precisely the expression \eqref{P1_P2_mainB}.

\section{An integral identity}\label{ap2_3}

\begin{quote}
In this appendix we prove that the result
\begin{equation*}
 I := \int_{0}^{T}\!\!\!\d t \left[2 \int_{0}^{t}\!\!\d t'
f(t-t')- \int_{0}^{T}\!\!\!\d t'f(t-t'-\kvs\cdot\Delta \xv)\right] =
0,
\end{equation*}
used in Section \ref{cdu}, holds for an \emph{arbitrary even
function} $f$
and in the limit $\kvs\cdot\Delta\xv / T \rightarrow 0$.\\
\end{quote}
For simplicity let $\Delta := \kvs\cdot\Delta \xv.$ Defining the
variable $\tau:= t-t'$ we have
\begin{equation*}
    \int_{0}^{t}\!\!\d t'
f(t-t') = \int_0^t\!\!\!\d\tau f(\tau),
\end{equation*}
while, with $\tau:= t-t'-\Delta$,
\begin{equation*}
    \int_{0}^{T}\!\!\!\d
t'f(t-t'-\Delta)=\int_{t-\Delta - T}^{t-\Delta}\!\!\!\d\tau
f(\tau)=\int_0^{t-\Delta}\!\!\!\d\tau f(\tau) - \int_0^{t-\Delta -
T}\!\!\!\d\tau f(\tau).
\end{equation*}
Introducing the primitive of $f(t)$
\begin{equation*}
F(t):=\int_0^t\!\!\!\d\tau f(\tau)
\end{equation*}
we can re-write $I$ as
\begin{equation*}
I=2\int_{0}^{T}\!\!\!{\d}t F(t)-\int_{0}^{T}\!\!\!{\d}t\,
F(t-\Delta) + \int_{0}^{T}\!\!\!{\d}t\, F(t-\Delta -T).
\end{equation*}
Performing a further change of variable $z:= t - \Delta$ the second
integral reads
\begin{eqnarray*}
 \int_{0}^{T}\!\!\!{\d}t\, F(t-\Delta) = \int_{-\Delta}^{T -
\Delta}\!\!\!dz\, F(z) := \int_{0}^{T - \Delta}\!\!\!dz\, F(z) -
\int_{0}^{- \Delta}\!\!\!dz\, F(z).
\end{eqnarray*}
Performing a similar operation on the third integral we obtain
\begin{equation*}
 I=2\int_{0}^{T}\!\!\!{\d}t F(t)- \int_{0}^{T - \Delta}\!\!\!dz\,
F(z) + \int_{0}^{- \Delta}\!\!\!dz\, F(z) + \int_{0}^{-
\Delta}\!\!\!dz\, F(z) - \int_{0}^{-T - \Delta}\!\!\!dz\, F(z).
\end{equation*}
Now we use the information $T \gg \Delta$ and $f(t)=f(-t)$. As an
elementary consequence we have that $F(t)=-F(-t)$, implying that
\begin{equation*}
\int_{0}^{\chi}\!\!\!\d\tau F(\tau)=\int_{0}^{-\chi}\!\!\!\d\tau
F(\tau).
\end{equation*}
Approximating $T \pm \Delta \approx T$ and swapping the sign of the
upper integration bound appropriately we have
\begin{eqnarray*}
 I=2\int_{0}^{T}\!\!\!{\d}t F(t)- \left[\int_{0}^{T}\!\!\!dz\,
F(z) - \int_{0}^{\Delta}\!\!\!dz\, F(z)\right] - \left[
\int_{0}^{T}\!\!\!dz\, F(z) - \int_{0}^{\Delta}\!\!\!dz\,
F(z)\right].
\end{eqnarray*}
The integrals from 0 to $\Delta$ can all be neglected exploiting
again the fact that $T \gg \Delta$ and we obtain
\begin{eqnarray*}
I \approx 2\int_{0}^{T}\!\!\!{\d}t F(t)-2\int_{0}^{T}\!\!\!{\d}t
F(t)=0.
\end{eqnarray*}
The result is exact in the limit $\Delta / T \rightarrow 0.$

\section{Fourth order term in the Dyson expansion}\label{ap2_4}

The fourth order propagator in the Dyson expansion \eqref{P2_Xe22}
would be given by
\begin{equation}
 \Ko_4(T):=\left(\frac{-i}{\hbar}\right)^4\!\!
\int_{0}^{T}\!\!{\d}t^{(1)}\!\!\int_{0}^{t^{(1)}}\!\!\!{\d}t^{(2)}\!\!\int_{0}^{t^{(2)}}\!\!\!{\d}t^{(3)}\!\!
\int_{0}^{t^{(3)}}\!\!\!\!{\d}t^{(4)}{\Ho}(t^{(1)}){\Ho}(t^{(2)}){\Ho}(t^{(3)}){\Ho}(t^{(4)}),
\end{equation}
implying a fourth order term in the expression for the density
matrix evolution given by
\begin{small}
\begin{eqnarray*}
\m{\Ko_4 \rho_0} \sim
\int_{0}^{T}\!\!{\d}t^{(1)}\!\!\int_{0}^{t^{(1)}}\!\!\!{\d}t^{(2)}\!\!\int_{0}^{t^{(2)}}\!\!\!{\d}t^{(3)}\!\!
\int_{0}^{t^{(3)}}\!\!\!\!{\d}t^{(4)} \int \d^3 y^{(1)}
\Pr{\yv^{(1)}}\cdots\int \d^3 y^{(4)}\Pr{\yv^{(4)}}\times\\
\times\m{V(\yv^{(1)},t^{(1)})V(\yv^{(2)},t^{(2)})V(\yv^{(3)},t^{(3)})V(\yv^{(4)},t^{(4)})}\rho_0.
\end{eqnarray*}
\end{small}
Since the potential is $V = Mc^2 (\cI A + \cII A^2)$ the average
yields one term proportional to $A_0^4$:
\begin{eqnarray*}
\m{V(\yv^{(1)},t^{(1)})V(\yv^{(2)},t^{(2)})V(\yv^{(3)},t^{(3)})V(\yv^{(4)},t^{(4)})}
\Rightarrow\\
\hspace{-1cm}\left(\frac{Mc^2 A_0}{\hbar}\right)^4\int\!
\d\kvs^{(1)}\cdots\int\!
\d\kvs^{(4)}\m{f_{\kvs^{(1)}}(\tau^{(1)})f_{\kvs^{(2)}}(\tau^{(2)})f_{\kvs^{(3)}}(\tau^{(3)})f_{\kvs^{(4)}}(\tau^{(4)})},
\end{eqnarray*}
where $\tau^{(i)} := t^{(i)} - \yv^{(i)}\cdot\kvs^{(i)}$. This
requires knowledge of the four-point correlation function, involving
the average of the product of four directional components evaluated
at different points. For a real random process having a zero mean
and gaussian distribution the four-points function reduces to
\cite{adler81,erkisen02}:
\begin{eqnarray*}
\m{f_{\kvs^{(1)}}(\tau^{(1)})f_{\kvs^{(2)}}(\tau^{(2)})f_{\kvs^{(3)}}(\tau^{(3)})f_{\kvs^{(4)}}(\tau^{(4)})}
=
\m{f_{\kvs^{(1)}}(\tau^{(1)})f_{\kvs^{(2)}}(\tau^{(2)})}\m{f_{\kvs^{(3)}}(\tau^{(3)})f_{\kvs^{(4)}}(\tau^{(4)})}\\
+\m{f_{\kvs^{(1)}}(\tau^{(1)})f_{\kvs^{(3)}}(\tau^{(3)})}\m{f_{\kvs^{(2)}}(\tau^{(2)})f_{\kvs^{(4)}}(\tau^{(4)})}\\
+\m{f_{\kvs^{(1)}}(\tau^{(1)})f_{\kvs^{(4)}}(\tau^{(4)})}\m{f_{\kvs^{(2)}}(\tau^{(2)})f_{\kvs^{(3)}}(\tau^{(3)})}.
\end{eqnarray*}
We can now use equation \eqref{P1_e17} to express the 2-point
correlations:
\begin{eqnarray*}
\m{f_{\kvs^{(1)}}(\tau^{(1)})f_{\kvs^{(2)}}(\tau^{(2)})f_{\kvs^{(3)}}(\tau^{(3)})f_{\kvs^{(4)}}(\tau^{(4)})}
=\\
\hspace{0.3cm}\delta(\kvs^{(1)},\kvs^{(2)})R(\tau^{(1)}-\tau^{(2)})\delta(\kvs^{(3)},\kvs^{(4)})R(\tau^{(3)}-\tau^{(4)})\\
+\delta(\kvs^{(1)},\kvs^{(3)})R(\tau^{(1)}-\tau^{(3)})\delta(\kvs^{(2)},\kvs^{(4)})R(\tau^{(2)}-\tau^{(4)})\\
+\delta(\kvs^{(1)},\kvs^{(4)})R(\tau^{(1)}-\tau^{(4)})\delta(\kvs^{(2)},\kvs^{(3)})R(\tau^{(2)}-\tau^{(3)}).
\end{eqnarray*}
This implies that the term $T_4[A_0^4]$ deriving from $\m{\Ko_4
\rho_0}$ has the structure:
\begin{equation*}
 T_4[A_0^4] \sim 3 \left[\int\! \d\kvs^{(1)}\!\!\int\!\!
\d\kvs^{(2)}\!\!
\int_{0}^{T}\!\!\!{\d}t^{(1)}\!\!\int_{0}^{T}\!\!\!{\d}t^{(2)}
\delta(\kvs^{(1)},\kvs^{(2)})R(\tau^{(1)}-\tau^{(2)}) \right]^2,
\end{equation*}
where all upper bounds in the time integration can be set equal to
$T$ by appropriate normal ordering \cite{roman65}. By carrying out
one of the two angular integrations we have:
\begin{equation*}
 T_4[A_0^4] \sim 3 \left[\int\! \d\kvs^{(1)}\!\!
\int_{0}^{T}\!\!\!{\d}t^{(1)}\!\!\int_{0}^{T}\!\!\!{\d}t^{(2)}
R[t^{(1)} - t^{(2)} - \kvs^{(1)}\cdot(\yv^{(1)}-\yv^{(2)})]
\right]^2.
\end{equation*}
The double time integral can be simplified using the general result
\eqref{P1_TTtt01} and we have
\begin{equation*}
T_4[A_0^4] \sim 3 \left[\int\! \d\kvs^{(1)} \mathfrak{F}\left[R(t +
\tau)\right](0) T \right]^2,
\end{equation*}
where $\mathfrak{F}$ denotes Fourier transform and $\tau :=
-\kvs^{(1)}\cdot(\yv^{(1)}-\yv^{(2)}). $ The Fourier transform can
be evaluated using that $ R(t + \tau)= C(t+\tau) / C_0.$ Then
\begin{small}
\begin{eqnarray*}
\mathfrak{F}\left[R(t + \tau)\right](\omega) =
\frac{1}{C_0}\int_{-\infty}^{\infty}\d t C(t+\tau) e^{-i\omega t}
\nonumber\\
= \frac{1}{C_0(2\pi c)^3} \int_0^{\omega_c}\!\!\d
\omega'\int_{-\infty}^{\infty}\d t\, \omega'^2 S(\omega')
e^{-i\omega t} \cos \omega' (t+\tau)\nonumber\\
= \frac{1}{2C_0(2\pi c)^3} \int_0^{\omega_c}\!\!\d
\omega'\int_{-\infty}^{\infty}\d t\, \omega'^2 S(\omega')
e^{-i\omega t} \left[ e^{i\omega'(t+\tau)} + e^{-i\omega'(t+\tau)}\right]\nonumber\\
= \frac{1}{2C_0(2\pi c)^3} \int_0^{\omega_c}\!\!\d \omega' \omega'^2
S(\omega') \int_{-\infty}^{\infty}\d t\, \left[
e^{i(\omega'-\omega)t}e^{i\omega'\tau} +
e^{i(-\omega'-\omega)t}e^{-i\omega'\tau}\right].
\end{eqnarray*}
\end{small}
Integrating with respect to $t$ gives
\begin{eqnarray*}
\mathfrak{F}\left[R(t + \tau)\right](\omega) =\frac{\pi}{C_0(2\pi
c)^3} \int_0^{\omega_c}\!\!\d \omega' \omega'^2 S(\omega') \left[
\delta(\omega'-\omega)e^{i\omega'\tau} +
\delta(-\omega'-\omega)e^{-i\omega'\tau}\right].
\end{eqnarray*}
This vanishes for $\omega > \omega_c$. For $\omega = 0$ we have:
\begin{eqnarray*}
\mathfrak{F}\left[R(t + \tau)\right](0) =\frac{\pi}{C_0(2\pi c)^3}
\int_0^{\omega_c}\!\!\d \omega' \omega'^2 S(\omega') \left[
\delta(\omega')e^{i\omega'\tau} +
\delta(-\omega')e^{-i\omega'\tau}\right].
\end{eqnarray*}
Carrying out the frequency integral and using the properties of the
$\delta$ function we have:
\begin{eqnarray*}
\mathfrak{F}\left[R(t + \tau)\right](0) &=\frac{2\pi}{C_0(2\pi c)^3}
\lim_{\omega\rightarrow 0} \omega^2 S(\omega).
\end{eqnarray*}
In the interesting case $S \propto 1/\omega$ this tends to 0, in
such a way that $T_4[A_0^4] \rightarrow 0$ and the fourth order
Dyson expansion term doesn't give any contribution.

\newpage
\thispagestyle{empty}


\singlespacing
\chapter{Zero point energy density and pressure of matter fields
in the free field approximation}\label{ap3}

\begin{quote}
\begin{small}
In this Appendix we estimate the zero point energy density and
pressure of a typical matter field in its vacuum state as described
within the random gravity framework. The special case of a scalar
massive field with a Klein-Gordon stress energy tensor is considered
as a simple model and used to derive the expression for the vacuum
power spectral density introduced in Section \ref{P1_pcf}.
\end{small}
\end{quote}
\singlespacing

\section{Modeling the zero point energy of a massive scalar field}

Estimating the total amount of vacuum energy in a rigorous way is a
nontrivial QFT problem: within the Standard Model, vacuum energy
density is estimated to be given by, at least, three main
contributions \cite{rugh02}: (i) vacuum zero-point energy plus
virtual particles fluctuations, (ii) QCD gluon and quark condensates
(iii) Higgs field. With the following calculation we want to get a
first approximation estimate, by neglecting fields interactions and
by describing the free-field configuration as a collection of
decoupled harmonic oscillators of frequency
\begin{equation}\label{wkm}
\omega_{k}=\sqrt{c^2k^2 +m^2c^4},
\end{equation}
where $k$ is the norm of the spatial wave vector $\kv$ and $m$ is
the mass of the field quanta under examination. This is quite
accurate for the EM field but it is likely to be quite a crude
approximation for e.g. the QCD sector. By doing so we are neglecting
higher order contributions to the vacuum energy density as well as
the nonlinear and strong coupling effects of QCD and a detailed
treatment of the Higgs fields. Nonetheless we expect to obtain a
meaningful lower bound estimate to the vacuum energy.

We consider an arbitrary matter field of mass $m$ within the
Standard Model. We calculate a lower bound to the associated vacuum
energy density by estimating the overall energy density resulting
from the summation of the zero point energies of all frequency
modes, up to the random scale cutoff set by $\ell = \lambda L_P.$

We model each independent field component as a scalar field $\phi$
with mass $m$ and associated Klein-Gordon stress-energy tensor
\begin{equation}\label{strs}
\TKG[\phi]=\phi_a\phi_b-\frac12\eta_{ab}
\left(\phi^c\phi_c+\frac{m^2c^2}{\hbar^2}\phi^2\right).
\end{equation}
From this we see that $\phi^2$ has the dimension of force. Regarding
$\phi$ as in its zero-point fluctuating state, we shall assume it to
be stationary over spacetime and having an isotropic spectral
density $\Sphi(k)$. In Appendix \ref{ap1} we have shown that the
mean squared field is given by
\begin{equation}\label{mphi}
\m{\phi^2}= \frac{1}{(2\pi)^3}\int d^3 k\,\Sphi(k) .
\end{equation}
Furthermore, the mean squared time derivative $\phi_{,t}=c\phi_{,0}$
satisfies
\begin{equation}\label{mphid}
\ms{\phi_{,t}^{\;2}}= \frac{1}{(2\pi)^3}\int d^3 k\,\omega_k^2
\Sphi(k)
\end{equation}
and the mean squared gradient $(\nabla\phi)_i:=\phi_{,i}$ for
$i=1,2,3$ satisfies
\begin{equation}\label{mphip}
\m{\abs{\nabla\phi}^2}= \frac{1}{(2\pi)^3}\int d^3 k\,k^2 \Sphi(k).
\end{equation}

It follows from \eqref{wkm}, \eqref{mphi}, \eqref{mphid},
\eqref{mphip} that
\begin{equation}\label{rel1}
\m{\phi^a\phi_a}=-\frac1{c^2}\ms{\phi_{,t}^{\;2}}+\m{\abs{\nabla\phi}^2}
=-\frac{m^2c^2}{\hbar^2}\m{\abs{\phi}^2}.
\end{equation}
Using \eqref{mphid}, \eqref{mphip} and \eqref{rel1} we see that the
mean stress-energy tensor given by \eqref{strs} becomes
\begin{equation}\label{strsB}
\m{\TKG}=\m{\phi_a\phi_b},
\end{equation}
which yields the effective energy density
\begin{equation}\label{phirho1}
\rho=\m{T_{00}}=\frac1{c^2}\ms{\phi_{,t}^{\;2}}
=\frac{1}{(2\pi)^3}\int d^3 k\,\frac{\omega_k^2}{c^2} \Sphi(k)
\end{equation}
and the effective pressure
\begin{equation}\label{phiP1}
p= \frac13\ms{T^i{}_i} =\frac13\m{\abs{\nabla\phi}^2}
=\frac{1}{3(2\pi)^3}\int d^3 k\,k^2 \Sphi(k).
\end{equation}
A useful combination follows from \eqref{wkm}, \eqref{phirho1} and
\eqref{phiP1} as
\begin{equation}\label{phirhoP2}
\rho-3p=\frac{1}{(2\pi)^3}\frac{m^2c^2}{\hbar^2}\int d^3 k\,
\Sphi(k).
\end{equation}

The spectral density $\Sphi(k)$ itself can be determined through the
well-known zero point energy density expression
\begin{equation}\label{phirho2}
\rho=\frac{1}{(2\pi)^3}\int d^3 k\,\frac{\hbar\omega_k}{2},
\end{equation}
that adds up contributions from all wave modes of $\phi$. This
amounts to neglecting, in a first approximation, the virtual
particles non linear terms due to the matter field interactions.
Comparing \eqref{phirho1} and \eqref{phirho2} we see that the
spectral density $\Sphi(k)$ takes the form
\begin{equation}\label{phiSDk}
\Sphi(k)=\frac{\hbar c^2}{2\omega_k}.
\end{equation}

Substituting \eqref{wkm} and \eqref{phiSDk} into \eqref{phirho1} and
\eqref{phiP1} and integrating using $\int d^3 k = 4\pi\int dk k^2$
up to the cutoff value $k_\lambda=\frac{2\pi}{\lambda L_P}$, we
obtain
\begin{equation}\label{phirho4}
\rho =\frac{\hbar}{4\pi^2}\int_0^{k_\lambda}d k\,k^2
\sqrt{c^2k^2+\frac{m^2c^4}{\hbar^2}},
\end{equation}
and
\begin{equation}\label{phiP4}
p =\frac{\hbar c^2}{12\pi^2}\int_0^{k_\lambda}d k\,
{k^4}\Big/{\sqrt{c^2k^2+\frac{m^2c^4}{\hbar^2}}}.
\end{equation}
Introducing the dimensionless variable $y:= \frac{k\hbar}{mc}$, we
can rewrite \eqref{phirho4} and \eqref{phiP4} as
\begin{equation}\label{phirho5}
\rho = \frac{\rho_P}{4\pi^2}\left(\frac{m_\lambda}{\lambda}\right)^4
 \int_0^{\frac{2\pi}{m_\lambda}}\!\!\!dy\,y^2\sqrt{1+y^2},
\end{equation}
and
\begin{equation}\label{phiP5}
p =\frac{\rho_P}{12\pi^2}\left(\frac{m_\lambda}{\lambda}\right)^4
\int_0^{\frac{2\pi}{m_\lambda}}d y\, \frac{y^4}{\sqrt{1+y^2}},
\end{equation}
in terms of the Planck energy density $\rho_P$ and the effective
mass of the field in units of $M_P / \lambda$, i.e. $ m_\lambda :=
\frac{m}{(M_P/\lambda)}$.

For $m_\lambda \ll 1$ we can approximate \eqref{phirho5},
\eqref{phiP5} by
\begin{equation}\label{phirho6}
\rho = \frac{\pi^2 \rho_P}{\lambda^4} +
    \frac{\rho_P}{4\lambda^2}\left(\frac{m}{M_P}\right)^2
\end{equation}
and
\begin{equation}\label{phiP6}
p = \frac{\pi^2 \rho_P}{3\lambda^4} -
    \frac{\rho_P}{12\lambda^2}\left(\frac{m}{M_P}\right)^2,
\end{equation}
up to $\mathcal{O}([m\lambda/M_P]^4)$ terms. This approximation is
physically well justified. The heaviest particles in the Standard
Model are the quark top, with $m_t \approx 173$ GeV, the weak bosons
$W^{\pm}$, with $m_{W^{\pm}} \approx 80$ GeV and the $Z$ boson with
$m_Z \approx 91$ GeV. All the other particles have $m \lesssim 1$
GeV. Being the Planck mass of the order of $10^{19}$ GeV, it will be
$m_\lambda \ll 1$ as long as the cutoff parameter satisfies $\lambda
\lesssim 10^{15}$. This is a safe upper bound, as the stochastic
classical conformal fluctuations are expected to have a cutoff which
should not exceed $\lambda \approx 10^2 - 10^5$
\cite{powerpercival00,wang2006}. Within this approximation, by
subtracting \eqref{phirho6} and \eqref{phiP6} we get
\begin{equation}\label{phirhosubP}
\rho - 3p = \frac{\rho_P}{2\lambda^2}\left(\frac{m}{M_P}\right)^2.
\end{equation}
This relation can also be obtained directly from the right hand side
of \eqref{phirhoP2} by following through the steps leading from
\eqref{phirho4} to \eqref{phiP6}.

\newpage
\thispagestyle{empty}


\singlespacing
\chapter{Autoconsistent theory of metric perturbations over a curved background}\label{ap4}

\begin{quote}
\begin{small}
In this Appendix we review the theory of small perturbation over a
general curved background geometry. We start by the standard results
holding for linear perturbations upon a Minkowski background and
then we look in detail at the more general case, originally studied
by Isaacson \cite{isaacson1968I,isaacson1968II}, in which fast
varying small perturbations propagate upon a smooth slow varying
background that they themselves contribute to set up through their
energy density backreaction. This material is at the base of the
material presented in Chapter \ref{ch2} and should be read prior or
in conjunction to it.
\end{small}
\end{quote}
\singlespacing

\section{Perturbation theory on a flat background}\label{ptfb}

\subsection{Linear approximation}

We start by reviewing the standard linearized gravity theory over a
flat background \cite{wald84,weinberg72,flanagan2005}. In a
situation where gravity is week one can write the metric as
\begin{equation}
    g_{ab} = \eta_{ab} + {h}_{ab}
\end{equation}
where the metric perturbation ${h}_{ab}$ is small, i.e. it is
possible to find some global inertial coordinate system of
$\eta_{ab}$ where its components satisfy $|{h}_{\mu\nu}| \ll 1.$ To
first order in the perturbation the inverse metric is given by
$g^{ab} = \eta^{ab} - {h^{ab}} + O(2),$ where indices are raised by
$\eta_{ab}$, e.g. $h^{ab} = \eta^{ac}\eta^{bd}{h}_{cd}.$

The Christoffel symbol and the Ricci tensor are in general given by
\begin{equation}
{\Gamma^c}_{ab} = \frac{1}{2}g^{cd}\left\{ \pd_a g_{bd} + \pd_b
g_{ad} - \pd_d g_{ab} \right\},
\end{equation}
\begin{equation}\label{ricci}
R_{ab}=\pd_c {\Gamma^c}_{ab} - \pd_a {\Gamma^c}_{cb} +
{\Gamma^c}_{ab}{\Gamma^d}_{cd} - {\Gamma^c}_{db}{\Gamma^d}_{ca}.
\end{equation}
Then to linear order in ${h}_{ab}$ it is
\begin{equation}
{{\Gamma^c}_{ab}}^{(1)} := \frac{1}{2}\eta^{cd}\left\{ \pd_a
{h}_{bd} + \pd_b {h}_{ad} - \pd_d {h}_{ab} \right\},
\end{equation}
\begin{equation}\label{R1}
R_{ab}^{(1)}:=\pd_c {{\Gamma^c}_{ab}}^{(1)} - \pd_a
{{\Gamma^c}_{cb}}^{(1)} = \frac{1}{2}\pd^c\pd_b {h}_{ac} +
\frac{1}{2}\pd^c \pd_a {h}_{bc} -\frac{1}{2} \pd^c\pd_c {h}_{ab} -
\frac{1}{2}\pd_a \pd_b {h},
\end{equation}
where ${h} := \eta^{ab}{h}_{ab}$ is the trace of the perturbation.

The linearized Ricci scalar is given by
$R^{(1)}:=\eta^{ab}R_{ab}^{(1)}$ and the linear part of the Einstein
tensor follows as
\begin{align}\label{G1}
G^{(1)}_{ab}&:= R_{ab}^{(1)} - \frac{1}{2}\eta_{ab}R^{(1)}\nonumber\\
&=\frac{1}{2}\pd^c\pd_b {h}_{ac} + \frac{1}{2}\pd^c\pd_a {h}_{bc} -
\frac{1}{2}\pd^c \pd_c {h}_{ab} - \frac{1}{2}\pd_a \pd_b {h} -
\frac{1}{2}\eta_{ab}\left( \pd^c\pd^d {h}_{cd} - \pd^c\pd_c {h}
\right).
\end{align}
This defines a linear operator whose action on an arbitrary
symmetric tensor $\chi_{ab}$ of type $(0,2)$ (i.e. with two
covariant indices) will be indicated through the notation
$G^{(1)}_{ab}[\chi]$.

This framework is appropriate to describe the propagation of GWs
whose self-energy is so small that it doesn't produce appreciable
curvature. In empty space \ee reads $G_{ab} = 0$. In the linear
approximation one has the usual equation for the small metric
perturbation ${h}_{ab}$, $G^{(1)}_{ab}[{h}] = 0.$ By defining the
trace reversed perturbation ${\bar{h}}_{ab}:= {h}_{ab} -
\frac{1}{2}\eta_{ab}{h}$, with ${\bar{h}}=-{h}$, this simplifies to
\begin{equation}
G^{(1)}_{ab}[{h}] = -\frac{1}{2}\bx{\bar{h}}_{ab} + \frac{1}{2}\pd^c
\pd_b {\bar{h}}_{ac} + \frac{1}{2}\pd^c \pd_a {\bar{h}}_{bc} -
\frac{1}{2}\eta_{ab}\pd^c\pd^d{\bar{h}}_{cd} = 0.
\end{equation}

As it is well known this simplifies to the free wave equation when
the Lorentz gauge condition $\pd^b {\bar{h}}_{ab} = 0$ is imposed.
This can always be done by exploiting the gauge freedom of GR
related to the group of diffeomorphisms. In the linear approximation
the gauge freedom is given by
\begin{equation}
    {h}_{ab} \rightarrow {h'}_{ab}:= {h}_{ab} + \pd_a v_b + \pd_b v_a,
\end{equation}
where $v_a$ is an arbitrary vector field that generates the
coordinate transformation. The tensors ${h}_{ab}$ and ${h'}_{ab}$
represent the same physical perturbation since their components are
related, to first order, by a coordinate transformation. The
linearized classical vacuum \ee in the Lorentz gauge reads
\begin{equation}\label{flatlin}
    \bx {\bar{h}}_{ab} = 0.
\end{equation}
As it is well known, this equation admits a class of homogeneous
solutions which are superpositions of elementary plane waves, i.e.
\begin{equation}
{\bar{h}}_{ab}(x;\kv) := A_{ab}(\kv)\,e^{i(\kv\cdot\xv - \omega t)}
\equiv  A_{ab}(\kv)\,e^{i k_a x^a},
\end{equation}
where $k_a = (\omega, \kv)$ is the wave vector. The wave equation
implies that this is a null vector, i.e. $\abs{\kv} = \omega$. The
complex amplitude $A_{ab}(\kv)$ satisfies the constraint $k^a A_{ab}
= 0$, due to the Lorentz gauge.

\subsection{\texttt{TT} gauge}

With the metric perturbation put in the Lorentz gauge it looks as if
all the degrees of freedom of the metric perturbation are radiative,
as they all obey a wave equation. However this is an artifact due to
the choice of gauge. It can be shown that only the traceless and
transverse part of the metric perturbation involves radiative,
physical degrees of freedom. The remaining components obey Poisson
like equations and are thus not radiative. For a proof see e.g.
\cite{flanagan2005}. In particular, for a globally classically
vacuum spacetime, it is always possible to exploit the residual
gauge freedom to impose the \texttt{TT} gauge, in which the metric
perturbation satisfies the conditions:
\begin{equation}
    {h}_{0\mu} = 0, \quad(\mu = 0,1,2,3);\quad\quad
    {h} = {h^{i}}_i = 0; \quad\quad
    \pd^i {h}_{ij} = 0.
\end{equation}
Because of the traceless condition, in the \texttt{TT} gauge, it is
${h}_{ab} \equiv {\bar{h}}_{ab}$. This gauge shows clearly that GWs
only have two degrees of freedom, linked to their two possible
polarization components. This can be seen clearly, e.g. by
considering a plane wave propagating in the $z$ direction, for which
the \texttt{TT} metric perturbation has the form
\begin{equation}
    {h}_{ab} = \left(
               \begin{array}{cccc}
                 0 & 0 & 0 & 0 \\
                 0 & h_+ & h_{\times} & 0 \\
                 0 & h_{\times} & -h_+ & 0 \\
                 0 & 0 & 0 & 0 \\
               \end{array}
             \right),
\end{equation}
where the two quantities $h_+ := {h}_{xx}(t-z)$ and $h_{\times} :=
{h}_{xy}(t-z)$ represent the two polarization components of the
wave.

\subsection{Second order corrections}

The inclusion of second order terms in the Ricci tensor becomes
necessary when one wants to assess the backreaction energy on
spacetime due to the presence of GWs. This is not a trivial problem,
the main reason being that in GR the spacetime metric plays both
roles of `stage', or background for physical processes, and `actor'.
Any formalism in which one wishes to identify the actor with GWs
only, then will also require to identify somehow a background. A
satisfactory answer to this problem was given by Isaacson
\cite{isaacson1968I,isaacson1968II}, whose formalism is presented in
Section \ref{isaacson}.

We notice that the approach of this section, based on the linear
approximation $g_{ab} = \eta_{ab} + {h}_{ab},$ has a flat Minkowski
background. As we now show, one can still carry the analysis to
second order, thus identifying a background \emph{correction} due to
the presence of linear GWs. Since these however satisfy the flat
spacetime wave equation, this approach gives a good approximation
only when this correction is so small that the related curvature
effects upon the GWs propagation can be neglected.

Substituting $g_{ab} = \eta_{ab} + {h}_{ab}$ into \eqref{ricci} and
retaining terms up to second order yields \cite{wald84,weinberg72}
\begin{align}\label{G2}
R_{ab}^{(2)}:=&\,\frac{1}{2}\,h^{cd}\pd_a\pd_b {h}_{cd} -
h^{cd}\pd_c \pd_{(a}{h}_{b)d} + \frac{1}{4}\, (\pd_a {h}_{cd})\pd_b
h^{cd} + (\pd^d
{h^c}_b)\pd_{[d}{h}_{c]a} + \frac{1}{2}\, \pd_d (h^{dc}\pd_c {h}_{ab})\nonumber\\
& - \frac{1}{4}\, (\pd^c h)\pd_c {h}_{ab} - (\pd_d h^{cd} -
\frac{1}{2}\, \pd^c h)\pd_{(a}{h}_{b)c},
\end{align}
where $( \,\, )$ and $[ \,\, ]$ denote respectively symmetrization
and antisymmetrization.

The classical vacuum \ee $R_{ab}[g] = 0$ now would read
\begin{equation}\label{2ndno}
 R_{ab}^{(1)}[h] + R_{ab}^{(2)}[h] + O(3) = 0.
\end{equation}
Since the first order equation gives $R_{ab}^{(1)}[h] = 0$,
\eqref{2ndno} cannot be satisfied but for the trivial case ${h}_{ab}
= 0.$ This simply means that one is forced to introduce a second
order metric perturbation $h^{(2)}_{ab}$ and consider a metric of
the form
\begin{equation}
    g_{ab} = \eta_{ab} + {h}_{ab} + h^{(2)}_{ab}.
\end{equation}
Then the second order expansion of the Ricci tensor yields
\begin{equation}\label{2nd}
 R_{ab} = R_{ab}^{(1)}[h] + R_{ab}^{(1)}[h^{(2)}] + R_{ab}^{(2)}[h] + O(3) =
 0,
\end{equation}
which implies that the second order metric perturbation satisfies
\begin{equation}\label{R2}
R_{ab}^{(1)}[h^{(2)}] = -R_{ab}^{(2)}[h].
\end{equation}

\subsubsection{General second order expansion of the Einstein tensor}

To put equation \eqref{R2} into a form that resembles \ee we can
find the trace of equation \eqref{2nd} for the Ricci tensor:
\begin{align}
R:=g^{ab}R_{ab} &\approx (\eta^{ab} - {h}^{ab})(R_{ab}^{(1)}[h] +
R_{ab}^{(1)}[h^{(2)}] +
R_{ab}^{(2)}[h])\nonumber\\
\nonumber\\
&= \eta^{ab}R_{ab}^{(1)}[h] + \eta^{ab}R_{ab}^{(1)}[h^{(2)}] +
\eta^{ab}R_{ab}^{(2)}[h] - {h}^{ab}R_{ab}^{(1)}[h] + O(3)\nonumber\\
\nonumber\\
&= R^{(1)}[h] + R^{(1)}[h^{(2)}] + R^{(2)}[h] -
{h}^{ab}R_{ab}^{(1)}[h] + O(3),
\end{align}
where we defined the $n$-th order Ricci scalars by
$R^{(n)}:=\eta^{ab}R^{(n)}_{ab}$. By using $G_{ab} = R_{ab} -
\frac{1}{2}g_{ab}g^{cd}R_{cd}$ and keeping terms up to second order
the Einstein tensor can be expanded as:
\begin{equation}\label{Gexp}
G_{ab} = G^{(1)}_{ab}[h] + G^{(1)}_{ab}[h^{(2)}] + G^{(2)}_{ab}[h] +
G^{(hh)}_{ab}[h] + O(3),
\end{equation}
where the \emph{linear tensor} $G^{(1)}_{ab}$ is defined as
\begin{equation}
    G^{(1)}_{ab}[\,\cdot\,]:= R^{(1)}_{ab}[\,\cdot\,] -
    \frac{1}{2}\eta_{ab}R^{(1)}[\,\cdot\,],
\end{equation}
and the \emph{quadratic tensors} $G^{(2)}_{ab}$ and $G^{(hh)}_{ab}$
by
\begin{equation}\label{G2def}
    G^{(2)}_{ab}[\,\cdot\,]:= R^{(2)}_{ab}[\,\cdot\,] -
    \frac{1}{2}\eta_{ab}R^{(2)}[\,\cdot\,],
\end{equation}
\begin{equation}\label{Ghhdef}
    G^{(hh)}_{ab}[h]:= \frac{1}{2}\eta_{ab}h^{cd}R^{(1)}_{cd}[h] -
    \frac{1}{2}{h}_{ab}R^{(1)}[h].
\end{equation}
The linear Einstein tensor general expression is given in equation
\eqref{G1} in terms of the trace reversed linear metric
perturbation, while the second order part can be built starting from
equation \eqref{G2} for the second order terms of the Ricci tensor.
For convenience we re-write those expressions below:
\begin{equation}\label{G1n}
G^{(1)}_{ab}[h] = -\frac{1}{2}\bx{\bar{h}}_{ab} + \frac{1}{2}\pd^c
\pd_b {\bar{h}}_{ac} + \frac{1}{2}\pd^c \pd_a {\bar{h}}_{bc} -
\frac{1}{2}\eta_{ab}\pd^c\pd^d{\bar{h}}_{cd}
\end{equation}
\begin{align}\label{G2n}
R_{ab}^{(2)}[h]:=&\,\frac{1}{2}\,h^{cd}\pd_a\pd_b {h}_{cd} -
h^{cd}\pd_c \pd_{(a}{h}_{b)d} + \frac{1}{4}\, (\pd_a {h}_{cd})\pd_b
h^{cd} + (\pd^d
{h^c}_b)\pd_{[d}{h}_{c]a} + \frac{1}{2}\, \pd_d (h^{dc}\pd_c {h}_{ab})\nonumber\\
& - \frac{1}{4}\, (\pd^c h)\pd_c {h}_{ab} - (\pd_d h^{cd} -
\frac{1}{2}\, \pd^c h)\pd_{(a}{h}_{b)c}.
\end{align}

It is important to note that the above expansion \eqref{Gexp} for
the Einstein tensor is general. It also holds in the general case
where matter enters \ee on the r.h.s. through its stress energy
tensor $T_{ab}$. However, in the case of classical vacuum it is
$T_{ab} = 0$, so that $G_{ab}[g] = 0$ implies the following set of
equations:
\begin{align}
 &G^{(1)}_{ab}[h] = 0 \quad \Leftrightarrow \quad R^{(1)}_{ab}[h] = 0,\nonumber\\
\\
 &G^{(1)}_{ab}[h^{(2)}] = -G^{(2)}_{ab}[h].\nonumber
\end{align}
Note that, in this case, it is $G^{(hh)}_{ab}[h]=0$ because
${h}_{ab}$ satisfies the classical vacuum linearized \eep.

Note that, up to second order in perturbation, the second order
equation can be cast into the more suggestive form
\begin{equation}
    G_{ab}[\eta_{ab} + h^{(2)}_{ab}] = 8\pi t_{ab}[h],
\end{equation}
where we have defined the `source' tensor
\begin{equation}\label{tten}
t_{ab}[h] := -\frac{1}{8\pi}G^{(2)}_{ab}[h].
\end{equation}
This suggests that one may interpret $t_{ab}[h]$ as the GWs
backreaction stress energy tensor causing a correction to the
background geometry. Once again, it is important to note that GWs
propagate on the flat background according to the linear equation,
in such a way that this scheme is satisfactory only when the metric
correction $h^{(2)}_{ab}$ is very small and its influence on the
propagation of the GWs can be ignored.

\subsubsection{Properties of the backreaction effective stress energy tensor}

The main properties of $t_{ab}[h]$ that would suggest considering it
as the effective stress energy tensor of GWs are: (1) it is
\emph{quadratic} in ${h}_{ab}$, (2) \emph{symmetry} and (3)
\emph{conservation} with respect to the flat background, i.e. $\pd^a
t_{ab}[h] = 0.$ In particular this second property is true if
${h}_{ab}$ satisfies the classical vacuum linearized \eep.

However the main problem towards such an interpretation is that
$t_{ab}[h]$ is \emph{not} gauge invariant, i.e. under a gauge
transformation ${h}_{ab} \rightarrow {h}_{ab} + 2\pd_{(a}v_{b)}$,
the expression for $t_{ab}[h]$ does not remain unchanged. This means
that, for two observers whose coordinates and metric perturbation
components are respectively $x^{\mu}$, ${h}_{\mu\nu}$ and
${x^{\mu}}' := x^{\mu} + \delta v^{\mu}$, ${h'}_{\mu\nu} :=
{h}_{\mu\nu} + 2\pd_{(\mu}\delta v_{\nu)}$, where $\delta v_a$ is an
infinitesimal vector field, the quantities $t_{\mu\nu}[h]$ and
$t_{\mu\nu}[{h'}]$ are \emph{not} the components of the same tensor.

While some \emph{global} quantities such as the total energy $E =
\int_{\Sigma} t_{00}[h] d^3 x$, where $\Sigma$ is a spacelike
hypersurface, can be shown to be gauge invariant for asymptotically
flat metrics $\eta_{ab} + {h}_{ab}$ \cite{wald84}, one would in
general wish to be able to define some meaningful \emph{local}
notion of stress energy tensor associated with GWs.

This can be achieved, e.g., by defining a suitably averaged tensor
as it is done, among other authors, by Isaacson in his important
work of 1968 on \emph{Gravitational radiation in the limit of high
frequency} \cite{isaacson1968I,isaacson1968II}. This is the object
of Section \ref{isaacson}. Before that, we review in the next
section a more general formalism for an arbitrarily large
perturbation over Minkowski spacetime.

\section{Non covariant exact definition of gravity `stress energy tensor'}

Further insight into the nonlinear structure of GR and the inherent
backreaction effect can be obtained by writing the exact \ee in a
non manifestly covariant form \cite{weinberg72}. This is done by
choosing a quasi-Minkowskian coordinates system, in which the metric
components are written as
\begin{equation}
g_{\mu\nu} = \eta_{\mu\nu} + H_{\mu\nu}.
\end{equation}
The perturbation component $H_{\mu\nu}$ are \emph{not} assumed to be
small but they fall off to zero at infinity. The part of the Ricci
tensor linear in $H_{\mu\nu}$ is given of course by
$R^{(1)}_{\mu\nu}[H]$, where $R^{(1)}_{\mu\nu}[\,\cdot\,]$ is the
operator defined in \eqref{R1}. In this coordinate system, the exact
\ee with a matter source, $G_{\mu\nu}[g] = 8\pi T_{\mu\nu}$, can be
written as
\begin{equation}\label{eex}
G^{(1)}_{\mu\nu}[H] = 8\pi \left\{ T_{\mu\nu} + t_{\mu\nu}[H]
\right\} := 8\pi \tau_{\mu\nu}[H],
\end{equation}
where we defined
\begin{equation}\label{tH}
t_{\mu\nu}[H] := -\frac{1}{8\pi}\left\{ G_{\mu\nu}[\eta + H] -
G^{(1)}_{\mu\nu}[H] \right\}
\end{equation}
and with the linear operator $G^{(1)}_{\mu\nu}[\,\cdot\,]$ defined
in \eqref{G1}.

Equation \eqref{eex} has the expected form for a spin 2 field
$H_{\mu\nu}$ suffering a self-interaction as dictated by the
`source' term $\tau_{\mu\nu}[H] := T_{\mu\nu} + t_{\mu\nu}[H]$,
which depends nonlinearly upon $H_{\mu\nu}$. As Weinberg suggests,
if one decided to view $H_{\mu\nu}$ as \emph{the gravitational
field}, then the tensor $t_{\mu\nu}[H]$ would be the candidate to
represent gravitation's stress energy tensor. We remark that the
above dynamical equation is \emph{exact}, even though not
\emph{manifestly} covariant. Indeed it is based upon an a-priori
choice of one coordinate system in which the metric components
$g_{\mu\nu}$ are simply re-written as $\eta_{\mu\nu} + H_{\mu\nu}$,
thereby \emph{defining} the \emph{observer-related} gravitational
field $H_{\mu\nu}$.

The main properties that the matter plus gravitation `stress energy
tensor' $\tau_{\mu\nu}[H]$ enjoys are: (1) \emph{symmetry}, (2)
\emph{conservation}, i.e. $\pd_{\sigma}\tau_{\mu\nu}[H] = 0$, as
implied by the linearized Bianchi identity obeyed by
$G^{(1)}_{\mu\nu}[H]$, (3) if one computes $t_{\mu\nu}[H]$ as a
formal power series, the first term is \emph{quadratic in
$H_{\mu\nu}$} and given by
\begin{equation}
t_{\mu\nu}[H] = -\frac{1}{8\pi}\left\{  G^{(2)}_{\mu\nu}[H] +
G^{HH}_{\mu\nu}[H] \right\} + O(H^3),
\end{equation}
where are the operators $G^{(2)}_{\mu\nu}[H]$ and
$G^{HH}_{\mu\nu}[H]$ are defined by equations \eqref{G2def} and
\eqref{Ghhdef}. Of course the third and higher order terms in
$H_{\mu\nu}$ would account for gravity self interaction. A fourth
important property enjoyed in general by $\tau_{\mu\nu}[H]$ is that
of being (4) \emph{Lorentz invariant} \cite{weinberg72}.

Though not manifestly covariant the above formalism is exact. The
link to the perturbation approach over a flat non-dynamical
background of Section \ref{ptfb} can be found when gravity is weak
and $|H_{\mu\nu}| \ll 1$, by expanding
\begin{equation}
H_{\mu\nu} = h_{\mu\nu} + h^{(2)}_{\mu\nu} + h^{(3)}_{\mu\nu} +
\ldots
\end{equation}
Since $G^{(1)}_{\mu\nu}[\,\cdot\,]$ is linear, equation \eqref{eex}
reads
\begin{equation}
G^{(1)}_{\mu\nu}[h] + G^{(1)}_{\mu\nu}[h^{(2)}] + O(h^3) = 8\pi
\left\{ T_{\mu\nu} - \frac{1}{8\pi}\left\{  G^{(2)}_{\mu\nu}[h] +
G^{hh}_{\mu\nu}[h] \right\} + O(h^3) \right\},
\end{equation}
and where the metric tensor expansion also appears implicitly in the
expression for $T_{\mu\nu}$. In the classical vacuum case
$T_{\mu\nu} = 0$ and $G^{hh}_{\mu\nu}[h] = 0$. By equating equal
order terms, the above equation yields of course
\begin{align}
&G^{(1)}_{\mu\nu}[h] = 0,\\
&G^{(1)}_{\mu\nu}[h^{(2)}] = -G^{(2)}_{\mu\nu}[h].
\end{align}

\section{Isaacson's perturbation theory on a curved background geometry}\label{isaacson}

The material in this section is based upon Isaacson papers
\cite{isaacson1968I,isaacson1968II} as well as the review paper by
Flanagan \cite{flanagan2005} and \emph{Gravitation} \cite{MTW} by
Wheeler and co-authors.

\subsection{Slow varying vs. fast varying components: the main assumptions}

The perturbation approach over Minkowski spacetime given in the
previous sections suffers of a serious limitation problem: in the
weak field limit and for classical vacuum, the GWs as described by
$h_{ab}$ and satisfying the linear wave equation $G^{(1)}_{ab}[h] =
0$ propagate on a flat background; the formalism is therefore
suitable only to describe situations in which the spacetime geometry
is essentially flat to 0-th order.

However, there are situations in which classical vacuum spacetime
geometry may be highly curved by itself, even at the lowest order of
approximation. This can be the case for, e.g., the neighborhood of
black holes, a neutron stars or a collapsing supernovae. If GWs are
also present, one would require a formalism allowing to study their
propagation on an arbitrary curved background geometry. This also
seems to be desirable to study the vacuum random \ee (Section
\ref{rga}) in situations where the background curvature due to the
net vacuum energy amount plus standard non-vacuum matter in a given
spacetime region cannot be ignored.

Isaacson devised a suitable method that works well in the case of
high frequency GWs and which is based upon a suitable \emph{split}
of the spacetime geometry $g_{ab}$ into a background $\gb_{ab}$ and
a perturbation $h_{ab}$. The split \emph{defines} the background
geometry as the smooth, slow varying part of $g_{ab}$, while the GWs
are defined as a high frequency, fast varying superimposed
perturbation.

A meaningful definition for the local energy content of the high
frequency perturbation can be given by means of a suitable spacetime
averaging procedure, originally introduced by Brill and Hartle
\cite{brillhartle}. Isaacson shows that the resulting stress energy
tensor is quadratic in $h_{ab}$, gauge invariant and can be
identified with that of massless spin-2 field if the perturbation
$h_{ab}$ satisfies the GWs equation on the curved background. This
result is important for the work presented in this thesis and the
details are reported in \ref{HB}.

The key assumption behind Isaacson paper is that the GWs
perturbation must have high frequency, in such a way that their
typical small wavelength $\ell$ can be used as a formal expansion
parameter.
A perturbation is defined to have a high frequency whenever its
wavelength is much shorter than the typical radius of curvature $L$
of the background geometry. It is important to note that the case
$\gb_{ab} = \eta_{ab}$ implies an infinite radius of curvature, so
that the standard flat background expansion scheme presented in
Section \ref{ptfb} is a very special case of Isaacson more general
framework. In that case `high frequency' simply means `all
frequencies'!

To help visualize the physical situation involved, one may think of
an orange's skin: the overall shape represents the curved background
geometry, while the small scale ripples represent the GWs. The
amplitude of the GWs is assumed to be small but this doesn't imply
that their energy content also must be small: in fact the formalism
is also suited to describe situations in which the GWs energy
content is the \emph{only} cause of the background geometry
curvature. We refer to this as to the \emph{autoconsistent} case.

\subsection{GWs propagating on a curved background: autoconsistent case}

Isaacson framework can be implemented by splitting the spacetime
metric as
\begin{equation}
g_{ab} = \gb_{ab} + \varepsilon h_{ab},
\end{equation}
where the slow varying background geometry $\gb_{ab}$ varies on a
typical scale $L$, while the high frequency component $h_{ab}$,
representing the GWs, varies on a typical scale $\ell \ll L$. The
fact that the GWs have a small amplitude is embodied in the choice
$\varepsilon \ll 1$. This ensures that the laboratory geometry has
only microscopic fluctuations.

\subsubsection{Orders of magnitude estimates}

The background metric components are assumed to be of order
$\gb_{ab} = O(1).$\footnote{By definition, $f = O(\varepsilon^n)$
means that one can find a constant $0 < C < \infty$ such that $f <
C\varepsilon^n$ as $\varepsilon$ goes to zero. Then $f = O(1)$
simply means that $f$ has a finite value that does not depend on the
smallness parameter $\varepsilon$.} The metric derivatives then have
the typical magnitudes $\pd \gb \sim \gb/L$ and $\pd h \sim h /
\ell$. To lowest order the effective energy connected to the GWs and
acting as a source in \ee is of order $(\varepsilon / \ell)^2$,
while the background curvature is of order $(1/L)^2$. Because of \ee
and the fact that other sources beyond GWs in general may curve the
background one has
\begin{equation}
(1/L)^2 \geq (\varepsilon / \ell)^2,
\end{equation}
implying that
\begin{equation}
\varepsilon \leq \ell / L \ll 1.
\end{equation}
The case in which no other causes of curvature are present beside
GWs implies the equality $\varepsilon = \ell / L$. This relation can
be used to estimate the order of magnitude of the various quantities
involving derivatives that define the Ricci tensor. Since it is $L =
O(1)$, $\gb_{ab} = O(1)$ and $h_{ab} = O(1)$, it follows that $\ell$
and $\varepsilon$ are small parameters of the same order and one can
deduce the following orders of magnitudes estimates:
\begin{equation}\label{orders}
\pd_{a}\gb_{bc} = O(1),\quad \pd_a\pd_b \gb_{cd} = O(1), \quad
\pd_{a}h_{bc} = O(1/\varepsilon)\quad \pd_a\pd_b h_{cd} =
O(1/\varepsilon^2).
\end{equation}

The autoconsistent framework implies that we want to study a
situation in which \emph{all} of the background curvature is
produced precisely by the backreaction energy of the GWs. We thus
set $\varepsilon = \ell / L$.

\subsubsection{Ricci tensor expansion}

By expanding the Ricci tensor $R_{ab}[g] = R_{ab}[\gb + \varepsilon
h]$ in powers of $\varepsilon$ one obtains, in a similar way to what
happens in the flat background case,
\begin{equation}\label{Rabexp}
R_{ab}[\gb + \varepsilon h] = R_{ab}^{(0)} + \varepsilon
R_{ab}^{(1)} + \varepsilon^2 R_{ab}^{(2)} + \varepsilon^3
R_{ab}^{(3+)},
\end{equation}
where
\begin{equation}
R_{ab}^{(0)} := R_{ab}[\gb]
\end{equation}
is the full Ricci tensor of the smooth, slow varying background
geometry while
\begin{equation}\label{ord1}
R_{ab}^{(1)} :=\, \frac{1}{2}\left(\cd^c\cd_b {h}_{ac} + \cd^c \cd_a
{h}_{bc} - \cd^c\cd_c {h}_{ab} - \cd_b \cd_a {h}\right)
\end{equation}
and
\begin{align}\label{ord2}
R_{ab}^{(2)}:=&\,\frac{1}{2}\,h^{cd}\cd_b\cd_a {h}_{cd} -
h^{cd}\cd_c \cd_{(a}{h}_{b)d} + \frac{1}{4}\, (\cd_a {h}_{cd})\cd_b
h^{cd} + (\cd^d
{h^c}_b)\cd_{[d}{h}_{c]a} \nonumber\\
& + \frac{1}{2}\, \cd_d (h^{cd}\cd_c {h}_{ab}) - \frac{1}{4}\,
(\cd^c h)\cd_c {h}_{ab} - (\cd_d h^{cd} - \frac{1}{2}\, \cd^c
h)\cd_{(a}{h}_{b)c}.
\end{align}
In the above expressions $\cd_a$ is the covariant derivative of the
background geometry $\gb_{ab}$, which is also used to raise or lower
all indices, in such a way that, e.g., $\cd^c = \gbu^{ca}\cd_a$ and
$h = \gbu^{ab}h_{ab}.$ The linear operator $R_{ab}^{(1)}$ reduces to
the flat background expression \eqref{R1} when $\gb_{ab} \rightarrow
\eta_{ab}$ and $\cd \rightarrow \pd$ and the same happens for the
quadratic operator $R_{ab}^{(2)}$ which reduces to the expression
\eqref{G2}. Note however that in the general background case the
order in which second covariant derivatives appear is important.
Finally, the higher order term $R_{ab}^{(3+)}$ is simply defined by
equation \eqref{Rabexp}.

The fact that the formal expansion parameter $\varepsilon$ appears
in \eqref{Rabexp} should not deceive. Indeed the various
$R^{(n)}_{ab}$ terms are \emph{not} of the same magnitude, and this
must be carefully assessed before one can write down the approximate
\ee order by order. By using $\varepsilon = \ell / L,$ the estimates
\eqref{orders}, and by inspecting the structure of the various Ricci
tensor components it is found:
\begin{equation}
    R^{(0)}_{ab}\sim {\gb}^{-1}\pd^2\gb\sim 1/L^2 = O(1),
\end{equation}
\begin{equation}
    \varepsilon R^{(1)}_{ab}\sim {\gb}^{-1}\pd^2(\varepsilon h)\sim \varepsilon/\ell^2 =
    O(1/\varepsilon),
\end{equation}
\begin{equation}
    \varepsilon^2 R^{(2)}_{ab}\sim \varepsilon h{\gb}^{-2}\pd^2(\varepsilon h)\sim \varepsilon^2/\ell^2 =
    O(1).
\end{equation}
Moreover it is possible to verify that $\varepsilon^3 R^{(3+)}_{ab}
= O(\varepsilon)$ so that it truly represents a small correction.
The important point about the above estimates is that they show how
the background curvature and $\varepsilon^2 R^{(2)}_{ab}$ which will
be connected to the GWs local energy density have the same order of
magnitude. In particular, even though $\varepsilon h_{ab}$ is small,
the associated energy needs not be small. Note also that, correctly
$\varepsilon^2 R^{(2)}_{ab}$ is smaller than $\varepsilon
R^{(1)}_{ab}$ by a factor $\varepsilon.$

\subsubsection{\ee expansion}

\ee in classical vacuum reads $R_{ab}[g] = 0.$ Collecting together
terms of the same order in the Ricci tensor expansion \eqref{Rabexp}
yields:
\begin{equation}\label{isac1}
R_{ab}^{(1)}[h] = 0,
\end{equation}
governing the GWs propagation upon the background $\gb_{ab}$ which
they themselves produce according to
\begin{equation}\label{isac2}
R_{ab}^{(0)}[\gb] = -R^{(2)}_{ab}[h],
\end{equation}
where for convenience we have set $\varepsilon = 1.$

Isaacson showed that, in the high frequency limit, the theory enjoys
gauge invariance similar to that of the linearized weak field limit
and given by $h_{ab} \rightarrow h_{ab} - \cd_{(a}v_{b)}.$ In order
to exploit the gauge freedom the linear equation \eqref{isac1} is
re-expressed in terms of the trace reversed perturbation
${\bar{h}}_{ab}:= {h}_{ab} - \frac{1}{2}\gb_{ab}{h}$. By exploiting
the following properties of the Riemann tensor and of the covariant
derivative, holding for any scalar function $f$ and any tensor
$T_{ab}$:
\begin{align}
&\cd_a\cd_b f = \cd_b \cd_a f,\nonumber\\
\nonumber\\
&(\cd_a\cd_b - \cd_b\cd_a) T_{cd} = {R_{abc}}^e T_{ed} + {R_{abd}}^e
T_{ce},\\
\nonumber\\
&R_{abcd} = R_{cdab},\quad R_{abcd} = -R_{bacd},\quad R_{abcd} =
-R_{abdc},\nonumber
\end{align}
equation \eqref{ord1} yields
\begin{equation}
\cd^c\cd_c {\bar{h}}_{ab} - \frac{1}{2}\gb_{ab}\cd^c\cd_c {\bar{h}}
-2\cd_{(b} \cd^c{\bar{h}}_{a)c} - 2R^{(0)}_{cbad}{\bar{h}}^{dc} -
2R^{(0)}_{c(a}{\bar{h}_{b)}}^{\hspace{2mm}c} = 0.
\end{equation}
The gauge freedom can now be exploited to obtain a traceless and
transverse perturbation satisfying $\bar{h} = 0$ and
$\cd^c{\bar{h}}_{ac} = 0$. In this gauge $h_{ab} \equiv
\bar{h}_{ab}$ and the wave equation simplifies to:
\begin{equation}
\cd^c\cd_c {{h}}_{ab} - 2R^{(0)}_{cbad}{{h}}^{dc} -
2{R^{(0)}}_{c(a}{{h}_{b)}}^{\hspace{2mm}c} = 0.
\end{equation}
This obviously reduces to equation \eqref{flatlin} in the flat
background limit.

The curved background GWs propagation equation couples the
perturbation to the background curvature. This causes a gradual
evolution in the properties of the wave. In the high frequency limit
this evolution can be described using the formalism of geometric
optics, showing that GWs travel along null geodesics with slowly
evolving amplitudes and polarizations \cite{isaacson1968I}.

\subsection{GWs stress energy tensor}\label{HB}

\subsubsection{Second order equations}

The analysis of the second order equation \eqref{isac2} is central
to this thesis since it provides information on the local energy
content of GWs. It can be re-written as
\begin{equation}\label{finally}
G_{ab}^{(0)}[\gb] = 8\pi
\left(-\frac{1}{8\pi}G^{(2)}_{ab}[h]\right),
\end{equation}
where $G^{(2)}_{ab}:=
R_{ab}^{(2)}[h]-\frac{1}{2}\gb_{ab}{\gb}^{cd}R^{(2)}_{cd}[h]$, with
$R_{ab}^{(2)}[h]$  given in \eqref{ord2}, and where we have used the
linear equation $R_{ab}^{(1)}[h] = 0$. The tensor in between
brackets on the r.h.s. is the curved background generalization of
$t_{ab}[h]$ defined in \eqref{tten}. Even though it enjoys various
properties as observed above, the main obstacle in interpreting it
as the stress energy tensor of GWs is that it is not invariant under
a general change of coordinates. This implies that one can in
principle always find a suitable system of coordinate where, locally
at some spacetime event $\P$, the \emph{total} physical metric
$g_{ab}$ is Minkowski. For such an observer
$t_{ab}[h]\left|_{\P}\right.$ would vanish. Such considerations lead
to expect that in order to define a meaningful concept of local
energy density for GWs one should recur to some kind of spacetime
averaging over a region containing many wavelengths.

Another reason for this emerges by looking at \eqref{finally}. The
background Einstein tensor is smooth and varies on a typical scale
$L \gg \ell.$ This means that it contains no fluctuations. Yet, on
the r.h.s. we have a tensor built upon the high frequency
perturbation $h_{ab}$. This also leads to think that it is some
suitable smooth average of $G^{(2)}_{ab}[h]$ that should act as a
source for the smooth background curvature.

To see this better we can recall that $R_{ab}[g] = R_{ab}^{(0)} +
\varepsilon R_{ab}^{(1)} + \varepsilon^2 R_{ab}^{(2)} +
\varepsilon^3 R_{ab}^{(3+)}, = 0$. Once the linear equation is
imposed by $R_{ab}^{(1)} = 0$, the classical vacuum \ee implies $
R_{ab}^{(0)} + R_{ab}^{(2)} = 0.$ Setting again $\varepsilon = 1$
for convenience, we can re-write this as
\begin{equation}\label{temp}
R_{ab}^{(0)}[\gb] + \m{R_{ab}^{(2)}[h]} + \left\{R_{ab}^{(2)}[h] -
\m{R_{ab}^{(2)}[h]} \right\} = 0,
\end{equation}
where $\m{\,\cdot\,}$ denotes some average to be defined later. The
first two terms are now both smooth and slow varying so that we can
set
\begin{equation}
R_{ab}^{(0)}[\gb] = -\m{R_{ab}^{(2)}[h]}.
\end{equation}
The term in curly bracket however defines a high frequency
perturbation with zero average. In order to maintain a solution of
\ee at second order one must introduce a higher order metric
perturbation $g^{(2)}_{ab}$ by writing from the start $g_{ab} =
\gb_{ab} + h_{ab} + g^{(2)}_{ab}.$ This way an extra term
$R_{ab}^{(1)}[g^{(2)}]$ appears in \eqref{temp} on the l.h.s. and we
get the extra equation
\begin{equation}
R_{ab}^{(1)}[g^{(2)}] = \m{R_{ab}^{(2)}[h]} - R_{ab}^{(2)}[h].
\end{equation}

Summarizing we have the following interpretation: the GWs are
represented by a high frequency perturbation $h_{ab}$ of a smooth,
slow varying background $\gb_{ab}$. Their propagation is described
by the linear equation $R_{ab}^{(1)}[h] = 0$ which, in an arbitrary
gauge, reads
\begin{equation}\label{1gws}
\cd^c\cd_c {h}_{ab} + \cd_b \cd_a {h} - \cd^c\cd_b {h}_{ac} -\cd^c
\cd_a {h}_{bc} = 0.
\end{equation}
The smooth background is produced by a suitable average of a
backreaction tensor quadratic in the perturbation, according to the
effective \ee
\begin{equation}
G_{ab}^{(0)}[\gb] = 8\pi \Tgw_{ab},
\end{equation}
where the GWs stress energy tensor will be defined by an expression
like
\begin{equation}
\Tgw_{ab} := -\frac{1}{8\pi}\left\{\m{R_{ab}^{(2)}[h]} - \frac{1}{2}
\gb_{ab} \m{R^{(2)}[h]}\right\} \equiv
-\frac{1}{8\pi}\m{G_{ab}^{(2)}[h]}.
\end{equation}
Finally, due to the the fact that gravity auto-interacts with
itself, the linear GWs are themselves source of extra, higher order
perturbations that obey a wave equation with a source given by
\begin{equation}
R_{ab}^{(1)}[g^{(2)}] = \m{R_{ab}^{(2)}[h]} - R_{ab}^{(2)}[h].
\end{equation}
This can account for phenomena such as e.g. wave-wave scattering or
higher harmonics \cite{MTW}.

\subsubsection{General definition and properties of the GWs stress energy tensor}

The second order part of the Einstein tensor, giving the microscopic
structure of what will become the GWs stress energy tensor, can be
found starting from \eqref{ord2}. It is straightforward to show that
this can be written as \cite{isaacson1968I}:
\begin{equation}\label{gen1}
G^{(2)}_{ab} = \frac{1}{2}\left( \cd_c {S_{ab}}^c - Q_{ab} \right),
\end{equation}
where
\begin{equation}\label{gen2}
{S_{ab}}^c := {\delta_b}^c h^{ed}\cd_a h_{ed} + h^{cd}( \cd_d h_{ab}
- \cd_{b} h_{da} - \cd_{a} h_{db} ) + \gb_{ab}\left[ h^{cd}(\cd^e
h_{de} - \frac{1}{2}\cd_d h) - \frac{1}{2}h_{ed} \cd^c h^{ed}
\right],
\end{equation}
and
\begin{align}\label{gen3}
Q_{ab} := &\frac{1}{2} \cd_a h^{cd} \cd_b h_{cd} - \cd^c {h_b}^d
(\cd_c h_{da} - \cd_d h_{ca} ) - \frac{1}{2}\cd^d h (\cd_b h_{da} +
\cd_a h_{db} - \cd_d h_{ab}) \nonumber\\
&+ \frac{1}{2}\gb_{ab}\left[ \frac{1}{2}\cd^e h^{cd} \cd_e h_{cd} -
\cd^c h^{ed} \cd_d h_{ec} + \cd^d h ( \cd^e h_{de} - \frac{1}{2}
\cd_d h ) \right].
\end{align}
Symbolically the simple structure emerges $G^{(2)} \sim (\pd h)^2 +
\pd (h \pd h), $ which implies that only the part quadratic in $\pd
h$ will survive an averaging procedure over extended spacetime
regions.

Similarly to what is done to study \emph{macroscopic} electric
fields inside a dielectric, one can define the energy content of GWs
by neglecting the fine details due to the fast fluctuations and
recovering an average expression which is suitable to represent the
local energy content of GWs. The procedure used by Isaacson is the
same as that introduced by Arnowitt, Deser and Misner in
\cite{ADM1961}. The notation $\m{\,\cdot\,}$ then denotes a
spacetime integral average over a region whose characteristic size
$\Lav$ contains many fluctuations wavelengths: i.e. it must be small
compared to the scale $L$ over which the background varies, yet
still of order $O(1)$ and much larger of $\ell$, the typical
fluctuations scale, and we have
\begin{equation}
    [\ell = O(\epsilon)] \ll [\Lav = O(1)] \ll [L = O(1)].
\end{equation}
The procedure is defined in such a way that the average of a tensor
still yields a tensor. Further details are given at the end of this
section.

The GWs stress energy tensor is then given by
\begin{equation}\label{Tgw}
\Tgw_{ab} = -\frac{1}{8\pi}\m{G_{ab}^{(2)}[h]} \equiv
\frac{1}{16\pi}\m{Q_{ab} - \cd_c {S_{ab}}^c}.
\end{equation}
When performing the average many terms can be simplified or
neglected thanks to the following rules \cite{isaacson1968II,MTW}:
\begin{enumerate}\label{rules}
  \item covariant derivatives of $h_{ab}$ commute as $\varepsilon \rightarrow
  0$ and one can use $\m{h \cd_d\cd_c h_{ab}}  = \m{h \cd_c\cd_d h_{ab}} $;
  \item averages of a divergence, e.g. $\m{\cd_c {S_{ab}}^c}$, are
  reduced by a factor $\varepsilon$ and can also be neglected;
  \item for the same reason it is possible to integrate by parts under
  integrals so that, ignoring extra terms smaller by a factor
  $\varepsilon$,
  one can use e.g. $\m{h \cd_c\cd_d h_{ab}} = -\m{\cd_d h \cd_c
  h_{ab}}$.
\end{enumerate}
Using these prescriptions \emph{as well as} \cite{MTW}:
\begin{quote}
  \hspace{-0.5cm}4. the fact that $h_{ab}$ satisfies the linearized GWs
  equation $G^{(1)}_{ab}[h] = 0$ given in \eqref{1gws}
\end{quote}
it is possible to show that \cite{MTW,flanagan2005}
\begin{equation}\label{finally2}
\Tgw_{ab} = \frac{1}{32\pi}\m{ \cd_a \bar{h}_{cd} \cd_b \bar{h}^{cd}
- \frac{1}{2} \cd_a \bar{h}\cd_b \bar{h} - 2\cd_d \bar{h}^{cd}
\cd_{(a}\bar{h}_{b)c} } + O(\varepsilon),
\end{equation}
which is an expression valid in \emph{any} gauge.

Isaacson studied the gauge invariance properties of this quantity.
He showed that, under a general gauge transformation $h_{ab}
\rightarrow h_{ab} + \cd_{(a} v_{b)}$, it is
\begin{equation}
G^{(2)}_{ab} \rightarrow G'^{(2)}_{ab} = G^{(2)}_{ab} + \cd_c
{U_{ab}}^c[v] + O(\varepsilon).
\end{equation}
The extra term $\cd_c {U_{ab}}^c[v]$ is of order $O(\varepsilon)$ if
the coordinate transformation vector $v_a$ contains only low
frequency components and it could be neglected as such. When $v_a$
includes high frequency modes it is instead $\cd_c {U_{ab}}^c[v] =
O(1)$, showing that the un-averaged $G^{(2)}_{ab}$ is not indeed
gauge invariant. However, since divergences are reduced under
averages, it follows that the averaged tensor \emph{is} gauge
invariant up to smaller terms of order $\varepsilon$, i.e.
\begin{equation}
\Tgw_{ab} \rightarrow T'^{\text{\scriptsize{GW}}}_{ab} = \Tgw_{ab} +
O(\varepsilon).
\end{equation}

The gauge invariance of $\Tgw_{ab}$ being shown, one can always go
to the traceless-transverse gauge for $\bar{h}_{ab}$, where it takes
the very simple form
\begin{equation}\label{finally3}
\Tgw_{ab} = \frac{1}{32\pi}\m{ \cd_a {h}_{cd} \cd_b {h}^{cd} } +
O(\varepsilon).
\end{equation}
Note from the background \ee $G_{ab}^{(0)}[\gb] = 8\pi \Tgw_{ab}$
that this is conserved with respect to $\gb_{ab}$, i.e.
\begin{equation}
\cd^a \Tgw_{ab} = 0 + O(\varepsilon).
\end{equation}

\subsubsection{GWs stress energy tensor in the geometric optic approximation}

In the geometric optic approximation, suitable to describe GWs
propagation on the curved background, one has an explicit
representation of $h_{ab}$ as
\begin{equation}
h_{ab} = \mathcal{A} \, e_{ab} e^{i\phi},
\end{equation}
where the polarization $e_{ab}$ is defined to satisfy $ e_{ab}e^{ab}
=1$, while $\phi$ is a rapidly fluctuating phase with large first
derivatives but negligible higher derivatives and $k_a := \pd_a
\phi$ represents ray vectors normal to the surfaces of constant
phase. The effect of the curvature is to induce a slow variation in
the amplitude $\mathcal{A}$, polarization $e_{ab}$ and propagation
vector $k_a$ of the wave. Then imposition of the
transverse-traceless gauge yields
\begin{equation}
{\gb}^{ab} e_{ab} = 0,
\end{equation}
\begin{equation}
k_a e^{ab} = 0,
\end{equation}
while imposing the wave equation yields:
\begin{equation}
k_a k^a = 0
\end{equation}
and
\begin{equation}
\cd_a \left(\mathcal{A}^2 k^a\right) = 0,\quad \cd_c e_{ab} k^c = 0.
\end{equation}
The first condition gives the variation of the amplitude once the
integral null curves along which waves propagate are known, while
the second shows that the polarization tensor is
parallel-transported along such curves. By using this information
the GWs stress energy tensor \eqref{finally2} reduces to the very
simple form
\begin{equation}
\Tgwgo_{ab} := \frac{\mathcal{A}^2}{64\pi}\, k_a k_b.
\end{equation}
Note that, as it happens for a radiation fluid, it is traceless
\begin{equation}
\Tgwgo = {\gb}^{ab} \Tgwgo_{ab} = \frac{\mathcal{A}^2}{64\pi}\, k^a
k_a = 0.
\end{equation}

We summarize now some important points which will be crucial to the
development of this thesis:
\begin{enumerate}
  \item the non-averaged expressions \eqref{gen1}-\eqref{gen3} are general
  and simply define the non linear operator:
\begin{equation}
G^{(2)}_{ab}[\,\cdot\,]= \frac{1}{2}\left( \cd_c
{S_{ab}}^c[\,\cdot\,] - Q_{ab}[\,\cdot\,] \right);
\end{equation}
  \item if acting on a symmetric tensor $h_{ab}$ describing GWs, i.e. satisfying
  the wave equation \eqref{1gws}, its average yields the expression
  \begin{equation}
  \Tgw_{ab} = \frac{1}{32\pi}\m{ \cd_a \bar{h}_{cd} \cd_b \bar{h}^{cd}
- \frac{1}{2} \cd_a \bar{h}\cd_b \bar{h} - 2\cd_d \bar{h}^{cd}
\cd_{(a}\bar{h}_{b)c} };
  \end{equation}
  \item finally, if $h_{ab}$ also satisfies the
  geometric optics approximation, this further reduces to
  \begin{equation}
  \Tgwgo_{ab} = \frac{\mathcal{A}^2}{64\pi}\, k_a k_b.
  \end{equation}
\end{enumerate}

\subsection{Definition of the spacetime averaging procedure}

In general, averages of tensors over a curved geometry do not yield
a tensor as the integration involves tensors defined at different
spacetime points which, therefore, have different transformation
properties. To defined a meaningful average whose result is a true
tensor one must devise a method to transport the tensors entering
the average at the same spacetime point, where they can be summed in
the ordinary manner \cite{isaacson1968II,MTW}.

This can be done in a unique manner by defining the \emph{bivector
of geodesic parallel displacement} ${{\gb}_{a}}^{a'}$, discussed by
De Witt \cite{dewitt60} and Synge \cite{synge60}. Its definition
follows by observing that, in a small macroscopic region containing
many perturbation's wavelengths, there exists a unique geodesic
$\gamma_{\P\P'}$ connecting any two events $\P$ and $\P'.$ Then, if
$E$ denotes an arbitrary tensor field, one can define
\begin{equation}
E(\P')_{\rightarrow \P} := \text{parallel transport of } E(\P')
\text{ to } \P \text{ along $\gamma_{\P\P'}$.}
\end{equation}
Next one defines a scalar weighting function $f(\P,\P')$ such that
\begin{align}
&f(\P,\P') \rightarrow 0 \text{ when }d(\P,\P') \geq
\Lav,\\
\nonumber\\
&\int f(\P,\P') \sqrt{-\gb} d^4 x' = 1,
\end{align}
where $\Lav$, such that $\ell \ll \Lav \ll L$, defines the averaging
scale containing many wavelengths and $d(\P,\P')$ is the distance
between $\P$ and $\P'$. The average of the tensor field $E(\P')$
about event $\P$ is then defined as
\begin{equation}\label{av}
\m{E}_{\P} := \int E(\P')_{\rightarrow \P} f(\P,\P') \sqrt{-\gb} d^4
x'.
\end{equation}

Given these definition it is possible to show that \cite{MTW} an
object ${{\gb}_{a}}^{a'}$ can be defined such that it transform as a
tensor at $\P'$ with respect to the index $a'$ and as a tensor at
$\P$ with respect to the index $a$. This is the \emph{bivector of
geodesic parallel displacement} mentioned above. In the case of $E$
having two covariant indices, this allows to express its parallel
transport as
\begin{equation}
E_{ab}(\P')_{\rightarrow \P} = {{\gb}_{a}}^{a'} {{\gb}_{b}}^{b'}
E_{a'b'}(\P').
\end{equation}
Then, when expressed explicitly in the coordinate system, the
average \eqref{av} yields
\begin{equation}
\m{E_{\mu \nu}}_{x} = \int {{\gb}_{\mu}}^{\mu'}(x,x')
{{\gb}_{\nu}}^{\nu'}(x,x') E_{\mu'\nu'}(x') f(x,x') \sqrt{-\gb(x')}
d^4 x'.
\end{equation}
Finally Isaacson paper \cite{isaacson1968II} shows how to derive the
rules 1., 2. and 3. listed at page \pageref{rules} starting from
this expression.

\newpage
\thispagestyle{empty}


\singlespacing
\chapter{Quantum physics background}\label{ap5}

\begin{quote}
\begin{small}
In this Appendix we give a very synthetic summary of some quantum
mechanics concepts that we employ in Chapter \ref{ch1}. In
particular we illustrate how the density matrix formalism can be
introduced as an alternative to the well known state vector
formalism. In the last section we illustrate some ideas related to
decoherence theory.
\end{small}
\end{quote}
\singlespacing

\section{State vector and unitary evolution}

In quantum mechanics the state of a system is described by a
normalized vector $\psi$ in an abstract complex Hilbert space. In
the \emph{\schr picture} dynamical variables are represented by
fixed hermitian operators while the state vector evolves in time
according to \schr equation
\begin{equation}\label{1e11b}
\Ho \psi_t = i\hbar\frac{d\psi_t}{dt},
\end{equation}
where $\Ho$ is the hamiltonian operator. The fact that the operators
are hermitian guarantees that they admit a complete set of
orthogonal eigenvectors with real eigenvalues. One important
postulate of quantum mechanics states the possible results of an
experiment devoted to measure, say, an observable $A$ can be only
the eigenvalues of the corresponding quantum operator $\Ao$. If $\Ao
v_n = a_n v_n$ is the eigenvalue equation for the operator $\Ao$
then the set $\{v_n\}_n$ of orthogonal eigenvectors allows to expand
a generic state $\psi$ as
\begin{equation*}
    \psi = \sum_n c_n v_n,
\end{equation*}
where $c_n = (v_n,\psi)$ and $(\,,\,)$ denotes the Hilbert space
scalar product. The components $c_n$ are related to the
probabilities that a single measurement of the observable $A$ yields
the value $a_n$, i.e.\footnote{For simplicity we assume the
eigenvalues to be not degenerate.}
\begin{equation}\label{Pan}
    P(A= a_n) = \abs{c_n}^2 = \abs{(v_n,\psi)}^2.
\end{equation}
By the usual properties of probability this implies that it must be
\begin{equation*}
    \sum_n \abs{c_n}^2 = 1.
\end{equation*}

The expectation or average value of the observable $A$ after many
identical experiments on identically prepared systems is given by
\begin{equation*}
\m{A}=(\psi,\Ao\psi) = \sum_n a_n \abs{c_n}^2.
\end{equation*}
Since the state vector changes in time, the expectation values of an
observable also change in general. From
\begin{equation}\label{1e17}
\brk{A}_t=(\psi_t,\Ao\psi_t),
\end{equation}
using the equation of motion \eqref{1e11b}, and if $\Ao$ doesn't
depend on time, we have
\begin{equation}\label{1e18}
\frac{d}{dt}\brk{A}_t=\frac{i}{\hbar}(\Ho\psi,\Ao\psi)-\frac{i}{\hbar}(\psi,\Ao
\Ho\psi)=(\psi,[\Ho,\Ao]\psi),
\end{equation}
where $[\,\,,\,\,]$ denotes the commutator and where we have
exploited the fact that the $\Ho$ is hermitian. Thus
\begin{equation}\label{1e19}
\frac{d}{dt}\brk{A}_t=\frac{i}{\hbar}\left\langle[\Ho,\Ao]\right\rangle_t.
\end{equation}

The spectrum of an operator can be discrete or continuous as, e.g.,
it is the case for the position operator $\xo$. Considering for
simplicity a single particle in one-dimension the eigenvalue
equation reads $\xo u_x = x u_x$, where $x \in \mathbb{R}$. An
arbitrary state $\psi$ could then be expanded as
\begin{equation*}
\psi = \int_{\mathbb{R}} \psi(x) u_x dx,
\end{equation*}
where
\begin{equation}
\psi(x) := (u_x,\psi).
\end{equation}
The complex function $\psi(x)$ describes the quantum state in the
well known \emph{position representation}. In this case the scalar
product has the explicit representation
\begin{equation*}
(\psi,\phi) = \int_{\mathbb{R}}dx \psi^*(x)\phi(x),
\end{equation*}
where $^*$ denotes complex conjugation, and the Hilbert space can be
identified with $L^2(\mathbb{R})$. In the position representation
$\abs{\psi(x)}^2$ can be interpreted as the position probability
density to measure the particle position at the location $x$.

Going back to the abstract level, the state vector evolution in the
Hilbert space can be explicitly written in terms of the
\emph{temporal evolution operator} $\Uo(t,t_0)$ as
\begin{equation}\label{1e12}
\psi_t=\Uo(t,t_0)\psi_0,
\end{equation}
where $\psi_0$ represents the initial state of the system at
$t=t_0$. The operators $\Uo(t,t_0)$ are unitary and form a group
\begin{equation}\label{1e13}
\Uo(t,t_0)=\Uo(t,t_1)\Uo(t_1,t_0),
\end{equation}
\begin{equation}\label{1e13b}
\Uo^{-1}(t,t_0)=\Uo(t_0,t),
\end{equation}
\begin{equation}\label{1e14}
\Uo\,\Uo^{\dag}=\Uo^{\dag}\Uo=\I,
\end{equation}
where $^{\dag}$ denotes the adjoint operator and $\I$ is the
identity operator. Substituting equation \eqref{1e12} into equation
\eqref{1e11b} we find
\begin{equation}\label{1eq15}
\frac{\hbar}{i}\frac{\partial\,\Uo(t,t_0)}{\partial t} + \Ho
\Uo(t,t_0) = 0,\,\,\,\,\,\,\Uo(t_0,t_0)=\I.
\end{equation}
Equation \eqref{1eq15} can be cast in the equivalent integral form
\begin{equation}\label{1eq16}
\Uo(t,t_0)=\I-\frac{i}{\hbar}\int_{t_0}^{t}\Ho\Uo(t',t_0)dt'.
\end{equation}
A formal solution of \eqref{1eq15}, valid when $\Ho$ \emph{does not}
depend itself on time, is
\begin{equation*}
\Uo(t) =  e^{-\frac{i}{\hbar}\Ho t},
\end{equation*}
where we set $t_0 = 0.$

In many situations, and as it also happens for the problem we study
in Chapter \ref{ch1}, the Hamiltonian operator can have an explicit
time dependence. This is typically the case when $\Ho$ represents
some time dependent small perturbation acting on the system. Then a
solution to equation \eqref{1eq16} can be attempted through a
\emph{Dyson's perturbation series}. As the 0-th approximation we
have
\begin{equation}\label{1e35}
\Uo^{(0)}(t,t_0)=\I.
\end{equation}
Substituting back into equation \eqref{1eq16} we find the first
order approximation
\begin{equation}\label{1e36}
\Uo^{(1)}(t,t_0)=\I-\frac{i}{\hbar}\int_{t_0}^{t}\Ho(t')dt'.
\end{equation}
In the same manner, the second order approximation results in
\begin{equation}\label{1e37}
\Uo^{(2)}(t,t_0)=\I-\frac{i}{\hbar}\int_{t_0}^{t}\Ho(t')dt'+\left(\frac{-i}{\hbar}\right)^2
\int_{t_0}^{t}\int_{t_0}^{t'}\Ho(t')\Ho(t^{\prime\prime})dt^{\prime\prime}dt'.
\end{equation}
If the procedure converges\footnote{It is known that whenever $\Ho$
is a bounded operator the series does indeed converge.} the
approximation process can be continued in the same fashion and we
get the series expansion
\begin{equation}\label{1e38}
\Uo(t,t_0)=\sum_{n=0}^{\infty}\Uo_n (t,t_0),
\end{equation}
where
\begin{equation}\label{1e39}
\Uo_n(t,t_0):=\left(\frac{-i}{\hbar}\right)^n\int_{t_0}^{t}\int_{t_0}^{t'}\cdots\int_{t_0}^{t^{(n-1)}}
\Ho(t')\Ho(t^{\prime\prime})\cdots \Ho(t^{(n)})dt^{(n)}\cdots dt'.
\end{equation}
Then $N$-th order truncated series for the evolution operator is
\begin{equation}\label{1e40}
\Uo^{(N)}(t,t_0)=\sum_{n=0}^{N}\Uo_n (t,t_0).
\end{equation}
If the hamiltonian depends upon some small parameter $\varepsilon$
then $\Uo^{(N)}(t,t_0)$ will be of order $\varepsilon^N$

Going back to the case of time independent Hamiltonian for
simplicity, an important feature of quantum mechanics is that the so
called unitary evolution described by
\begin{equation*}
    \psi_t = e^{-\frac{i}{\hbar}\Ho t} \psi_0
\end{equation*}
is deterministic and continuous. The indeterminism comes in only
when a measurement is performed. In that case only one of
eigenvalues $a_n$ of the observable $A$ under exam results
unpredictably. The usual view is that if the state \emph{before} the
measurement was $\psi = \sum_n c_n v_n$, then the state \emph{after}
the measurement is $\psi' = v_n$. The transition $\psi \rightarrow
v_n$ goes under the name of \emph{state vector reduction} or
\emph{collapse} of the state vector. Within the standard framework
of quantum mechanics it is supposed to happen instantly and it thus
represents a non-unitary, discontinuous element in the system
evolution. It is important to realize that the transition $\psi
\rightarrow v_n$ cannot happen by virtue of unitary evolution alone.
This is because \schr equation is linear and the operators of
quantum mechanics are linear. In fact if the system initial state is
$\psi_0 = \sum_n c_n^{(0)} v_n$, where the $v_n$ are the
eigenvectors of some hermitian operator $\Ao$, then the evolved
state at time $t$ is
\begin{equation*}
\psi_t = \Uo(t)\left( \sum_n c_n^{(0)} v_n \right) = \sum_n
c_n^{(0)} \Uo(t) v_n,
\end{equation*}
which is still given by a superposition of states.

The unitary evolution maintains the correlations that may be present
in the state vector describing some system. This is important for
phenomena such as, e.g., quantum interference to happen. We can
consider for simplicity a superposition of two states $\psi =
c_1\psi_1 + c_2\psi_2$, which may describe the wave function of a
single particle emerging behind two slits like in the standard Young
interference experiment. Until the particle hits the photographic
plate and a position measurement is performed, unitary evolution
maintains the superposition state intact. This implies that the
position probability density at the screen location is given by
$\abs{\psi_1 + \psi_2}^2 = \abs{c_1}^2\abs{\psi_1}^2 +
\abs{c_2}^2\abs{\psi_2}^2 + c_1^* c_2 \psi^*_1\psi_2 + c_1
c_2^*\psi_1\psi^*_2$ and it is well known that interference is due
to the cross terms. It is only after the measurement that the
particle wave functions collapses to some position eigenstate
according to $\psi \rightarrow u_x$. The important feature of a so
called \emph{pure state} like the one considered above is that,
prior to measurement, the particle is neither in the state $\psi_1$
nor $\psi_2$. It is really in a superposition of the two. The two
states are there simultaneously and interact giving rise to the
typical interference pattern. This particular pure state could be
described in words as $\psi_1$ \textsc{AND} $\psi_2$.

In some situations one may however encounter a situation in which it
is $\psi_1$ \textsc{OR} $\psi_2$. This would corresponds to the
particle state being really either given by $\psi_1$ or $\psi_2$.
The `doubt' standing solely in our lack of knowledge on the `real'
state of the system. In this case one usually speaks of a
\emph{mixed state} or \emph{statistical mixture}. Notice that no
interference is possible because the two states $\psi_1$ and
$\psi_2$ cannot interact; they simply are \emph{not} there
simultaneously! In the next section we introduce the density matrix
formalism which allows to treat the case of mixed states in a very
natural way.

\section{The density operator}

A quantum system which is not in pure state can be described by a
statistical mixture of pure states $\psi^{(i)}$, each one having a
probability $w^{(i)}$ to occur. The $\psi^{(i)}$ are normalized but
do not have to be necessarily orthogonal. Choosing any maximal set
of commuting operators, any vector of the Hilbert space can be
expanded in the orthonormal basis of their common eigenvectors
${u}_n$. In particular
\begin{equation}\label{1e1}
\psi^{(i)}=\sum_n c_n^{(i)}{u}_n,
\end{equation}
where
\begin{equation}\label{1e2}
({u}_n,{u}_m)=\delta_{nm},\,\,\,\,\,\,\sum_n |c_n^{(i)}|^2=1.
\end{equation}
The expectation value of an observable $\Ao$ in the pure state
$\psi^{(i)}$ is
\begin{equation}\label{1e3}
\brk{A}_i=(\psi^{(i)},\Ao\psi^{(i)})=\sum_{n,\,
m}c_n^{(i)*}c_m^{(i)}({u}_n,\Ao{u}_m)=\sum_{n,\,m}c_n^{(i)*}c_m^{(i)}A_{nm},
\end{equation}
where $A_{nm}:=({u}_n,\Ao{u}_m)$ are the matrix elements of the
operator $\Ao$. Since the system has a statistical probability
$w^{(i)}$ to be described by the pure state $\psi^{(i)}$, the
statistical \emph{grand average} of the observable in the mixed
state can be defined as
\begin{equation}\label{1e4}
\brk{A}:=\sum_i
w^{(i)}\brk{A}_i=\sum_{nm}A_{nm}\sum_iw^{(i)}c_n^{(i)*}c_m^{(i)}.
\end{equation}
Defining the \emph{density matrix} elements by
\begin{equation}\label{1e5}
\rho_{mn}:=\sum_i w^{(i)}c_n^{(i)*}c_m^{(i)}
\end{equation}
then
\begin{equation}\label{1e6}
\brk{A}=\sum_{nm}\rho_{mn}A_{nm}=\mbox{Tr}(\rho\Ao),
\end{equation}
where $\rho$ indicates the abstract density matrix operator and
$\mbox{Tr}$ denotes the trace.

Dirac's \emph{bra-ket} notation allows to get an useful
representation for the density matrix. Denoting the base states
${u}_n$ by the \emph{kets} $\ket{n}$ and the pure states
$\psi^{(i)}$ by $\ket{i}$ then by definition
\begin{equation}\label{1e7}
c_n^{(i)}=\brk{n|i},
\end{equation}
so that
\begin{equation}\label{1e8}
\rho_{mn}=\sum_i \brk{m|i}w^{(i)}\brk{i|n}
\end{equation}
and the density operator itself can be written as
\begin{equation}\label{1e9}
\rho=\sum_i w^{(i)}\ket{i}\bra{i}.
\end{equation}
Thus $\rho$ appears as a statistical sum of the \emph{projector
operators} $\ket{i}\bra{i}$ upon the pure states $\ket{i}$.

In the special case in which the system happens to be in the pure
state $\ket{k}$ then $w^{(i)}=\delta_{ik}$ and $\rho$ reduces to be
a simple projection operator
\begin{equation}\label{1e10}
\rho^P=\ket{k}\bra{k},
\end{equation}
where $^P$ stands for pure. Equation \eqref{1e6} for the expectation
value would read
\begin{equation}\label{1e11}
\brk{A}=\sum_{m}(\rho^P\Ao)_{mm}=\sum_m\brk{m|k}\brk{k|\Ao|m}=\sum_m
c_m^{(k)}\brk{k|\Ao|m}=\brk{k|\Ao\sum_m c_m^{(k)}|m}=\brk{k|\Ao|k},
\end{equation}
as it would result from the usual quantum mechanics pure states
formalism.

This example shows that the density matrix formalism includes as a
special case the more usual pure states formalism. At a formal
level, equation \eqref{1e6} can be used as a general definition of
the density operator, in the sense that \emph{$\rho$ is some
operator characterizing the system (whether mixed or pure), allowing
to get the expectation value of any observable $A$ through
$\brk{A}=\mbox{\emph{Tr}}(\rho\Ao)$}.

In the density matrix formalism the rule \eqref{Pan} for the
probability that the measurement of $A$ yields the value $a_n$
reads:
\begin{equation}\label{eq4}
P(A=a_n) = \tr [ \rho \hat{P}(a_n) ],
\end{equation}
Here the projector operator $\hat{P}(a_n)$ into the subspace spanned
by the eigenvectors of $\Ao$ corresponding to the eigenvalue $a_n$
is given by $\hat{P}(a_n) = \ket{a_n}\bra{a_n}$. Equation
\eqref{eq4} is general and it can be easily checked that it reduces
to \eqref{Pan} in the special case in which $\rho^P =
\ket{\psi}\bra{\psi}$ describes a pure state:
\begin{align*}
\tr [ \rho^P \hat{P}(a_n) ] &= \sum_{m}
\bra{m}\kett{\psi}\bra{\psi}\kett{a_n}\bra{a_n}\kett{m}=\abs{\bra{a_n}\kett{\psi}}^2,
\end{align*}
where we have used the completeness relation $\sum_m
\ket{m}\bra{m}=\I.$

The following formal properties of a density operator are important.
For a proof we refer to \cite{espagnat76}:
\begin{enumerate}
  \item $\rho$ is Hermitian;
  \item $\rho$ is positive definite, i.e. $\bra{u}\rho\ket{u}\geq 0$ for any
  $\ket{u}$;
  \item $\tr\rho = 1$, where $\tr\rho = \sum_n \bra{n}\rho\ket{n}$ denotes the Trace;
  \item The diagonal elements of $\rho$ in any representation are
  nonnegative;
  \item The eigenvalues $p_n$ satisfy $0 \leq p_n \leq 1;$
  \item If $\rho$ is a projection operator it projects into a one
  dimensional subspace;
  \item $\tr (\rho^2)\leq 1$, and the equality holds only if $\rho$ is
  a projector operator;
  \item With $\rho$ written as $ \rho=\sum_i w^{(i)}\ket{i}\bra{i}, $ a
necessary and sufficient condition for it to be a projection
operator is that all the $\ket{i}$ should be identical up to a phase
factor. In this case the above sum reduces to only one term.
\end{enumerate}
Thus we have the important property that $\tr (\rho^2) = 1$ only
when $\rho$ describes a pure state. Statistical mixtures are thus
characterized by the important property $\tr (\rho^2) < 1$.

An elementary but instructive example that shows the utility of the
density matrix formalism in relation to the problem of decoherence
is the following. Let $\ket{\psi} := c_1 \ket{1} + c_2 \ket{2}$ be a
pure state and let us consider the special case in which the two
states $\ket{1}$ and $\ket{2}$ are orthogonal and make part of
complete set $\{\ket{n}\}_n$. As it has been discussed above the ket
$\ket{\psi}$ is a superposition of states that can interfere with
each other. In the density matrix formalism the same system would be
described by the projector operator
\begin{equation*}
    \rho^P = \ket{\psi}\bra{\psi} = \abs{c_1}^2 \ket{1}\bra{1} + \abs{c_2}^2
    \ket{2}\bra{2} + c_1^*c_2 \ket{2}\bra{1} + c_1 c_2^*
    \ket{1}\bra{2}.
\end{equation*}
The matrix elements would follow form $\rho_{nm} = \bra{n}\rho
\ket{m}$. The only non vanishing matrix element would be
\begin{equation}
\rho^P_{11}=\abs{c_1}^2,\quad \rho^P_{22}=\abs{c_2}^2,\quad
\rho^P_{12}=c_1 c_2^*, \quad \rho^P_{21}=c_2 c_1^*.
\end{equation}
In particular the possibility of interference between the pure state
two components is expressed in the matric element formalism by the
fact that the \emph{off-diagonal terms are not zero}.

On the other hand a mixed state describing a system which is either
$\ket{1}$ or $\ket{2}$ and where interference cannot happen would be
described by the density matrix
\begin{equation*}
    \rho = w^{(1)} \ket{1}\bra{1} + w^{(2)}
    \ket{2}\bra{2}.
\end{equation*}
Using as above the set of orthonormal vectors $\{\ket{n}\}_n$ as a
base, in this case only the diagonal elements are present, while the
off-diagonal components vanish. Quite in general we have that in the
density matrix formalism the absence of off-diagonal components
implies the inability of the system to show interference.

\subsection{Unitary evolution of the density operator}

We now derive the equation for the time evolution of the density
matrix. To this end we can consider an arbitrary observable $A$ and
write the time derivative of its expectation value as
\begin{equation}\label{1e20}
\frac{d}{dt}\brk{A}=\frac{d}{dt}\mbox{Tr}(\rho\Ao)=\mbox{Tr}(\dot{\rho}\Ao).
\end{equation}
Moreover we have
\begin{equation}\label{1e21}
\frac{i}{\hbar}\left\langle[\Ho,\Ao]\right\rangle
=\frac{i}{\hbar}\mbox{Tr}(\rho[\Ho,\Ao])=\frac{i}{\hbar}\mbox{Tr}(\rho
\Ho \Ao - \rho \Ao \Ho)=\frac{i}{\hbar}\mbox{Tr}(\rho \Ho \Ao - \Ho
\rho \Ao) = \mbox{Tr}(\frac{i}{\hbar}[\rho,\Ho]\Ao),
\end{equation}
where we have used the fact that the trace is addictive and
unchanged under cyclic permutation of the factors. Comparing
equations \eqref{1e19}, \eqref{1e20}, \eqref{1e21} we get
\begin{equation}\label{1e22}
\dot{\rho}=\frac{i}{\hbar}[\rho,\Ho],
\end{equation}
which is the analogue of the \schr equation and describes unitary
evolution.

This equation is general but we are also interested in having an
explicit formal expression for the evolved density matrix $\rho_t$
of an initial density operator $\rho_0$ through some kind of
temporal evolution operator as it has been done for the state vector
in equation \eqref{1e12}. If the system is in a mixed state then it
can conveniently be described by a density operator given by
\begin{equation}\label{1e42}
\rho_t=\sum_i  w^{(i)} \ket{i}_t\bra{i}_t
\end{equation}
in terms of the time dependent pure states $\ket{i}_t$. The time
evolution of the density operator stems from the time evolution of
the individual pure states. For these we have
\begin{equation}\label{1e43}
\ket{i}_t=\Uo(t,t_0)\ket{i}_0.
\end{equation}
The \emph{bra} $\bra{i}_t$ corresponding to the \emph{ket}
$\ket{i}_t$ must evolve according to
\begin{equation}\label{1e44}
\bra{i}_t=\bra{i}_0\Uo^{\dag}(t,t_0).
\end{equation}
This guarantees indeed the normalization of the pure states at every
time
\begin{equation}\label{1e45}
\brk{i|_t i}_t=\brk{i|_0\Uo^{\dag}(t,t_0)\Uo(t,t_0)|i}_0=\brk{i|_0
i}_0=1.
\end{equation}
The time evolution of the density matrix follows as
\begin{equation}\label{1e46}
\rho_t=\sum_i \Uo(t,t_0)\ket{i}_0 w^{(i)}
\bra{i}_0\Uo^{\dag}(t,t_0)=\Uo(t,t_0)\rho_0 \Uo^{\dag}(t,t_0).
\end{equation}
Then the expansion \eqref{1e38} can be used to express $\Uo(t,t_0)$
and obtain an explicit expression for the evolved density matrix.

Let us think now of $\Ho$ as a small perturbation (which is indeed
the case in our case of interest where it represents the tiny
fluctuations in the gravitational field). In the expansion series
for $\Ko$ we retain then only the terms up to second order and the
evolved density matrix is
\begin{equation}\label{1e47}
\rho_t=[\I+\Ko_1(t,0)+\Ko_2(t,0)]\rho_0
[\I+\Ko_1^{\dag}(t,0)+\Ko_2^{\dag}(t,0)].
\end{equation}
Keeping terms up to second order gives
\begin{equation}\label{1e48}
\rho_t=\rho_0 + \Ko_1(t,0)\rho_0+\rho_0
\Ko_1^{\dag}(t,0)+\Ko_2(t,0)\rho_0+\Ko_1(t,0)\rho_0
\Ko_1^{\dag}(t,0)+\rho_0 \Ko_2^{\dag}(t,0),
\end{equation}
with
\begin{equation}\label{1e49}
\Ko_1(t,0)=-\frac{i}{\hbar}\int_{0}^{t}\Ho(t')dt',
\end{equation}
\begin{equation}\label{1e50}
\Ko_2(t,0)=-\frac{1}{\hbar^2}
\int_{0}^{t}\int_{0}^{t'}\Ho(t')\Ho(t^{\prime\prime})dt^{\prime\prime}dt'.
\end{equation}
These are the expressions \eqref{dy} used in chapter \ref{ch1}.

\subsection{Proper and improper mixtures}

In many important situations a quantum system can be decomposed into
various subsystems. The state vector (or density matrix) describing
the whole system encodes information about all the subparts and
their correlations. This applies e.g. to the relatively simple case
of a system composed of $N$ particles or, which is relevant to the
issue of decoherence, to the case of one particle and a complex
quantum environment with which it interacts. One of the key and
striking features of QM is that, because there is only one
\emph{overall} state vector for the whole system, its subparts are
often entangled; i.e. systems that interacts once and get correlated
maintain their correlation even at later times. A typical simple
example includes EPR-like systems \cite{epr35}, e.g. when a particle
with, say, zero spin decays into two particles that then propagate
far from each other: the time evolution of the whole system is
unitary until a spin measurement is performed on one of the
particles, in which case the part of the wave function describing
the second particle, possibly causally separated from the first,
also collapses, \emph{instantaneously} to a new appropriate state.
This example shows the non-local nature of QM, as it is also
implicit in the violation of Bell inequalities \cite{bell87} and as
it has also been confirmed by experiments \cite{aspect82}.

For sake of illustration we can consider e.g. a composite quantum
system $\Sigma = S + E$, where the subsystems $S$ and $E$ could be
described individually through their relevant Hilbert spaces
$\H^{(S)}$ and $\H^{(E)}$. The total system Hilbert is given by the
tensor product $\H = \H^{(S)} \otimes \H^{(E)}$ and a pure state
could be generally written as
\begin{equation}\label{kp}
\ket{\psi} = \sum_{ij}c_{ij}\ket{v_i}\ket{u_j},
\end{equation}
where $\{\ket{v_i}\}$ and $(\{\ket{u_i}\})$ are two complete sets of
orthonormal kets in $\H^{(S)}$ and $\H^{(E)}$. Such a situation
could typically occur as a result of some past interaction between
$S$ and $E$. The specific example of the two particles having
opposite spins could e.g. be described as
\begin{equation*}
\ket{\psi} = \frac{1}{\sqrt{2}}\ket{\uparrow}\ket{\downarrow} +
\frac{1}{\sqrt{2}}\ket{\downarrow}\ket{\uparrow}.
\end{equation*}
If a measurement, say of particle 1, yields a spin up alignment as
described by $\ket{\uparrow}$, then the entanglement would force the
wave function related to particle 2 to collapse to a spin down
alignment, as described by $\ket{\downarrow}$.

Let us go back now to the general case \eqref{kp} and suppose we are
interested in describing the subsystem $S$ only, which could be a
quantum particle propagating through an environment described by
$E$. Let $A$ be an observable pertaining to $S$. The expectation
value of $A$ is
\begin{equation*}
\m{A}= \bra{\psi}\Ao\ket{\psi} = \sum_{ijrs}
c^*_{ij}c_{rs}\bra{v_i}\bra{u_j} \Ao \ket{v_r}\ket{u_s}.
\end{equation*}
Since $A$ belongs to the system $S$, the operator $\Ao$ does not
affect the vectors in $\H^{(E)}$ and the previous equation can be
rewritten as
\begin{align*}
\m{A} &= \bra{\psi}\Ao\ket{\psi} = \sum_{ijr}
c^*_{ij}c_{rj}\bra{v_i} A\ket{v_r} = \sum_{ir}\rho_{ri}A_{ir} = \tr
(\rho \Ao),
\end{align*}
where we used $\bra{u_j}\kett{u_s}=\delta_{js}$ and where $A_{ir} =
\bra{v_i} A\ket{v_r}$ are the matrix elements of the operator $\Ao$
in the base $\{\ket{v_i}\}$ while
\begin{equation}\label{eq5}
\rho_{ri} = \sum_j c_{rj}c^*_{ij}
\end{equation}
are the matrix elements of the operator
\begin{equation}\label{eq6}
\rho = \sum_{ri} \ket{v_r} \rho_{ri} \bra{v_i}.
\end{equation}
The density operator
\begin{equation*}
\rho^{(\Sigma)} = \sum_{rsij} \ket{v_r}\ket{u_s}
\rho^{(\Sigma)}_{rs;ij} \bra{v_i}\bra{u_j}
\end{equation*}
describing the whole system $\Sigma = S + E$ has matrix elements
given by $\rho^{(\Sigma)}_{rs;ij} = c_{rs}c^*_{ij}$. Equations
\eqref{eq5} and \eqref{eq6} can thus be rewritten as
\begin{equation*}
\rho_{ri} = \sum_j \rho^{(\Sigma)}_{rj;ij},
\end{equation*}
which corresponds to
\begin{equation*}
\rho = \tr^{(E)}\rho^{(\Sigma)} := \sum_n \bra{u_n}\rho\ket{u_n}
\end{equation*}
where the \emph{partial trace} $\tr^{(E)}$ means that the operation
of taking the trace of the operator is carried over only with
respect to $\H^{(E)}$.

It is possible to show that the operator $\rho$ defined above
satisfies all the property enjoyed by any density operator as listed
above. Moreover it can be verified that
\begin{equation*}
\tr \rho^2 < 1,
\end{equation*}
implying that, as long as it is considered on its own, the subsystem
$S$ resembles a statistical mixture. This fact is remarkable and
motivates the convention which extends the name of mixtures also to
the systems like $S$ considered here. Such an \emph{emerging}
mixture should be more properly called an \emph{improper mixture};
the reason being that the full system $\Sigma = S + E$ is still
described by a pure state. The system $S$ gains mixtures properties
only due to the fact that in tracing out the properties of the
environment $E$ one loses information about the correlation that are
still present in the overall system.

It is important to note that the fact that the density matrix
related to system $S$ turns into the appropriate form to describe
mixtures \emph{cannot} stem out of unitary evolution. Indeed, from
$\rho_t=\Uo(t,t_0)\rho_0 \Uo^{\dag}(t,t_0)$, we can observe that:
\begin{align*}
\rho^2_t &= \Uo(t,t_0) \rho_0 \Uo^{\dag}(t,t_0) \Uo(t,t_0) \rho_0
\Uo^{\dag}(t,t_0) = \Uo(t,t_0) \rho^2_0 \Uo^{\dag}(t,t_0).
\end{align*}
We see that, if $\rho^2_0 = \rho_0$, then $\rho^2_t = \rho_t$ for
every $t$. \emph{Thus a pure state can never evolve into a mixture
if the time evolution is governed by Schr\"{o}dinger equation.}

\section{Basic concepts in decoherence theory}

Many excellent reviews articles analyzing decoherence, its roots,
its meaning, and its consequences can be found in the literature
\cite{zeh06,zeh96,zureck03,joss99,kiefer98}. We refer to these for a
comprehensive review. With the following we aim to describe briefly
some essential features that are important in relation to the
present work.

As succinctly expressed by Joss in \cite{joss99}
\begin{quote}
Decoherence is the irreversible formation of quantum correlations of
a system with its environment. These correlations lead to entirely
new properties and behavior compared to that shown by isolated
objects.
\end{quote}
The motivation behind this statement is expressed in a nutshell by
the example of the previous section. Quantum mechanics unitary
evolution applies to the closed system $\Sigma = S + E$. However
because of the entanglement from interaction, as expressed by
$\ket{\psi} = \sum_{ij}c_{ij}\ket{v_i}\ket{u_j}$, the system of
interest $S$ appears to gain new properties: namely it becomes
analogue to a statistical mixture, thus losing the coherence
properties that are needed to show genuine quantum effects such as,
e.g., interference. Whenever the environment $E$ has many degrees of
freedom the entanglement is practically irreversible and the effects
of decoherence on $S$ can be quite dramatic.

Decoherence is an ubiquitous quantum phenomenon and it is important
to understand that it represents the rule more than the exception.
Indeed a perfectly isolated and closed system does not exist,
exception made, perhaps, for the whole universe. Of course the
degree of decoherence suffered by a quantum system coupled to its
environment will depend upon the complexity of the environment and
the strength of the coupling. Decoherence is strongly believed to
play an important role in determining the quantum to classical
transition and helping alleviating the measurement problem of
quantum mechanics. Starting from the 90s it received important
experimental verifications in \cite{brune96}, where decoherence of
mesoscopic superposition of quantum states involving radiation
fields was observed, and in \cite{arndt99}, when interference with
fullerene molecules was first obtained.

According to Von Neumann's measurement process description, in the
interaction with a macroscopic quantum apparatus, the state
$\ket{n}$ of a quantum system gets correlated with the so called
macroscopic \emph{pointer state} $\ket{\Phi(n)}$. If $\ket{\Phi(0)}$
represents the initial pointer position prior to interaction, an
ideal measurement should look like
\begin{equation*}
\ket{n}\ket{\Phi(0)} \rightarrow \ket{n}\ket{\Phi(n)}.
\end{equation*}
The roots of the measurement problem are in the fact that if the
particle is initially in a pure superposition state then, by
linearity, we have:
\begin{equation*}
\left(\sum_n c_n \ket{n}\right)\ket{\Phi(0)} \rightarrow \sum_n c_n
\ket{n}\ket{\Phi(n)},
\end{equation*}
i.e. because of the interaction the system and the apparatus become
correlated in such a way that the superposition state also affects
the macroscopic apparatus. Clearly macroscopic superpositions of
classical objects have never been observed and the problem stands in
asking why this is so and why the measurement always yields
\emph{one} definite answer, corresponding to one definite pointer
state. The problem is somehow alleviated if one takes the
environment into account and considers the resulting decoherence.
The coupling of the apparatus to the environment dislocalizes the
phase relations to the enlarged total system made up of quantum
particle + apparatus + environment according to
\begin{equation*}
\left(\sum_n c_n \ket{n}\ket{\Phi(n)}\right)\ket{E_0} \rightarrow
\sum_n c_n \ket{n}\ket{\Phi(n)}\ket{E_n} := \psi,
\end{equation*}
where the kets $\ket{E_n}$ denote orthogonal environment states.
Taking the partial trace of the total density matrix $\rho:=
\ket{\psi}\bra{\psi}$ with respect to the environment states yields
the reduced density matrix describing the particle + apparatus as
\begin{equation*}
    \rho^{(S+A)} = \sum_n \abs{c_n}^2 \ket{n} \bra{n} \otimes
    \ket{\Phi_n}\bra{\Phi_n}.
\end{equation*}
This density matrix describes an improper mixture with \emph{no}
correlations; off-diagonal terms are absent and, in this sense, the
system $S+A$ acquires classical behavior.

Another example is the localization process of macroscopic objects.
In this case one can consider the scattering of microscopic
environmental particles off a body that represents the system of
interest. It can be shown that, as a result of many scattering
processes, the reduced density matrix describing the object suffers
an exponential damping of spacial coherence, i.e. in the position
representation one finds \cite{joss85}
\begin{equation*}
    \rho_t(x,x') = \rho_0(x,x')\exp{\left[-\Lambda t
    (x-x')^2\right]},
\end{equation*}
where the localization rate $\Lambda$ depends on the flux and
momenta of the environment particles and on scattering cross
section. Numerical estimates indicate that macroscopic particles
such as dust grains or even relatively large molecules ($d \approx
10^{-6}$ cm) have a very large localization rate (and thus suffer
fast decoherence) even for a typical laboratory vacuum environment,
estimated as containing roughly $10^3$ particles per cm$^3$.

We conclude this short summary by remarking that the reduced density
matrix describing a system that suffers decoherence usually obeys
some master equation that doesn't describe unitary evolution such as
in \eqref{1e22}. In this localization example the appropriate master
equation takes the form:
\begin{equation}
\dot{\rho}=\frac{i}{\hbar}[\rho,\Ho] +
\left.\dot{\rho}\right|_{\scriptsize{\text{scattering}}},
\end{equation}
where $\Ho$ describes in this case the \emph{free} \schr evolution
while the extra term results as an effect of the scattering process.
In the position representation and in one dimension this equation
reads ($\hbar = 1$):
\begin{equation*}
i \dot{\rho}_t(x,x') = \frac{1}{2m}\left( \frac{\pd^2}{\pd x'^2} -
\frac{\pd^2}{\pd x^2} \right) \rho - i\Lambda (x-x')^2\rho.
\end{equation*}

\newpage
\thispagestyle{empty}


\newpage

\bibliographystyle{amsplain}

\addcontentsline{toc}{chapter}{Bibliography}

\bibliography{references}

\newpage
\thispagestyle{empty}

\end{document}